\pdfoutput=1
\documentclass[12pt,twoside,openright]{book}
\usepackage{vmargin}

\usepackage[english]{babel}
\usepackage[latin2]{inputenc}

\usepackage{indentfirst}
\setpapersize{A4}
\usepackage{amsmath}
\usepackage{amssymb}
\usepackage{graphicx}

\usepackage{fancyheadings}

\renewcommand{\ln}{\,\textrm{ln}\,}
\newcommand{\Tr}{\,\textrm{Tr}\,}
\renewcommand{\det}{\,\textrm{det}\,}
\renewcommand{\exp}{\,\textrm{exp}\,}
\newcommand{\arth}{\,\textrm{arth}\,}
\renewcommand{\arctan}{\,\textrm{arctan}\,}
\newcommand{\perm}{\,\textrm{perm.}\,}
\newcommand{\BPHZ}{\,\textrm{BPHZ}\,}
\newcommand{\inte}{\,\textrm{int}\,}
\newcommand{\kaon}{\,\textrm{kaon}\,}

\newcommand{\clearemptydoublepage}{\newpage{\pagestyle{empty}\cleardoublepage}}

\newcommand{\diag}{\,\textrm{diag}\,}
\newcommand{\diagrams}{\,\textrm{diagrams with at least 2-loops}\,}

\newcommand{\num}{\,\textrm{numerator: }\,}

\newcommand{\rt}[1]{{}}     
\newcommand{\be}{\begin{equation}}      
\newcommand{\ee}{\end{equation}}      
\newcommand{\bea}{\begin{eqnarray}}      
\newcommand{\eea}{\end{eqnarray}}    

\newcommand{\ket}[1]{{\left|#1\right\rangle}}
\newcommand{\bra}[1]{{\left\langle #1\right|}} 

\renewcommand{\k}{{\bf k}}

\newcommand{\vac}{\textrm{(v)}}

\newcommand{\Gv}{G^{\,\textrm{(v)}}}
\newcommand{\Gm}{G^{\,\textrm{(m)}}}
\newcommand{\Gr}{G^{\,\textrm{(r)}}}
\newcommand{\Dv}{\Delta^{\textrm{(v)}}}
\newcommand{\Dm}{\Delta^{\textrm{(m)}}}
\newcommand{\Dr}{\Delta^{\textrm{(r)}}}
\newcommand{\Sigmar}{\Sigma^{\textrm{(r)}}}
\newcommand{\SigmaO}{\Sigma^{\textrm{(0)}}}
\newcommand{\Sigmav}{\Sigma^{\textrm{(v)}}}

\title{\vspace{-3.0cm}{\bf Resummed perturbative series of scalar quantum field theories in two-particle-irreducible formalism} \vspace{1.0cm}} 
\date{\LARGE 2011}
\author{\Large Ph.D. thesis  \vspace {1.75cm} \\
\LARGE Gergely Fejős \vspace {2.0cm} \\ \Large Advisor: Prof. András Patkós, D.Sc. \vspace{1.0cm}\\ Eötvös University, Budapest
\\ Physics Doctoral School\\ Particle Physics and Astronomy Programme \\ Department of Atomic Physics \vspace{0.4cm} \\  Doctoral school and programme leader: \\ Prof. Ferenc Csikor, D.Sc. \vspace{1.5cm}}

\begin{document}
\maketitle

\thispagestyle{empty}
\begin{titlepage}
\end{titlepage}

\pagenumbering{roman}

\renewcommand{\headrulewidth}{0.3pt}
\renewcommand{\chaptermark}[1]{
\markleft{\chaptername\ \thechapter.\ #1}{}}
\renewcommand{\sectionmark}[1]{
\markright{\sectionname\ \thesection.\ #1}}
\lhead[\let\uppercase\relax\bf{\small{\thepage}}]{\let\uppercase\relax\bf{\footnotesize{\rightmark}}}
\rhead[\let\uppercase\relax\bf{\footnotesize{\leftmark}}]{\let\uppercase\relax\bf{\small{\thepage}}}
\cfoot{}
\pagestyle{fancy}

\tableofcontents

\chapter*{Introduction}
\addcontentsline{toc}{chapter}{Introduction}
\pagenumbering{arabic}
\setcounter{page}{1} 

\lhead[\let\uppercase\relax\bf{\small{\thepage}}]{\let\uppercase\relax\bf{\footnotesize{Introduction}}}
\rhead[\let\uppercase\relax\bf{\footnotesize{Introduction}}]{\let\uppercase\relax\bf{\small{\thepage}}}

Quantum field theory (QFT) is an essential tool in order to understand phenomena occuring from atomic ($10^{-10}$ m) to the Planck scale ($10^{-35}$ m). It provides a theoretical framework to calculate scattering cross sections, particle lifetimes, and other observables of processes of the quantum world. Its most powerful strength is demonstrated by the fact that every elementary interaction known in particle physics can be described by a corresponding relativistic quantum field theory. The main reason of the introduction of the {\it field} description is that
the processes of particle physics changing the particle number can not be described by the simple application of quantum mechanics (e.g. see the Klein-paradox \cite{Peskin-book}). Furthermore, QFT combines quantum mechanics and special relativity naturally: it explains the relation between spin and statistics, and solves the causality problem of (relativistic) quantum mechanics. Quantum field theories also play an important role in statistical- and condensed-matter physics, e.g. in the description of critical phenomena and quantum phase transitions. It is extensively used in atomic and nuclear physics as well.

In field theory we have infinitely many degrees of freedom: one assigns dynamical variables (typically scalars, vectors, spinors and/or matrices) to every point of the three (or more generally $d$) dimensional space and follow for the time evolution of them. The classical dynamics is determined through the Lagrange formalism as a generalization of classical mechanics. In the quantum version of the theory, the variables are considered to be linear operators acting on the Hilbert space of the quantum states in question (in QFT this is the so-called Fock space). The key of the transition to QFT from the classical version is the so-called canonical equal-time commutation relation.

In order to obtain measurable quantities, one has to compute $n$-point functions, which are vacuum expectation values of the time ordered product of the fields. The most successful method for obtaining them is perturbation theory, in which every quantity is expressed through power series (perturbative series) of small parameters of the theory, i.e. the coupling constants. Although general analyses proved
that perturbative series are asymptotic \cite{collins}, therefore the convergence radius is zero, very valuable results can be obtained within this framework, e.g. the perturbatively solved quantum electrodynamics (QED) has extremely accurate predictive power. For example its prediction for the anomalous magnetic moment of the electron agrees with the experimentally measured value to more than 10 significant digits (this is the most accurately verified prediction in the history of physics) \cite{Peskin-book}. However, for strongly coupled theories perturbation theory completely breaks down; the series do not seem to show even practical convergence at all.
At low enough energy, this is the case of the fundamental theory of the strong interaction, the quantum chromodynamics (QCD) \cite{zinn-justin02}. One of the most challenging task of contemporary particle physics is to solve QCD in
different setups relevant for phenomena from accelerator experiments to
internal structure of compact stellar objects.

The most productive and mathematically well controlled method to attack the problem of large couplings is lattice field theory. It is a fully non-perturbative
approach, in which the degrees of freedom are reduced in a way that the quantum equations of motion are solved in a lattice, instead of the continuum formulation \cite{montvay}. The method typically requires very huge computational resources, nevertheless, it seems to be the most competitive method in order to describe non-perturbative features of the spectra of strongly coupled theories. 

A more traditional alternative way to deal with the problem is the application of functional methods. The best known technical framework is offered by the infinite set of Dyson-Schwinger equations. In the first chapter, we will discuss the main properties of this approach, in particular a variant called {\it two-particle-irreducible} (2PI) formalism, which is the central topic of this thesis.

Before we proceed to review the literature on 2PI formalism,
let us shortly discuss a central problem of every quantum field theory. QFT in its original formulation is ill-defined: divergences are obtained even at the lowest orders of perturbation theory. These divergences come from the ultra-violet regime of the momenta, or in other words
from spacetime points which are very close to each other. This
implies that QFT at very short distances must break down, which
gives no surprise; we should not expect that in an arbitrary small neighborhood of a given point {\it infinitely many} physically relevant degrees of freedom can appear. We should think of QFT as an effective model of some more fundamental theory (e.g. string theory), and
expect that the details of the dynamics at this scale are not relevant for obtaining results to a much larger scale. This expectation leads us to the idea of renormalization.

In order to obtain well-defined quantities, first we have to regularize QFT. The most natural regularization is to build up equations on a lattice, or the use of a cutoff in the momenta.
On the regularized theory, we can perform the so-called {\it renormalization programme}. This means that our lack of knowledge of the short range dynamics is hidden into a finite number of physical quantities determined by appropriate measurements or fixed by some conditions. If this can be done in a consistent way, where the cutoff can really be thought as a large quantity, therefore formally(!) can be sent to infinity with obtaining finite results, then the theory is said to be renormalizable.

Renormalization and renormalizability have an extended literature and in
perturbation theory it is worked out in great details. Usually the way we treat
the theory is to think of the model parameters as cutoff (defined as the highest allowed momentum) dependent
objects, which can be separated into renormalized parts and
counterterms. In perturbation theory we treat also the counterterms as
perturbations, which can be determined from the requirement of
the cancellation of the appearing divergences.
Pioneers of perturbative renormalizaton with rigorous proofs built upon Feynman's diagram technique were Bogoliubov, Parasiuk, Hepp and Zimmermann \cite{bogoliubov57,zimmermann69}, together with Weinberg \cite{weinberg60}. A very detailed description of their work and the general theory of perturbative renormalization can be found in \cite{collins}.

2PI formalism generalizes the idea of the 1PI effective action \cite{rivers}. The so-called {\it 2PI effective action} has two types of variables: mean fields and propagators. Stationary conditions of this action lead to self-consistent equations for these quantities. The solution induces a partial resummation of the perturbative series (i.e. resums certain infinite classes of Feynman diagrams), therefore it is a rather nontrivial question, how the perturbative renormalizability is carried over into certain 2PI approximations. One of the goals of this thesis is to show various methods, which are capable to formulate 2PI renormalizability.

The formalism was developed in the 1960s for non-relativistic field theories of condensed matter physics \cite{ward60}, and was generalized to relativistic theories in the 1970s \cite{Cornwall:vz} by Cornwall, Jackiw and Tomboulis. Referring to the latter paper, the method is sometimes called {\it CJT formalism}. The {\it $\Phi$-derivable approximation} appellation is also used with reference to the paper \cite{baym62b}, but these mean the very same formulation. Although the technique is known for a long time, the applications have become an especially active research topic in the last 10 years placing emphasis on non-equilibrium simulations. This is due to the fact, that 2PI formalism is the only known solution to the so-called secular time-evolution problem. Considering scalar fields for instance, the simplest non-local contribution to the 2PI effective action leads to non-secular time evolution showing thermalization at late times \cite{Berges:2000ur}. Same observations were made in theories coupling scalars to fermions \cite{Berges:2002wr}.
Some of the non-perturbative effects of 2PI approximations can be captured by combining the approach with the large-N expansion. In out-of-equilibrium phenomena, the 2PI 1/N expansion at next-to-leading order proved to be very valuable in studying various problems, including thermalization \cite{Berges:2001fi,aarts02,cooper03} and cosmological preheating \cite{Berges:2002wr,arrizab04}, non-thermal fixed points \cite{berges08} and decoherence \cite{giraud10}. The lowest ordered 2PI approximations in equilibrium also show good convergence of thermodynamical quantities, such as the pressure \cite{Berges:2004hn,Borsanyi:2008ar}. Discussions of finite temperature gauge theories can be found in \cite{blaizot01,blaizot05}. There were attempts to determine transport coefficients in \cite{aarts03,aarts04,aarts05}. Phenomenological studies have also appeared using various 2PI approximations, see \cite{petropoulos,andersen04,andersen08,Roder:2005vt}. Very recently promising applications appeared for the evaluation of critical exponents for continuous phase transition \cite{saito11}.

The renormalization of approximations obtained from the 2PI technique was investigated first for scalar theories at finite temperature in real time formalism \cite{hees02,VanHees:2001pf} and also in imaginary time formalism \cite{blaizot04,berges05}. The method was then extended to fermions \cite{Reinosa:2005pj} and to QED \cite{Reinosa:2006cm}.
Renormalization with non-vanishing field expectation values and therefore the determination of the effective potential was discussed in \cite{berges05,vanHees:2002bv,cooper05,arrizabalaga06,patkos08}. In case of theories with more complicated global symmetries renormalization techniques were developed in \cite{fejos08}. Although many papers discussed 2PI renormalization and renormalizability, there are still some open questions, in particular the standardized qualification of its numerical realizations is missing.

The present thesis deals with quantum field theories {\it in equilibrium}. Its goal is to map and solve some problems of renormalization both from an analytic and a numerical point of view, together with applications of some newly developed techniques to scalar quantum field theories. It is important to stress that publications discussing the accuracy of numerical realizations of 2PI renormalization in a reproducible manner are virtually missing from the literature. One of the aims of this thesis is to fill the gap caused by this absence of interest.

The greater part of the models examined here are used as effective theories of strong interactions. With this, the thesis would like to contribute to developing the treatments of models of phenomenological importance beyond perturbation theory, which - as already discussed - is crucial when large coupling constants appear. It is important to stress that this thesis has no task to do phenomenology. We wish to search for appropriate methods applicable for phenomenological investigations.

The structure of the thesis is as follows. In Chapter 1, we provide
a discussion of the functional methods of QFT. The task of this part is to present pedagogically the usual generator functionals of the 
Green's functions in order and end up at 2PI formalism. As an intermediate step, we will also discuss the Dyson-Schwinger equations and establish connections to two-particle-irreducibility. In Chapter 2, we introduce the models investigated in the thesis. At this point we will discuss some fundamental aspects of the large-N technique, which will also be a central subject of our investigations during the next chapters. In Chapter 3, we deal with the O(N) model in 2PI formalism at next-to-leading order of the large-N expansion. We will use the so-called auxiliary field formulation, and demonstrate its renormalizability with constructing appropriate counterterms explicitly. The equivalence between the original and
the auxiliary field formulation will also be discussed. The task of Chapter 4 is to present a careful numerical study of 2PI renormalization, in which the one component $\phi^4$ theory was employed. The first part of this chapter nevertheless deals with a theoretical issue: we show that the minimal subtraction procedure used in Chapter 3 is equivalent to imposing appropriate renormalization conditions on some quantities. This will allow us in one hand to obtain explicitly finite 2PI equations, and on the other hand the opportunity to compare the convergence of the numerical solutions of these equations with the ones containing counterterms. We will discuss the features of various numerical algorithms and produce very accurate results. In Chapter 5 we turn again to a phenomenologically more important theory, the $U(N)\times U(N)$ symmetric meson model. We will present an approximate large-N solution in the broken phase, which turns out to be renormalizable and fulfills Goldstone's theorem. Various symmetry breaking patterns will be investigated, and the renormalized effective potential will be constructed. We will investigate the vacuum structure of the theory which reveals new extrema of the effective potential not observed in previous perturbative computations.

The thesis is based on the following four publications:
\newline
\begin{itemize}
\item
G. Fej\H{o}s, A. Patk\'os, Zs. Sz\'ep: {\it Renormalized effective actions for the O(N) model at next-to-leading order of the 1/N expansion} \newline Phys. Rev. D{\bf 80}, 025015 (2009), arXiv:0902.0473 [hep-ph]
\item
G. Fej\H{o}s, A. Patk\'os: {\it A renormalized large N solution of the U(N)$\times$U(N) linear sigma model in the broken symmetry phase} \newline Phys. Rev. D{\bf 82}, 045011 (2010), arXiv:1005.1382 [hep-ph]
\item
G. Fej\H{o}s, A. Patk\'os: {\it Spontaneously broken ground states of the U(N)$\times$U(N) linear sigma model at large N} \newline Phys. Rev. D{\bf 84}, 036001 (2011), arXiv:1103.4799 [hep-ph]
\item
G. Fej\H{o}s, Zs. Sz\'ep: {\it Broken symmetry phase solution of the $\phi^4$ model at two-loop level of the $\Phi$-derivable approximation} \newline Phys. Rev. D{\bf 84}, 056001 (2011), arXiv:1105.4124 [hep-ph]
\end{itemize} 
\newpage
\pagestyle{plain}
\vspace{4cm}
{\huge{{\bf Acknowledgements}}}
\newline

First of all I would like express my gratitude to my advisor, Prof. Andr\'as Patk\'os for his continuous support and useful advises
over the last five years. His painstaking and most valuable guidance
has been an indispensable help during my undergraduate and PhD years.

I am very greatful to Zsolt Sz\'ep for his precision and endurance. He introduced me to the numerical work, which was an enormous help. Without him, the thesis in its current form could not have been completed.

I thank Urko Reinosa and Julien Serreau for the illuminating discussions during my three months visit in France.

I benefited a lot from the discussions and lectures of Antal Jakov\'ac.

I thank my fellow students Gergely Mark\'o and Istv\'an Sz\'ecs\'enyi for various discussions related to the topics of my thesis.

Travel grants from the Doctoral School of E\"otv\"os University are greatfully acknowledged. My work was supported by the Hungarian Research Fund (OTKA) under contract No. K77534 and No. T068108.

\clearemptydoublepage

\chapter{Functional techniques in quantum field theory}
\pagestyle{fancy}

\lhead[\let\uppercase\relax\bf{\small{\thepage}}]{\let\uppercase\relax\bf{\footnotesize{\rightmark}}}
\rhead[\let\uppercase\relax\bf{\footnotesize{\leftmark}}]{\let\uppercase\relax\bf{\small{\thepage}}}
In this chapter we review the standard functional techniques of quantum field theory. We will go through the ordinary generator functionals and show explicitly the properties of these quantities. The motivation for such a summary is that our goal at the end of this chapter is to arrive at the so-called two-particle-irreducible (2PI) formalism, which is a generalization of the standard one-particle-irreducible (1PI) formulation. The best way to achieve this is to build up the usual generator functionals from the beginning to see the correspondence between the different approaches. In this introductory chapter we present the functional techniques in a less formal but nevertheless expressive way. The discussion is built in a way from which the 2PI formalism can be obtained very naturally. Derivations will be presented mainly graphically, using the language of the Feynman-diagram technique. For a greater transparency we illustrate the general line of thinking within the one component $\phi^4$ theory in the symmetric phase, nevertheless all steps can be performed easily in arbitrary theories.

In this thesis we work in $\hbar=c=1$ units and this chapter will contain calculations only in Minkowski space. The metric is defined as $x^2\equiv x_{\mu}x^{\mu}=x_0^2-\vec{x}^{\,2}$. In this thesis only scalar quantum fields will appear in $3+1$ space-time dimensions, with Lagrangians parametrised as
\bea
{\cal L}[\partial_{\mu} \phi, \phi]=-\frac12 \phi (\partial_{\mu} \partial^{\mu}+m^2)\phi-U(\phi),
\eea
where $\phi$ is a multicomponent $c$-number scalar field and $U(\phi)$ describes the self-interaction. The $n$-point Green function is defined as a time-ordered vacuum expectation value of Heisenberg field operators
\bea
G_n(x_1,x_2,...,x_n)=\bra{0}T\big(\hat{\phi}(x_1)\hat{\phi}(x_2)...\hat{\phi}(x_n)\big)\ket{0}.
\eea
In functional integral formalism these functions can be obtained as \cite{rivers}
\bea
G_n(x_1,x_2,...,x_n)=\frac{\int {\cal D}\phi \phi(x_1)\phi(x_2)...\phi(x_n)e^{iS[\phi]}}{\int {\cal D}\phi e^{iS[\phi]}}
\eea
where $S[\phi]$ refers to the action of the configuration $\phi(x)$: $S[\phi]=\int d^4x {\cal L}[\partial_{\mu} \phi, \phi]$.

The Green functions are not measurable quantities, but with the help of the so-called reduction formulae (Lehmann, Symanzik and Zimmermann, 1955), they can be turned into physical transition amplitudes, see details in \cite{Peskin-book}. Due to this result, one can prove that the Green functions incorporate all information which can be extracted from a quantum field theory. Therefore, it is worthwhile to concentrate on building up equations for these quantities, which we eventually do in the next sections: the corresponding Dyson-Schwinger and 2PI equations will be derived. Before turning to this task, let us discuss the generator functionals of the Green functions. The properties of these functionals summarized here are quite general and can be obtained in arbitrary theories.

\section{Generator functionals}

The generator functional of the scalar Green functions is a functional of a source $J$, which is an arbitrary $c$-number function:
\bea
Z[J]=\sum_{n=0}^{\infty}\frac{i^n}{n!}\int d^4x_1... d^4x_n J(x_1)...J(x_n)G_n(x_1...x_n),
\eea
therefore the Green functions can be obtained by functional differentiation as
\bea
G_n(x_1,x_2,...,x_n)=(-i)^n\frac{\delta^n Z[J]}{\delta J(x_1)...\delta J(x_n)}\bigg|_{J=0}\equiv (-i)^n \frac{\delta^n Z[J]}{\delta J^n}\bigg|_{J=0}.
\eea
We will often not write down explicitly the spacetime arguments and use the short-hand notation as shown by the second equality. $Z[J]$  can be represented with functional integrals:
\bea
Z[J]=\frac{\int {\cal D}\phi e^{i(S[\phi]+\int J \phi)}}{\int {\cal D} \phi e^{iS[\phi]}},
\label{1-Z_func}
\eea
as it can be easily seen by expanding $\exp (i \int J \phi)$ in Taylor series. To calculate $Z[J]$ and eventually the Green functions we split the action into two parts:
\bea
S[\phi]=S_0[\phi]+S_{I}[\phi],
\eea
where $S_0$ refers to the gaussian part (quadratic in $\phi$), while $S_I$ to the interaction:
\bea
S_0[\phi]=-\frac12\int d^4x \phi(x)(\partial_\mu \partial^\mu +m^2)\phi(x), \qquad S_I[\phi]=-\int d^4x U[\phi(x)].
\eea
We write $Z[J]$ as
\bea
\label{1-Z_av}
Z[J]=\frac{\int {\cal D}\phi e^{i(S[\phi]+\int J \phi)}/\int {\cal D} \phi e^{iS_0[\phi]}}{\int {\cal D} \phi e^{iS[\phi]}/\int {\cal D} \phi e^{iS_0[\phi]}}\equiv \frac{<\!\!\exp(iS_I+i\int J\phi)\!\!>}{<\!\!\exp(iS_I)\!\!>},
\eea
where $<.>$ refers to the operation of the gaussian averaging. The evaluation of the numerator and the denominator can be made by expanding the exponentials in the brackets, since Green functions of the gaussian theory can be easily calculated \cite{Peskin-book}. Here we implicitly assumed that $m^2>0$ in order to have a positive definite quadratic form for the gaussian part, therefore the expansion makes sense. For each term in the expansion one can associate a diagram with defining model specific Feynman rules. For $U[\phi]=\lambda \phi^4/4!$ of a one component field these are the following \cite{Peskin-book}: \vspace{0.3cm}
\bea
\includegraphics[bb=145 482 334 431,scale=0.7]{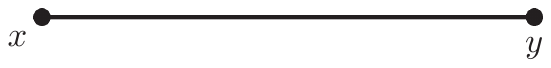} =\qquad \Delta_F(x-y) \nonumber
\eea
\bea
\includegraphics[bb=145 482 334 431,scale=0.7]{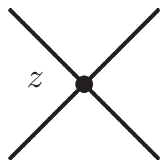} =\qquad -i\lambda\int d^4z \nonumber
\eea
\bea
\includegraphics[bb=135 482 334 441,scale=0.7]{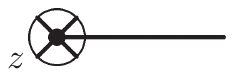} =\qquad i\int d^4z J(z), \nonumber
\eea
\newline where $\Delta_F(x-y)$ is the Feynman (or perturbative) propagator:
\bea 
\Delta_F(x-y)=-i(\partial^2+m^2)^{-1}(x,y)=\int \frac{d^4k}{(2\pi)^4}\frac{ie^{ik(x-y)}}{k^2-m^2+i\epsilon}.
\eea
Every diagram must be divided by a symmetry factor, which is defined as the dimension of the graph's symmetry group.
Note that the $1$-point vertex appears only in the expansion of the numerator, due to the term proportional to the source. In the averages every possible diagram without external legs appear, these are the so-called {\it vacuum graphs}. Using the Feynman rules (keeping also the symmetry factors in mind), in $\phi^4$ theory the first few terms of the denominator are \newline
\bea
\label{1-denom}
<\!\!e^{iS_I}\!\!>=1+\includegraphics[bb=106 540 200 631,scale=0.21]{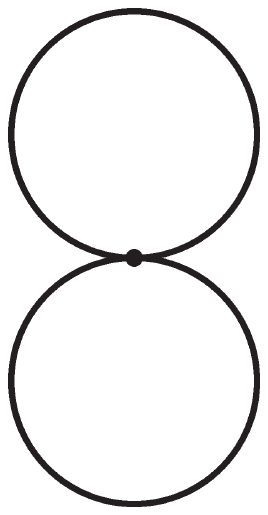}+ \includegraphics[bb=106 540 200 631,scale=0.21]{graf1.pdf} \includegraphics[bb=106 540 200 631,scale=0.21]{graf1.pdf} + \includegraphics[bb=63 540 234 631,scale=0.21]{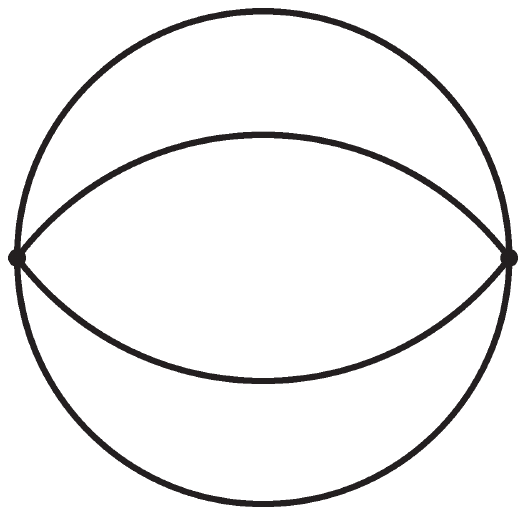} + \includegraphics[bb=106 585 200 631,scale=0.21]{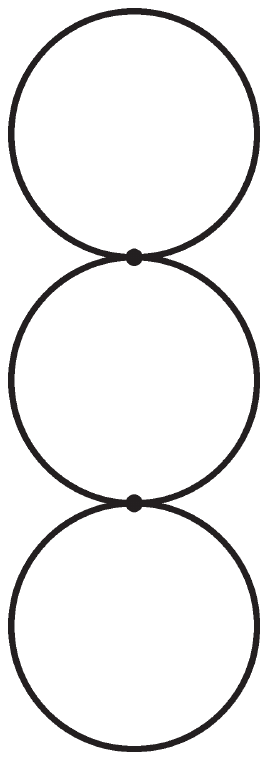} +  \includegraphics[bb=106 540 200 631,scale=0.21]{graf1.pdf} \includegraphics[bb=63 540 234 631,scale=0.21]{graf2.pdf} + \includegraphics[bb=63 540 234 631,scale=0.21]{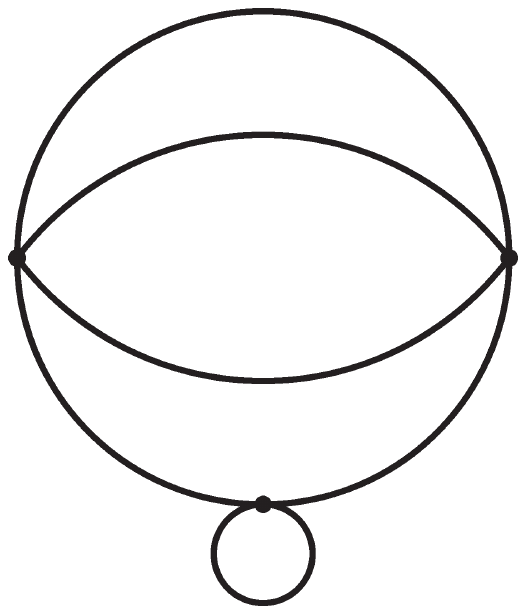} +  ...
\eea
\newline
We can observe that some terms contain disconnected diagrams, which offers a possibility for a simplification of the expression. Let $V$ be the set of the internally connected diagrams:
\bea
V = \Bigg\{\includegraphics[bb=106 540 200 631,scale=0.21]{graf1.pdf} \quad \includegraphics[bb=63 540 234 631,scale=0.21]{graf2.pdf} \quad \includegraphics[bb=106 585 200 631,scale=0.21]{graf4.pdf} \quad \includegraphics[bb=63 540 234 631,scale=0.21]{graf3.pdf} \quad ... \Bigg\}.\nonumber
\eea
We can check that every term in (\ref{1-denom}) can be written in the following form:
\bea
\prod_{i} \frac{1}{n_i!}(V_i)^{n_i}, \nonumber
\eea
where $V_i$ is the $i$th element of $V$ and $n_i$ equals to the number of occurrences of this connected diagram in the term in question (the factorials are symmetry factors of the interchanging of disconnected diagrams). The denominator of (\ref{1-Z_av}) is therefore
\bea
<\!\!e^{iS_I}\!\!>=\sum_{\{n_i\}} \prod_{i} \frac{1}{n_i!}(V_i)^{n_i},
\eea
where the summation goes through all ordered sets $\{n_1,n_2,...\}$. It is easy to show that this expression can be factored:
\bea
<\!\!e^{iS_I}\!\!>=\Big(\sum_{n_1}\frac{1}{n_1!}V_1^{n_1}\Big)\Big(\sum_{n_2}\frac{1}{n_2!} V_2^{n_2}\Big)\Big(\sum_{n_3}\frac{1}{n_3!}V_3^{n_3}\Big)...=\prod_i \sum_{n_i}\frac{1}{n_i!}V_i^{n_i}.
\eea
We recognize the definition of the exponential function and obtain
\bea
\label{1-denom_diag}
<\!\!e^{iS_I}\!\!>=\prod_i \!\exp V_i=\!\!\exp\!\!\sum_i V_i=\!\exp\Big(\includegraphics[bb=106 540 200 631,scale=0.21]{graf1.pdf}
+ \includegraphics[bb=63 540 234 631,scale=0.21]{graf2.pdf}+ \includegraphics[bb=106 585 200 631,scale=0.21]{graf4.pdf} +\includegraphics[bb=63 540 234 631,scale=0.21]{graf3.pdf}+...\Big).
\eea
This is the {\it linked cluster theorem}, which states that the sum of all possible vacuum diagrams is equal to the exponential of the sum of all the connected ones. The numerator of (\ref{1-Z_av}) can be evaluated the very same way. We have to keep in mind that there we also have a $1$-point vertex, therefore in addition to the diagrams of $V$, we have new vacuum graphs. Let $\tilde{V}$ be the extended set of the vacuum diagrams:
\bea
\tilde{V}=\Bigg\{\includegraphics[bb=106 540 200 631,scale=0.21]{graf1.pdf} \quad \includegraphics[bb=63 540 234 631,scale=0.21]{graf2.pdf} \quad \includegraphics[bb=106 585 200 631,scale=0.21]{graf4.pdf} \quad \includegraphics[bb=63 540 234 631,scale=0.21]{graf3.pdf} \quad ... \includegraphics[bb=165 482 318 441,scale=0.35]{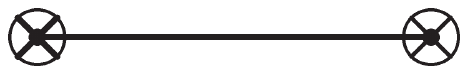}\quad \includegraphics[bb=165 482 318 441,scale=0.35]{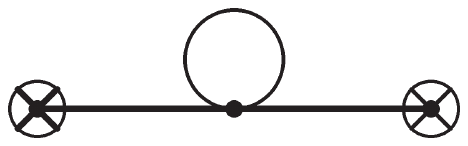}\quad \includegraphics[bb=165 482 318 441,scale=0.35]{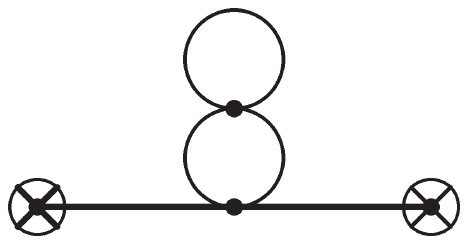} \nonumber\\
\includegraphics[bb=165 482 318 441,scale=0.35]{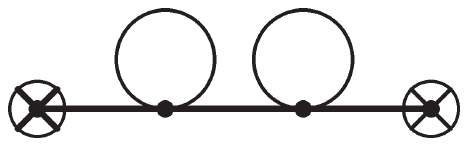}\quad \includegraphics[bb=165 482 318 441,scale=0.35]{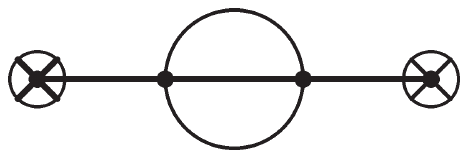}\quad ... \includegraphics[bb=165 482 318 441,scale=0.355]{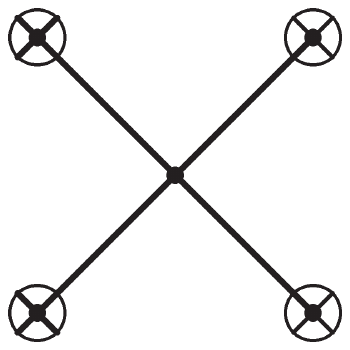}\quad \includegraphics[bb=195 482 298 441,scale=0.355]{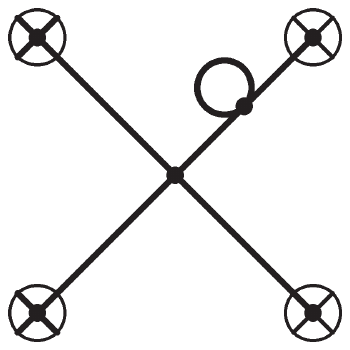}\quad ...\includegraphics[bb=180 482 338 441,scale=0.355]{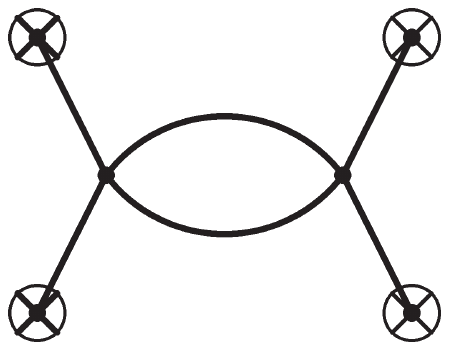}\quad ...
 \Bigg\}.\nonumber
\eea 
Note that the crosses do not correspond to external legs, these are $1$-point vertices, therefore every diagram of $\tilde{V}$ is indeed a vacuum graph. Following the method described above we arrive at
\bea
\label{1-nom_diag}
<\!\!e^{iS_I+i\int J \phi}\!\!>=\!\exp\Big(\includegraphics[bb=106 540 200 631,scale=0.21]{graf1.pdf}
+ \includegraphics[bb=63 540 234 631,scale=0.21]{graf2.pdf}+ \includegraphics[bb=106 585 200 631,scale=0.21]{graf4.pdf} +\includegraphics[bb=63 540 234 631,scale=0.21]{graf3.pdf}+...\includegraphics[bb=165 482 333 441,scale=0.35]{graf5.pdf}+ \includegraphics[bb=165 482 333 441,scale=0.35]{graf6.pdf}+ \nonumber 
\eea
\vspace{0.2cm}
\bea
\!\!\!+ \includegraphics[bb=165 482 333 441,scale=0.35]{graf7.pdf}+\includegraphics[bb=174 482 328 441,scale=0.35]{graf8.pdf}+ \includegraphics[bb=175 482 328 441,scale=0.35]{graf9.pdf}+... \includegraphics[bb=206 482 298 441,scale=0.355]{graf10.pdf}+ \includegraphics[bb=206 482 298 441,scale=0.355]{graf11.pdf}+\includegraphics[bb=205 482 338 441,scale=0.355]{graf12.pdf}...\Big).
\eea
\newline
Since $Z[J]$ is the quotient of (\ref{1-nom_diag}) and (\ref{1-denom_diag}), only diagrams involving the source remain \cite{Peskin-book}:
\bea
\label{1-Z_final}
\vspace{0.5cm}
Z[J]=\exp \Bigg(\!\!\!\includegraphics[bb=165 482 333 441,scale=0.35]{graf5.pdf}+ \includegraphics[bb=165 482 333 441,scale=0.35]{graf6.pdf}+ \includegraphics[bb=165 482 333 441,scale=0.35]{graf7.pdf}+\includegraphics[bb=174 482 328 441,scale=0.35]{graf8.pdf}+ \includegraphics[bb=175 482 328 441,scale=0.35]{graf9.pdf}...\nonumber
\eea
\newline
\vspace{-1.2cm}
\bea
+\includegraphics[bb=186 482 328 441,scale=0.355]{graf10.pdf}+ \includegraphics[bb=186 482 328 441,scale=0.355]{graf11.pdf}+...+\includegraphics[bb=186 482 348 441,scale=0.355]{graf12.pdf}+...\Bigg).
\eea

If now we are interested in the value of the various $n$-point functions, then we should re-expand the exponential function and read off the coefficients of the power series in the source $J$. We immediately recognize that in our concrete example only even number of sources could appear in every term, therefore when we take the source to zero only Green functions of even variables are non-zero. (In the broken phase of $\phi^4$ theory also Green functions of odd variables are present.) For example the $2$- and $4$-point functions read as
\newline
\vspace{-0.4cm}
\bea
\label{1-G_2}
G_2(x_1,x_2)=\!\!\includegraphics[bb=186 482 328 441,scale=0.355]{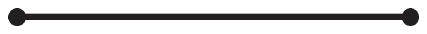}+\includegraphics[bb=186 482 328 441,scale=0.355]{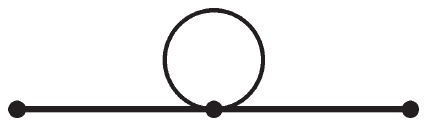}+
\includegraphics[bb=186 482 328 441,scale=0.355]{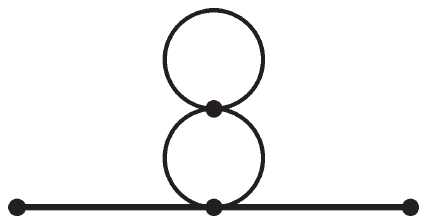}+\includegraphics[bb=186 482 328 441,scale=0.355]{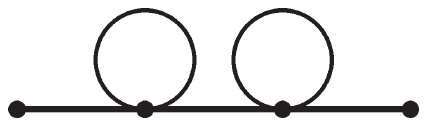}+\includegraphics[bb=186 482 328 441,scale=0.355]{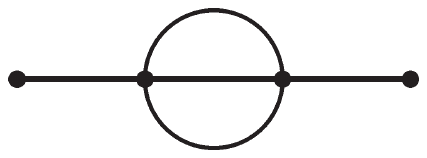}...,
\eea
\bea
\!\!\!\!\!\!\!\!\!\!\!\!\!\!\!\!G_4(x_1,x_2,x_3,x_4)=\includegraphics[bb=186 482 328 441,scale=0.355]{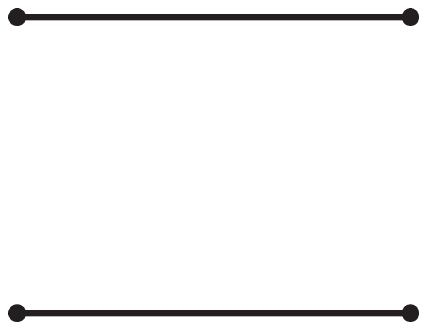}+\includegraphics[bb=186 482 328 441,scale=0.355]{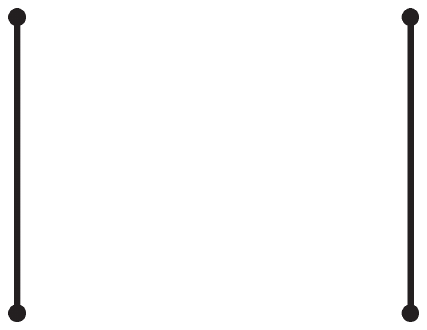}+\includegraphics[bb=186 482 328 441,scale=0.355]{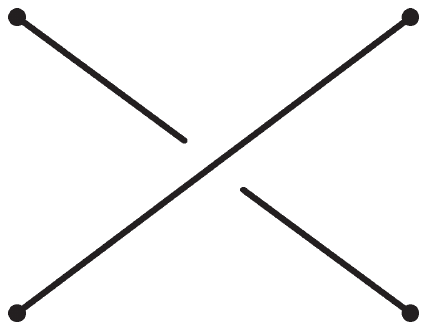}+\includegraphics[bb=186 482 328 441,scale=0.355]{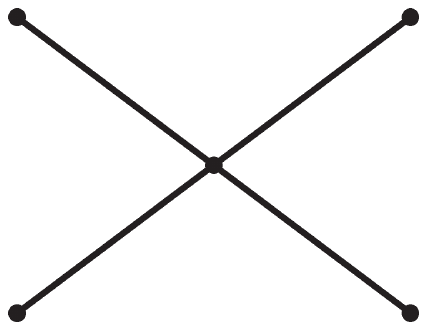}+ \nonumber
\eea
\newline
\vspace{-0.7cm}
\bea
+\includegraphics[bb=186 482 328 441,scale=0.355]{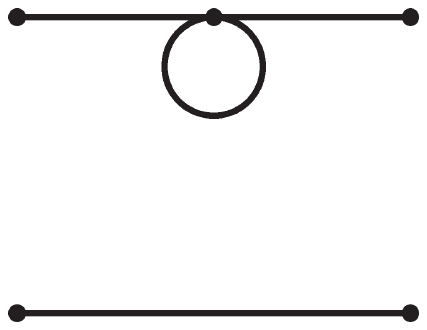}+\includegraphics[bb=186 482 328 441,scale=0.355]{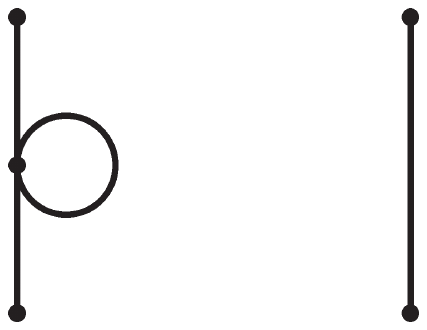}+...\includegraphics[bb=186 482 328 441,scale=0.355]{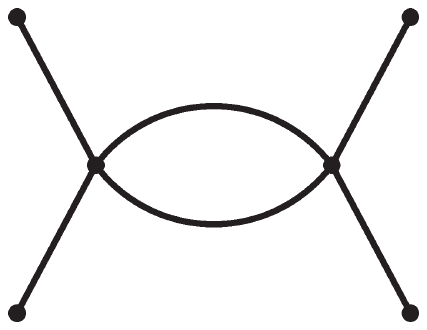}+...\includegraphics[bb=186 482 328 441,scale=0.355]{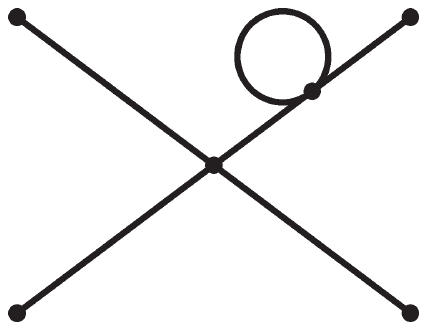}+...
\includegraphics[bb=186 482 328 441,scale=0.355]{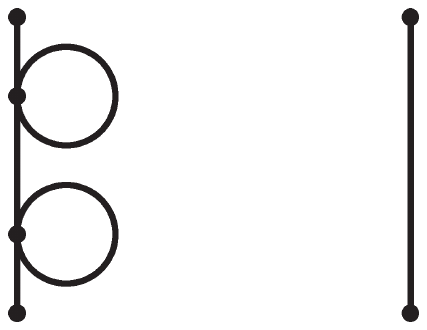}+...
\eea
\newline
The crosses from the endpoints disappeared, they indicated the presence of the sources (therefore neither multiplicative factors nor integration is needed corresponding to these endpoints). This means that these diagrams are not vacuum graphs anymore but have $2$ and $4$ external legs, respectively.

We recognize that there are lots of disconnected diagrams appearing in the series of the Green functions. (We note that in the symmetric
phase of the $\phi^4$ theory $G_2$ contains only connected diagrams - see (\ref{1-G_2}). This would not be the case in the broken phase, since then there would be a $1$-point vertex in the original Lagrangian, which could also produce nonzero contribution to $G_1$.) Recalling (\ref{1-Z_final}) it is possible to build up a generator functional which generates only the connected pieces, which provides great simplifications in further calculations. We introduce $W[J]$ as
\bea
iW[J]=\log Z[J].
\eea
Factoring an $i$ is just a convention, which will prove to be useful later. Using (\ref{1-Z_final}) we have
\bea
iW[J]=\includegraphics[bb=165 482 333 441,scale=0.35]{graf5.pdf}+ \includegraphics[bb=165 482 333 441,scale=0.35]{graf6.pdf}+ \includegraphics[bb=165 482 333 441,scale=0.35]{graf7.pdf}+\includegraphics[bb=174 482 328 441,scale=0.35]{graf8.pdf}+ \includegraphics[bb=175 482 328 441,scale=0.35]{graf9.pdf}...\nonumber
\eea
\newline
\vspace{-1.2cm}
\bea
+\includegraphics[bb=186 482 328 441,scale=0.355]{graf10.pdf}+ \includegraphics[bb=186 482 328 441,scale=0.355]{graf11.pdf}+...+\includegraphics[bb=186 482 348 441,scale=0.355]{graf12.pdf}+...
\eea
\newline
\vspace{-0.4cm}
If we think of $W[J]$ as a power series of the source:
\newline
\bea
iW[J]=\sum_n \frac{i^n}{n!}\int d^4x_1...d^4x_n J(x_1)...J(x_n) W_n(x_1,x_2,...,x_n),
\eea
it is obvious that the $W_n$ quantities refer to sums of {\it connected} diagrams. These are to be called as connected Green functions or {\it cumulants} and can be calculated as
\bea
W_n(x_1,...,x_n)=(-i)^m\frac{\delta^n (iW[J])}{\delta J^n}\bigg|_{J=0}.
\eea
In our $\phi^4$ example due to the absence of nonzero $G_1$, we have $W_1=0$ and $W_2=G_2$, however $W_4$ and further Green functions are greatly simplified:
\newline
\bea
\label{1-W_2}
W_2(x_1,x_2)=\!\!\includegraphics[bb=186 482 328 441,scale=0.355]{graf5b.pdf}+\includegraphics[bb=186 482 328 441,scale=0.355]{graf6b.pdf}+
\includegraphics[bb=186 482 328 441,scale=0.355]{graf7b.pdf}+\includegraphics[bb=186 482 328 441,scale=0.355]{graf8b.pdf}+\includegraphics[bb=186 482 328 441,scale=0.355]{graf9b.pdf}...,
\eea
\newline
\vspace{-0.6cm}
\bea
W_4(x_1,x_2,x_3,x_4)=\includegraphics[bb=186 482 328 441,scale=0.355]{graf10b.pdf}+\includegraphics[bb=186 482 328 441,scale=0.355]{graf18.pdf}+...\includegraphics[bb=186 482 328 441,scale=0.355]{graf19.pdf}+\includegraphics[bb=186 482 328 441,scale=0.355]{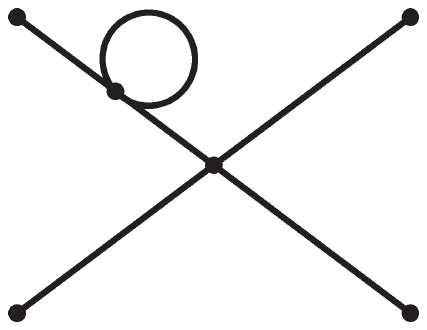}+...
\eea
\newline
\vspace{-0.4cm}

It turns out that even more simplification can be made, which leads us to the idea of the {\it quantum effective action}. Let us introduce the so-called {\it semi-classical} or {\it mean-field} $\bar{\phi}$ as
\bea
\label{1-phibar}
\bar{\phi}=\frac{\delta W[J]}{\delta J},
\eea
which is the vacuum expectation value of $\phi$ in the presence of the source $J$. We assume that this relation can be inverted uniquely: for a given $J$ only one $\bar{\phi}$ exists and vice-versa. The quantum effective action is the functional Legendre-transform of the generator of the connected Green functions $W[J]$ \cite{rivers}:
\bea
\Gamma[\bar{\phi}]=W[J]-\int J\bar{\phi}.
\eea
Using (\ref{1-phibar}) we have the usual relation
\bea
\label{1-gamma_deriv}
\frac{\delta \Gamma[\bar{\phi}]}{\delta \bar{\phi}}=\int \frac{\delta W}{\delta J} \frac{\delta J}{\delta \bar{\phi}}-\int \frac{\delta J}{\delta \bar{\phi}}\bar{\phi}-J=-J.
\eea
Let us try to calculate the diagrams contributing to this quantity. We have
\bea
e^{iW[J]}=\frac{\int {\cal D}\phi e^{i(S[\phi]+\int J \phi)}}{\int {\cal D}\phi e^{iS[\phi]}},
\eea
therefore
\bea
\label{1-expgamma}
e^{i\Gamma[\bar{\phi}]}=\frac{\int {\cal D}\phi e^{i\big(S[\phi]+\int J (\phi-\bar{\phi})\big)}}{\int {\cal D}\phi e^{iS[\phi]}}=\frac{\int {\cal D}\phi e^{i(S[\bar{\phi}+\phi]+\int J \phi)}}{\int {\cal D}\phi e^{iS[\phi]}},
\eea
where in the second equality we performed the following change of the integration variable in the numerator: $\phi \rightarrow \bar{\phi}+\phi$. The shifted action in the numerator can be expanded around $S[\bar{\phi}]$:
\bea
S[\bar{\phi}+\phi]=S[\bar{\phi}]+S^{'}[\bar{\phi}]\phi+\frac12 S^{''}[\bar{\phi}]\phi^2+...,
\eea
which gives new vertices.

We are now equipped to calculate such a quantity like (\ref{1-expgamma}). As we have already seen, the denominator is the exponential of all possible connected vacuum diagrams with couplings of the original Lagrangian. The numerator contains the same type of graphs but with vertices of the shifted Lagrangian extended with an additional term for the $1$-point vertex coming from the source. Diagrams of the denominator cancel the diagrams of the numerator which do not contain at least one of the new vertices or the source. This is a consequence of the normalization $Z[0]=1$, which implies $\Gamma[0]=0$.

\begin{figure}
\label{fig1_1}
\includegraphics[bb=20 445 254 516,scale=0.85]{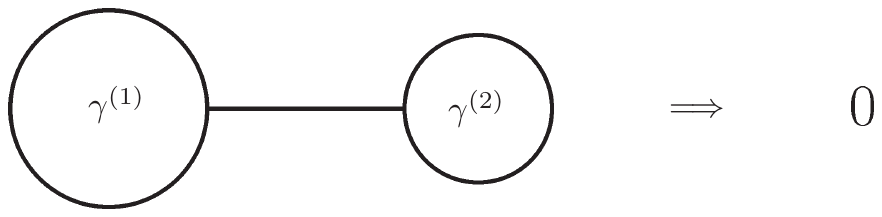}
\caption{1-particle-reducible diagrams do not count in the effective action.}
\end{figure}

Before starting to calculate the remaining diagrams by brute force, let us point out an important simplification taking place. The key observation is that the $1$-point function of the shifted action is zero by construction \cite{jakovac}:
\bea
\frac{\int {\cal D}\phi \phi e^{i(S[\bar{\phi}+\phi]+\int J\phi)}}{\int {\cal D}\phi e^{iS[\phi]}}&=&\frac{\int {\cal D} \phi (\phi-\bar{\phi})e^{i{\big(S[\phi]+\int J (\phi-\bar{\phi})\big)}}}{\int {\cal D} \phi e^{iS[\phi]}}=-\bar{\phi}e^{i\Gamma[\bar{\phi}]} \nonumber\\
&+&\frac{\int {\cal D}\phi \phi e^{i\big(S[\phi]+\int J (\phi-\bar{\phi})\big)}}{\int {\cal D}\phi e^{iS[\phi]}}=-\bar{\phi}e^{i\Gamma[\bar{\phi}]}+e^{-i\int J\bar{\phi}}\frac{1}{i}\frac{\delta}{\delta J}e^{iW[J]} \nonumber \\
&=&e^{i\Gamma[\bar{\phi}]}\Big(\frac{\delta W[J]}{\delta J}-\bar{\phi}\Big)=0.
\eea
This gives a great simplicity of the structure of $\Gamma[\bar{\phi}]$. Consider a diagram of the numerator in which two subdiagrams $\gamma^{(1)}$ and $\gamma^{(2)}$ are connected with one single line. This is to be called $1$-particle-reducible (1PR). Then find every possible diagram where $\gamma^{(2)}$ is replaced with some other subdiagram. In the sum of these diagrams, the exact $1$-point function will appear hanging on $\gamma^{(1)}$. Since we saw that the $1$-point function is identically zero, we may forget about diagrams with the previously indicated property (see Figure \ref{fig1_1}). In other words, we may omit all 1PR diagrams and consider only one-particle-irreducible (1PI) ones. These do not fall into two parts if one internal line is cut. The expressions corresponding to these diagrams do not refer to the $1$-point vertices at all, therefore the source does not appear in $\Gamma[\bar{\phi}]$, as it should. We may write
\bea
\label{1-exp_gamma}
e^{i\Gamma[\bar{\phi}]}=\frac{\int {\cal D}\phi e^{iS[\bar{\phi}+\phi]}}{\int {\cal D}\phi e^{iS[\phi]}}\bigg|_{1PI}.
\eea
When we are interested in the physical case (i.e. $J\rightarrow 0$), from (\ref{1-gamma_deriv}) we have
\bea
\frac{\delta \Gamma[\bar{\phi}]}{\delta \bar{\phi}}=0,
\eea
which means that the semi-classical field is an extremal configuration of $\Gamma$. This is the reason for calling it quantum effective action. In agreement with the expectations, after expanding the shifted action in (\ref{1-exp_gamma}) around $S[\bar{\phi}]$ and ignoring fluctuations, $\Gamma[\bar{\phi}]$ becomes the classical action: $\Gamma[\bar{\phi}]\approx S[\bar{\phi}]$. 

In $\phi^4$ theory the shifted action introduces the following new $2$ and $3$-point vertices (the $1$-point vertex is irrelevant due to the previous analysis):
\vspace{0.3cm}
\bea
\includegraphics[bb=127 481 228 512,scale=0.8]{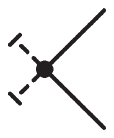}=\quad(-i\lambda)\frac{1}{2}\int d^4z \bar{\phi}^2(z), \nonumber
\eea
\bea
\includegraphics[bb=143 481 228 512,scale=0.8]{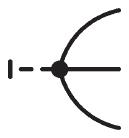}=\quad(-i\lambda) \int d^4z \bar{\phi}(z). \nonumber 
\eea

\vspace{0.6cm}
Note, that some sticks appeared at the end of the dashed lines. These refer to the multiplication(s) of $\bar{\phi}$ and to the integration over spacetime. The first few terms in the series of $\Gamma[\bar{\phi}]$ turns out to be: 
\newpage
\bea
\label{1-gamma_diags}
\!\!\!\!\!\!\!\!\!\!\!\Gamma[\bar{\phi}]=S[\bar{\phi}]-i\Bigg[ \includegraphics[bb=192 492 246 371,scale=0.65]{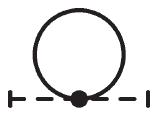}+\includegraphics[bb=192 492 246 371,scale=0.65]{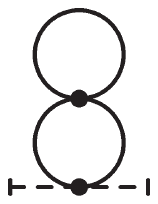}+\includegraphics[bb=192 482 291 371,scale=0.65]{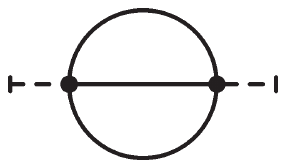} 
\eea
\vspace{0.1cm}
\bea
\qquad\quad +\includegraphics[bb=177 482 279 371,scale=0.65]{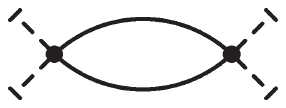}+\includegraphics[bb=172 482 279 331,scale=0.65]{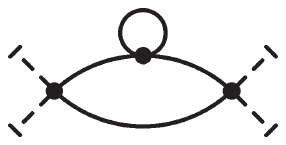}+...\includegraphics[bb=172 482 259 331,scale=0.65]{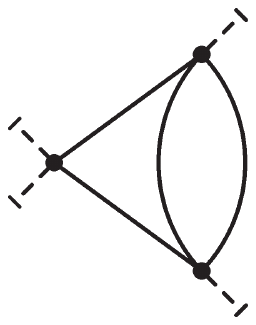}...\Bigg] \nonumber
\eea
\vspace{0.5cm}
\newline
We may think of $\Gamma[\bar{\phi}]$ as a power series of $\bar{\phi}$:
\bea
\label{1-gamma_series}
\Gamma[\bar{\phi}]=\sum_n \frac{1}{n!}\int d^4x_1...d^4x_n \Gamma_n(x_1,...,x_n)\bar{\phi}(x_1)...\bar{\phi}(x_n),
\eea
where the $\Gamma_n$ coefficients are the so-called {\it proper vertices}. From (\ref{1-gamma_diags}) we can read for example that
\newline
\bea
\label{1-gamma2}
\Gamma_2(x_1,x_2)=-(\partial^2+m^2)\delta(x_1-x_2)-i\Bigg[\includegraphics[bb=192 492 246 371,scale=0.65]{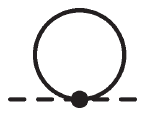}+\includegraphics[bb=192 492 246 371,scale=0.65]{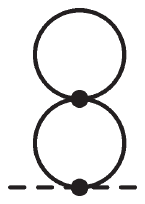}+\includegraphics[bb=192 482 281 371,scale=0.65]{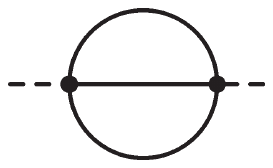}...\Bigg], \nonumber
\eea
\vspace{-0.8cm}
\bea \label{1-gamma2_diag} \eea
\vspace{-0.9cm}
\bea
\!\!\!\!\!\!\!\!\!\!\!\!\!\!\Gamma_4(x_1,x_2,x_3,x_4)=\lambda\delta(x_1-x_2)\delta(x_1-x_3)\delta(x_1-x_4)-i\Bigg[\includegraphics[bb=177 482 269 371,scale=0.65]{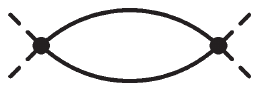} \nonumber
\eea
\bea
\label{1-gamma4_diag}
\qquad\qquad+\includegraphics[bb=208 482 259 331,scale=0.65]{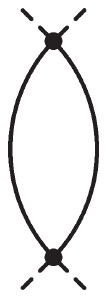}+...\includegraphics[bb=172 482 264 331,scale=0.65]{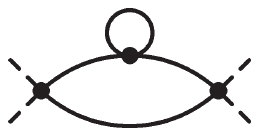}+...\includegraphics[bb=172 482 259 331,scale=0.65]{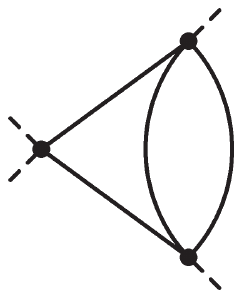}+...\quad\Bigg].
\eea
\vspace{0.3cm}
\newline
Proper vertices with odd number of variables are zero in the symmetric phase of the $\phi^4$ theory.
Note that diagrams in (\ref{1-gamma2_diag}) and (\ref{1-gamma4_diag}) do not have sticks on their ``legs'', as the $\bar{\phi}$ multiplicative factors (their integrals more precisely) were not included in the definition of the proper vertices. One also has to note that the factorials in (\ref{1-gamma_series}) were ``eaten up'' by the symmetry factors of diagrams still containing these sticks (i.e. diagrams in (\ref{1-gamma_diags})).

$\Gamma_2$ is strongly related to the connected $2$-point function, since
\bea
\label{1-derivJ}
\frac{\delta}{\delta J(y)}=\int_z \frac{\delta \bar{\phi}(z)}{\delta J(y)}\frac{\delta}{\delta \bar{\phi}(z)}=\int_z \frac{\delta^2 W[J]}{\delta J(y)\delta J(z)}\frac{\delta}{\delta \bar{\phi}(z)},
\eea
which applied to the function $J(x)$ gives
\bea
\delta(x-y)=\int_z \frac{\delta^2W[J]}{\delta J(y)\delta J(z)}\frac{\delta J(x)}{\delta\bar{\phi}(z)}=-\int_z \frac{\delta^2W[J]}{\delta J(y)\delta J(z)}\frac{\delta^2 \Gamma[\bar{\phi}]}{\delta \bar{\phi}(z)\delta \bar{\phi}(x)}.
\eea
This shows that in functional sense
\bea
\label{1-inverse}
\Big(\frac{\delta^2 W}{\delta J^2}\Big)^{-1}=-\frac{\delta^2\Gamma}{\delta \bar{\phi}^2},
\eea
which implies that
\bea
iW_2^{-1}=\Gamma_2.
\eea
The squared bracket of (\ref{1-gamma2_diag}) including the factor $i$ is called the {\it self-energy}, and is denoted by $\Sigma(x_1,x_2)$. With the help of (\ref{1-inverse}) we arrive at the {\it Dyson-equation} of the propagator
\bea
\label{1-Dyson}
iW_2^{-1}(x_1,x_2)=-(\partial^2+m^2)\delta(x_1-x_2)-\Sigma(x_1,x_2),
\eea
to which we will refer several times in the forthcoming sections.

It is possible to organize the series of $\Gamma$ in terms of the number of loops. When one works in ordinary units, it turns out to be a series in $\hbar$. Introducing the notations
\bea
\Gamma[\bar{\phi}]=S[\bar{\phi}]+\Gamma_{\hbar}[\bar{\phi}]+\Gamma_{\hbar^2}[\bar{\phi}]+...,
\eea
the leading order piece (beyond the classical value) $\Gamma_{\hbar}$ can be resummed in a rather compact form, in which one has to take into account every 1-loop diagram:
\newline
\vspace{-0.35cm}
\bea
i\Gamma_{\hbar}&=&\includegraphics[bb=151 498 211 331,scale=0.63]{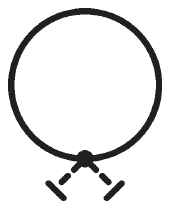}+\includegraphics[bb=131 498 211 331,scale=0.63]{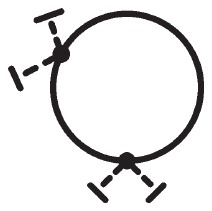}+
\includegraphics[bb=131 498 221 331,scale=0.63]{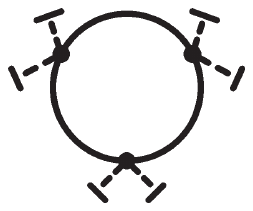}+...
\eea
\vspace{-0.1cm}
\newline
Taking into account the symmetry factors, the sum of these diagrams is
\bea
\!\!\!\!\!\!\!i\Gamma_{\hbar}&\!\!\!=\!\!\!&\frac12\Big(-i\frac{\lambda}{2}\Big)\int_x \Delta_F(x,x)\bar{\phi}^2(x)\nonumber\\
\!\!\!\!\!\!\!&\!\!\!+\!\!\!&\frac14\Big(-i\frac{\lambda}{2}\Big)^2\int_x\int_y \Delta_F(x,y)\Delta_F(y,x)\bar{\phi}^2(x)\bar{\phi}^2(y)\nonumber\\
\!\!\!\!\!\!\!&\!\!\!+\!\!\!&\frac16 \Big(-i\frac{\lambda}{2}\Big)^3 \int_x \int_y \int_z \Delta_F(x,y)\Delta_F(y,z)\Delta_F(z,x)\bar{\phi}^2(x)\bar{\phi}^2(y)\bar{\phi}^2(z)+...,
\eea
where the integrals are over the whole space-time with respect to the corresponding letters.
One can identify the series of the logarithm function:
\bea
i\Gamma_{\hbar}&=&\frac12 \Tr \sum_{n=1}^{\infty} \frac{1}{n}\Big[\frac{(-i\lambda)}{2}\bar{\phi}^2\Delta_F\Big]^n = -\frac12 \Tr \log \Big(1+\frac{\lambda}{2}\bar{\phi}^2i\Delta_F\Big).
\eea
The trace has to be taken in functional sense. Recalling that $U^{''}(\bar{\phi})=\frac{\lambda}{2}{\bar{\phi}}^2$, for $\Gamma$ we have
\bea
\label{1-gamma_1loop}
\!\!\!\Gamma[\bar{\phi}]=S[\bar{\phi}]+\frac{i}{2}\Tr \log \Big(1+U^{''}(\bar{\phi})i\Delta_F\Big)-i(\diagrams).
\eea
This equation serves as a general formula applicable in arbitrary scalar theories, not just in the $\phi^4$ model. In the case when $\phi$ is a multicomponent field, the trace has to be taken over the internal indices as well.

From the proper vertices every Green function can be built up, therefore it is sufficient to determine the $\Gamma_n$ functions to solve fully a quantum field theory. We wish to mention that it is often more convenient to work with the Fourier transform of the proper vertices, denoted by $\tilde{\Gamma}_n$. Due to the translation invariance, one can always factor out a delta function, therefore $\tilde{\Gamma}_n(p_1,...,p_n)$ can be always written as
\bea
\label{1-Gamma_Fourier}
\tilde{\Gamma}_n(p_1,...,p_n)=(2\pi)^4\delta(p_1+...+p_n)\Gamma_n(p_1,...,p_n).
\eea

\section{Dyson-Schwinger equations}

As already announced at the beginning of the chapter, in order to obtain all the information contained in a quantum field theory, it is sufficient to determine all the Green functions. In practice we need to calculate the $\Gamma_n$ proper vertex functions generated by the effective action. In this section we derive an infinite tower of coupled equations from which any $\Gamma_n$ can be obtained. We begin with recalling the translation invariance of the measure of the functional integral \cite{rivers}:
\bea
0=\int {\cal D}\phi \frac{\delta }{\delta \phi}(...).
\eea
Let us apply this identity to the function $e^{i(S[\phi]+\int J \phi)}$:
\bea
0=\int {\cal D}\phi \frac{\delta}{\delta \phi} e^{i(S[\phi]+\int J \phi)}=\int {\cal D} e^{i(S[\phi]+\int J \phi)} \Big(i\frac{\delta S[\phi]}{\delta \phi}+iJ\Big).
\eea
Since under the integral a multiplication with $\phi$ is equivalent to differentiating with respect to the source, we may write the formal expression
\bea
\label{1-DS_G}
0=\int {\cal D} \phi \Bigg(\frac{\delta S}{\delta \phi}\bigg|_{\phi\rightarrow -i\frac{\delta }{\delta J}}+J\Bigg)e^{i(S[\phi]+\int J \phi)}.
\eea
Now neither of the terms in the bracket contain $\phi$ therefore it can be taken outside the integral. Dividing (\ref{1-DS_G}) by $\int {\cal D}\phi e^{iS[\phi]}$ we obtain
\bea
0=\Bigg(\frac{\delta S}{\delta \phi}\bigg|_{\phi\rightarrow -i\frac{\delta }{\delta J}}+J\Bigg)Z[J].
\eea
Since $Z[J]=\exp{iW[J]}$, this equation is equivalent to
\bea
\label{1-DS_W}
0=e^{-iW[J]}\frac{\delta S}{\delta \phi}\bigg|_{\phi\rightarrow -i\frac{\delta}{\delta J}}e^{iW[J]}+J.
\eea
Let $T[J]$ be a test functional, then
\bea
\Big[e^{-iW[J]}\Big(-i\frac{\delta}{\delta J}\Big)e^{iW[J]}\Big]T[J]=\Big(\frac{\delta W[J]}{\delta J}-i\frac{\delta}{\delta J}\Big)T[J].
\eea
This means that (\ref{1-DS_W}) can be also written as
\bea
\Big(\frac{\delta S}{\delta \phi}\bigg|_{\phi\rightarrow \frac{\delta W[J]}{\delta J}-i\frac{\delta}{\delta J}}+J\Big){\bf 1}=0,
\eea
where ${\bf 1}$ is the unit functional. Recalling (\ref{1-phibar}) and (\ref{1-gamma_deriv}), we have
\bea
\frac{\delta \Gamma[\bar{\phi}]}{\delta \bar{\phi}}=\frac{\delta S}{\delta \phi}\bigg|_{\phi\rightarrow \bar{\phi}-i\frac{\delta}{\delta J}}{\bf 1}.
\eea
Using (\ref{1-derivJ}) and (\ref{1-inverse}) we arrive at the generator equation of the {\it Dyson-Schwinger equations} for $\Gamma_n$:
\bea
\label{1-genDS}
\frac{\delta \Gamma[\bar{\phi}]}{\delta \bar{\phi}}=\frac{\delta S}{\delta \phi}\bigg|_{\phi\rightarrow \bar{\phi}+i\int \frac{\delta^2 \Gamma}{\delta\bar{\phi}^2}\frac{\delta}{\delta \bar{\phi}}}{\bf 1},
\eea
which shows how to compute the derivatives of the effective action and builds up a tower of coupled equations for the proper vertices.

In $\phi^4$ theory with the classical potential $U(\phi)=\lambda \phi^4/4!$, the first two derivatives are:
\bea
\label{1-gamma_deriv_1}
\frac{\delta \Gamma[\bar{\phi}]}{\delta \bar{\phi}(x)}&=&-\Big(\partial^2+m^2+\frac{\lambda}{3!}(\bar{\phi}^2(x)+3 W_2(x,x))\Big)\bar{\phi}(x)\nonumber\\
&-&i\frac{\lambda}{3!}\int_{uvw} W_2(x,u) W_2(x,v) W_2(x,w) \Gamma_3(u,v,w)\\
\label{1-gamma_deriv_2}
\frac{\delta^2 \Gamma}{\delta\bar{\phi}(x)\delta\bar{\phi}(y)}&=&-\Big(\partial^2+m^2+\frac{\lambda}{2}(\bar{\phi}^2(x)+W_2(x,x))\Big)\delta(x-y)\nonumber\\
&-&i\frac{\lambda}{2}\bar{\phi}(x)\int_{uv} W_2(x,u)W_2(x,v)\Gamma_3(u,v,y)\nonumber\\
&+&\frac{\lambda}{2}\int_{stuvw} W_2(x,s) W_2(x,u) W_2(x,v) \Gamma_3(u,v,w)W_2(w,t)\Gamma_3(s,t,y)\nonumber\\
&-&i\frac{\lambda}{3!}\int_{uvw} W_2(x,u) W_2(x,v) W_2(x,w)\Gamma_4(u,v,w,y).
\eea
For a given $n$-point proper vertex, higher $\Gamma_m$ functions always appear in their Dyson-Schwinger equations, therefore practically at some point we have to make an approximation and truncate the tower of the coupled equations. When the configuration is set to the physical one (i.e. $J\rightarrow 0$), we know that in the symmetric phase of $\phi^4$ theory only proper vertices with even variables are nonzero. In this case (\ref{1-gamma_deriv_1}) becomes trivial, and for (\ref{1-gamma_deriv_2}) we have (note that $\Gamma_2=iW_2^{-1}$)
\bea
\label{1-Dyson2}
\Gamma_2(x,y)&=&-\big(\partial^2+m^2+\frac{\lambda}{2}W_2(x,x)\big)\delta(x-y)\nonumber\\
&-&i\frac{\lambda}{3!}\int_{uvw} W_2(x,u) W_2(x,v) W_2(x,w)\Gamma_4(u,v,w,y).
\eea
In agreement with the previously introduced Feynman rules, diagrammatically it can be written as
\bea
\label{1-Dyson2_graph}
\Gamma_2(x,y)=-(\partial^2+m^2)\delta(x-y)-i\bigg[\includegraphics[bb=179 492 235 331,scale=0.7]{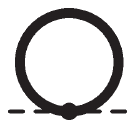}+\includegraphics[bb=199 508 281 331,scale=0.73]{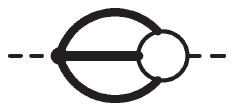}\bigg],
\eea
where the thick lines do not refer to Feynman propagators, but to full connected $2$-point functions ($W_2$) and in the second term of the squared bracket the circle represents $-i\Gamma_4$ (one should keep also the symmetry factors in mind).
(\ref{1-Dyson2_graph}) is very similar to (\ref{1-gamma2}). In fact, this is the same equation, but in a different form. In (\ref{1-Dyson2_graph}), diagrams of (\ref{1-gamma2}) are resummed into $W_2$ and $\Gamma_4$. If we insert the perturbative expansion of $W_2$ and $\Gamma_4$ into the right hand side of (\ref{1-Dyson2_graph}), we obtain (\ref{1-gamma2}). In the next section, we will derive this equation in a third form using 2PI formalism.

What makes the tower of Dyson-Schwinger equations (and therefore the form (\ref{1-Dyson2}) of the Dyson-equation) in particular important is that they are valid and make sense also non-perturbatively (i.e. without referring to any Feynman diagrams). However, closing the hierarchy using appropriate approximation(s) typically lead to partially resummed perturbative series of the vertices yet to determine. This will also be a key feature of the 2PI formalism, which is introduced in the next section. Due to this, Dyson-Schwinger and 2PI formalisms are related in certain approximations (this will be highlighted later).

\section{2PI formalism}

Two-particle-irreducible (2PI) formalism generalizes the idea of the effective action: it treats also the propagator as a variable. The so-called 2PI effective action can be obtained easily in the framework showed before.

We introduce into the usual $Z$ generator functional of the Green functions a bilocal source $K(x,y)$ with respect to the product of the fields \cite{Cornwall:vz}:
\bea
Z[J,K]=\frac{\int{\cal D}\phi e^{i(S[\phi]+\int J \phi+\frac12\int\int K \phi \phi)}}{\int {\cal D}\phi e^{iS[\phi]}}.
\eea
The quadratic term including $K$ can be viewed formally as an additional ``mass term'', therefore our previous analysis can be directly applied. This means that
\bea
iW[J,K]\equiv \log Z[J,K]
\eea
contains the connected diagrams and in
\bea
\label{1-gammaK_1PI}
\Gamma_K[\bar{\phi}]\equiv W[J,K]-\int J \bar{\phi}
\eea
we find the 1PI diagrams, where the $\bar{\phi}$ semi-classical field in the presence of sources $J$ and $K$ is defined as before:
\bea
\label{1-2PI-phi_def}
\frac{\delta W[J,K]}{\delta J(x)}=\bar{\phi}(x).
\eea
Since we have two sources, it is possible to perform one more Legendre-transformation. Therefore in addition we introduce $G(x,y)$ (which will refer to the connected $2$-point function) with defining the derivative with respect to $K$ as
\bea
\label{1-2PI-G_def}
\frac{\delta W[J,K]}{\delta K(x,y)}=\frac12\Big(\bar{\phi}(x)\bar{\phi}(y)+G(x,y)\Big).
\eea
The $J$ and $K$ dependence of $\bar{\phi}$ and $G$ are not indicated explicitly, as usual. The 2PI effective action is the double Legendre-transform of $W[J,K]$:
\bea
\Gamma[\bar{\phi},G]=W[J,K]-\int J \bar{\phi}-\frac12\int\int K (G+\bar{\phi}\bar{\phi}).
\eea
Using (\ref{1-2PI-phi_def}) and (\ref{1-2PI-G_def}) we have
\bea
\label{1-legendre_2PI}
\frac{\delta \Gamma[\bar{\phi},G]}{\delta \bar{\phi}(x)}=-J(x)-\int_y K(x,y)\bar{\phi}(y),\qquad\frac{\delta \Gamma[\bar{\phi},G]}{\delta G(x,y)}=-K(x,y).
\eea
In the physical case (i.e. $J\rightarrow 0$, $K\rightarrow 0$) $\bar{\phi}$ and $G$ are determined from the stationary conditions
\bea
\label{1-stat}
\frac{\delta \Gamma[\bar{\phi},G]}{\delta \bar{\phi}(x)}=0, \qquad \frac{\delta \Gamma[\bar{\phi},G]}{\delta G(x,y)}=0.
\eea
Similarly to the procedure of obtaining the 1PI effective action, we have
\bea
\label{1-eGamma_2PI}
e^{i\Gamma[\bar{\phi},G]}=\frac{\int {\cal{D}}\phi e^{i\big(S[\phi]+\int J (\phi-\bar{\phi})+\frac12\int\int K(\phi\phi-\bar{\phi}\bar{\phi}-G)\big)}}{{\cal D}\phi e^{iS[\phi]}}.
\eea
Performing the change of variable $\phi \rightarrow \bar{\phi}+\phi$,
\bea
e^{i\Gamma[\bar{\phi},G]}=\frac{\int {\cal{D}}\phi e^{i\big(S[\bar{\phi}+\phi]+\int J \phi+\frac12\int\int K(2\phi\bar{\phi}+\phi\phi-G)\big)}}{{\cal D}\phi e^{iS[\phi]}}.
\eea
We may now try to calculate both the numerator and the denominator with the method introduced in the beginning of the chapter (divide both with $\int {\cal D}\phi e^{iS_0[\phi]}$ and use perturbation theory), however there are simplifications which should be taken into account. First we see again, that the $1$-point function corresponding to the shifted action is zero:
\bea
&&\frac{\int {\cal D}\phi \phi e^{i\big(S[\phi+\bar{\phi}]+\int J \phi +\frac12 \int \int K(2\phi\bar{\phi}+\phi\phi-G)\big)}}{\int {\cal D}\phi e^{iS[\phi]}}=e^{-i\big(\int J\bar{\phi}+\frac12 \int\int K(\bar{\phi}\bar{\phi}+G)\big)}\frac{1}{i}\frac{\delta}{\delta J} e^{iW[J,K]}\nonumber\\
&&-\bar{\phi}e^{i\Gamma[\bar{\phi},G]}=e^{i\Gamma[\bar{\phi},G]}\Big(\frac{\delta W[J,K]}{\delta J}-\bar{\phi}\Big)=0.
\eea
In the first equality we performed the change of variable $\phi \rightarrow \phi-\bar{\phi}$. As it was already argued, the $1$-particle-reducible diagrams drop out. Furthermore, where the connected $2$-point function ($W_2$) appears, it can be substituted by the new variable $G$ \cite{jakovac}. It can be seen by the following argument. Let us calculate the $2$-point function of the shifted theory in two different ways. The first one is:
\bea
\label{1-2p_W2}
&&\frac{\int {\cal D}\phi \phi \phi e^{i\big(S[\phi+\bar{\phi}]+\int J \phi +\frac12 \int \int K(2\phi\bar{\phi}+\phi\phi-G)\big)}}{\int {\cal D}\phi e^{iS[\phi]}}=e^{-i\big(\int J\bar{\phi}+\frac12 \int\int K(\bar{\phi}\bar{\phi}+G)\big)}\frac{1}{i^2}\frac{\delta^2}{\delta J^2} e^{iW[J,K]}\nonumber\\
&&-\bar{\phi}\bar{\phi}e^{i\Gamma[\bar{\phi},G]}=e^{i\Gamma[\bar{\phi},G]}\frac{1}{i}\frac{\delta^2W[J,K]}{\delta J^2}=e^{i\Gamma[\bar{\phi},G]}W_2.
\eea
The other way is:
\bea
\label{1-2p_G}
&&\frac{\int {\cal D}\phi \phi \phi e^{i\big(S[\phi+\bar{\phi}]+\int J \phi +\frac12 \int \int K(2\phi\bar{\phi}+\phi\phi-G)\big)}}{\int {\cal D}\phi e^{iS[\phi]}}=e^{-i\big(\int J\bar{\phi}+\frac12 \int\int K(\bar{\phi}\bar{\phi}+G)\big)}2\frac{\delta}{\delta K}e^{iW[J,K]}\nonumber\\
&&-\bar{\phi}\bar{\phi}e^{i\Gamma[\bar{\phi},G]}=e^{i\Gamma[\bar{\phi},G]}\Big(\bar{\phi}\bar{\phi}+G-\bar{\phi}\bar{\phi}\Big)=e^{i\Gamma[\bar{\phi},G]} G.
\eea
The key part was to observe that the product of two $\phi$ fields could appear with differentiating with respect to $K$, or two times with respect to $J$. Comparing (\ref{1-2p_W2}) with (\ref{1-2p_G}) we have $W_2=G$. Consider then a typical diagram of the numerator of (\ref{1-eGamma_2PI})  which contains two subdiagrams $\gamma^{(1)}$ and $\gamma^{(2)}$ connected by two perturbative propagators. Let us treat the subdiagram $\gamma^{(2)}$ as a self-energy input. (We could have chosen $\gamma^{(1)}$, it does not matter.) Then find every possible diagram in which $\gamma^{(1)}$ is connected to an other self-energy insertion also with two lines. In other words, let us find all self-energy inputs, and connect them to $\gamma^{(1)}$ via two lines. In the sum of these diagrams, the connected $2$-point function ($W_2$) appears as it connects a part of $\gamma^{(1)}$ with another one (see Figure 1.2). Since $W_2$ can be replaced by $G$, we conclude that in $\Gamma[\bar{\phi},G]$ diagrams with properties described above do not appear, when propagator lines are set to $G$. 
\begin{figure}
\label{fig1_2}
\includegraphics[bb=45 445 254 516,scale=0.85]{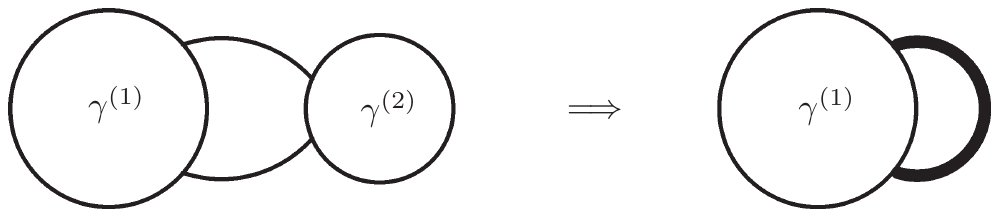}
\caption{Self-energy insertions are resummed and a full propagator appears (thick line). This produces the absence of 2-particle-reducible diagrams.}
\end{figure}
We say, that in $\Gamma[\bar{\phi},G]$ only those diagrams are included, which do not fall into two parts if two internal lines are cut. These are to be called {\it two-particle-irreducible} (2PI) diagrams and they have to be calculated with vertices of the shifted action and with propagator lines equal to G. This is the reason for calling $\Gamma[\bar{\phi},G]$ the {\it 2PI effective action}, which can be calculated as:
\bea
\label{Eq:1-Gamma2PI}
e^{i\Gamma[\bar{\phi},G]}=\frac{\int {\cal D}\phi e^{iS[\bar{\phi}+\phi]}}{\int {\cal D}\phi e^{iS[\phi]}}\bigg|_{\num 2PI,\hspace{0.1cm} W_2\rightarrow G}.
\eea
In (\ref{Eq:1-Gamma2PI}) there is no reference to the sources (since neither the $1$- nor the $2$-point functions are needed for the calculation), as it should.

Careful readers may complain about the denominator of (\ref{Eq:1-Gamma2PI}), which was used in 1PI formalism to cancel diagrams of the numerator including only the original vertices of the theory (i.e. before the shift). Since now in the numerator these diagrams were used to sum up the self-energy insertions, no cancellations occur at all. Therefore the denominator is just playing the role of normalization and is often neglected since it gives an additive constant for $\Gamma[\bar{\phi},G]$ and in practice we need only the derivatives of it.

From the 2PI effective action we can obtain the standard 1PI effective action. Using the definition, the 1PI action can be thought as $\Gamma[\bar{\phi},G]|_{K=0}$. In the language of the 2PI action, the $K=0$ condition refers to a $\tilde{G}[\bar{\phi}]$ propagator, which satisfies the second stationary condition of (\ref{1-stat}) as a function of $\bar{\phi}$:
\bea
\frac{\delta \Gamma [\bar{\phi},G]}{\delta G}\bigg|_{G={\tilde{G}[\bar{\phi}}]}=0.
\eea
Therefore the 1PI effective action is
\bea
\label{1-gamma-12PI}
\Gamma[\bar{\phi}]=\Gamma\big[\bar{\phi},G=\tilde{G}[\bar{\phi}]\big].
\eea
Note that the equality is true only if the normalization is appropriately chosen (see above), otherwise a constant shift may appear.

We can use the loop expansion of the 1PI effective action determined in (\ref{1-gamma_1loop}), since from (\ref{1-gammaK_1PI}) it is obvious that at leading order
\bea
\Gamma_K[\bar{\phi}]&\!\!\!=\!\!\!&S[\bar{\phi}]+\frac12\int_x\int_y K(x,y)\bar{\phi}(x)\bar{\phi}(y)+\frac{i}{2}\Tr \log (1+(U^{''}(\bar{\phi})-K)i\Delta_F), \nonumber\\
\eea
or equivalently
\bea
\label{1-loop_2PI_K}
\Gamma[\bar{\phi},G]=S[\bar{\phi}]+\frac{i}{2}\Tr \log\Big[\frac{i\Delta_F^{-1}-U^{''}(\bar{\phi})+K}{i\Delta_F^{-1}}\Big]-\frac12\Tr (KG).
\eea
This is not the final form we are looking for, since the source $K$ should be expressed with $G$ and $\bar{\phi}$. This can be achieved using the second equation of (\ref{1-legendre_2PI}). We find that at lowest order
\bea
iG^{-1}=i\Delta_F^{-1}-U^{''}(\bar{\phi})+K,
\eea
which is consistent if we plug this relation to (\ref{1-loop_2PI_K}). With this we obtained the loop expansion of the 2PI effective action at leading order \cite{berges}:
\bea
\label{1-loop_2PI}
\Gamma[\bar{\phi},G]=S[\bar{\phi}]+\frac{i}{2}\Tr \log\Big[\frac{G^{-1}}{\Delta_F^{-1}}\Big]+\frac{i}{2}\Tr (D^{-1}G-1),
\eea
where we introduced the short hand notation $iD^{-1}=i\Delta_F^{-1}-U^{''}(\bar{\phi})$, which is the classical (tree level) propagator at a given $\bar{\phi}$ background in the absence of the bilocal source. If we want to go further in the expansion, we have to add 2PI diagrams containing at least
$2$ loops. To be very precise and taking into account the normalization, we have
\bea
\label{1-2PI}
\Gamma[\bar{\phi},G]=S[\bar{\phi}]+\frac{i}{2}\Tr \log\Big[\frac{G^{-1}}{\Delta_F^{-1}}\Big]+\frac{i}{2}\Tr (D^{-1}G-1)+\Gamma^{(2)}-N,
\eea
where $iN=$ (every vacuum diagram with at least 2 loops built by perturbative propagators containing vertices of the original theory), and $i\Gamma^{(2)}=$ (2PI vacuum diagrams with at least 2 loops built by propagators $G$ with vertices of the shifted theory).

In the literature the $\bar{\phi}$ and $G$ independent parts are usually omitted, therefore the final form of the 2PI effective action is
\bea
\label{1-2PI_final}
\Gamma_{\textnormal{2PI}}[\bar{\phi},G]=S[\bar{\phi}]+\frac{i}{2}\Tr \log G^{-1}+\frac{i}{2}\Tr(D^{-1}G)+\Gamma^{(2)},
\eea
which is valid for every multicomponent scalar field theory.

From stationary conditions (\ref{1-stat}) we get equivalent equations already obtained in the Dyson-Schwinger formalism. The equivalence is demonstrated as follows. The first observation is that the partial derivative of the 2PI action with respect to $\bar{\phi}$ is equal to the first derivative of the 1PI action, if the propagator is set to $\tilde{G}$ (i.e. the solution of the second stationary condition):
\bea
\label{1-diff}
\frac{\delta \Gamma[\bar{\phi}]}{\delta\bar{\phi}}=\frac{\Delta \Gamma\big[\bar{\phi},\tilde{G}[\bar{\phi}]\big]}{\Delta \bar{\phi}}=\frac{\delta \Gamma[\bar{\phi},\tilde{G}]}{\delta \bar{\phi}}+\frac{\delta\Gamma[\bar{\phi},G]}{\delta G}\bigg|_{G=\tilde{G}}\frac{\delta\tilde{G}[\bar{\phi}]}{\delta\bar{\phi}}\equiv\frac{\delta \Gamma[\bar{\phi},\tilde{G}]}{\delta \bar{\phi}},
\eea
where $\frac{\Delta}{\Delta \bar{\phi}}$ refers to total differentiation. This shows that the condition of the physical field configuration being the extremal point of the 1PI effective action is equivalent to the first stationary condition of (\ref{1-stat}) obtained in 2PI formalism, if the propagator is set to the solution of the second stationary condition of (\ref{1-stat}). The other important observation is that the latter equation in turn is found to be equivalent to the Dyson-equation. We have
\bea
\label{1-2PI_prop}
0=\frac{\delta \Gamma_{\textnormal{2PI}}[\bar{\phi},G]}{\delta G(x,y)}=-\frac{i}{2}G^{-1}(x,y)+\frac{i}{2}D^{-1}(x,y)+\frac{\delta \Gamma^{(2)}}{\delta G(x,y)},
\eea
which leads to
\bea
\label{1-propeq}
iG^{-1}(x,y)\equiv\Gamma_2(x,y)=-(\partial^2+m^2+U^{''}(\bar{\phi}))\delta(x-y)+2\frac{\delta \Gamma^{(2)}}{\delta G(x,y)}.
\eea
In $\phi^4$ theory the diagrams of $\Gamma^{(2)}$ are
\newline \vspace{-0.5cm}
\bea
\label{1-2PIdiag}
i\Gamma^{(2)}=\includegraphics[bb=151 481 205 331,scale=0.63]{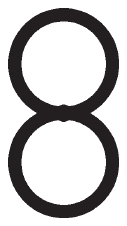}+\includegraphics[bb=120 479 219 331,scale=0.53]{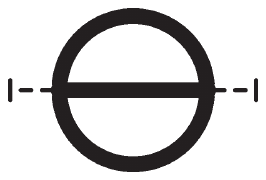}+\includegraphics[bb=138 481 188 331,scale=0.83]{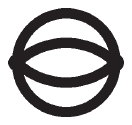}+...
\eea
\vspace{-0.3cm} \newline
from which in the symmetric phase
\bea
\label{1-2PI_Dyson}
\Gamma_2(x,y)=-(\partial^2+m^2)\delta(x-y)-i\Big(\includegraphics[bb=151 491 205 331,scale=0.63]{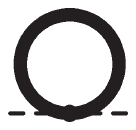}+\includegraphics[bb=125 479 214 331,scale=0.53]{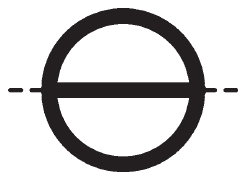}+...\Big).
\eea
(\ref{1-2PI_Dyson}) is reminiscent of (\ref{1-Dyson2_graph}) and (\ref{1-gamma2_diag}). At a first look it seems to be identical with (\ref{1-gamma2_diag}), since a differentiation with respect to $G$ removes a propagator line, therefore what remains is only 1-particle-irreducible, just as it can be seen on (\ref{1-gamma2_diag}) (the factor of $2$ is also correct since two lines can be cut to obtain the very same diagram). However, at a second look we may think to have an apparent contradiction, since in $\Gamma^{(2)}$ we have diagrams which contain no self-energy insertions. This can lead to confusion, since in (\ref{1-gamma2_diag}) every 1PI diagrams are present, but in (\ref{1-2PI_Dyson}) not. The resolution is that we should keep in mind that in the right hand side of (\ref{1-2PI_Dyson}) the diagrams are made of {\it full} propagators (thick lines), and not of perturbative ones, as on the right hand side of (\ref{1-gamma2_diag}). Indeed, we can organize the perturbative series on the right hand side of (\ref{1-gamma2_diag}) in a way that the self energy insertions are resummed and full propagators appear. With this we arrive exactly at (\ref{1-2PI_Dyson}), which is called the {\it self-consistent Dyson-equation}. This equation appears in 2PI formalism in a very natural way. Since we have already argued that the Dyson-equation is identical with the Dyson-Schwinger equation of $\Gamma_2$, we obtained the same relation in three different forms. The difference between them is that while the ordinary Dyson-equation generates the $2$-point function via perturbative propagators and vertices, the self-consistent Dyson equation and the Dyson-Schwinger formalism resum certain diagrams. In Dyson-Schwinger equations diagrams are resummed into proper vertices, while in 2PI formalism (leading to the self-consistent Dyson-equation) perturbative vertices appear and only the self-energy insertions are resummed.

Since $\Gamma^{(2)}$ contains infinitely many diagrams, analogously to the Dyson-Schwinger method, in practice we have to truncate the series. The truncation usually does not respect general identities of the $n$-point functions of the full theory. For instance we derive such an identity for the propagator (later we will mention a similar argument for the $4$-point function as well). Since the $2$-point proper vertex is $\Gamma_2=\frac{\delta^2 \Gamma[{\bar{\phi}]}}{\delta \bar{\phi}\delta \bar{\phi}}$, using (\ref{1-gamma-12PI}) and (\ref{1-propeq}), we have the relation
\bea
\label{1-propagators}
\frac{\Delta^2 \Gamma_{\textnormal{2PI}}\big[\bar{\phi},\tilde{G}[\bar{\phi}]\big]}{\Delta \bar{\phi}\Delta \bar{\phi}}=iD^{-1}+2\frac{\delta \Gamma^{(2)}\big[\bar{\phi},G\big]}{\delta G}\bigg|_{G=\tilde{G}}.
\eea
With the help of (\ref{1-diff}), the left hand side of (\ref{1-propagators}) can be extracted:
\bea
\label{1-propagators2}
\frac{\delta^2 \Gamma_{\textnormal{2PI}}\big[\bar{\phi},\tilde{G}[\bar{\phi}]\big]}{\delta \bar{\phi}\delta\bar{\phi}}+\frac{\delta^2 \Gamma_{\textnormal{2PI}}[\bar{\phi},G]}{\delta \bar{\phi}\delta G}\bigg|_{G=\tilde{G}}\frac{\delta \tilde{G}[\bar{\phi}]}{\delta \bar{\phi}} = iD^{-1}+2\frac{\delta \Gamma^{(2)}[\bar{\phi},G\big]}{\delta G}\bigg|_{G=\tilde{G}}.
\eea
In the symmetric phase the second term of the left hand side of (\ref{1-propagators2}) vanishes, since in this case functions made of an odd number of field derivatives of $\Gamma_{\textnormal{2PI}}$ vanish (this is true in every theory with $Z_2$ symmetry).
It is convenient to define $\Gamma_{\inte}[\bar{\phi},G]=\Gamma_{\textnormal{2PI}}[\bar{\phi},G]-S[\bar{\phi}]-\frac{i}{2}\Tr \log G^{-1}$, since when $\bar{\phi}=0$ (\ref{1-propagators2}) can be written as:
\bea
\label{1-propagators3}
iD^{-1}\big|_{\bar{\phi}=0}+\frac{\delta^2 \Gamma_{\inte}[\bar{\phi},\tilde{G}]}{\delta \bar{\phi}\delta\bar{\phi}}\bigg|_{\bar{\phi}=0}=iD^{-1}\big|_{\bar{\phi}=0}+2\frac{\delta\Gamma_{\inte}[\bar{\phi},\tilde{G}]}{\delta G}\bigg|_{\bar{\phi}=0},
\eea
which means that
\bea
\label{1-propagators4}
\frac{\delta^2 \Gamma_{\inte}[\bar{\phi},\tilde{G}]}{\delta \bar{\phi}\delta\bar{\phi}}\bigg|_{\bar{\phi}=0} = 2\frac{\delta\Gamma_{\inte}[\bar{\phi},\tilde{G}]}{\delta G}\bigg|_{\bar{\phi}=0}
\eea
is an exact relation in the full theory. However, in general, the truncation of $\Gamma^{(2)}$ does not respect it. It is possible then to define two propagators corresponding to the left- and right-hand side of (\ref{1-propagators3}). This plays an important role when one attempts to renormalize a specific 2PI approximation. (The left-hand side of (\ref{1-propagators}) is sometimes referred as the curvature of the effective potential.)

The truncation of $\Gamma^{(2)}$ gives a partial resummation of the perturbative series. To get a taste how a specific resummation works, let us see the most simple example, the Hartree approximation. In this only the first diagram of (\ref{1-2PIdiag}) is kept. This leads to the following:
\bea
\Gamma_2(x,y)=-(\partial^2+m^2)\delta(x-y)-i\includegraphics[bb=151 491 205 331,scale=0.63]{graf37.pdf},
\eea
whose iterative solution is
\bea
\label{1-Sdaisy}
\!\!\!\!\!\Gamma_2(x,y)=-(\partial^2+m^2)\delta(x-y)-i\Big(\includegraphics[bb=156 483 195 341,scale=0.63]{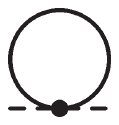}+\includegraphics[bb=156 483 195 341,scale=0.63]{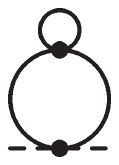}+\includegraphics[bb=156 483 195 341,scale=0.63]{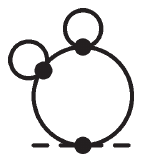}+\includegraphics[bb=156 483 195 341,scale=0.63]{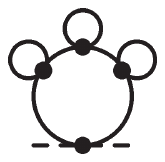}+\includegraphics[bb=156 483 195 341,scale=0.63]{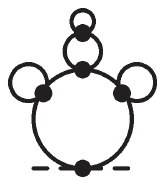}+...\Big).
\eea
(\ref{1-Sdaisy}) is the so-called {\it super-daisy} resummation.

\clearemptydoublepage

\chapter{Effective scalar theories of QCD}

In this thesis we are investigating scalar field theories, particularly those which are effective models of the fundamental theory of the strong interaction, the quantum-chromodynamics (QCD). The demand of building up such effective theories of QCD follows from the fact that it can not be solved at low energy scales due to a feature of the asymptotic freedom. This forces the running coupling of the theory to decrease monotonically with the energy, which is a nice effect at large enough scales, however around and below the scale of QCD ($\Lambda_{QCD}$) is rather inconvenient, since it breaks down perturbation theory. One way out is to give up this very successful method and apply lattice field theory, or we can build up effective models which are sensitive to specific features and based on the global symmetry properties of QCD. The spontaneously broken approximate chiral symmetry of QCD leads us to the idea of the {\it linear sigma models}, which we introduce in details in the following.

\section{The U(N)$\times$U(N) model}

At zero quark masses QCD with $N$ quark flavors has a global $U_L(N)\times U_R(N)$ {\it chiral symmetry}, where $L$ and $R$ refers to the left and right projections of the Dirac quark field, respectively \cite{Peskin-book}. This means that a $q=q_R+q_L$ quark field can be transformed without introducing changes into the Lagrangian as:
\bea
q = q_R + q_L \longrightarrow \exp(i\theta_R^a T^a) q_R + \exp(i\theta_L^a T^a) q_L,
\eea
where $R$ and $L$ refer to the left and right projections of the Dirac quark field, $T^a$ are the group generators and $\theta_R$ and $\theta_L$ are transformation parameters. The basic properties of the $U(N)$ group can be found in the Appendix. Alternatively we can write:
\bea 
q \longrightarrow \exp\Big(i(\theta_V^a+\theta_A^a\gamma_5)T^a\Big)q,
\eea
where $\theta_{V,A}:=(\theta_R\pm \theta_L)/2$ are the parameters of the vector and axialvector transformations. $\gamma_5$ is the product of the Dirac matrices: $\gamma_5=i\gamma_0\gamma_1\gamma_2\gamma_3$. This means that this global symmetry can be realized also in terms of vector and axialvector symmetry groups: $U_V(N)\times U_A(N)$, which can be decomposed into $SU_V(N)\times SU_A(N)\times U_V(1)\times U_A(1)$. This is convenient, since the explicit, spontaneous and anomalous breaking of the symmetry is expressed through the vector and axialvector spectra. For instance, one can argue that due to the axial vector anomaly, in the quantum version of the theory the symmetry is reduced to $SU_V(N)\times SU_A(N)\times U_V(1)$ \cite{lenaghan00}. Furthermore, nonzero, but equal quark masses reduce the symmetry further to $SU_V(N)\times U_V(1)$, and with unequal masses only a $U_V(1)$ symmetry remains. This refers to an arbitrary phase shift in the definition of the fields and is responsible for the baryon number conservation.

In the $U(N)\times U(N)$ linear sigma model this symmetry is realized in terms of a new dynamical variable, an $N\times N$ matrix field $M$. A chiral transformation is defined as
\bea
M \longrightarrow R M L^\dagger,
\eea
where $R=\exp(i\theta_R^aT^a)$ and $L=\exp(i\theta_L^aT^a)$. For infinitesimal transformations we have
\bea
\delta M=i(\theta^a_R T^a M-\theta_L^aMT^a)\equiv i(\theta_A^a[T^a,M]_++\theta_V^a[T^a,M]_-),
\eea
where $[\cdot,\cdot]_{\pm}$ refers to anticommutation and commutation, respectively. We have also introduced the vector and axial-vector transformation parameters as $\theta_{V,A}=(\theta_R\pm\theta_L)/2$. From this, the pure vector and axial-vector transformations read as
\bea
M \longrightarrow VMV^\dagger, \qquad M \longrightarrow A^\dagger MA^\dagger,
\eea
where $V=\exp(i\theta_V^aT^a)$ and $A=\exp(i\theta_A^aT^a)$. The $U(N)\times U(N)$ linear sigma model is a chirally symmetric field theory of the $M$ matrix. One can check that the most general renormalizable Lagrangian which preserves chiral symmetry can be written as
\bea
\label{2-Lagr}
{\cal L}=\Tr (\partial_{\mu}M^\dagger \partial^{\mu}M-m^2M^\dagger M)-g_1\Big(\Tr(M^\dagger M)\Big)^2-g_2\Tr\Big((M^\dagger M)^2\Big).
\eea
Since under pure vector and axial-vector transformations the determinant of $M$ transforms as
\bea
\det(M)\longrightarrow \det(VMV^\dagger)&=&\det(V^\dagger V M)=\det(M), \\
\det(M)\longrightarrow \det(A^\dagger M A^\dagger)&=&\big(\det(A^\dagger)\big)^2\det(M)=\exp(2i\theta_A^a\Tr T^a)\det(M), \nonumber
\eea
the scalar combination $\det(M)+\det(M^\dagger)$ changes only with respect to the $U_A(1)$ part of the transformation. This property is appropriate for describing the $U_A(1)$ anomaly of QCD by introducing the following piece into the Lagrangian:
\bea
{\cal L}_{U_A(1)}=c(\det M+\det M^\dagger).
\eea
We note that this term gives birth to a non-renormalizable coupling for $N>4$. 
However, since above three quark flavors, the chiral symmetry of QCD is broken explicitly such that we do not expect that an effective theory based on this symmetry can be applicable at all.

The simplest explicit symmetry breaking term is written as
\bea
{\cal L}_h=\Tr\big(H(M+M^\dagger)\big),
\eea
which is built in a way to have the same transformation properties for the chiral transformation as the mass term of QCD.

Introducing the scalar ($s$) and pseudoscalar ($\pi$) fields as
\bea
M=T^a(s^a+i\pi^a),
\eea
various pieces of the Lagrangian (\ref{2-Lagr}) can be written as follows \cite{lenaghan00}.
\bea
\label{2-L_ch}
{\cal L}&=&\frac12\big(\partial_{\mu}s^a\partial^{\mu}s^a+\partial_{\mu}\pi^a\partial^{\mu}\pi^a-m^2(s^as^a+\pi^a\pi^a)\big) \nonumber\\
&-&\frac13 F_{abcd}s^as^bs^cs^d-\frac13 F_{abcd}\pi^a\pi^b\pi^c\pi^d-2H_{ab,cd}s^as^b\pi^c\pi^d,
\eea
where
\begin{subequations}
\bea
\!\!\!\!\!\!\!\!\!\!\!\!\!\!\!\!\!F_{abcd}&=&\frac{g_1}{4}(\delta_{ab}\delta_{cd}+\delta_{ac}\delta_{bd}+\delta_{ad}\delta_{bc})+\frac{g_2}{8}(d_{abm}d_{mcd}+d_{adm}d_{mbc}+d_{acm}d_{mbd}), \\
\!\!\!\!\!\!\!\!\!\!\!\!\!\!\!\!\!H_{ab,cd}&=&\frac{g_1}{4}\delta_{ab}\delta_{cd}+\frac{g_2}{8}(d_{abm}d_{mcd}+f_{acm}f_{mbd}+f_{bcm}f_{mad}).
\eea
\end{subequations}
Alternatively, we can write
\bea
\label{2-L_ch2}
{\cal L}&=&\frac12\Big(\partial_{\mu}s^a\partial^{\mu}s^a+\partial_{\mu}\pi^a\partial^{\mu}\pi^a-m^2(s^as^a+\pi^a\pi^a)\Big)\nonumber\\
&-&\frac{g_1}{4}\Big(s^as^a+\pi^a\pi^a\Big)^2-\frac{g_2}{2}U^aU^a,
\eea
where
\bea
U^a=\frac12 d_{abc} (s^bs^c+\pi^b\pi^c)-f_{abc}s^b\pi^c.
\eea
The $d_{abc}$ totally symmetric and the $f_{abc}$ totally antisymmetric tensors are defined in the Appendix. Furthermore, the term describing the anomaly is:
\bea
{\cal L}_{U_A(1)}=
\begin{cases}
G^2_{ab}(s^as^b-\pi^a\pi^b), \qquad \qquad \qquad \qquad \qquad \qquad N=2 \\
G^3_{abc}(s^as^bs^c-3\pi^a\pi^bs^c), \qquad \qquad \qquad \qquad \qquad \!\! N=3 \\
\frac13G^4_{abcd}(6s^as^b\pi^c\pi^d-s^as^bs^cs^d-\pi^a\pi^b\pi^c\pi^d),\qquad \! N=4 \\
\end{cases}
\eea
where
\begin{subequations}
\bea
\!\!\!\!\!G^2_{ab}&=&\frac{c}{2}(\delta_{a0}\delta_{b0}-\delta_{a1}\delta_{b1}-\delta_{a2}\delta_{b2}-\delta_{a3}\delta_{b3}), \\
\!\!\!\!\!G^3_{abc}&=&\frac{c}{6}\Big(d_{abc}-\frac32(\delta_{a0}d_{0bc}+\delta_{b0}\delta_{a0c}+\delta_{c0}d_{ab0})+\frac92 d_{000}\delta_{a0}\delta_{b0}\delta_{c0}\Big),\\
\!\!\!\!\!G^4_{abcd}&=&-\frac{c}{16}\Big[\delta_{ab}\delta_{cd}+\delta_{ad}\delta_{bc}+\delta_{ac}\delta_{bc}-d_{abm}d_{mcd}-d_{adm}d_{adm}d_{mbc}-d_{acm}d_{mbc}\nonumber\\
&+&16\delta_{a0}\delta_{b0}\delta_{c0}\delta_{d0}-4(\delta_{a0}\delta_{b0}\delta_{cd}+\delta_{a0}\delta_{c0}\delta_{bd}+\delta_{a0}\delta_{d0}\delta_{bc}+\delta_{b0}\delta_{c0}\delta_{ad}\nonumber\\
&+&\delta_{b0}\delta_{d0}\delta_{ac}+\delta_{d0}\delta_{c0}\delta_{ab})+\sqrt{8}(\delta_{a0}d_{bcd}+\delta_{b0}d_{cda}+\delta_{c0}d_{dab}+\delta_{d0}d_{abc})\Big].
\eea
\end{subequations}
As it was already mentioned, the term containing the determinant only makes sense for $N=2,3,4$ from the renormalizability point of view. Finally, with the parametrization $H=h_a T^a$, we have
\bea
{\cal L}_h=h_as^a.
\eea
Although only low $N$ values are physically relevant, we can try to attack the theory with the large-$N$ technique and obtain quantitative results with continuing the results to these rather low values. In the end of this chapter, the reader can find a review of the results obtained in the past using large-$N$ technique.

\section{The O(N) model}

Let us consider the $U(N)\times U(N)$ model with $N=2$ and introduce the following multiplets:
\bea
\Phi=(s_0,\pi_1,\pi_2,\pi_3)\equiv (s,\vec{\pi}), \qquad \chi=(\pi_0,s_1,s_2,s_3)\equiv (\eta',\vec{a}_0).
\eea 
With these notations, after some calculations, (\ref{2-L_ch}) extended with an explicit symmetry breaking term can be written as
\bea
\label{2-SU2}
L&=&\frac12(\partial_{\mu}\Phi^T\partial^{\mu}\Phi-m_{\Phi}^2\Phi^T\Phi)-\frac{\lambda}{4!}(\Phi^T\Phi)^2\nonumber\\
&+&\frac12(\partial_{\mu}\chi^T\partial^{\mu}\chi-m_{\chi}^2\chi^T\chi)-\frac{\lambda}{4!}(\chi^T\chi)^2+L_I(\Phi,\chi)+h_as^a,
\eea
where $L_I(\Phi,\chi)$ describes the interaction between the multiplets and
\bea
m_{\Phi}^2=m^2-c, \qquad m_{\chi}^2=m^2+c, \qquad \lambda=6g_1 + 3g_2.
\eea
Since in the $N=2$ case the strange quark mass is formally infinite, for consistency reasons $\eta'$ and $a_0$ must also be treated as infinitely heavy particles, since for example from experiments we know that $\eta'$ is heavier than the strange kaon: $m_K<m_{\eta'}$. We note that it is the axial anomaly which pushes up the mass of the $\chi$ multiplet, therefore for consistency reasons we should treat the combination $m^2+c$ as infinity, while $m^2-c$ remains finite. This means that $\chi$ should be omitted from (\ref{2-SU2}) and the $SU(2)\times SU(2)$ linear sigma matrix model transfigures into the $O(4)$ symmetric vector model: ($h_0\equiv h$)
\bea
\label{2-O4}
L&=&\frac12(\partial_{\mu}\Phi^T\partial^{\mu}\Phi-m_{\Phi}^2\Phi^T\Phi)-\frac{\lambda}{24}(\Phi^T\Phi)^2+hs.
\eea
The fields of $\Phi$ are usually called as sigma ($s$) and pion ($\pi$).
The $O(N)$ model is the generalization of (\ref{2-O4}), where instead of $3$ pions, we have a number of $(N-1)$. This extension is motivated by the need of the application of the non-perturbative large-$N$ expansion. In this case we factor out an $N$ from the coupling: ($m^2_{\Phi}\equiv m^2$) 
\bea
\label{2-ON}
L&=&\frac12(\partial_{\mu}\Phi^T\partial^{\mu}\Phi-m^2\Phi^T\Phi)-\frac{\lambda}{24N}(\Phi^T\Phi)^2+hs,
\eea
since it can be argued that the coupling must scale with $1/N$ in order to have the free energy as an extensive quantity. In the large-$N$ expansion we formally treat $N\gg 1$ and use $1/N$ as an expansion parameter. Surprisingly, the large-$N$ results evaluated at $N=4$ provide phenomenologically quite interesting informations even at leading order.

\section{The large-N expansion}
Large-$N$ expansion is a rather old classical non-perturbative tool of quantum
field theory \cite{dolan74,schnitzer74,coleman74}, in which instead of
building up perturbative series in powers of the coupling, we treat $1/N$ as the small parameter. This leads to a certain partial resummation to all orders in the ordinary Feynman diagrams.

Leading order solution of the large-$N$ expansion of the $O(N)$ model
has been applied to interesting problems of finite temperature phenomena,
in particular the study of the restoration of the $SU(2)\times SU(2)
\sim O(4)$ chiral symmetry of QCD \cite{meyer-ortmanns93,bocskarev96,patkos02}.
One of its attractive points is that it preserves Goldstone's theorem at every
order of the expansion, a feature not shared by all resummations of
the original perturbative series. Being a resummation to all orders,
it has some features which are absent at any given order of the
perturbation theory, but which are believed to be true for the exact
solution.  On the other hand, related to the now well established triviality
of scalar theories (see {\it e.g.} \cite{kuti} for a review), its spectra
shows the presence of a tachyonic pole (Landau ghost) in the renormalized
propagators.

In early publications \cite{schnitzer74,coleman74,root74} the
appearance of tachyons was considered as an inconsistency of the
large-$N$ approximation.  In next-to-leading order (NLO)
investigations of the expansion, started already in 1974
\cite{root74}, the tachyonic problem seemed to be aggravated, because
the renormalized effective potential appeared to be complex for all
values of the field. This led to the claim that the large-$N$
expansion breaks down, when one is going beyond leading-order.
However, extensive studies of the $O(N)$ symmetric model
\cite{abott76,moshe83} established the consistency of the $1/N$
expansion for the effective potential and revealed its rich phase
structure.  Considered in a restricted sense, as a renormalized
effective theory, the large-$N$ expansion turned out to be a valuable
nonperturbative tool, when applied to phenomena dominated by scales
much lower than the cut-off.  A strict cut-off version of the model
was considered in \cite{schnitzer95} showing that with restrictions on
the value of the background field the model has a phase with
spontaneously broken symmetry, free of tachyons. A different solution
to the tachyonic problem, called the tachyonic regularization, was
proposed in \cite{binoth98,binoth99a}, where the tachyonic pole is
minimally subtracted.

Next-to-leading order results were also presented for the pressure
of the relativistic $O(N)$ model and applied to the physical
($N=4$) sigma-pion gas \cite{andersen04,andersen08}. Some time ago the the method was initiated in the same model for the description of the chiral symmetry restoration \cite{patkos02,jakovac04}.
Further applications in modeling chiral symmetry breaking can be found even in technicolor theories \cite{appelquist95,kikukawa08}. Applications to non-equilibrium phenomena, e.g. to the quantum dynamics of disoriented chiral condensates \cite{blaizot92,rajagopal93,bjorken97} also benefit from extending the
number of flavors. Most of the studies were realized in a reformulation of the model in which the quartic self-coupling of the $N$-component scalar field is replaced by an auxiliary field mediated interaction.

An analogous development for the three-flavor linear sigma model
model is hindered, because even the leading order solution of  
the large-$N$ limit of the $U(N)\times U(N)$ symmetric model is
unknown. To the author's best knowledge, no published attempts exist, which would go beyond the partial large-$N$ treatments of the $O(2N^2)$-symmetric part of the symmetry group (i.e. corresponding to the term proportional to $s^as^a+\pi^a\pi^a$). The results obtained with such an approach are questionable in light of the rather different finite temperature renormalization group behavior of the $O(2N^2)$ and $U(N)\times U(N)$ symmetric models for $N\geq 3$ \cite{pisarski84,paterson81}. In case of $O(2N^2)$, one expects continuous (second order) transition, while in the $U(N)\times U(N)$ symmetric model it is argued to be first order for $N\geq 3$ \cite{pisarski84}.

\chapter{Renormalization of the O(N) model in the large-N approximation}

In this chapter we investigate renormalizability of the $O(N)$ model in the large-$N$ limit using 2PI formalism. Since most of the studies of the $O(N)$ model described in Chapter 2 were realized in a reformulation of the model in which the quartic self-coupling of the $N$-component scalar field is replaced by an auxiliary field mediated interaction, we also start our investigation this way. 

We should keep in mind, that the renormalization of the $O(N)$ model was performed at NLO in \cite{root74}. However, while pointing out all the divergences, this analysis did not attempt the explicit calculation of the counterterms, which were determined only at LO. Analyses of the renormalization of the self-coupling of the model showed that at NLO the $\beta$-function is corrected by terms of order $1/N$ \cite{smekal94}. The success of the renormalization program in 2PI formalism \cite{hees02,blaizot04,berges05} revived also the study of the renormalizability of the NLO corrections of the large-$N$ expansion \cite{cooper05,cooper04}. Despite all these efforts, the detailed and explicit knowledge of the counterterms was missing in the literature. Another, perhaps more serious motivation for the study is that recent publications even raised doubts on the renormalizability of the NLO approximation of the model formulated with an auxiliary (composite) field for arbitrary values of the field expectation values \cite{andersen04,andersen08}.

In this chapter we provide an explicit construction of the renormalized effective action at NLO in the auxiliary field formulation of the $O(N)$ model. The leading order part of the effective action is proportional to the number of degrees of freedom $N$, therefore our goal is to construct and renormalize it with ${\cal O}(N^0)$ accuracy. This goal will be achieved by separating divergences in the derivatives of the effective action and then integrating them to obtain
the counterterms of the action and eventually get a finite effective action. The identification of the divergences will take place by expanding the propagators in inverse powers of $N$ and omitting all terms which contribute beyond NLO. We note, that this step ``kills'' the self-consistent nature of the 2PI formalism, most of the propagators equal their tree level value at
leading order, where by tree level we mean the expression which can be derived from the classical action {\it including} the auxiliary field. The construction of the counterterm functional will respect Goldstone's theorem, obviously obeyed in the NLO $1/N$-expansion of the bare theory. 

The fact that the 2-point functions will be determined by following the $1/N$
hierarchy and not self-consistently has the definite backward effect
of implying secular behavior in time dependent applications
\cite{mihaila01,yaffe00}. It is however essential for ensuring
Goldstone's theorem, since an ${\cal O}(1/N)$ expansion on the level
of the 2PI generating functional was shown not to cure the violation
of Goldstone's theorem by the self-consistent propagator
\cite{cooper05}.  Proposition for the preservation of Goldstone's
theorem for the self-consistent (2PI) propagators exists at present
only at LO (Hartree-level) \cite{ivanov05} and the proposed solution 
follows efforts initiated in the framework of non-relativistic
many-body theory \cite{hohenberg65}.

By explicit construction of the counterterm functional in the auxiliary
field formulation we demonstrate that the model is renormalizable at
NLO in the large-$N$ expansion for {\it arbitrary} vacuum expectation values of the fields. Substitution of the solution of the auxiliary field and the related propagators at the level of the ${\cal O}(N^0)$ accurate functional leads to the recovery of ${\cal O}(N)$ and ${\cal O}(N^0)$ terms of the NLO 2PI effective action of the $O(N)$ model in its original formulation \cite{dominici93}, this time completed with all appropriate counterterms of the same accuracy. Alternatively, 
elimination of the NLO pion and LO sigma propagators produces the 
renormalized effective potential for the sigma and the auxiliary 
fields \cite{andersen04}, again with correct counterterms.

We remark that in the ``complete'' 2PI-1/N approximation, where one is looking for a self-consistent solution of the stationary conditions of the 2PI effective action with respect to the propagators (i.e. after building up the 2PI effective action, no further expansion takes place in $1/N$) represents a different resummation strategy. For the sake of completeness this will also be presented in the last section of this chapter.

\section{Auxiliary field formulation}

At the level of the generating functional $Z[J]$ (\ref{1-Z_func}), we introduce 
the auxiliary field $\alpha$ into the Lagrangian (\ref{2-ON}) of
the $O(N)$ model through the functional Hubbard-Stratonovich
transformation \cite{HST}:
\bea
&&\!\!\!\!\!\!\!\!\!\!\!\!\!\!\!\!\!\!\!\!\!\!\!\!\!\!\!\!\! \int{\cal D}\alpha \exp\bigg\{i\int d^4 x\bigg[
-\frac{1}{2}\alpha^2(x)+\frac{i}{2}\sqrt{\frac{\lambda}{3N}}
\alpha(x)\Phi^2(x)\bigg]\bigg\}\nonumber\\
&&\qquad \qquad \qquad \qquad\propto
\exp\left\{i\int d^4 x\left[
-\frac{\lambda}{24 N}\big(\Phi^2(x)\big)^2
\right]
\right\},
\eea
where $\Phi^2=s^2+\pi_n^2.$
Then, the extended action without the explicit symmetry breaking term and shifted in $s$ by the background $\sqrt{N}v$ reads:
\bea
S[s,\pi_n,\alpha;v]&=&\int
d^4x\bigg[\frac{1}{2}\big(\partial_\mu s(x)\big)^2
+\frac{1}{2}\big(\partial_\mu\pi_n(x)\big)^2-\frac{1}{2}\alpha^2(x)
\nonumber\\
&-&\frac{m^2}{2}\Big(s^2(x)+\pi^2_n(x)
+2\sqrt{N}v s(x)+N v^2\Big)
\nonumber\\
&+&\frac{i}{2}\sqrt{\frac{\lambda}{3N}}\alpha(x)\Big(s^2(x)+\pi_n^2(x)+2\sqrt{N}v s(x)+N v^2\Big)\bigg].
\label{4-Eq:S_aux}
\eea

\subsection{Approximate 2PI effective potential}

\begin{figure}
\includegraphics[bb=30 475 224 551,scale=0.85]{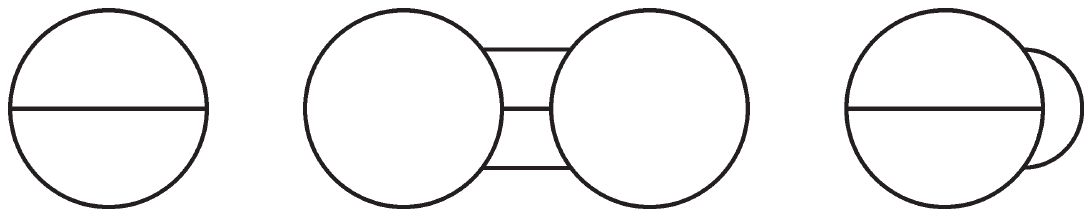}
\caption{The first few 2PI diagrams contributing to $V_{2PI}$. The lines can represent various propagators allowed by the vertices. Only the first type (setting sun) can contribute at NLO of the large-$N$ expansion.}
\label{fig3_0}
\end{figure}

From the modified action (\ref{4-Eq:S_aux}), we construct an approximate 2PI effective action. Since the $4$-point couplings were changed to $3$-point ones, the most simple truncation is to include only the 2-loop setting sun diagrams (check Fig. \ref{fig3_0}) containing 3 propagators. As we have multicomponent fields and propagators, we have three of this type of diagram (see (\ref{4-Phi-funct}) below). Because of translational invariance, the mean fields are spacetime independent and in this case we can always factor out a four-volume and work with the effective potential instead of the effective action: 
\bea
V_{\textnormal 2PI}=-\frac{1}{\int d^4x}\Gamma_{\textnormal 2PI}.
\eea
Allowing the auxiliary field to have a saddle point value $<\!\alpha\!>\equiv\sqrt{\frac{3N}{\lambda}}\hat\alpha$, using (\ref{1-2PI_final}) Fourier representation of the propagators we have (the $\pi$ and the shifted $s$ fields have no expectation values):
\bea
V_{2PI}[\hat\alpha,v,G_\pi,{\cal G}]&=&\frac{1}{2}\left
(m^2-i\hat\alpha\right)N v^2+\frac{3N}{2\lambda}\hat\alpha^2
-\frac{i}{2}\int_k\bigg[(N-1) \big(\ln G^{-1}_\pi(k) \nonumber\\
&+& D^{-1}_\pi(k)G_\pi(k)\big)
+\textrm{Tr}\ln {\cal G}^{-1}(k)
+\textrm{Tr}\left( {\cal D}^{-1}(k) {\cal G}(k)\right)
\bigg]
\nonumber\\
&+&i\frac{\lambda}{12N}\int_k\int_p\bigg[G_{\alpha\alpha}(k)
\Big((N-1)G_\pi(p)G_\pi(p+k)\nonumber\\
&+&G_{ss}(p)G_{ss}(p+k)\Big)
+2G_{\alpha s}(p)G_{ss}(k)G_{s\alpha}(p+k)\bigg]
\nonumber\\
&+&\Delta V[\hat\alpha,v,G_\pi,{\cal G}].
\label{4-Phi-funct}
\eea
Note that the momentum integrals are defined as $\int_k=\int \frac{d^4k}{(2\pi)^4}$. In (\ref{4-Phi-funct}) $m^2$ and $\lambda$ are the renormalized couplings, and
$\Delta V[\hat\alpha,v,G_\pi,{\cal G}]$ is the yet undetermined
counter-term functional.  ${\cal G}$ and ${\cal D}$ are the variational and tree-level symmetric $2\times 2$ propagator matrices with components
$G_{ss},G_{s\alpha},$ $G_{\alpha\alpha}$ and
$D_{ss},D_{s\alpha},$ $D_{\alpha\alpha},$ respectively.
Their inverse matrices are denoted by ${\cal G}^{-1}$ and 
${\cal D}^{-1}$. The elements of the inverse of the tree-level propagators appearing here are the following:
\bea
\label{Eq:tree_prop}
i(D^{-1})_{\alpha\alpha}(k)=-1&\!\!\!\!\!\!\!,\!\!\!& \qquad i(D^{-1})_{\alpha s}(k)=i\sqrt{\frac{\lambda}{3}}v, \nonumber\\
i(D^{-1})_{ss}(k)&\!\!\!=\!\!\!&i D^{-1}_\pi(k)=k^2-M^2,
\eea
where we introduced the shorthand notation $M^2=m^2-i\hat\alpha$. 

It is important to mention that the 2PI approximation defined by
(\ref{4-Phi-funct}) is identical to a specific truncation of the
tower of the Dyson-Schwinger equations. We introduce the so-called
Bare Vertex Approximation (BVA) as we close the Dyson-Schwinger hierarchy
at the level of the $3$-point function. In other words we reduce the
tower to two class of equations (i.e. equations of the $1$- and $2$-point
functions). The exact $3$-point vertices are denoted by $\Gamma_{A B C}(x,y,z)=\delta^3\Gamma/\delta\phi_A(x)\delta\phi_B(y)\delta\phi_C(z)$ (here $\phi_i$ can refer to any of the fields $\pi,s,\alpha$) are approximated with their classical value:
\bea
\Gamma_{ss\alpha}(x,y,z)&\!\!\!\approx\!\!\!&\frac{\delta^3 S}{\delta s(x)\delta s(y)\alpha(z)}=i\sqrt{\frac{\lambda}{3N}}
\delta(x-y)\delta(x-z), \nonumber\\
\Gamma_{\pi_n\pi_m\alpha}(x,y,z)&\!\!\!\approx\!\!\!&\frac{\delta^3 S}{\delta \pi(x)\delta \pi(y)\alpha(z)}=\delta_{n m}
\Gamma_{ss\alpha}(x,y,z).
\label{Eq:BVAvertices}
\eea
The other $3$-point vertices and all the higher ones are zero at
classical level. Without going into details, we just state that when
we plug these expressions into the Dyson-Schwinger equations (actually
they appear only in the equation of the propagator, check (\ref{1-genDS})
for obtaining the derivatives of the effective action),
we obtain the very same equations for the propagators
and for the $1$-point functions, as the ones which come from (\ref{4-Phi-funct})
after differentiating with respect to the propagators or the fields.
This means that in this certain 2PI approximation it is possible to find the
corresponding truncation of the tower of Dyson-Schwinger equations (or
vice versa). Then, if we renormalized the 2PI equations, we would
immediately prove the renormalizability of the DS equations in the
given truncation.

We can easily figure out, that beyond the setting sun, no 2PI
diagrams are present which would contribute to the NLO effective action (see Fig. \ref{fig3_0}). In agreement, it reproduces the NLO 2PI effective action presented in (44) and (55) of \cite{aarts02}. Having in mind the ${\cal O}(N^0)$ (NLO) determination of the effective action we realize that only the setting sun involving pion propagators should be kept in (\ref{4-Phi-funct}) at NLO. Therefore we truncate further  the functional in (\ref{4-Phi-funct}) and will work with the approximate effective potential
\bea
V_{2PI}[\hat\alpha,v,G_\pi,{\cal G}]&\!\!\!=\!\!\!&\frac{1}{2}\left
(m^2-i\hat\alpha\right)N v^2+\frac{3N}{2\lambda}\hat\alpha^2
-\frac{i}{2}\int_k\bigg[(N-1) \big(\ln G^{-1}_\pi(k) \nonumber\\
&\!\!\!+\!\!\!& D^{-1}_\pi(k)G_\pi(k)\big)
+\textrm{Tr}\ln {\cal G}^{-1}(k)
+\textrm{Tr}\left( {\cal D}^{-1}(k) {\cal G}(k)\right)
\bigg]
\nonumber\\
&\!\!\!+\!\!\!&i\frac{\lambda}{12}\int_k\int_pG_{\alpha\alpha}(k)
G_\pi(p)G_\pi(p+k)+\Delta V[\hat\alpha,v,G_\pi,{\cal G}].
\label{4-Phi-funct2}
\eea
This truncated form of the functional influences the propagator
equations of the $\alpha - s$ sector (at LO) and of the pion
fields (at NLO). We emphasize, that even if we preserved the
omitted terms, the ${\cal O}(1/N)$ solution of that approximation would
not correspond to the complete NLO order solution in the 
$\alpha-s$ sector, since there are more 2PI diagrams proportional to $1/N$.

As already announced, renormalizability of this approximation will be
investigated at the level of the derivatives of (\ref{4-Phi-funct2}),
then we attempt the construction of an appropriate counterterm functional
$\Delta V[\hat\alpha,v,G_\pi,{\cal G}]$. This counterterm functional 
allows for a uniform treatment of the counterterms and also makes possible 
to keep track of the effect a counterterm determined from one equation 
has in the renormalization of another equation. Some new insights will be
offered when the results are compared to other approaches, where the
propagators are not considered as variational variables
\cite{andersen04,andersen08}, or when the quantities related to the
auxiliary field are eliminated \cite{dominici93}. We shall demonstrate
the validity of Goldstone's theorem also for the renormalized NLO pion
propagators. Before we start to determine the counterterm functional,
we have to fix the renormalization scheme we are using.

\subsection{Renormalization scheme}
Following the spirit of Ref.~\cite{patkos08} we will expand the
propagators around appropriately chosen infrared safe auxiliary
propagators (which has nothing to do with the auxiliary field!)
 and subtract the divergent pieces defined through these
quantities (it reminds us to the minimal subtraction scheme).
These subtractions define our renormalization scheme.

An important fact for finding appropriate auxiliary propagators is that
we see from (\ref{4-Phi-funct}) that only the $\alpha-\alpha$ component
of the ${\cal G}$ propagator receives a correction at leading order:
\bea
\label{Eq:bub}
i(G^{-1})_{\alpha \alpha}(k)=-1-i\frac{\lambda}{6}\int_k D_{\pi}(k)D_{\pi}(k+p)\Big|_F\equiv -1+\frac{\lambda}{6}I^F_{\pi}.
\eea
Note that we took the finite part of the $I_{\pi}$ integral immediately, since an appropriate counterterm of $\Delta V$ should kill the divergent piece. We will come to its construction later. The corresponding inverse is therefore (the $0$ index corresponds to the leading order expression):
\bea
\label{Eq:G_LO_matrix}
G^{(0)}_{ss}(k)=\left(1-\frac{\lambda}{6} I_\pi^F(k)\right) 
\tilde G(k)&\!\!\!\!\!,\!\!\!&\qquad
G^{(0)}_{\alpha\alpha}(k)=-i D_\pi^{-1}(k)\tilde G(k),\nonumber\\
G^{(0)}_{\alpha s}(k)&\!\!\!=\!\!\!&i \sqrt{\frac{\lambda}{3}}v\tilde G(k),
\label{Eq:D_matrix}
\eea
where 
\bea
i\tilde G^{-1}(k)=(k^2-M^2)\left(1-\frac{\lambda}{6}
I_\pi^F(k)\right) -\frac{\lambda}{3} v^2.
\label{Eq:Gt}
\eea
It will turn out that at NLO the asymptotics is determined basically by
integrals of the LO propagator $D_\pi(p)$ and of the propagator
$\tilde G(p)$ given in (\ref{Eq:Gt}) which incorporates the effect of
the resummation of the chain of pion bubbles (Hint: expand the inverse of
(\ref{Eq:Gt}) in terms of $\lambda$!), we need two auxiliary propagators.
The first one is
\be
G_0(p)=\frac{i}{p^2-M_0^2},
\ee
where $M_0$ is an arbitrary mass scale. With this propagator we 
define the quadratically divergent integral
\be
T_d^{(2)}=\int_p G_0(p),
\label{Eq:Td2}
\ee
and the following one-loop bubble integral
\be
I_0(p)=-i \int_k G_0(k) G_0(k+p)=T_d^{(0)}+I_0^F(p),
\ee
where the logarithmically divergent part of the integral above is defined as
\be
T_d^{(0)}=-i \int_k G_0^2(k),
\label{Eq:Td0}
\ee
and $I_0^F$ is the finite part of the bubble integral.
The explicit cutoff dependence of the introduced divergent quantities are:
\begin{subequations}
\label{Eq:Tds}
\bea
T_d^{(0)}&=&\frac{1}{16\pi^2}\left[
\ln\left(\frac{\Lambda^2}{M_0^2}+1\right)
-\frac{\Lambda^2}{\Lambda^2+M_0^2}
\right],\\
T_d^{(2)}&=&\frac{1}{16\pi^2}\left[
\Lambda^2+M_0^2\ln\frac{M_0^2}{\Lambda^2+M_0^2}
\right],
\eea
\end{subequations}
which were calculated in Euclidean space with $p_0\to i p_E^0$ and a 4d cutoff $\Lambda$. 
The finite part of the bubble integral behaves asymptotically as
$I_0^F(p)\sim\ln\frac{p^2}{M_0^2}-2-i\pi+{\cal O}\left(p^{-2}\ln p\right)$,
and defines together with $G_0(p)$ the second
auxiliary propagator:
\be
G_a(p)=\frac{i}{(p^2-M_0^2)(1-\lambda I_0^F(p)/6)}.
\ee
Integrals involving combinations of $G_0(p),$ $G_a(p)$ and
$I_0^F(p)$  will be fully included in the counterterms. 
\subsection{Leading order (LO) construction of the counterterms}
We start our program with a short description of the leading order
(pieces proportional to $N$) analysis. Since the pion propagator equals
its tree level value at LO, first we will discuss the saddle
point equation and obtain the corresponding LO counterterm $\Delta V^{\alpha,N}$.
Then we turn to the equation of state and the propagator equations in the
$\alpha-s$ sector. At this level of the approximation, corresponding to this
sector only one piece of $\Delta V$ will be necessary, which is denoted by $\Delta V^{\alpha\alpha}$ and is proportional to the $\alpha-\alpha$ propagator.
\newline \newline
{\bf{Saddle point equation (SPE)}} \newline \newline
Taking the derivative of the functional in (\ref{4-Phi-funct2}) 
with respect to $\hat\alpha$ we arrive at the expression:
\be
\frac{\delta V_{2PI}}{\delta\hat\alpha}[\hat\alpha,v,G_\pi,{\cal G}]=
\frac{3N}{\lambda}\hat\alpha-i\frac{N}{2}
\left(v^2+\int_k D_\pi(k)\right)+\frac{\delta \Delta V^{\alpha,N}}{\delta \hat{\alpha}},
\label{Eq:LO-SPE}
\ee
where we replaced $G_\pi,$ originally appearing in the integral above
by $D_\pi$ introduced in (\ref{Eq:tree_prop}). The last term is the
contribution of the counterterm functional which has to be
constructed to ensure the finiteness of this equation. As announced, the structure of the tadpole integral is given in the third section of this chapter.
Since we have:
\bea
\int_k D_\pi(k)&\!\!\!=\!\!\!&
T_d^{(2)}+\left(M^2-M_0^2\right)T_d^{(0)}+T_\pi^F,\nonumber\\
T_\pi^F&\!\!\!=\!\!\!&\frac{1}{16\pi^2}\left(
M^2\ln\frac{M^2}{M_0^2}-M^2+M_0^2\right),
\label{Eq:tadpole}
\eea
we can see that the finiteness of (\ref{Eq:LO-SPE}) ($T_{\pi}^F$ is the
finite part of the pion tadpole integral) for {\it any} value of
$\hat\alpha$ is ensured by the following ${\cal O}(N)$ piece of the
counterterm functional:
\be
\Delta V^{\alpha,N}=i\hat\alpha\frac{N}{2}\left[T_d^{(2)}+\left(m^2-
M_0^2-i\frac{1}{2}\hat\alpha\right)T_d^{(0)}\right].
\label{Eq:ct_aN}
\ee
Here, $T_d^{(2)}$ and $T_d^{(0)}$ are the quadratic and logarithmic
divergences of the pion tadpole given by the propagator $G_0(p)$,
see (\ref{Eq:Td2}) and (\ref{Eq:Td0}). Then, from (\ref{Eq:LO-SPE})
we obtain the finite saddle point equation:
\be
\frac{3N}{\lambda}\hat\alpha-i\frac{N}{2}\left(v^2+T_{\pi}^F\right)=0.
\ee
\newline \newline \newline
{\bf{Equation of state and Goldstone's theorem}} \newline \newline
The leading order term in the derivative of the
functional $V_{2PI}$ with respect to $v$ is of order $N$
\be
\frac{\delta V_{2PI}}{\delta v}[\hat\alpha,v,G_\pi,{\cal G}]=
N v M^2.
\label{LO-EOS}
\ee
The right hand side is finite in itself, it does not necessitate the
introduction of any ${\cal O}(N)$ counterterm.

Since the equation of state is obtained by equating the r.h.s. of
(\ref{LO-EOS}) to zero and the leading order inverse pion
propagator is of the form (\ref{Eq:tree_prop}), we immediately see that
Goldstone's theorem ($D_\pi^{-1}(\hbox{$k=0$})=0$) is obeyed.
\newline \newline
{\bf{Leading order propagator matrix in the $\alpha - s$-sector}}
\newline \newline
As it was already stressed, the only entry of the $2\times 2$ inverse propagator matrix which receives LO correction is $(G^{-1})_{\alpha\alpha}:$
\be
\label{Eq:alpha-alpha}
i(G^{-1})_{\alpha\alpha}(k)=-1+\frac{\lambda}{6}I_\pi(k)-2\frac{\delta \Delta V^{\alpha\alpha}}{\delta G_{\alpha\alpha}},
\ee
where $I_\pi$ is defined in (\ref{Eq:bub}). Its finite part $I_\pi^F$ can be obtained in terms of $D_\pi$ as follows. Recall that the divergence of the one-loop bubble integral built by propagators $G_0$ is chosen to be $T_d^{(0)}$ given in (\ref{Eq:Td0}). Power counting shows that changing the mass of the propagators in this integral leads only to a finite difference. In other words, the divergence of a perturbative bubble integral does not depend on the masses of the propagators. Then we define $I^F_{\pi}$ as:
\bea
\label{Eq:I_pi^F}
I_\pi(p)&\!\!\!=\!\!\!&-i\int_k D_\pi(k) D_\pi(k+p)\equiv T_d^{(0)}+I_\pi^F(p),\nonumber\\
I_\pi^F(p)&\!\!\!=\!\!\!&\frac{1}{16\pi^2}\left[-2+\ln\frac{M^2}{M_0^2}+L(p^2,M)\right].
\label{Eq:mid1}
\eea
Here, the momentum dependent function $L(p^2,M)$ is
\bea
\label{Eq:Lfunc}
L(p^2,M)=
\begin{cases}
2\sqrt{1-\frac{4M^2}{p^2}}\arth\Big(\sqrt{1-\frac{4M^2}{p^2}}\Big)-i\sqrt{1-\frac{4M^2}{p^2}}\pi \qquad \!\! p^2>4M^2, \\
2\sqrt{\frac{4M^2}{p^2}-1}\arctan\Big(\sqrt{\frac{4M^2}{p^2}-1}\Big)^{-1} \qquad \qquad \quad \!\! 4M^2>p^2>0, \\
2\sqrt{1-\frac{4M^2}{p^2}}\arth\Big(\sqrt{1-\frac{4M^2}{p^2}}\Big)^{-1} \qquad \qquad \qquad \qquad \quad p^2<0.
\end{cases}
\eea
It can be shown that $\underset{p\to 0}{\lim}$ $L(p^2,M)=2$, which results that
\be
I_\pi^F(p=0)=\frac{1}{16\pi^2}\ln\frac{M^2}{M_0^2}.
\label{Eq:Bp0}
\ee
Since $I_0^F(p)$ has exactly the same form as 
$I_\pi^F(p),$ but with $M^2$ replaced by $M_0^2,$ for large $p^2$ we have
\be
I_\pi^F(p)-I_0^F(p)=\frac{1}{8\pi^2}\frac{M_0^2-M^2}{p^2}\ln\frac{p^2}{M_0^2}
+{\cal O}\left(\frac{1}{p^2}\right).
\label{Eq:mid2a}
\ee
We see now that in order to make (\ref{Eq:alpha-alpha}) finite we have to introduce another piece into the counterterm functional:
\be
\Delta V^{\alpha\alpha}=\frac{\lambda}{12}T_d^{(0)}
\int_k G_{\alpha\alpha}(k).
\label{phi-alpha-alpha}
\ee
With this the LO renormalization is complete. Note, that
in the broken symmetry phase all elements of the
LO propagator matrix have common pole structure determined by
$\tilde G(k)$.  This is the manifestation of the hybridization
for $v\ne 0$ of the longitudinal field component $s$ and of the
composite field $\alpha\sim s^2+\pi_n^2$ \cite{szepfalusy1974}.
This feature is relevant when studying dynamical aspects of
the phase transition at finite temperature.

Using the first entry of (\ref{Eq:G_LO_matrix}), we can derive 
the following relation between the LO sigma and pion propagators: 
\be
G_{ss}^{(0)}(k)=D_\pi(k)-i\frac{\lambda}{3}v^2
\frac{G_{ss}^{(0)}(k) D_\pi(k)}{1-\lambda I_\pi^F(k)/6}.
\label{Eq:LO_relation}
\ee
This relation will be very useful for the divergence analysis at NLO.

\subsection{Next-to-leading order (NLO) construction of the counterterms}

The construction of the NLO counterterm starts with discussing the
$1/N$ expansion of the pion propagator. Its asymptotics will be
completely determined by certain integrals of the LO propagators. We
shall see that the counterterm functional can be determined by the
asymptotics of the LO propagators and of the NLO self-energy of the
pions.  In order to explicitly demonstrate the NLO renormalizability
for all values of $v$ and $\hat\alpha$ we investigate the derivatives
of the functional (\ref{4-Phi-funct2}) with respect to these
variables. 

First we analyze the divergence structure of the pion propagator equation, and determine the corresponding piece of the counterterm functional, denoted by $\Delta V^{\pi}$. Then we turn again to the equation of state (now at NLO). To have a finite version, we will determine the next piece: $\Delta V^v$. At this point we have the unrenormalized NLO saddle point equation of $\hat{\alpha}$, and propagator equations in the $\alpha-s$ sector. It will turn out that both $\Delta V^{\pi}$ and $\Delta V^v$ depend on $\hat{\alpha}$, therefore they give contribution also to the SPE. The mutual consistency of the counterterms renormalizing the SPE, EoS and the pion propagator equation is fundamental for the consistent outcome of our analysis. During the renormalization of the SPE, this consistency is checked and also a new counterterm piece $\Delta V^{\alpha,0}$ will be introduced, which completes the renormalization.

We do not present here the renormalization of the partial NLO
corrections which would occur in the $\alpha-s$ sector when
we used the 2PI effective action (\ref{4-Phi-funct}). All
counterterms necessary for the renormalization of these pieces are of
${\cal O}(1/N)$, not contributing to the renormalized effective action
at NLO (i.e. ${\cal O}(N^0)$) in the $1/N$ expansion.
\newline \newline
{\bf{Pion propagator}}
\newline \newline
The pion propagator at NLO in the $1/N$ expansion is given by 
\be
i G^{-1}_\pi(k)=i D^{-1}_\pi(k)-i\frac{\lambda}{3N}\int_p
G_{\alpha\alpha}^{(0)}(p)D_\pi(p+k)-2\frac{\delta \Delta V^{\pi
}}{\delta G_\pi}.
\label{Eq:Gp_NLO}
\ee
Since we need the pion self-energy to ${\cal O}(1/N)$ accuracy, 
we replaced $G_{\alpha\alpha}$ with $G_{\alpha\alpha}^{(0)}$
and $G_\pi$ with $D_\pi$ in the above integral. 

In order to determine the counterterm contribution in
(\ref{Eq:Gp_NLO}), we have to study the divergence of the NLO
self-energy.  Using the first two entries of (\ref{Eq:G_LO_matrix}) together with 
(\ref{Eq:LO_relation}) we can write
\be
G^{(0)}_{\alpha\alpha}(p)=-\frac{i}{1-\lambda I_\pi^F(p)/6}
-\frac{\lambda v^2}{3} \frac{G_{ss}^{(0)}(p)}
{\left(1-\lambda I_\pi^F(p)/6\right)^2}.
\label{Eq:noZ1}
\ee
First we show that the divergence of (\ref{Eq:Gp_NLO}) is momentum
independent, i.e. there is no need for infinite wave function
renormalization, in accordance with \cite{binoth98}.
It is clear by dimensional reasons that we have to analyze that
piece of the second term of (\ref{Eq:Gp_NLO}), which is proportional
to
\be
\int_p \frac{D_\pi(p+k)}{1-\lambda I_\pi^F(p)/6}.
\label{Eq:noZ2}
\ee
To study this one, we use the following expansion
\be
-i D_\pi(p+k)=\frac{1}{(p+k)^2-M^2}=\frac{1}{p^2-M^2}
+\frac{1}{p^2-M^2}\sum_{n=1}^\infty
\left(-\frac{k^2+2 p\cdot k}{p^2-M^2}\right)^n.
\label{Eq:noZ3}
\ee
In order to prove that there is no infinite wave function renormalization, it
is enough to look at the appearance of $k^2$ in the numerator, that is
at terms $n=1,2$ in the sum. Keeping only terms up to and including 
${\cal O}(1/p^4),$ but throwing away those which vanish upon
symmetrical integration
(note that $I_\pi^F(p)$ depends on $p^2$) amounts to the following 
replacement at the level of the integrand in (\ref{Eq:noZ2}):
\be
\frac{1}{(p+k)^2-M^2}\longrightarrow \frac{1}{p^2-M^2}+
\frac{4 (p\cdot k)^2 - k^2 p^2}{(p^2-M^2)^3}.
\label{Eq:noZ4}
\ee
The second term gives vanishing contribution in
(\ref{Eq:noZ2}) due to the following
property which holds for any integrable function $f(p^2)$
upon the use of a Lorentz-invariant
regulator: 
\be
\int d^4 p\, p_\mu p_\nu f(p^2)=\frac{g_{\mu\nu}}{4}\int d^4 p\, p^2 f(p^2).
\ee

It is clear that the other piece of the second term of (\ref{Eq:Gp_NLO}),
which is coming from the second term of (\ref{Eq:noZ1}) and proportional to
$v^2$ does not have a divergent part with respect to the external momenta.
In fact it does not contribute to the divergences at all, since after
setting $k=0$ and iterating (\ref{Eq:LO_relation}) we recognize that, by
logarithmic(!) power counting the following integral is actually convergent:
\be
\label{Eq:conv-int}
\int_p\frac{D_\pi^2(p)}{\big(1-\lambda I_\pi^F(p)/6\big)^2}.
\ee
Due to this fact we do not encounter any divergence proportional to
$v^2$ in the NLO pion self-energy. 

It is instructive to point out here a peculiarity of the resummation
procedure as compared to the order-by-order perturbative
renormalization.  Namely, when in the above integral the denominator
is expanded in powers of $\lambda$ then at $n$th order of the
expansion, we find a $\lambda^n (\log \Lambda)^{n+1}$ type
divergence.  Through formal resummation of this divergent series a
finite result is obtained. This argument explicitly shows that in
resummed perturbation theory the structure of the counterterms can be
different from that seen at any given order of the perturbation
theory. The same effect was noticed in \cite{binoth98} in connection
with the wave function renormalization constants of pion and sigma
fields which arise from imposing renormalization conditions on the
residua of their propagators. At NLO in the $1/N$ expansion they are
finite whereas in an expansion to any given order in the coupling they
appear divergent. 

We see now that the momentum independent divergence is determined by the
first term of (\ref{Eq:noZ1}):
\be
i\int_p G_{\alpha\alpha}^{(0)}(p) D_\pi(p)\bigg|_{\textnormal{div}}=
\int_p\frac{D_\pi(p)}{1-\lambda I_\pi^F(p)/6}
\bigg|_{\textnormal{div}}=:\tilde T_\textnormal{div}(M^2).
\label{pion-p0}
\ee
To find the expression of $ \tilde T_\textnormal{div}(M^2)$ using the
auxiliary propagators we start by taking into account that
the one-loop bubble integral behaves logarithmically for large momentum. 
In the asymptotic momentum region the above expression allows us to write 
\bea
\nonumber
\!\!\!\!\!\!\!\!\!\!\!\!\!\frac{1}{1-\lambda I_\pi^F(p)/6}&\!\!\!=\!\!\!&\frac{1}{1-\lambda I_0^F(p)/6}+
\frac{\lambda}{6}\frac{I_\pi^F(p)-I_0^F(p)}{(1-\lambda I_0^F(p)/6)^2}
+{\cal O}\left(\frac{1}{p^4 \ln p}\right)\\
\!\!\!\!\!\!\!\!\!\!\!\!\!&\!\!\!=\!\!\!&\frac{1}{1-\lambda I_0^F(p)/6}+
\frac{M_0^2-M^2}{p^2-M_0^2}
\frac{\lambda I_0^F(p)/3}{(1-\lambda I_0^F(p)/6)^2}+{\cal O}
\left(\frac{1}{p^2\ln^2 p}\right),
\label{Eq:mid2b}
\eea
where in the last line we used
$I_0^F(p)\sim\ln\frac{p^2}{M_0^2}-2-i\pi+{\cal O}\left(p^{-2} \ln p\right).$
The neglected terms give finite contribution in the integral of
(\ref{pion-p0}). Using there (\ref{Eq:mid2b}) and  
\bea
\frac{1}{p^2-M^2}=\frac{1}{p^2-M_0^2}+\frac{M^2-M_0^2}{(p^2-M_0^2)^2}+
{\cal O}\left(\frac{1}{p^6}\right),
\label{Eq:mid3}
\eea
we obtain
\be
\tilde T_\textnormal{div}(M^2)=T_a^{(2)}-(M^2-M_0^2) i\int_p G_a^2(p)
\left(1-\frac{\lambda}{6}I_0^F(p)\right)\bigg|_\textnormal{div}
-\frac{\lambda}{3} (M^2-M_0^2) T_a^{(I)},
\label{Eq:mid4}
\ee
where the following divergent integrals were defined:
\be
T_a^{(2)}=\int_p G_a(p),\qquad T_a^{(I)}=-i \int_p G_a^2(p) I_0^F(p).
\label{Eq:Ta2I}
\ee
Using in the remaining integral of (\ref{Eq:mid4})
that $\int_p G_a^2(p)$ is finite, we obtain for 
$\tilde T_\textnormal{div}$ the expression
\be
\tilde T_\textnormal{div}(M^2)=T_a^{(2)}-\frac{\lambda}{2}(M^2-M_0^2)T_a^{(I)}.
\label{tilde-div}
\ee
Since the necessary counterterm in (\ref{Eq:Gp_NLO}) compensates
$\tilde T_\textnormal{div},$ we readily find the required counterterm
functional piece upon functional integration of (\ref{tilde-div}) with
respect to $G_\pi$ and multiplying it by $\lambda/6$:
\be
\Delta V^{\pi}=-
\frac{\lambda}{6}
\left[T_a^{(2)}-\frac{\lambda}{2}(M^2-M_0^2)
T_a^{(I)}\right]\int_k G_\pi(k).
\label{Eq:ct_Gp}
\ee
We conclude this part by giving the finite pion propagator at NLO in
the $1/N$ expansion, including also the contribution of the
counter-functional $\Delta V^{\pi}.$ It reads as
\be
i G^{-1}_\pi(k)=k^2-M^2-\frac{\lambda}{3N}\Sigma^{F}_\pi(k),
\qquad 
\Sigma^{F}_\pi(k)=i\int_p G_{\alpha\alpha}^{(0)}(p) D_\pi(k+p)
-\tilde T_\textnormal{div}(M^2).
\label{pi-prop-1}
\ee 
It is instructive to evaluate explicitly the cutoff dependence of the
divergent integrals $T_a^{(2)}$ and $T_a^{(I)}$. This can be also
of some practical interest when we proceed to the numerical solution
of the renormalized equations. As we did during the calculation of
$T_d^{(0)}$ and $T_d^{(2)}$, we go to Euclidean space with
$p_0\to i p_E^0$ and use a 4d cutoff $\Lambda$, however we
limit ourselves to an asymptotic analysis and expand the integrands
for large momentum (check the asymptotics of the second entry of
(\ref{Eq:I_pi^F}) for $M=M_0$). Exploiting the freedom to omit those
contributions to $T_a^{(I)}$ which are formally finite for
$\Lambda\to \infty,$ we choose the scheme in which
\be
T_a^{(I)}=-\frac{2}{(4\pi)^4}\int^\Lambda \frac{d k}{k}
\frac{\ln(k^2/M_0^2)}{(1+2 a -a \ln(k^2/M_0^2))^2}.
\label{Eq:TaI_calc}
\ee
Here, we introduced $a=\lambda/(96\pi^2).$ We notice that
for $k=M_0\exp(1+48\pi^2/\lambda)$ the denominator of the integral
above vanishes. To avoid this non-integrable singularity to occur in
the range of integration, that is for $k<\Lambda,$ we need
$\Lambda<\Lambda_\textnormal{max}=M_0\exp(1+48\pi^2/\lambda).$ That
means that for $\lambda\ne 0$ the cutoff cannot be sent to infinity,
there is a maximal value for it, which reflects the triviality
of the theory. Performing the integral in (\ref{Eq:TaI_calc}), and obeying this 
restriction, we find
\be
\label{Eq:TaI}
T_a^{(I)}=-\frac{36}{\lambda^2}\ln\left(
-\frac{\lambda}{96\pi^2}\ln\frac{\Lambda^2}{M_0^2}+1+\frac{\lambda}{48\pi^2}
\right).
\ee
With the same strategy we have
\bea
\label{Eq:Ta2}
T_a^{(2)}=\frac{3 M_0^2}{8\pi^2\lambda}
\Bigg[&\!\!\!-\!\!\!&e^{2+96\pi^2/\lambda}\textnormal{Ei}\left(
\ln\frac{\Lambda^2}{M_0^2}-2-\frac{96\pi^2}{\lambda}\right)\nonumber\\
&\!\!\!+\!\!\!&3 \ln\left(
-\frac{\lambda}{96\pi^2}\ln\frac{\Lambda^2}{M_0^2}+1+\frac{\lambda}{48\pi^2}
\right)
\Bigg],
\eea
where Ei$(x)$ refers to the exponential integral function:
\bea
\textnormal{Ei}(x)={\cal P}\int_{-\infty}^x \frac{e^{t}}{t}dt.
\eea
An asymptotic series representation of this function looks as 
\bea
\textnormal{Ei}(x)=\frac{\exp(x)}{x}\sum_{n=0}^\infty \frac{n!}{x^n},
\eea
from which it is obvious that for large cutoffs $T_a^{(2)} \sim \Lambda^2/\log\Lambda$. In the light of the previous two equations, it is important to
mention that a similar consideration applies to the divergence
appearing in (\ref{pion-p0}) to what was already stressed below
(\ref{Eq:conv-int}): resummed perturbation theory can change the nature of
the counterterms compared to order-by-order perturbation theory. Expanding
the denominator of the middle expression of  (\ref{pion-p0}) into powers of
$\lambda$,  we find the usual quadratic and logarithmic divergences
characteristic for the perturbative contributions. Resummation however
modifies these divergences as can be  explicitly seen in the formulas
(\ref{Eq:TaI}) and (\ref{Eq:Ta2}).
\newline \newline
{\bf Equation of state and Goldstone's theorem revisited}
\newline \newline
Next, we investigate the renormalization of the derivative of the
2PI effective potential with respect to the background $v.$ At NLO in the
$1/N$ expansion this is given by
\bea
\frac{\delta V_{2PI}}{\delta v}[\hat\alpha,v,G_\pi,{\cal G}]
&=&
N v M^2-i\sqrt{\frac{\lambda}{3}}\int_k G_{\alpha s}^{(0)}(k)
+\frac{\delta \Delta V^{v}}{\delta v}.
\nonumber
\\
&=&N v\left[M^2+\frac{\lambda}{3N}\int_k \tilde G(k)\right]++\frac{\delta \Delta V^{v}}{\delta v},
\label{Eq:v_derivalt}
\eea
where for the second equality we used the last entry of
(\ref{Eq:G_LO_matrix}). The counterterm functional $\Delta V^{\pi}$
determined above does not contribute (neither $\Delta V^{\alpha \alpha}$ nor $\Delta V^{\alpha,N}$ do), since its derivative with respect to $v$ is zero.

The equation of state is obtained by equating the r.h.s. of
(\ref{Eq:v_derivalt}) to zero. Its unrenormalized expression obviously
implies when confronted with (\ref{Eq:Gp_NLO}) the validity of
Goldstone's theorem with ${\cal O}(1/N)$ accuracy.  Note that, as it
is well known, Goldstone's theorem is not followed if we proceed in
strict 2PI sense which requires the self-consistent determination of
the full propagator without expansion in $1/N.$

In a renormalization procedure which preserves Goldstone's theorem we
need to construct a counterterm which does not depend on $G_\pi,$
therefore does not interfere with its already renormalized equation.
Since the divergence in (\ref{Eq:v_derivalt}) coincides with the
divergence of the NLO pion self-energy, the new contribution to the
counterterm functional is obtained upon integrating with respect to
$v$ the expression given in (\ref{tilde-div}) multiplied by 
$\lambda v/3.$ We find
\be
\Delta V^{v}=-\frac{\lambda}{6}v^2\left[
T_a^{(2)}-\frac{\lambda}{2}(M^2-M_0^2)T_a^{(I)}\right].
\label{Eq:ct_v}
\ee
A remarkable feature of this ${\cal O}(1/N)$ renormalized solution is
that it satisfies Goldstone's theorem for {\it arbitrary values} of
$\hat\alpha$~!
\newline \newline
{\bf{Saddle point equation}}
\newline \newline
In writing down the derivative of the effective potential with respect
to $\hat\alpha$, we must not forget about the contributions of the
$\hat\alpha$-dependent counterterms $\Delta V^{\alpha,N},$
$\Delta V^{\pi},$ and $\Delta V^{v}$ constructed above (see
(\ref{Eq:ct_aN}), (\ref{Eq:ct_Gp}), and (\ref{Eq:ct_v})):
\bea
\frac{\delta V_{2PI}}{\delta \hat\alpha}[\hat\alpha,v,G_\pi,{\cal G}]
&=&
\frac{3N}{\lambda}\hat\alpha-i\frac{N}{2}
\left(v^2+\int_k G_\pi(k)\right)-\frac{i}{2}
\int_k \Big(G_{ss}(k)-G_\pi(k)\Big)\nonumber\\
&+&i\frac{N}{2}\left[T_d^{(2)}+(M^2-M_0^2)T_d^{(0)}\right]
-i\frac{\lambda^2}{12}T_a^{(I)}\left(v^2+\int_k G_\pi(k)\right)\nonumber\\
&+&\frac{\delta\Delta V^{\alpha, 0}(\hat\alpha)}{\delta\hat\alpha}.
\label{counter-spe-div}
\eea
The contribution of the counterterms determined by the renormalization
of the equation of the inverse pion propagator  (\ref{Eq:Gp_NLO}) and
of the equation of state is displayed in the last but one term. 
\rt{Such type of ``counterterm cross-contribution'' would be more
difficult to recognize on the level of the Dyson-Schwinger equations.}

The last, yet undetermined part of the NLO counterterm functional, 
{\it i.e.} $\Delta V^{\alpha, 0},$ provides the NLO completion to
$\Delta V^{\alpha, N}$.  In order to determine it, first we have to
evaluate the pion tadpole in the second term of 
the r.h.s. of (\ref{counter-spe-div}) with NLO accuracy. Taking the inverse of $G_\pi^{-1}$
given in (\ref{pi-prop-1}) and expanding it to ${\cal O}(1/N)$, we obtain
\bea
\int_k G_\pi(k)
&=&\int_k D_\pi(k)+\frac{\lambda}{3N}\int_k D_\pi^2(k)
\int_p G_{\alpha\alpha}^{(0)}(p) D_\pi(k+p) + \frac{\lambda}{3N}
i \tilde T_\textnormal{div}(M^2) \int_k D_\pi^2(k)
\nonumber
\\
&=&\int_k D_\pi(k)-i\frac{\lambda}{3N}\left[
\tilde J(M^2)-\tilde T_\textnormal{div}(M^2)\int_k D_\pi^2(k)
\right]
-\frac{\lambda^2 v^2}{9N} J(M^2),
\label{pion-tadpole}
\eea
where for the second equality we used (\ref{Eq:noZ1}) and introduced the 
following functions
\begin{subequations}
\label{Eq:J_ints}
\bea
\tilde J(M^2)&=&\int_k D_\pi^2(k)\int_p 
\frac{D_\pi(p+k)}{1-\lambda I_\pi^F(p)/6},\\
J(M^2)&=&\int_k D_\pi^2(k)
\int_p\frac{D_\pi(p+k)G_{ss}^{(0)}(p)}{(1-\lambda I_\pi^F(p)/6)^2}.
\eea
\end{subequations}
Collecting all ${\cal O}(N^0)$ (NLO) divergent terms in 
(\ref{counter-spe-div}) we see that 
$\delta\Delta V^{\alpha,0}(\hat\alpha)/\delta\hat\alpha$ is determined by
\bea
\frac{\delta\Delta V^{\alpha, 0}(\hat\alpha)}{\delta\hat\alpha}&\!\!\!=\!\!\!&
\frac{\lambda}{6}\left[\tilde J_\textnormal{div}(M^2)
-\tilde T_\textnormal{div}(M^2)\int_k D_\pi^2(k)
\right]
-i\frac{\lambda^2}{18}v^2 J_\textnormal{div}(M^2)
\nonumber
\\
&\!\!\!+\!\!\!&\frac{i}{2}\int_k 
\Big(G_{ss}^{(0)}(k)-D_\pi(k)\Big)\bigg|_\textnormal{div}
+i\frac{\lambda^2}{12}T_a^{(I)}\left(v^2+\int_k D_\pi(k)\right),
\label{Eq:Act1}
\eea
where $\tilde J_\textnormal{div}(M^2)$ and $J_\textnormal{div}(M^2)$
denote the divergences of the integrals defined in equations (\ref{Eq:J_ints}).
Note that to the order of interest it was
again allowed to replace in the last two terms the original 
$G_\pi$ and $G_{ss}$ by $D_\pi$ and $G_{ss}^{(0)},$
respectively.

The important question of consistency inquires whether the previously
constructed counterterms cancel all subdivergences of $\tilde J(M^2)$
and the $v^2$-dependent divergence of second and third terms in
(\ref{Eq:Act1}). Note that the double integral $J(M^2)$ has only overall
divergence, since both $k$ and $p$ integrals are individually
finite.

We start with investigating the second double integral given in
(\ref{Eq:J_ints}). Changing the order of integration we use
(\ref{Eq:LO_relation}) and the following relation which can be
derived from (\ref{Eq:mid1}):
\be
\int_k D_\pi^2(k) D_\pi(p+k)=-\frac{1}{2}
\frac{d }{d M^2} I_\pi(p)=\frac{1}{p^2-4 M^2}\Big[
I_\pi^F(p)-\frac{1}{16\pi^2}\ln\frac{M^2}{M_0^2}+\frac{1}{8\pi^2}
\Big],
\label{Eq:v2_div3}
\ee
to find 
\be
J_\textnormal{div}(M^2)=
\int_p \frac{D_\pi(p)}{(1-\lambda I_\pi^F(p)/6)^2}
\frac{I_\pi^F(p)}{p^2-4 M^2}\Bigg|_\textnormal{div}
=-i\int_p G_a^2(p) I_0^F(p)=T_a^{(I)}.
\label{Eq:v2_div4}
\ee
To obtain the second equality above we replaced $I_\pi^F$ with $I_0^F$
in view of (\ref{Eq:mid2a}) and used (\ref{Eq:mid3}).

We continue with the differences of the tadpoles involving LO sigma and pion propagators. Using (\ref{Eq:LO_relation}) iteratively we find
\be
\frac{i}{2}\int_k\big(G_{ss}^{(0)}(k)
-D_\pi(k)\big)\bigg|_\textnormal{div}=
\frac{\lambda}{6}v^2 \int_k \frac{D_\pi^2(k)}{1-\lambda I_\pi^F(k)/6}
\bigg|_\textnormal{div}.
\label{Eq:v2_div1}
\ee
In view of (\ref{Eq:mid2b}), we can replace $I_\pi^F(k)$ with
$I_0^F(k)$ in the denominator above. Then using (\ref{Eq:mid3}) we
obtain
\be
\frac{i}{2}\int_k\big(G_{ss}^{(0)}(k)
-D_\pi(k)\big)\bigg|_\textnormal{div}=
\frac{\lambda}{6}v^2 
\int_k G_a^2(k)
\left(1-\frac{\lambda}{6}I_0^F(k)\right)\bigg|_\textnormal{div}=
-i\frac{\lambda^2}{36}v^2T_a^{(I)}.
\label{Eq:v2_div2}
\ee
With this term the $v^2$-dependent divergences are canceled by the
counterterm contribution proportional to $v^2$ appearing in the
last term of (\ref{Eq:Act1}).

Let us now go through the detailed analysis of the divergence structure
of $\tilde J(M^2)$, which is the first double integral given in
(\ref{Eq:J_ints}). This one contains an overall divergence as well as
subdivergences. By changing the order of integration and using
(\ref{Eq:v2_div3}), we have
\bea
\tilde J_\textnormal{div}(M^2)=\frac{6}{\lambda}\int_p
\frac{-1}{p^2-4M^2}\Bigg|_\textnormal{div}
+\Bigg(\frac{6}{\lambda}&\!\!\!+\!\!\!&\frac{1}{8\pi^2}-\frac{1}{16\pi^2}
\ln\frac{M^2}{M_0^2} \Bigg)\nonumber\\
&\!\!\!\times\!\!\!&\int_p \frac{1}{p^2-4M^2}\frac{1}{1-\lambda I_\pi^F(p)/6}
\Bigg|_\textnormal{div},\ 
\label{Eq:odiv1}
\eea
where we have separated a divergence independent of $I_\pi^F(p),$
which by using a relation similar to (\ref{Eq:mid3}) can be expressed
as a linear combination of $T_d^{(2)}$ and $T_d^{(0)}$. 
Since the form of the second integral in (\ref{Eq:odiv1}) is similar to
the last one in (\ref{pion-p0}),  the calculation follows very closely 
the determination of $\tilde T_\textnormal{div}$. 
Using (\ref{tilde-div}) and (\ref{Eq:Bp0}) the result is
\bea
\nonumber
\!\!\!\!\!\!\!\!\!\frac{\lambda}{6}\tilde J_\textnormal{div}(M^2)&\!\!\!=\!\!\!&
i T_d^{(2)}+(4M^2-M_0^2)i T_d^{(0)}\\
\!\!\!\!\!\!\!\!\!&\!\!\!+\!\!\!&i\left(\frac{\lambda}{2} M^2 T_a^{(I)}-\tilde T_\textnormal{div}(M^2)\right)
\left(1+\frac{\lambda}{6}T_d^{(0)}+\frac{\lambda}{48\pi^2}-
\frac{\lambda}{6}I_\pi(p=0)\right).
\label{Eq:odiv2}
\eea
The presence of subdivergences
is reflected by the appearance of divergent terms proportional to
$\ln(M^2/M_0^2)$ (coming from $I_\pi(p=0)$). These should cancel if
the approximation is renormalizable, since in this case only
divergences proportional to powers of $\hat\alpha$ (that is $M^2$)
are allowed. The result given in (\ref{Eq:odiv2}) for
$\tilde J_\textnormal{div}(M^2)$ shows the cancellation of the
subdivergence $\tilde T_\textnormal{div}(M^2) I_\pi(p=0)$,
when it is combined with the second term of the square bracket of
(\ref{Eq:Act1}). The cancellation of this self-energy type subdivergence
of the double integral $\tilde J(M^2)$ is expected in view of
(\ref{pi-prop-1}).

\begin{figure}[!t]
\centerline{ 
\includegraphics[bb=116 630 440 701,scale=1.0]{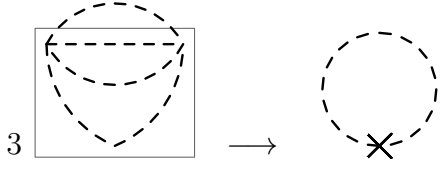}}
\caption{
The appearance of vertex type subdivergences in the first 
double integral of (\ref{Eq:J_ints}) as illustrated 
at leading $\lambda$ 
order in the expansion of the denominator of the integrand. 
The factor of three indicates the possible ways in which 
two lines of the setting-sun diagram form the bubble which corrects 
the vertex at 2-loop level. The cross denotes the associated
lowest order counterterm.
}
\label{Fig:vertex}
\end{figure}

The double integral $\tilde J_\textnormal{div}(M^2)$ has also a
vertex type subdivergence as illustrated in Fig.~\ref{Fig:vertex} at
leading order of the expansion in $\lambda.$ This is cancelled
by the last integral of (\ref{Eq:Act1}). We can see this
analytically by separating in the difference of the two terms of the
square bracket of (\ref{Eq:Act1}) a contribution proportional to
$T_a^{(I)}T_\pi^F$ which on its turn cancels the contribution
of the last tadpole integral in (\ref{Eq:Act1}).

With all the subdivergences and $v^2$-dependent divergences of
(\ref{Eq:Act1}) cancelled, 
$\delta\Delta V^{\alpha, 0}(\hat\alpha)/\delta\hat\alpha$ 
is determined by the overall divergence of $\tilde J(M^2)$ 
and that of the last tadpole integral. Its expression reads:
\bea
\frac{\delta\Delta V^{\alpha, 0}(\hat\alpha)}{\delta\hat\alpha}&\!\!\!=\!\!\!&
i T_d^{(2)}+i(4M^2-M_0^2)T_d^{(0)}-i T_a^{(2)}
\left(1+\frac{\lambda}{6}T_d^{(0)}+\frac{\lambda}{48\pi^2}\right)+i T_a^{(I)}
\nonumber\\
&\!\!\!\times\!\!\!&\Bigg[\lambda\left(M^2-\frac{1}{2}M_0^2\right)
+\frac{\lambda^2}{6} T_d^{(0)}\left(M^2-M_0^2\right)+\frac{\lambda^2}{12} T_d^{(2)}+\frac{\lambda^2}{12}\frac{3M^2-M_0^2}{16\pi^2}\Bigg].\nonumber\\
\label{Eq:Act2}
\eea
Since the above expression depends only on $\hat\alpha$ it will have
no ``back-reaction'' neither on the propagator equations nor on the
derivative of the effective potential with respect to the
background.

In conclusion, only divergences proportional to zeroth or first powers
of $M^2$ remained which upon integration over $\hat\alpha$ determine
the $\hat\alpha$-dependent counterterm functional
$\Delta V^{\alpha,0}(\hat\alpha).$ The counterterms induced by the
renormalization of the NLO pion propagator and of $\delta 
V_{2PI}/\delta v$ played an essential role in the cancellation of
subdivergences and of $v^2$-dependent divergences present in the
expression of $\delta V_{2PI}/\delta\hat\alpha.$ No limitations
whatsoever showed up on the value of $M^2$ and/or $v^2,$ in contrast
to the findings in \cite{andersen04,andersen08}. We shall return to
the discussion of this difference after reducing the 2PI effective potential
to the ordinary 1PI one depending only on the background
fields.
 \newline \newline
{\bf The explicit form of the counterterm functional}
\newline \newline
Now we collect into a unique expression $\Delta V$ all
individual pieces determined in Eqs.~(\ref{Eq:ct_aN}),
(\ref{phi-alpha-alpha}), (\ref{Eq:ct_Gp}), (\ref{Eq:ct_v}),
(\ref{Eq:Act2}), and
express it in a conventional form, in which we associate them with
the renormalization of different couplings appearing in the terms
of $V_{2PI}[\hat\alpha,v,G_\pi,{\cal G}]$ (Eq. (\ref{4-Phi-funct2})). The
counterterm functional reads:
\bea
\Delta V[\hat\alpha,v,G_\pi,{\cal G}]&=&
\frac{1}{2}\left(\delta m^2-i\delta g\hat\alpha\right) v^2+
i\delta\kappa_1\hat\alpha+\delta\kappa_2\hat\alpha^2
\nonumber\\ 
&+&\frac{1}{2}(\delta m^2-i\delta g\hat\alpha)\int_k
G_\pi(k)  
+\frac{1}{2}\delta\kappa_0\int_k G_{\alpha\alpha}(k),
\label{Eq:Phi_ct-funct}
\eea
where the counter-couplings are given by the following expressions:
\bea
\delta m^2&=&-\frac{\lambda}{3}\left[T_a^{(2)}-
\frac{\lambda}{2}\left(m^2-M_0^2\right)T_a^{(I)}\right],\qquad
\delta g=\frac{\lambda^2}{6} T_a^{(I)}, \nonumber\\ 
\qquad
\delta\kappa_2&=&\frac{N+8}{4} T_d^{(0)}+\frac{\lambda}{2} T_a^{(I)}
\left(1+\frac{\lambda}{6} T_d^{(0)}+\frac{\lambda}{64\pi^2}\right),
\qquad
\delta\kappa_0=\frac{\lambda}{6} T_d^{(0)},
\nonumber\\
\delta\kappa_1&=&\frac{N}{2}\left[T_d^{(2)}+(m^2-M_0^2)T_d^{(0)}\right]
+T_d^{(2)}+(4 m^2-M_0^2) T_d^{(0)}\nonumber\\ 
&&-T_a^{(2)}\left(1+\frac{\lambda}{6} T_d^{(0)}+\frac{\lambda}{48\pi^2}\right)
+\lambda T_a^{(I)} \Bigg[
m^2-\frac{1}{2}M_0^2\nonumber\\
&&+\frac{\lambda}{12}\left(
2 T_d^{(0)} (m^2-M_0^2) + T_d^{(2)} 
+\frac{1}{16\pi^2}(3m^2-M_0^2)\right)
\Bigg].
\eea 
It is interesting to note that the term proportional to
$\delta\kappa_1$ in (\ref{Eq:Phi_ct-funct}) has no correspondent in
(\ref{4-Phi-funct2}). Moreover, the terms proportional to
$\delta\kappa_0$ and $\delta\kappa_2$ correspond in
(\ref{4-Phi-funct2}) to terms not proportional to any coupling of the
original theory. However, the symmetry of the auxiliary field formulated model and the general concept of renormalization allow the appearance of these terms. $\delta \kappa_0$ and $\delta \kappa_2$ correspond to different possibilities of defining the two-point function of the auxiliary field (see (\ref{1-propagators3})). They are calculated to different orders in $1/N$ since the corresponding
terms contribute at different levels of the expansion, but after 
scaling back the fluctuating part of the field $\hat\alpha$ to $\alpha$, the two counter couplings do agree to leading order in $1/N$. This
feature is reminiscent of the renormalization of the
operators corresponding to two different definitions of the $n$-point
functions in 2PI-approximations. The terms
with $\delta g$ can be regarded as renormalizing the coupling
$g=\sqrt{\lambda/(3N)}$ through which the auxiliary field couples to
fields of the $O(N)$ multiplet. We interpret the renormalization of
all these operators not occurring in the original model as renormalizations
of the two variants of the 2-point functions of the composite field
$\phi^2$ and of the $\phi^2-s-s$ vertex (see ch. 30 of 
\cite{zinn-justin02}).

It is notable that $\delta\kappa_2$ determines the renormalization of the
coupling $\lambda$ following the formula
\be
\frac{6}{\lambda_B}=\frac{6}{\lambda}+\frac{4}{N}\delta\kappa_2
=\frac{6}{\lambda}+\frac{N+8}{N}T_d^{(0)}+
2\frac{\lambda}{N} T_a^{(I)}\left(1+\frac{\lambda}{6}T_d^{(0)}+
\frac{\lambda}{64\pi^2}\right).
\label{Eq:lambdaB1}
\ee
This can be seen by looking at the terms proportional to
$\hat\alpha^2$ in the classical part of the functional. This
relation is only intermediary, since it will change in the course of
the elimination of $\hat\alpha$. The part of the counterterm
$\delta\kappa_2$ proportional solely to $T_d^{(0)}$ appears already in
the literature and follows the one-loop $\beta$-function of the model
\cite{kleinert}.  The entire NLO part of $\delta\kappa_2$
proportional to $T_a^{(I)}$ is missing in the analysis of
\cite{andersen04,andersen08} (see e.g. Eq.~(23) of \cite{andersen08}
for the expression of their counterterm $b_1$).

Introducing the following notations
\be
\delta\kappa_2=N\delta\kappa_2^{(0)}+ \delta\kappa_2^{(1)},\qquad
\lambda_B=\lambda+\delta\lambda_\alpha,\qquad
\delta\lambda_\alpha=\delta\lambda_\alpha^{(0)}+
\frac{1}{N} \delta\lambda_\alpha^{(1)},
\label{Eq:lambdaB2}
\ee
we readily obtain
\bea
\delta\lambda_\alpha^{(0)}=-\frac{\lambda^2}{6}
\frac{T_d^{(0)}}{1+\lambda T_d^{(0)}/6},\qquad
&\!\!\!\lambda_B^{(0)}\!\!\!&=\lambda+\delta\lambda_\alpha^{(0)}
=\frac{\lambda}{1+\lambda T_d^{(0)}/6}, \nonumber\\
\delta\lambda_\alpha^{(1)}&\!\!\!\!\!=\!\!\!\!\!&
-\frac{2}{3}\big(\lambda_B^{(0)}\big)^2 \delta\kappa_2^{(1)}.
\label{Eq:lambdaB3}
\eea
Comparing with Eq.~(25) of \cite{fejos08} we observe that
$\delta\lambda_\alpha^{(0)}$ is the counter-coupling of the $O(N)$
model formulated with its original variables and considered at LO in the
large-$N$ expansion.  Likewise $\lambda_B^{(0)}$ is the bare coupling of
the model in the large-$N$ limit. We shall see in the next section,
where the auxiliary field will be eliminated, that at NLO in the 
large-$N$ expansion the bare coupling of the $O(N)$ model differs from
$\lambda_B,$ it turns out to be a combination of $\lambda_B$ and
$\delta g$ (see (\ref{Eq:cc_sep1}) and (\ref{Eq:cc_sep2})).

With the counterterm functional $\Delta V$ explicitly determined by
(\ref{Eq:Phi_ct-funct}), we can give now in a compact form the
functional introduced in (\ref{4-Phi-funct2}):
\bea
\!\!\!\!\!\!\!V_\textnormal{2PI}[\hat\alpha,v,G_\pi,{\cal G}]&=&
\frac{N}{2}(m_B^2-i\hat c\hat\alpha)v^2
+i\delta\kappa_1\hat\alpha+\frac{3N}{2\lambda_B}\hat\alpha^2\nonumber\\
\!\!\!\!\!\!\!&-&\frac{i}{2}\int_k[\textnormal{Tr}\ln{\cal G}^{-1}(k)+(N-1)\ln G_\pi^{-1}(k)]
\nonumber\\
\!\!\!\!\!\!\!&-&\frac{1}{2}\int_k \big[k^2-m_B^2+i\hat c\hat\alpha\big]\big[
(N-1) G_\pi(k)+G_{ss}(k)\big]
\nonumber\\
\!\!\!\!\!\!\!&-&iv\sqrt{\frac{\lambda}{3}}\int_k G_{\alpha s}(k)
+\frac{c}{2}\int_k G_{\alpha\alpha}(k)
-\frac{\lambda}{12}\int_k G_{\alpha\alpha}(k)\Pi(k),
\label{eq:sum-phi}
\eea
where $\lambda_B$ is defined in (\ref{Eq:lambdaB1}) and we introduced 
the following notations:
\bea
m_B^2=m^2+\frac{1}{N}\delta m^2,\qquad 
&\!\!\!\hat{c}\!\!\!&=1+\frac{1}{N}\delta g, \qquad c=1+\delta \kappa_0,\nonumber\\
\Pi(k)&\!\!\!=\!\!\!&-i\int_p G_\pi(p) G_\pi(k+p).
\label{Eq:mBhc}
\eea

\section{Elimination of the auxiliary field}

In this section the ${\cal O}(N^0)$ accurate renormalized
functional will be established for the original formulation of
the $O(N)$ model by eliminating the auxiliary field $\hat\alpha$ and
the propagator components related to it (i.e. $G_{\alpha s}$ and
$G_{\alpha\alpha}$).  In order to achieve this, we substitute
the expressions of $\hat\alpha, G_{\alpha s}^{(0)},$ and
$G_{\alpha\alpha}^{(0)}$ as found from their respective
equations into (\ref{eq:sum-phi}).

\subsection{Determination of $V_\textnormal{2PI}[v,G_\pi,G_s]$}

For rewriting the terms depending on $G_{\alpha\alpha}^{(0)}$ and
$G_{\alpha s}^{(0)}$ we exploit their representations which allow
the expression of the result fully in terms of $\Pi(k)$ and
$G_{ss}^{(0)}.$ The latter will be replaced with $G_s,$ 
the exact sigma propagator of the $O(N)$ model.
In this way we find
\bea
 -\frac{i}{2}\int_k\textnormal{Tr}\ln{\cal G}^{-1}(k)&=&
-\frac{i}{2}\int_k\ln\textnormal{det}{\cal G}^{-1}(k)=
-\frac{i}{2}\int_k\ln\left[\left(1-\frac{\lambda}{6}I_\pi^F(k)\right)
i G_s^{-1}(k)\right]\nonumber\\
&=&
-\frac{i}{2}\int_k \ln G_s^{-1}(k)
-\frac{i}{2}\int_k \ln\left(1-\frac{\lambda}{6 c}\Pi(k)\right)
-\frac{i}{2}\int_k \ln(i c) \nonumber\\
&+& {\cal O}\left(\frac{1}{N}\right).
\label{Eq:Phi_S1}
\eea
In going from the first to the second line above we used that 
to ${\cal O}(1/N)$ accuracy 
\be
-i\int_p G_\pi(p) G_\pi(k+p)\Big|_\textnormal{div}=
-i \int_p D_\pi(p) D_\pi(k+p)\Big|_\textnormal{div}= T_d^{(0)},
\label{Eq:Gaa4}
\ee 
and therefore we have
\be
1-\frac{\lambda}{6} I_\pi^F(k)=c-\frac{\lambda}{6} I_\pi(k)=
c-\frac{\lambda}{6} \Pi(k)+ {\cal O}\left(\frac{1}{N}\right),
\label{Eq:I_Pi_relation}
\ee
where the neglected ${\cal O}(1/N)$ contribution is finite.

Using (\ref{Eq:noZ1}) for $G_{\alpha\alpha}^{(0)}$ we write the
following chain of expressions for the last term of (\ref{eq:sum-phi}):
\bea
\!\!\!\!\!\!\!\!\!\!\!\!\!-\frac{\lambda}{12}\int_k G_{\alpha\alpha}^{(0)}(k)\Pi(k)&\!\!\!=\!\!\!&
-\frac{c}{2}\int_k G_{\alpha\alpha}^{(0)}(k)+
\frac{1}{2}\int_k G_{\alpha\alpha}^{(0)}(k)
\left(c-\frac{\lambda}{6}\Pi(k)\right)
\nonumber\\
\!\!\!\!\!\!\!\!\!\!\!\!\!&\!\!\!=\!\!\!&
-\frac{c}{2}\int_k G_{\alpha\alpha}^{(0)}(k)
-\frac{\lambda}{6 c}v^2\int_k\frac{G_s(k)}{1-\frac{\lambda}{6c}\Pi(k)}
-\frac{i}{2}\!\int_k \!1+{\cal O}\!\!\left(\frac{1}{N}\right)\!.
\label{Eq:Phi_S2}
\eea
In writing the second line above we use again (\ref{Eq:I_Pi_relation}).
The first term on the r.h.s. above is cancelled against the last but one term 
of (\ref{eq:sum-phi}). Finally, for the last term in the second line of  
(\ref{eq:sum-phi}) we use (\ref{Eq:I_Pi_relation}) to write:
\be
\displaystyle
-i v\sqrt{\frac{\lambda}{3}}\int_k G_{\alpha s}^{(0)}(k)=
\frac{\lambda}{3 c}v^2\int_k\frac{G_s(k)}{1-\frac{\lambda}{6c}\Pi(k)}
+{\cal O}\left(\frac{1}{N}\right).
\label{Eq:Phi_S3}
\ee
As a short digression from our current task we mention that we could
proceed by further eliminating also the pion and sigma propagators
using their respective NLO and LO equations, then we would obtain the
renormalized version of the ordinary (1PI) effective potential as function of
$\hat\alpha$ and $v$ studied in \cite{andersen04,andersen08}. We will
come to that point later.

Now we proceed instead with the elimination of $\hat\alpha$ from
(\ref{eq:sum-phi}) keeping the variables of the original theory, i.e.
$v,G_\pi,$ and $G_s$.  The simplest way to do this
is to complete to a full square the functional depending quadratically on $\hat\alpha$ and then use the saddle point equation $\delta
V_{2PI}/\delta\hat\alpha=0.$ Replacing $G_{ss}^{(0)}$ by the
exact propagator $G_s$ of the $O(N)$ model to obtain from the
$\hat\alpha$-dependent part
\be
\frac{\lambda_B}{24 N}\left[
N\hat c v^2+\hat c \int_k\big[(N-1)G_\pi(k)+G_s(k)\big]
-2\delta\kappa_1 \right]^2.
\label{Eq:Phi_Sa}
\ee
Putting together all above pieces we also use that in view of 
(\ref{Eq:lambdaB3}) $\lambda/c=\lambda_B^{(0)}$ and obtain
\bea
V_{2PI}[v,G_\pi,G_s]&=&
\frac{N}{2}\left(m_B^2-\frac{\lambda_B\hat c \delta\kappa_1}{3 N}\right)v^2
+N\frac{\lambda_B \hat c^2}{24} v^4\nonumber\\
&&-\frac{i}{2}\int_k\big[(N-1)\ln G_\pi^{-1}(k)+\ln G_s^{-1}(k)\big]
\nonumber\\
&&
-\frac{1}{2}\int_k \left[
k^2-m_B^2+\frac{\lambda_B\hat c \delta\kappa_1}{3 N}
-\frac{\lambda_B \hat c^2}{6} v^2 \right]
\big[(N-1) G_\pi(k)+G_s(k)\big]
\nonumber\\
&&
+\frac{\lambda_B^{(0)}}{6}v^2\int_k\frac{G_s(k)}
{1-\lambda_B^{(0)}\Pi(k)/6}
-\frac{i}{2}\int_k\ln\bigg(1-\frac{\lambda_B^{(0)}}{6}\Pi(k)\bigg)
\nonumber\\
&&
+\frac{\lambda_B \hat c^2}{24} (N-2)\left(\int_k G_\pi(k)\right)^2
+\frac{\lambda_B \hat c^2}{12} \int_k G_\pi(k) \int_p G_s(p),
\label{eq:sum-phi2}
\eea
where we omitted terms of order ${\cal O}(1/N)$ and a divergent
constant $\sim \delta\kappa_1^2$ coming from (\ref{Eq:Phi_Sa}) as well
as the irrelevant last but one terms of 
(\ref{Eq:Phi_S1}) and (\ref{Eq:Phi_S2}).

The bare couplings appearing in (\ref{eq:sum-phi2}) have to be
given with an accuracy corresponding to that of the functional 
(\ref{4-Phi-funct2}), therefore we write the counterterms as a sum of
the LO and NLO contributions:
\begin{alignat}{4}
& \lambda_B \hat c^2=\lambda+\delta\lambda,\qquad
& &\delta\lambda =\delta\lambda^{(0)}+\frac{1}{N}\delta\lambda^{(1)},
& &
\nonumber \\ 
& m_B^2-\frac{\lambda_B\hat c \delta\kappa_1}{3 N}=m^2+\delta m^2, \qquad
& &\delta m^2 =\delta {m^2}^{(0)}+\frac{1}{N}\delta {m^2}^{(1)}.
& &
\label{Eq:cc_sep1}
\end{alignat}

With help of (\ref{Eq:lambdaB1}), (\ref{Eq:lambdaB2}), (\ref{Eq:lambdaB3}), 
(\ref{Eq:mBhc}) and the separation 
$\delta\kappa_1=N\delta\kappa_1^{(0)}+ \delta\kappa_1^{(1)},$
the counter-couplings above can be given in terms of the counter-couplings 
of the model in the auxiliary field formalism as
\bea
\delta\lambda^{(0)}&\!\!\!=\!\!\!&\delta\lambda_\alpha^{(0)},\qquad
\delta\lambda^{(1)}=\delta\lambda_\alpha^{(1)}+
2\lambda_B^{(0)}\delta g, \qquad
\delta {m^2}^{(0)}=-\frac{1}{3}\lambda_B^{(0)}\delta\kappa_1^{(0)},\qquad
\nonumber\\
&&\delta {m^2}^{(1)}=\delta m^2-\frac{1}{3}\left[
\delta\lambda_\alpha^{(1)}\delta\kappa_1^{(0)}+\lambda_B^{(0)}
\left(\delta\kappa_1^{(1)}+\delta\kappa_1^{(0)}\delta g\right)
\right].
\label{Eq:cc_sep2}
\eea
Using (\ref{Eq:cc_sep1}) in (\ref{eq:sum-phi2}) we obtain 
at NLO in the $1/N$ expansion the renormalized functional 
of the original $O(N)$ model, that is without the auxiliary field, 
in the following form:
\bea
V_{2PI}[v,G_\pi,G_s]&\!\!\!=\!\!\!&
-\frac{i}{2}\int_k \Big[
(N-1)\left(\ln G_\pi^{-1}(k)+{D}^{-1}_\pi(k) G_\pi(k)\right)\Big]
\nonumber\\
&\!\!\!-\!\!\!&\frac{i}{2}\int_k \Big[\ln G_s^{-1}(k)+D^{-1}_s(k) G_s(k)\Big] \nonumber \\
&\!\!\!+\!\!\!& \frac{N}{2} m^2 v^2+N\frac{\lambda}{24} v^4
+N\frac{\lambda}{24}\left(\int_k G_\pi(k)\right)^2
+\frac{\lambda}{12} \int_k G_\pi(k) \int_p G_s(p)
\nonumber\\
&\!\!\!-\!\!\!&
\frac{\lambda_B^{(0)}}{6}v^2\int_k G_s(k)
+\frac{\lambda_B^{(0)}}{6}v^2\int_k\frac{G_s(k)}
{1-\lambda_B^{(0)} \Pi(k)/6}
\nonumber\\
&\!\!\!-\!\!\!&\frac{\lambda_B^{(0)}}{12}\left(\int_k G_\pi(k)\right)^2
-\frac{i}{2}\int_k\ln\bigg(1-\frac{\lambda_B^{(0)}}{6}\Pi(k)\bigg)
\nonumber\\
&\!\!\!+\!\!\!&
\frac{N}{2} \delta m^2 v^2+N\frac{\delta\lambda}{24} v^4
+\frac{\delta m^2}{2}\int_k\big[(N-1) G_\pi(k)+ G_s(k)\big]
\nonumber\\
&\!\!\!+\!\!\!&\frac{\delta\lambda}{12}\Big[
v^2\int_k\big[(N-1) G_\pi(k)+3 G_s(k)\big]
+\frac{N}{2}\left[\int_k G_\pi(k)\right]^2
\nonumber\\
&\!\!\!+\!\!\!&\int_k G_\pi(k) \int_p G_s(p)
\Big],\ \ \ \ 
\label{Eq:Phi_ON}
\eea
where we introduced the usual tree-level propagators 
for the sigma and pion fields as
\be
i{{D}}^{-1}_s(k)=k^2-m^2-\frac{\lambda}{2} v^2,\qquad
i{{D}}^{-1}_\pi(k)=k^2-m^2-\frac{\lambda}{6} v^2.
\ee
If we forget about the counterterms, the expression in
(\ref{Eq:Phi_ON}) coincides with the 2PI effective potential obtained
in \cite{dominici93}. We also could have figured out directly the
diagrams present in the NLO large-$N$ 2PI effective potential of the
$O(N)$ model, the auxiliary field however indirectly did this job. The terms above can be combined in a way which makes explicit that in the
large-$N$ expansion there are only two bare
couplings, namely $m^2+\delta m^2$ and $\lambda+\delta\lambda$. When
we would attempt to get a self-consistent solution of the equations arising from
the variation of $V_{2PI}(v,G_\pi,G_s)$, this would not be true 
(see the corresponding section later).  We emphasize
also that since the above functional is ${\cal O}(N^0)$
accurate, in terms involving the sigma propagator $G_s$ we do
not need the NLO part of the counterterms, i.e.  $\delta m^{2^{(1)}}$
and $\delta\lambda^{(1)}.$ When differentiating with respect to
$G_\pi$, we should remember that also $\Pi(k)$ is a functional of
$G_\pi$!

The interpretation of the terms in (\ref{Eq:Phi_ON}), which makes explicit
the infinite series of diagrams summed up in the present treatment is as
follows. The first three and the last three lines represent the 2PI effective
potential of the $O(N)$ model at Hartree level of the truncation and at NLO
in its large-$N$ expansion. The remaining four terms represent the NLO
contribution of the 2PI vacuum diagrams beyond Hartree level. The
$v^2$-dependent part of these terms can be rewritten as
\bea
\frac{\lambda+\delta\lambda^{(0)}}{6} v^2 \int_k &\!\!\!\!G_s\!\!\!\!&(k)
\frac{\big(\lambda+\delta\lambda^{(0)}\big)\Pi(k)/6}
{1-\big(\lambda+\delta\lambda^{(0)}\big) \Pi(k)/6}
=\nonumber\\
\frac{\lambda+\delta\lambda^{(0)}}{6} &\!\!\!\!\!\!\!\!v^2\!\!\!\!\!\!\!\!&
\int_k G_s(k)\sum_{n=1}^\infty
\left(\frac{\lambda+\delta\lambda^{(0)}}{6}\Pi(k)\right)^n.
\label{Eq:sum_1}
\eea
The $v^2$-independent part can also be written as a sum:
\bea
\!\!\!\!-\frac{i}{2}\Bigg[\frac{\lambda+\delta\lambda^{(0)}}{6}\int_k \Pi(k)
+\int_k\ln\bigg(1-\frac{\lambda+\delta\lambda^{(0)}}{6}\Pi(k)\bigg)\Bigg]=
\nonumber\\
\frac{i}{2}\int_k \sum_{n=2}^\infty\frac{1}{n}
\Big(\frac{\lambda+\delta\lambda^{(0)}}{6}\Pi (k)\Big)^n.
\label{Eq:sum_2}
\eea
\begin{figure}[!t]
\centerline{ 
\includegraphics[bb=96 630 440 701,scale=1.0]{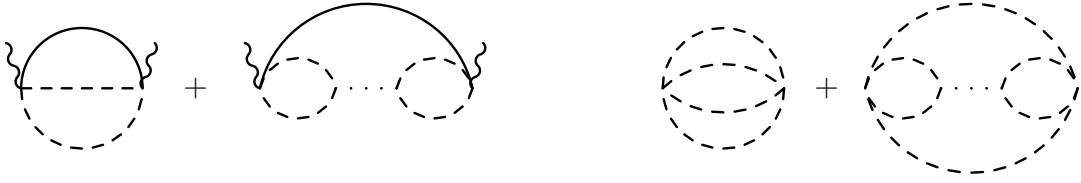}}
\caption{The two sets of vacuum diagrams which contribute beyond Hartree level 
and at NLO to the 2PI functional. Solid (dashed) line represents 
sigma (pion) propagator, while wiggly line represents the background $v.$ 
The dots indicate any number of pion bubbles.}
\label{Fig:2sets}
\end{figure}

\begin{figure}[!t]
\centerline{ 
\includegraphics[bb=172 610 440 701,scale=0.85]{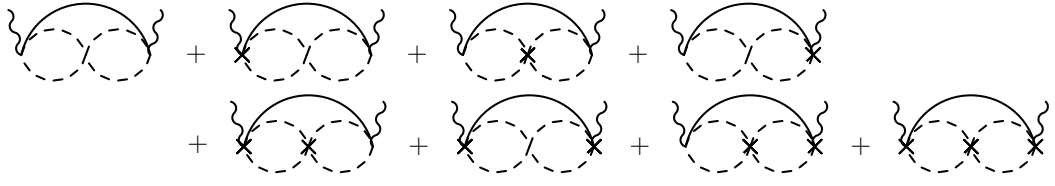}}
\caption{Diagrams corresponding to the $n=2$ term in the sum on the r.h.s. of
(\ref{Eq:sum_1}). A cross represents the counter-coupling
 $\delta\lambda^{(0)}.$}
\label{Fig:ctset1}
\end{figure}

It is easy to show that disregarding counterterms, these terms correspond to the two sets of diagrams given in Fig.~\ref{Fig:2sets}. Counterterm diagrams corresponding to the $n=2$ term of the sum in (\ref{Eq:sum_1}) are displayed in Fig.~\ref{Fig:ctset1}. Similar diagrams with different number of pion bubbles can be drawn for the other terms appearing in the sums in 
(\ref{Eq:sum_1}) and (\ref{Eq:sum_2}). A direct way to obtain 
(\ref{Eq:Phi_ON}) consists of summing up all these diagrams with the 
associated combinatorial factors determined by the Feynman rules.

\subsection{Renormalizability checks on $V_{\textnormal{2PI}}[v,G_\pi,G_s]$}

The significance of (\ref{Eq:Phi_ON}) is that it displays all 
counterterms which guarantee the renormalizability of the resummation
of the perturbative series, a resummation induced by the large-$N$ expansion. 
The finiteness of the equation of
state and the self-energies obtained from its respective variations is
ensured automatically, since this feature is ``inherited'' from the
finiteness of the same quantities achieved in the formulation with the
auxiliary field. Still, it is an instructive exercise to check this
feature directly. Exploiting the structure of our previous analysis 
done in the auxiliary formulation of the model we shall show the 
finiteness of the equations directly obtained from (\ref{Eq:Phi_ON}). 

The inverse pion propagator at NLO in the $1/N$ expansion is given then by
\be
i G_\pi^{-1}(k)=k^2-M^2-\frac{\lambda}{3N}\Sigma_\pi^F(k),
\ee
where the nonlocal $(\Sigma_\pi^F)$ and local $(M^2)$ parts of the self-energy are:
\bea
\nonumber
\Sigma_\pi^F(k)=\int_p\left[
-\frac{i}{1-\lambda \Pi_F(p)/6}
-\frac{\lambda v^2}{3} \frac{G_s(p)}
{\left(1-\lambda \Pi_F(p)/6\right)^2}
\right]G_\pi(k+p)-\tilde T_\textnormal{div}(M^2),\\
\!\!\!\!M^2=m_B^2+\frac{\lambda_B}{6}\left(v^2+\int_p G_\pi(p)\right)
+\frac{\lambda_B}{6 N}\int_p \Big(G_s(p)-G_\pi(p)\Big)
+\frac{\lambda}{3 N}\tilde T_\textnormal{div}(M^2).
\eea
For the nonlocal part we used that 
$6/\lambda_B^{(0)}-\Pi=6/\lambda-\Pi_F/6$ and $\tilde T_\textnormal{div}$ 
is given in (\ref{tilde-div}).

Since the local part has LO and NLO contributions we write 
$M^2={M^2}^{(0)}+{M^2}^{(1)}/N$ and expand the pion 
propagator to ${\cal O}(1/N)$. With help of the integrals defined in 
(\ref{Eq:J_ints}) we obtain 
\bea
\!\!\!\!\!{M^2}^{(0)}&=&m^2+\delta {m^2}^{(0)}+\frac{\lambda_B^{(0)}}{6}
\left(v^2+\int_k D_\pi(k)\right), \nonumber\\
\!\!\!\!\!3 i \frac{{M^2}^{(1)}}{\lambda_B^{(0)}}\left(
1-\frac{\lambda_B^{(0)}}{6} I_\pi(0)\right)&=&
\frac{3i}{\lambda_B^{(0)}}\left[\delta {m^2}^{(1)}+
\frac{\lambda}{3}\tilde T_\textnormal{div}\big({M^2}^{(0)}\big)\right]
\nonumber\\
\!\!\!\!\!&+&\frac{i}{2}\int_k \Big(G_s(k)-D_\pi(k)\Big)
\nonumber\\
\!\!\!\!\!&+&\frac{\lambda}{6}\left[\tilde J({M^2}^{(0)})
-\tilde T_\textnormal{div}({M^2}^{(0)})\int_k D_\pi^2(k) \right]\nonumber\\
\!\!\!\!\!&-&i\frac{\lambda^2}{18}v^2 J({M^2}^{(0)})
+\frac{i\delta\lambda^{(1)}}{2\lambda_B^{(0)}}
\left(v^2+\int_k D_\pi(k)\right),
\label{Eq:M2LO_NLO}
\eea
where in distinction to its definition given in (\ref{Eq:tree_prop}) 
$D_\pi$ depends now on ${M^2}^{(0)}.$ We used also that to leading order 
$\Pi(p)=I_\pi(p).$

The equation for ${M^2}^{(0)}$ in (\ref{Eq:M2LO_NLO}) is the usual 
gap equation at Hartree level of truncation of the effective action.
This was analyzed in \cite{fejos08} and the counterterms which
can be determined from this are $\delta \lambda^{(0)}$ and 
$\delta {m^2}^{(0)}$ given in (\ref{Eq:cc_sep2}).

Observing that the left hand side of the equation for ${M^2}^{(1)}$ 
in (\ref{Eq:M2LO_NLO}) is finite, we obtain the following relation 
between counterterms and divergences:
\bea
-\frac{3i}{\lambda_B^{(0)}}\left[\delta {m^2}^{(1)}+
\frac{\lambda}{3}\tilde T_\textnormal{div}\big({M^2}^{(0)}\big)\right]
&\!\!\!=\!\!\!&\frac{\lambda}{6}\left[\tilde J_\textnormal{div}({M^2}^{(0)})
-\tilde T_\textnormal{div}({M^2}^{(0)})\int_k D_\pi^2(k) \right] \nonumber\\
&\!\!\!-\!\!\!&i\frac{\lambda^2}{18}v^2 J_\textnormal{div}({M^2}^{(0)})
+\frac{i\delta\lambda^{(1)}}{2\lambda_B^{(0)}}\left(v^2+\int_k D_\pi(k)\right)
\nonumber\\
&\!\!\!+\!\!\!&\frac{i}{2}\int_k \Big(G_s(k)-D_\pi(k)\Big)\bigg|_\textnormal{div}.
\eea
Using the divergence analysis presented in the corresponding parts of the NLO renormalization, we have all divergences and integrals expressed in terms of ${M^2}^{(0)}$ for which we can substitute its finite gap equation 
${M^2}^{(0)}=m^2+\lambda(v^2+T_\pi^F)/6.$ Requiring the vanishing of 
the coefficient of $v^2+T_\pi^F$ determines $\delta\lambda^{(1)},$ 
while the remaining overall divergence determines 
$\delta {m^2}^{(1)}.$ Both are in accordance with (\ref{Eq:cc_sep2}).

The equation for the inverse sigma propagator obtained from 
(\ref{Eq:Phi_ON}) is
\be
i G_s^{-1}(k)=k^2-{M^2}^{(0)}-\frac{\lambda_B^{(0)}}{3}v^2 
\frac{1}{1-\lambda_B^{(0)} I_\pi(k)/6},
\ee
which is finite, since ${M^2}^{(0)}$ is finite and 
$6/\lambda_B^{(0)}-I_\pi=6/\lambda-I_\pi^F.$

Using the relation between the LO pion and sigma propagators we can show 
that the derivative of (\ref{Eq:Phi_ON}) with respect to $v$ reads as
\be
\frac{\delta V_{2PI}}{\delta v}[v,G_\pi,G_s]=
-N v i G_\pi^{-1}(k=0),
\ee
which is also finite, since we showed that $G_\pi^{-1}(k)$ is finite.
It also displays the validity of Goldstone's theorem.

We close this part by mentioning that the LO sigma propagator
equation, the NLO pion propagator equation and the equation of state
derived from (\ref{Eq:Phi_ON}) can be obtained also in the
Dyson-Schwinger formalism, similarly as it was demonstrated in the 
auxiliary field formulation.  Some details can be found in \cite{patkos06} 
(see also \cite{korpa} for the relation between the truncation of the 
Dyson-Schwinger equations and the $1/N$ expansion).  This means that 
our investigation implies also the renormalizability of the 
Dyson-Schwinger equations at NLO in the large-$N$ expansion. 

\section{Derivation of the 1PI effective potential $V[\hat\alpha,v]$}

Let us now go back to (\ref{Eq:Phi_S3}) and instead of eliminating the auxiliary
field, let us focus on substituting the solutions of the propagators into $V_{2PI}[\hat{\alpha},v,G_{\pi},{\cal G}]$ (\ref{eq:sum-phi}). In this case we obtain the ordinary 1PI effective potential (up to constants), which is defined as $V[\hat{\alpha},v]=-\frac{1}{\int d^4x}\Gamma[\hat{\alpha},v]$.

We start from (\ref{eq:sum-phi}) and after using (\ref{Eq:Phi_S2}) and 
(\ref{Eq:Phi_S3}) we find that in view of (\ref{Eq:LO_relation})
$G_{ss}^{(0)}\equiv G_s$ does not appear explicitly in the functional, 
which now becomes
\bea
V[\hat\alpha,v,G_\pi]&=&\frac{N}{2}(m_B^2-i\hat c\hat\alpha)v^2
+i\delta\kappa_1\hat\alpha+\frac{3N}{2\lambda_B}\hat\alpha^2\nonumber\\
&&-\frac{i}{2}(N-1)\int_k
\left(\ln G^{-1}_\pi(k) + D^{-1}_\pi(k)G_\pi(k)\right)
\nonumber
\\
&&-\frac{i}{2}\int_k \ln\bigg[\big(k^2-m^2+i\hat\alpha\big)
\bigg(1-\frac{\lambda_B^{(0)}}{6}\Pi(k)\bigg)
-\frac{\lambda_B^{(0)}}{3} v^2\bigg].
\label{Eq:V2a}
\eea
Here, the last term comes from (\ref{Eq:Phi_S1}).
Next, we use (\ref{pi-prop-1}), (\ref{tilde-div}), and the definitions in 
(\ref{Eq:mBhc}) to write the inverse pion propagator as 
\be
i G_\pi^{-1}(k)=k^2-m_B^2+i\hat c\hat\alpha-\frac{\lambda}{3 N} \Sigma_\pi(k),
\ee
where $\Sigma_\pi(k)$ is given by the integral of (\ref{pi-prop-1})
calculated with the expression $G_{\alpha\alpha}^{(0)}$ taken from
(\ref{Eq:noZ1}). Using this propagator in the first integral of 
(\ref{Eq:V2a}), we easily see that when expanding it to ${\cal O}(1/N)$, the contribution of the self-energy drops out and we are left with
\bea
V[\hat\alpha,v]&=&\frac{N}{2}(m_B^2-i\hat c\hat\alpha)v^2
+i\delta\kappa_1\hat\alpha+\frac{3N}{2\lambda_B}\hat\alpha^2
-N\frac{i}{2}\int_k \ln\big(k^2-m_B^2+i\hat c\hat\alpha\big)
\nonumber
\\
&&-\frac{i}{2}\int_k \ln\bigg(1-\frac{\lambda_B^{(0)}}{6}\Pi(k)
-\frac{\lambda_B^{(0)}}{3} v^2\frac{1}{k^2-m^2+i\hat\alpha}\bigg).
\label{Eq:V2b}
\eea

The radiative part of this functional has exactly the same form as the
effective potential given in \cite{root74,andersen04,andersen08}.  The
difference in the classical part corresponds to slightly different
ways of introducing the auxiliary field. More important is that the
authors of \cite{andersen04,andersen08} restrict their counterterm
functional only to terms proportional to pieces of the Lagrangian
which are present already in the original formulation of the model. In
their form of introducing the auxiliary field this restricts
the counterterms to those proportional to $\hat\alpha$ and
$\hat\alpha^2$. By allowing all independent counterterms to appear 
which have dimension less than or equal to 4 in the auxiliary field formulation 
we will have enough flexibility to ensure the renormalizability in arbitrary
background.

In order to demonstrate that 
(\ref{Eq:V2b}) contains all the NLO counterterms, we sketch the 
renormalization of the SPE obtained by differentiating (\ref{Eq:V2b}) 
with respect to $\hat\alpha.$ Using that $\Pi,$ to be taken only at LO 
in $1/N$ expansion, depends on $\hat\alpha$ through $D_\pi$ defined 
in (\ref{Eq:tree_prop}), we obtain 
\bea
0&=&\frac{3N}{\lambda_B}\hat\alpha-i\delta\kappa_1
-i\frac{N}{2}\left(v^2+\int_k\frac{i}{k^2-m_B^2+i\hat c\hat\alpha}\right)
\nonumber
\\
&&-\frac{i}{2}\int_k\big(G_{ss}^{(0)}(k)-D_\pi(k)\big)
-\frac{\lambda}{6} \tilde J(M^2)+i\frac{\lambda^2}{18}v^2 J(M^2).
\label{Eq:Asep1}
\eea
Here, we recognized the appearance of the expression of 
$G_{ss}^{(0)}$ which can be read from (\ref{Eq:G_LO_matrix}).
We used the relation (\ref{Eq:LO_relation}) 
and for the last two terms also (\ref{Eq:I_Pi_relation}) and 
(\ref{Eq:J_ints}).

All we have to do is to establish the connection between (\ref{Eq:Asep1})
and (\ref{counter-spe-div}), the latter being already renormalized. The last
three terms of (\ref{Eq:Asep1}) can be found in
(\ref{counter-spe-div}), if in that equation we take into account
(\ref{pion-tadpole}), so we have to work only on the first three terms
of (\ref{Eq:Asep1}). Using the definition of the couplings, 
we expand them to ${\cal O}(1/N)$ and obtains
\begin{subequations}
\bea
i\frac{N}{2}
\left(v^2+\int_k\frac{i}{k^2-m_B^2+i\hat c\hat\alpha}\right)&=&
i\left(\frac{N}{2}+\frac{\lambda^2}{12}T_a^{(I)}\right)
\left(v^2+\int_k D_\pi(k)\right)\nonumber\\
&&-\frac{\lambda}{6}\tilde T_\textnormal{div}(M^2)\int_k D_\pi^2(k),
\\
\frac{3N}{\lambda_B}\hat\alpha-i\delta\kappa_1&=&
\frac{3N}{\lambda}\hat\alpha
+i\frac{N}{2}\left[T_d^{(2)}+(M^2-M_0^2)T_d^{(0)}\right]\nonumber\\
&&+i \delta\kappa_1^{(1)}+2\delta\kappa_2^{(1)}\hat\alpha.
\eea
\end{subequations}
Since the last two terms of the second equality 
above coincide with
$\delta\Delta V^{\alpha,0}(\hat\alpha)/\delta\hat\alpha$ of 
(\ref{counter-spe-div}), the equivalence between (\ref{Eq:Asep1}) and
(\ref{counter-spe-div}) is demonstrated. 

\section{Case of the complete 2PI-1/N resummation}

In this section we present the renormalization of the complete 2PI-1/N approximation. By ``complete'' we mean that in the self-consistent 
equations derived from the approximate 2PI-1/N effective potential 
we do not expand further in 1/N for the propagators and the
condensate but renormalizate within the self-consistent framework.

Due to the complex self-consistent nature of the equations, we need a method in which the divergences resummed into self-consistent quantities can be extracted and eliminated. In the beginning of this section, we will review the method introduced in \cite{berges05}, in which these divergences are summed up by Bethe-Salpeter like ladder-equations in order to obtain counterterms. Although the method is elegant and well worked out, we will not have explicit, 
immediately calculatable counterterms as it was the case in the previous sections. However, we do not expect this at all, since we need to deal with more complex divergences as before.

\subsection{Theory}

Let us review briefly the results of \cite{berges05}. Mostly we will not go into details, just state some useful parts of the referred paper. For detailed proofs, we refer the reader to \cite{berges05}.

As we have seen, in 2PI approximations it is possible to define the connected $2$-point function in two independent ways. Considering the
full theory, in the symmetric phase we had two identical representations of the self energy, check (\ref{1-propagators4}).
If we denote the left hand side of (\ref{1-propagators4}) by $-\Sigma$, and the right hand side by $-\tilde{\Sigma}$, for approximate 2PI effective actions in general $\Sigma\neq \tilde{\Sigma}$.

\begin{figure}[!t]
\centerline{ 
\includegraphics[bb=319 502 336 569,scale=1.0]{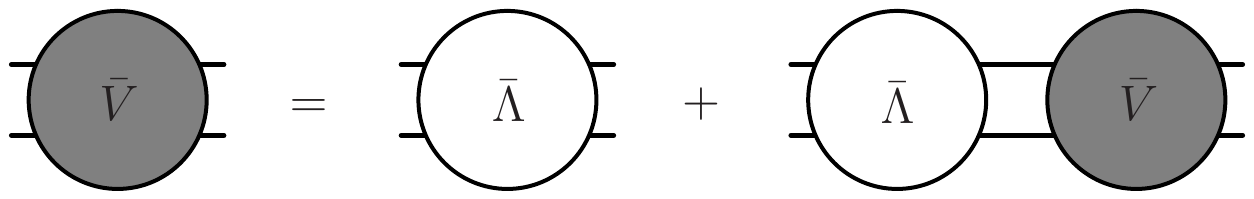}}
\caption{Bethe-Salpeter equation of $\bar{V}$. Ladder of $\bar{\Lambda}$'s are resummed.}
\label{Fig3_5}
\end{figure}

Considering a theory with $Z_2$ symmetry and staying in the symmetric phase, for the $4$-point function we have three possible definitions, which are introduced as follows. Let us define the $\bar{\Lambda}$ function as:
\bea
\bar{\Lambda}(x_1,x_2,x_3,x_4)=4\frac{\delta^2 \Gamma_{int}}{\delta G(x_1,x_2)\delta G(x_3,x_4)}\bigg|_{G=\tilde{G}},
\eea
where $\tilde{G}$ is the solution of the propagator equation, as before. Let us denote by $\bar{V}$ the infinite series of ladder graphs made by $\bar{\Lambda}$ (see Fig. \ref{Fig3_5}), which satisfies the following integral-equation (Bethe-Salpeter-like equation):
\bea
\label{V-BSp}
\bar{V}(x_1,x_2,x_3,x_4)=\bar{\Lambda}(x_1,x_2,x_3,x_4)+\frac{i}{2}\int_{yzwu} &\!\!\!\bar{\Lambda}\!\!\!\!\!&(x_1,x_2,y,z)\tilde{G}(y,w)\nonumber\\
&\!\!\!\times\!\!\!\!\!&\tilde{G}(u,z)\bar{V}(w,u,x_3,x_4).
\eea
$\bar{\Lambda}$ has the following symmetry properties 
\bea
\bar{\Lambda}(x_1,x_2,x_3,x_4)=\bar{\Lambda}(x_2,x_1,x_3,x_4)=\bar{\Lambda}(x_1,x_2,x_4,x_3)=\bar{\Lambda}(x_3,x_4,x_1,x_2),
\eea
which is inherited by $\bar{V}$. (\ref{V-BSp}) can be written in a short-hand notation as:
\bea
\bar{V}=\bar{\Lambda}+\frac{i}{2}\bar{\Lambda}\tilde{G}^2\bar{V}=\bar{\Lambda}+\frac{i}{2}\bar{V}\tilde{G}^2\bar{\Lambda},
\eea
where in the second equality we used the symmetry properties of the
functions $\bar{\Lambda}$ and $\bar{V}$. The first possible definition of the proper $4$-point vertex is $\bar{V}$.

Now we define the $V$ function as:
\bea
\label{Eq:V-def}
V(x_1,x_2,x_3,x_4)=-\frac{\delta \tilde{\Sigma}(x_3,x_4)}{\delta \bar{\phi}(x_1)\delta \bar{\phi}(x_2)}\equiv 2\frac{\Delta^2}{\Delta\bar{\phi}(x_1)\Delta \bar{\phi}(x_2)}\Bigg(\frac{\delta \Gamma_{int}}{\delta G(x_3,x_4)}\bigg|_{G=\tilde{G}}\Bigg),
\eea
where $\Delta/\Delta\bar{\phi}$ refers to total differentiation, as before. It can be shown that $V$ satisfies a very similar Bethe-Salpeter-like equation as $\bar{V}$. In the previously introduced short-hand notation it reads as:
\bea
\label{Eq:BS-for-V}
V=\Lambda+\frac{i}{2}V\tilde{G}^2\bar{\Lambda},
\eea
where
\bea
\Lambda(x_1,x_2,x_3,x_4)=2\frac{\delta^3 \Gamma_{int}}{\delta \bar{\phi}(x_1)\bar{\phi}(x_2)\delta G(x_3,x_4)}\bigg|_{G=\tilde{G}}.
\eea
It is possible to show that $V$ can be expressed in terms of $\bar{V}$:
\bea
V=\Lambda+\frac{i}{2}\Lambda\tilde{G}^2\bar{V}.
\eea
$V$ is the second possible definition for the $4$-point vertex.

The last variant for the $4$-point vertex is the most natural of the three possibilities:
\bea
\label{Eq:def-gamma4}
\Gamma^{(4)}=\frac{\delta^4 \Gamma[\bar{\phi}]}{\delta \phi^4}\equiv
\frac{\Delta^4 \Gamma[\bar{\phi},\tilde{G}[\bar{\phi}]]}{\Delta \phi^4}.
\eea
Working out carefully the total derivatives, after a short calculation we arrive at
\bea
\Gamma^{(4)}=\frac{\delta^4\Gamma_{int}}{\delta \bar{\phi}^4}\bigg|_{\tilde{G}}+\frac{i}{2}\Big(\Lambda\tilde{G}^2 V^\dagger+\perm\Big),
\eea
where $V^\dagger(x_1,x_2,x_3,x_4)=V(x_3,x_4,x_1,x_2)$, and the permutations must be taken in the last three arguments of $\Gamma^{(4)}$ (there are two more possibilities). The most important fact is that it can be showed that in the symmetric phase of the exact theory $\Gamma^{(4)}=\bar{V}=V$ (in the broken phase there are additional
terms in (\ref{Eq:BS-for-V}) and (\ref{Eq:def-gamma4})), therefore $\bar{V}$ and $V$ are as good definitions of the $4$-point vertex as $\Gamma^{(4)}$. Similarly as it happened to the $2$-point function, approximations of the 2PI effective action usually do not follow these relations.

In \cite{berges05} it is shown that with the renormalization of $\Sigma, \tilde{\Sigma}$ and $V, \bar{V}, \Gamma^{(4)}$, the theory is completely finite, also in the broken phase. However, since in general $\Sigma \neq \tilde{\Sigma}$ and $V\neq \bar{V} \neq \Gamma^{(4)}$, we have to introduce different mass and coupling counterterms corresponding to these quantities. Compared to ordinary perturbation theory, this is an unusual feature of 2PI approximations.

\begin{figure}[t]
\begin{center}
\includegraphics[bb=280 455 224 521,scale=0.95]{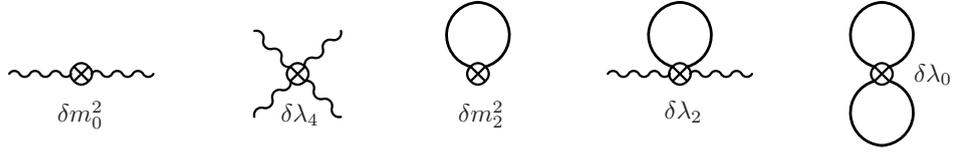}
\caption{The counterterm diagrams corresponding to different definitions of the $2$- and $4$-point functions. Neither wavefunction renormalization nor BPHZ counterterms are indicated. A wiggly line represents $v$, while a plain one corresponds to the full propagator $G$.}
\label{Fig:cts}
\end{center}
\end{figure}

The conventional names of the counterterms which cancel overall divergences of these quantities are as follows. The $\delta m_0^2$ mass counterterm with the $\delta Z_0$ wave function renormalization appear in the equation of $\Sigma$, therefore they can be found in the classical part of the effective action. $\delta m_2^2$ and $\delta Z_2$ are related to $\tilde{\Sigma}$, therefore they are included the tree level propagator of the $1$-loop part $\sim \Tr D^{-1}G$. For the $4$-point functions, we introduce three different coupling counterterms: $\delta \lambda_0, \delta \lambda_2, \delta \lambda_4$. The first one, $\delta \lambda_0$ is responsible for the renormalization of $\bar{V}$, which means that it is related to the Hartree part (``eight'' diagram). $\delta \lambda_2$ renormalizes $V$, therefore it can be found in the $\sim \Tr D^{-1}G$ piece, while $\delta \lambda_4$ makes $\Gamma^{(4)}$ finite, it is included in the classical part. 

Since $V$ and $\Gamma^{(4)}$ can be expressed in terms of $\bar{V}$, if the $2$-point functions are renormalized, subdivergences can arrive only from $\Lambda$ and $\bar{\Lambda}$. In order to cancel these, it is shown in \cite{berges05}, that one has to apply the BPHZ method \cite{bogoliubov57,zimmermann69}. This means that beyond the Hartree approximation we have to add corresponding counter-diagrams with {\it unique} counterterms related to $\bar{\Lambda}$, and another one related to $\Lambda$. These are usually denoted by $\delta \lambda_0^{\BPHZ}$ and $\delta \lambda_2^{\BPHZ}$, respectively. It is also shown that these counterterms are connected to $\delta \lambda_0$ and $\delta \lambda_2$, in general the latter ones include $\delta \lambda_0^{\BPHZ}$ and $\delta \lambda_2^{\BPHZ}$, therefore we can always introduce the following splitting:
\bea
\delta \lambda_0=\delta \lambda_0^{\BPHZ}+\Delta \lambda_0, \quad \delta \lambda_2=\delta \lambda_2^{\BPHZ}+\Delta \lambda_2.
\eea
In the next chapter, in a concrete example we will see the role of this splitting (see Fig. \ref{Fig:ren_with_cts}).

\newpage

\subsection{Effective potential revisited} 

TBA

\newpage

\subsection{Propagator- and Bethe-Salpeter equations}

TBA

\newpage

\subsection{Renormalization}

TBA

\newpage

TBA

\newpage

TBA

\newpage

TBA

\newpage

TBA

\newpage

TBA

\newpage

TBA

\newpage

\section{Concluding remarks}

We studied the renormalizability of the $O(N)$ model at
next-to-leading order in the large-$N$ expansion, at zero
temperature. In the main part of the chapter, we
constructed the ${\cal O}(N^0)$ counterterm 
functional of the model in auxiliary field formalism by 
studying the renormalization of the derivatives of the 2PI 
effective potential with respect to its variables. 
Expanding the propagator equation of the
pion to ${\cal O}(1/N)$ in the large-$N$ expansion we showed that the
renormalization can be achieved for arbitrary values of the background
and auxiliary fields, in a way that respects the internal symmetry of
the model (e.g. Goldstone's theorem). This is expected on
theoretical grounds, since divergences are determined only by the
asymptotic behavior of the propagators and because any consistent
resummation of the perturbation theory should resum also the
counterterm diagrams associated to the perturbative series.  Although,
one can anticipate the consistency of the auxiliary field technique
and of the large-$N$ expansion in dealing with perturbative series,
the difficulty we face when trying to infer the renormalizability of
the model in a given approximation from the fact that the model is
perturbatively renormalizable is that one cannot easily keep track of
what partial series of counterterm diagrams is actually resummed at a
given order of the large-$N$ expansion. In consequence, the actual
analytic check of the renormalization of a given approximation is
unavoidable.

We have also dealt with the elimination of the auxiliary field and
the related propagators with keeping the dynamical sigma and pion
propagators, which makes
transparent the classes of diagrams containing also counterterms which
are resummed in the original $O(N)$ theory at NLO of the large-$N$
expansion. The explicit form of the counterterms is given for the
first time for the theory using auxiliary field and also for the
original formulation. In the original theory, the 2PI effective potential
contains only two counterterms, a coupling and a mass counterterm,
both having LO and NLO parts. 

The two examples we worked out
(i.e. $V_\textnormal{2PI}[v,G_\pi,G_s]$ and
$V[\hat\alpha,v]$) demonstrate that the
renormalizability of the broadest functional
$V_{\textnormal{2PI}}[\hat\alpha,v,G_\pi,{\cal G}]$ implies the
renormalizability of the functionals arising after the elimination of
some subset of the variables. This result obtained at zero temperature
can makes us confident that the renormalization goes through for
$T\neq 0$ as well and that the counterterm functional determined here
will prove helpful for phenomenological studies in the $O(N)$ model.
Also, the method developed here can be used also to models with more
complicated global symmetries.

The final part of this chapter contained the complete 2PI-1/N renormalization
of the $O(N)$ model at NLO, i.e. when we looked for a self-consistent
solution of the propagator equations and the equation of state. In this case
no further expansion in 1/N was considered after deriving the equations for the
$1$- and $2$-point functions from the NLO 2PI effective potential. The 
renormalization was done in the symmetric phase of the model. In order to obtain
well-defined and finite results also at the broken phase, beyond the $2$-point
functions we had to renormalize the $4$-point functions as well (in the
previous case, we renormalized directly the equation of state). As expected
from \cite{berges05}, unlikely from our previous results, here we found that
in order to renormalize the complete theory, we had to include various
mass- and coupling counterterms regarding to different definitions of the
$2$- and $4$- point functions, which is an unusual feature of (self-consistent)
2PI approximations.

\chapter{Numerical implementation of 2PI renormalization methods}

In this chapter our goal is to investigate the numerical applicability of the 2PI formalism. This is motivated by the fact, that in an equilibrium setting and beyond the Hartree approximation (i.e. the ``eight'' diagram in theories with $4$-point coupling) there are relatively few papers reporting on numerical solutions of renormalized 2PI equations, even in scalar models. 

The application of the 2PI method did not went much beyond the
demonstration of its features in the simplest models (e.g. the
$N$-component real scalar model with $O(N)$ symmetry, in many cases
restricted to $N=1$).  The explicitly finite self-consistent
propagator equation of the $\phi^4$ model was solved including the
setting-sun diagram at vanishing field expectation value and at finite
temperature in \cite{VanHees:2001pf}.  The renormalized $O(N)$ model
was solved at zero temperature and at vanishing field expectation
value within the bare-vertex approximation of the auxiliary field
formalism in \cite{cooper04}.  The pressure of the one component
$\phi^4$ model was calculated in Euclidean space and finite
temperature in \cite{Berges:2004hn}, and of QED in \cite{Borsanyi:2008ar}. Solving the renormalized field and propagator
equation of this model in Minkowski space the phase transition was
studied in \cite{arrizabalaga06}, where it was found that including the
field-dependent two-loop skeleton diagram at the level of the 2PI
effective potential results in a second order phase transition. It
was proven analytically in
  \cite{Reinosa:2011ut} that in Hartree approximation the phase
  transition cannot be of second order.  The study of thermal
properties of the spectral function of the renormalized $\phi^4$
theory in the symmetric phase was reported in \cite{Jakovac:2006gi}.
In some phenomenologically oriented studies \cite{Roder:2005vt,roder05}
the necessity of renormalization was neglected, {\it e.g.} in
\cite{Roder:2005vt} the $O(N)$ model is solved in the Minkowski
space with some approximations
by taking into account momentum-dependent corrections only in the
imaginary part of the self-energy and treating the tadpole with a UV
cutoff.

Especially in the above listed finite temperature studies,
little technical details are given on the numerics, in particular the
rate of convergence of the iterative steps leading to self-consistent
solutions.  This fact does not allow us to easily infer the accuracy
of the methods used in obtaining the results. In our opinion, the lack
of standardized, well-tested algorithms might be the main reason
for the rather scarce applications of 2PI approximations in the
study of the thermodynamics of relevant field theoretical models. 

Our aim in this chapter is to quantify the accuracy of the iterative solutions of 2PI approximations and to compare different algorithms. We also would like to make connections between different renormalization methods.  Our highly accurate solutions could serve as a benchmark for methods to be used for a finite temperature solution of these equations.

In this chapter we stick to the $\phi^4$ theory. We will derive the usual
coupled set of self-consistent propagator equation and the equation of state. 
We will discuss in details their renormalization by adapting
the renormalization method of \cite{vanHees:2001ik,vanHees:2002bv} to the
broken symmetry phase (i.e. with nonvanishing field expectation value). 
We isolate the quantities which need to be renormalized, and imposing
renormalization conditions, we recover the set of counterterms derived using
the minimal subtraction method introduced in the previous chapter and which was
also used in \cite{patkos08}. On the other hand, we derive
explicitly finite equations as well.  We will present different
algorithms for solving the set of the finite equations and also the equations containing explicitly the counterterms. The solution will be given at zero temperature and in Euclidean space. We demonstrate that their numerical solutions coincide and compare the efficiency of the two methods. We show that in the case of
solving the equations containing explicitly the counterterms,
faster convergence of the iterative algorithm is obtained if the
counterterms themselves are derived iteratively and develop parallel to the
quantities to be determined (i.e. the propagator and the field expectation
value).

\section{The 2PI effective potential at two-loop field-dependent level}

The 2PI effective potential of the one-component real $\phi^4$ model
at the field ($v$) dependent two-loop truncation level is given in the following form 
\cite{berges05,arrizabalaga05,arrizabalaga06,patkos08}:
\bea
V_{\textnormal{2PI}}[v,G]&=&\frac{1}{2}m^2v^2+\frac{\lambda}{24}v^4
-\frac{i}{2}\int_p\big[\ln G^{-1}(p)+D^{-1}(p)G(p)\big]
+\frac{\lambda}{8}\left(\int_p G(p)\right)^2
\nonumber\\
&&-\frac{i\lambda^2}{12}v^2\int_k\int_P G(p)G(k)G(p+k)
+V_\textrm{ct}[v,G],
\label{Eq:l2-2PI_V}
\eea 
which can be obtained directly using the rules of 2PI formalism discussed in Chapter 1. The tree-level propagator is $D(p)=i/(p^2-m^2-\lambda v^2/2)$.
The counterterm functional is given by
\be
V_\textrm{ct}[v,G]=\frac{1}{2}\delta m^2_0 v^2+\frac{\delta\lambda_4}{24}v^4+
\frac{1}{2}\left(\delta m_2^2
+\frac{\delta\lambda_2}{2}v^2\right)\int_p G(p)
+\frac{\delta\lambda_0}{8}\left(\int_p G(p)\right)^2,
\label{Eq:l2-2PI-Vct}
\ee 
where the different terms are represented graphically in the previous
chapter on Fig. \ref{Fig:cts}.  As already discussed there, the
origin of two different mass counterterms and three coupling
counterterms stems from the fact that within the 2PI formalism it is
possible to define two two-point functions and three four-point
functions, which do not necessarily coincide within a given truncation
of the 2PI effective potential. 

The stationarity conditions $\delta V[v,G]/\delta G=0$ and  
$\delta V[v,G]/\delta v=0$ give self-consistent equations for the 
full two-point function and the equation of state: 
\bea
\label{Eq:full_prop}
i G^{-1}(p)&\!\!\!=\!\!\!&p^2-\Sigma(p),\\
\label{Eq:EoS}
0&\!\!\!=\!\!\!&v\left(m^2+\delta m_0^2+\frac{1}{6}(\lambda+\delta\lambda_4)v^2
+\frac{1}{2}(\lambda+\delta\lambda_2)T[G]+\frac{\lambda^2}{6}S(0,G)\right)\!,
\eea
where the self-energy $\Sigma(p)$ is given by
\be
\Sigma(p)=m^2+\delta m_2^2+\frac{1}{2}(\lambda+\delta\lambda_2)v^2
+\frac{1}{2}(\lambda+\delta\lambda_0) T[G]
+\frac{1}{2}\lambda^2 v^2 I(p,G).
\label{Eq:Sigma_def}
\ee
Note that we call $\Sigma(p)$ the self-energy, 
although its usual definition used in the 2PI
formalism does not contain the mass parameter and the corresponding
mass counterterm. The tadpole integral $T[G]$, the bubble integral $I(p,G)$ and the setting-sun integral at vanishing external momentum $S(0,G)$ which appear above are defined as follows:
\begin{subequations}
\bea
T[G]&=&\int_k G(k),
\label{Eq:T_def}
\\
I(p,G)&=&-i \int_k G(k)G(k+p),
\label{Eq:I_def}
\\
S(0,G)&=&-i \int_k\int_q G(k) G(q) G(k+q).
\label{Eq:SS_def}
\eea
\label{Eq:TBSS}
\end{subequations}
\vspace{-2mm}   

\noindent
Just as discussed in Chapter 2., \cite{patkos08} also calculates the divergences of these integrals by expanding the full propagator around the auxiliary propagator
\be
G_0(p)=\frac{i}{p^2-M_0^2}, 
\label{Eq:G0}
\ee
with the mass $M_0$ playing the role of the renormalization
scale. The counterterms absorbing these divergences were obtained in
\cite{patkos08} by requiring the separate cancellation of the
divergent coefficients of $v^0$, $v^2$ and of the environment (temperature and
the $v$ background) dependent finite function $T_F[G]$, representing the finite part of the tadpole integral, both in the propagator and field
equations.

\section{Renormalization conditions and counterterms}

In this section we adapt the renormalization method used at finite
temperature in \cite{vanHees:2001ik} to the broken symmetry phase at
zero temperature. We mention that the renormalization of scalar models
displaying spontaneously broken symmetries was discussed generally in
the context of the 2PI approximations in
\cite{vanHees:2002bv}. Here, we separate the field dependence from the
divergent quantities in the same way as the temperature dependence was
separated in \cite{vanHees:2001ik} and give the renormalization
conditions which, when imposed on the divergent quantities, lead on
the one hand to exactly the counterterms determined in \cite{patkos08},
and on the other hand
to explicitly finite equations, with no reference to any counterterm
at all.  We note that the counterterms in question have exactly the
same structure as those used in \cite{arrizabalaga06} and which were derived using the same minimal subtraction procedure introduced in Chapter 3.\!\!\! \footnote{Unfortunately the counterterms used in
  \cite{arrizabalaga06} appeared only on the corresponding poster, with no details on the calculations leading to their results.}
In the next section we will check explicitly that the solution of the
propagator equation and the equation of state obtained using explicit
counterterms agrees with the solution of the explicitly finite equations
to be derived below.

\subsection{Renormalization of the propagator equation}

We begin by splitting the propagator into ``vacuum'' and ``matter'' parts,
\be
G(p)=\Gv(p)+\Gm(p),
\label{Eq:G_split}
\ee
where the vacuum part refers to a propagator defined in the symmetric
phase ($v=0$), while the matter propagator includes all explicit and
implicit dependence on $v.$ The truncation of the 2PI effective
potential studied in this work leads to a momentum independent
self-energy for $v=0$ which means that the vacuum propagator can be
simply parametrized in terms of an effective positive mass $M_0$, as
in (\ref{Eq:G0}), so that $\Gv(p)\equiv G_0(p)$. Note that we need
this propagator with positive mass squared $M_0^2$ because in the
broken symmetry phase $m^2,$ the renormalized mass parameter of the
Lagrangian is negative.

As a consequence of the splitting (\ref{Eq:G_split}), 
the self-energy (\ref{Eq:Sigma_def}) is decomposed into three parts:
\bea
\Sigma(p)=\Sigmav+\SigmaO(p)+\Sigmar(p),
\label{Eq:sigma_decomp}
\eea
where the explicit expression of the different parts are given below.
The vacuum part is independent of the momentum, depends only on the 
vacuum propagator and has divergence degree 2. The last two pieces of 
the self-energy appearing in (\ref{Eq:sigma_decomp}) are $v$ dependent 
and are defined as follows. The part which does not result in any 
divergence when the full propagator is expanded around $\Gv$ is called 
regular part and is denoted by $\Sigmar$, while the remaining part, 
having divergence degree 0, is denoted by $\SigmaO$. The splitting of
the field-dependent part is somewhat arbitrary. Using (\ref{Eq:G_split})
in (\ref{Eq:Sigma_def}), the different parts of
(\ref{Eq:sigma_decomp}) are identified as:
\begin{subequations}
\label{Eq:sigmas}
\bea
\label{Eq:sigma-vac}
\!\!\!\!\Sigmav&\!\!\!=\!\!\!&M_0^2+\delta m_{2,A}^2
+\frac{\lambda+\delta \lambda_0}{2}\int_k \Gv(k),  \\
\label{Eq:sigma-0}
\!\!\!\!\SigmaO(p)&\!\!\!=\!\!\!&m^2-M_0^2+\delta m_{2,B}^2
+\frac{v^2}{2}\Lambda_2^{\vac}(0,p)
+\frac{\lambda+\delta \lambda_0}{2}\int_k \Gm(k),\ \  \\
\label{Eq:sigma-r}
\!\!\!\!\Sigmar(p)&\!\!\!=\!\!\!&-i\lambda^2 v^2\int_k \Gm(k) \Gv(k+p)
-i\frac{\lambda^2 v^2}{2}\int_k \Gm(k)\Gm(k+p),
\eea
\end{subequations}
where $\Lambda_2^{\vac}(0,p)$ is defined as
\be
\Lambda_2^{\vac}(0,p)=\lambda+\delta \lambda_2
-i\lambda^2\int_k \Gv(k)\Gv(k+p).
\label{Eq:Lambda2_vac}
\ee
We will see that with the above choice for the different pieces of the
self-energy, the expression of the counterterms
given in \cite{patkos08} can be obtained through simple
renormalization conditions. The mass counterterm $\delta m^2_2$ was
split into two parts: $\delta m^2_{2,A}$ is responsible for the
renormalization of the vacuum part, while $\delta m^2_{2,B}$ has to
remove the divergence generated by the $\SigmaO$ dependence of
$\Gm$. In order to see how this latter divergence proportional to
$\SigmaO$ emerges, we expand the propagator around the vacuum part:
\bea
G(p)&=&
\frac{i}{p^2-\Sigmav-\SigmaO(p)-\Sigmar(p)}=
\frac{i}{p^2-\Sigmav(p)}\left(1+\frac{\SigmaO(p)+
\Sigmar(p)}{p^2-\Sigmav}+\dots\right) \nonumber\\
&=&\Gv(p)-i\big(\Gv(p)\big)^2\SigmaO(p)+\Gr(p).
\label{Eq:prop}
\eea
The term proportional to $\SigmaO$ is ${\cal{O}}(1/p^4)$ (up to logs),
therefore it gives divergent contribution after integrating over the
momentum, while the regular part of the propagator goes with
${\cal{O}}(1/p^6)$ and leads to a finite contribution upon integration
over the momentum.  The sum of these two terms is the previously
introduced matter part
\be
\Gm(p)=-i\big(\Gv(p)\big)^2\SigmaO(p)+\Gr(p).
\label{Eq:G-mat}
\ee
As already announced at the beginning of this subsection, 
the first renormalization condition is 
\be
\Sigmav=M_0^2,
\label{Eq:ren_cond_sigma-vac}
\ee 
which determines $\delta m_{2,A}^2$ through the relation
\be
\delta m_{2,A}^2=-\frac{\lambda+\delta \lambda_0}{2}T_d^{(2)},
\label{Eq:delta-m22A}
\ee
where $T_d^{(2)}$ is defined in (\ref{Eq:Td2}). 

Now we turn to the renormalization of $\SigmaO$. 
After plugging (\ref{Eq:G-mat}) into (\ref{Eq:sigma-0}), 
we obtain the following integral equation:
\bea
\SigmaO(p)&\!\!\!=\!\!\!&m^2-M_0^2+\delta m_{2,B}^2+
\frac{v^2}{2}\Lambda_2^{\vac}(0,p)
-i\frac{\lambda+\delta \lambda_0}{2} 
\int_k\big(\Gv(k)\big)^2\SigmaO(k)\nonumber\\
&&+\frac{\lambda+\delta\lambda_0}{2}\int_k \Gr(k).
\label{Eq:Sigma0_def}
\eea
We can search for the solution of $\SigmaO$ in the following form: 
\be
\SigmaO(p)=\Gamma_m^{\vac}+\frac{v^2}{2}\Gamma_2^{\vac}(0,p)
+\frac12\Gamma^{\vac}\int_k \Gr(k).
\label{Eq:Sigma0}
\ee
Using (\ref{Eq:Sigma0}) in (\ref{Eq:Sigma0_def}) we find, that
$\Gamma_m^{\vac},$ $\Gamma_2^{\vac}(0,p),$ and $\Gamma^{\vac}$
fulfill the following Bethe-Salpeter type equations:
\begin{subequations}
\bea
\Gamma_m^{\vac}&=&m^2-M_0^2+\delta m_{2,B}^2
-i\frac{\lambda+\delta \lambda_0}{2} \Gamma_m^{\vac}
\int_k \big(\Gv(k)\big)^2,
\label{Eq:BS-c}
\\
\Gamma_2^{\vac}(0,p)&=&\Lambda_2^{\vac}(0,p)
-i\frac{\lambda+\delta \lambda_0}{2}\int_k\big(\Gv(k)\big)^2
\Gamma_2^{\vac}(0,k), 
\label{Eq:BS-a}
\\
\Gamma^{\vac}&=&\lambda+\delta \lambda_0
-i\frac{\lambda+\delta \lambda_0}{2}\Gamma^{\vac}
\int_k \big(\Gv(k)\big)^2,
\label{Eq:BS-b}
\eea
\end{subequations}
in which all integrals are divergent. To render these divergent
equations finite, we have to impose some renormalization conditions,
which will also determine the appropriate counterterms.  Due to our
introduction of the $M_0$ scale and the splitting of the mass
counterterm into to pieces, we need one more renormalization condition
in addition to (\ref{Eq:ren_cond_sigma-vac}) to fix also 
$\delta m_{2,B}^2.$  By using (\ref{Eq:sigma-vac}) we see that
$\displaystyle (m^2-M_0^2)\frac{\partial \Sigmav}{\partial M_0^2}$
satisfies 
\bea (m^2-M_0^2)\frac{\partial \Sigmav}{\partial M_0^2}
&=&m^2-M_0^2+(m^2-M_0^2)\frac{\partial (\delta m_{2,A}^2)}{\partial M_0^2}\nonumber\\
&&-i\frac{\lambda+\delta \lambda_0}{2} 
(m^2-M_0^2) \frac{\partial \Sigmav}{\partial M_0^2}
\int_k \big(\Gv(k)\big)^2,
\eea
which is the same equation fulfilled by $\Gamma_m^{\vac}$, up to counterterms.
Since 
\bea
\frac{\partial\Sigmav}{\partial M_0^2}=1,
\eea
a simple renormalization condition on $\Gamma_m^{\vac}$ is
\be
\Gamma_m^{\vac}=m^2-M_0^2.
\label{Eq:gamma-m_cond}
\ee
This fixes the $\delta m_{2,B}^2$ counterterm:
\be
\delta m_{2,B}^2=(m^2-M_0^2)\frac{\partial (\delta m_{2,A}^2)}{\partial M_0^2}=
-\frac{\lambda+\delta\lambda_0}{2}(m^2-M_0^2)T_d^{(0)},
\label{Eq:delta-m22B}
\ee
where $T_d^{(0)}$ is the familiar logarithmic divergent quantity defined in (\ref{Eq:Td0}). 
From (\ref{Eq:delta-m22A}) and (\ref{Eq:delta-m22B}) 
we obtain the complete $\delta m_2^2$ counterterm. As 
the sum of $\delta m^2_{2,A}$ and  $\delta m^2_{2,B}$, it leads to 
\bea
0=\delta m_2^2+\frac{\lambda+\delta \lambda_0}{2}\Big(T_d^{(2)}+(m^2-M_0^2)T_d^{(0)}\Big).
\label{4-ct1}
\eea
Before moving to the renormalization of (\ref{Eq:BS-a}) and
(\ref{Eq:BS-b}) we note that $\delta m_2^2$ can be equivalently fixed 
to the same value by keeping the condition (\ref{Eq:gamma-m_cond}) but  
replacing (\ref{Eq:ren_cond_sigma-vac}) with the condition
\be
\Sigma(P)\big|_{\textrm{overall}}=m^2,
\label{Eq:overall_cond1}
\ee 
where ``overall'' refers to the environment independent part of the
self-energy, which does not coincide with $\Sigma^{(v)}$ due to
additional terms coming from $\Sigma^{(0)},$ which contain some
overall divergences. We can identify this
part of the self-energy as the one which remains after $v$ and $\Gr$
are formally set to zero. Then, using the condition
(\ref{Eq:gamma-m_cond}) in the right hand side of (\ref{Eq:BS-c}), we obtain in (\ref{Eq:Sigma0}) 
\bea
\SigmaO(p)\big|_{\textrm{vac}}=m^2-M_0^2+\delta m^2_{2,B}+ 
(\lambda+\delta\lambda_0)(m^2-M_0^2) T_d^{(0)}/2,
\eea
so that upon adding it to $\Sigmav$ of (\ref{Eq:sigma-vac}), 
the condition (\ref{Eq:overall_cond1}) becomes
\be
m^2+\delta m_2^2+\frac{\lambda+\delta\lambda_0}{2} 
\left(T_d^{(2)}+(m^2-M_0^2) T_d^{(0)}\right)=m^2.
\ee

Next, we focus on the renormalization of $\Gamma_2^{\vac}$ and $\Gamma^{\vac}.$
As a renormalization condition we impose
\bea
\Gamma^{\vac}=\lambda,
\label{Eq:Gamma_cond}
\eea
which using (\ref{Eq:BS-b}) leads to
\bea
0=\delta\lambda_0+\frac12\lambda(\lambda+\delta \lambda_0)T_d^{(0)}.
\label{4-ct2}
\eea
Then, using (\ref{Eq:BS-a}) and (\ref{Eq:Lambda2_vac}), we see that the difference
\bea
\Gamma_2^{\vac}(0,p)-\Gamma_2^{\vac}(0,0)=-i\lambda^2\int_k 
\Big[\Gv(k+p)\Gv(k)&\!\!\!\!-\!\!\!\!&\big(\Gv(k)\big)^2\Big]\nonumber\\
&\!\!\!\!\equiv\!\!\!\!&\lambda^2 I_F^{\vac}(p)
\label{4-I-fin}
\eea
is finite, where we defined $I_F^{\vac}(p)$ as the finite part of the
bubble integral in the zero-momentum subtraction scheme in which
$I_F^\vac(0)=0$, as in the previous chapter. This shows that we only need to renormalize the momentum independent $\Gamma_2^{\vac}(0,0)$ which is achieved by imposing on it the renormalization condition
\be
\Gamma_2^{\vac}(0,0)=\lambda.
\label{Eq:Gamma2_cond}
\ee
Then, the explicitly finite $\Gamma_2^{\vac}(0,p)$ function is given by
\be
\Gamma_2^{\vac}(0,p)=\lambda+\lambda^2 I_F^{\vac}(p).
\label{Eq:fin-Gamma2}
\ee
Using (\ref{Eq:fin-Gamma2}) in (\ref{Eq:BS-a}), 
we obtain the equation
\bea
0=\delta \lambda_2+\frac12\lambda(\lambda+\delta\lambda_0)\Big(T_d^{(0)}+\lambda T_d^{(I)}\Big)+\lambda^2T_d^{(0)},
\label{4-ct3}
\eea
where
\bea
T_d^{(I)}=-i\int_p \big(G^{\vac}(p)\big)^2I^{\vac}_F(p).
\label{Eq:TdI}
\eea

In conclusion, the imposed renormalization conditions give explicitly
finite expressions for the functions $\Gamma_m^{\vac},$ $\Gamma_2^{\vac}(0,p),$
and $\Gamma^{\vac}$ appearing in (\ref{Eq:Sigma0}), and provide also 
the following finite equation for $\SigmaO(p)$: 
\be
\SigmaO(p)=m^2-M_0^2
+\frac{v^2}{2}\Big(\lambda+\lambda^2 I_F^{\vac}(p)\Big)
+\frac{\lambda}{2}\int_k \Gr(k).  
\label{Eq:Sigma0-final}
\ee

Two interesting remarks are in order at this point.
The first concerns the solution of (\ref{Eq:BS-a}) which, as we can check iteratively, is easily expressed in terms of $\Gamma^{\vac}$ as
\be
\Gamma_2^{\vac}(0,p)=\Lambda_2^{(v)}(0,p)-\frac{i}{2}\Gamma^{\vac}
\int_k\Lambda_2^{(v)}(0,k)\big(\Gv(k)\big)^2.
\label{Eq:Gamma-tilde-2}
\ee
The second comment refers to obtaining a different expression for the 
$\delta\lambda_2$ counterterm. Using (\ref{Eq:Gamma_cond}), 
(\ref{Eq:Gamma2_cond}), and (\ref{Eq:fin-Gamma2}) in 
(\ref{Eq:Gamma-tilde-2}) we obtain
\be
0=\delta\lambda_2+\frac{1}{2}\lambda(3\lambda+\delta\lambda_2)T_d^{(0)}+
\frac{1}{2}\lambda^3\left((T_d^{(0)})^2+T_d^{(I)}\right).
\label{Eq:delta-lambda-2b}
\ee
Note that (\ref{Eq:delta-lambda-2b}) does not coincide with
(\ref{4-ct3}), but the solutions for $\delta \lambda_2$
do. Indeed, using the result of (\ref{4-ct2}) for $\delta
\lambda_0$ in (\ref{4-ct3}), the same expression as that 
coming from (\ref{Eq:delta-lambda-2b}) is obtained. The equation
(\ref{Eq:delta-lambda-2b}) served in \cite{patkos08} 
as a consistency check of the renormalization procedure.

\begin{figure}[t]
\begin{center}
\includegraphics[bb=260 525 300 721,scale=0.95]{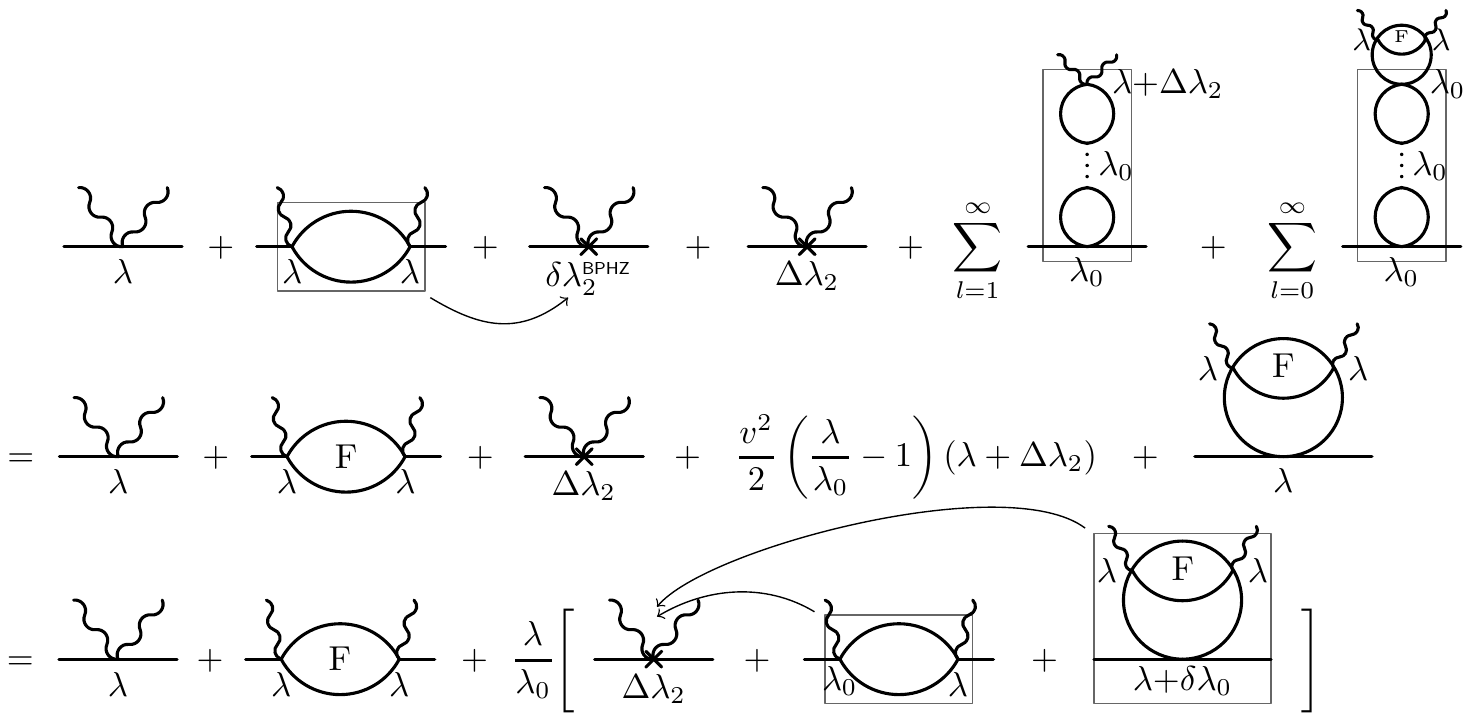}
\caption{Diagrammatic renormalization of the $v^2$-dependent part of
  the self-energy. See the text for details. The symmetry factors are
  not indicated and we used the shorthand notation
  $\lambda_0=\lambda+\delta\lambda_0.$ The boxes denote the divergence
  of the encircled part of the graph, the plain line denotes the
  propagator $\Gv(p)\equiv G_0(p),$ the wiggly one represents $v$, while 'F' denotes the finite part
  of the bubble integral. The index $l$ in the sums goes over the
  number of bubbles (for the two classes of diagrams the
  lowest number of bubbles is zero or one, as indicated).}
\label{Fig:ren_with_cts}
\end{center}
\end{figure}

Before we proceed to the renormalization of the equation of state, let us make a remark. So far in the thesis, renormalization was a more algebraic and formal, than a pictorial procedure, illustrated by diagrams. To help the reader getting closer to the more familiar BPHZ method, we also present a diagrammatic illustration of the cancellation of one of (sub)divergences. We choose the part proportional to $v^2$ in Eq.~(\ref{Eq:sigma-0}) for $\Sigma^{(0)},$ see
Fig.~\ref{Fig:ren_with_cts}. This example shows the
role the coupling counterterms play in the removal of both
subdivergences and overall divergences. The first line in Fig. {\ref{Fig:ren_with_cts}
shows the $v^2$-dependent diagrams generated upon iterating
Eq.~(\ref{Eq:sigma-0}) after the BPHZ subtraction procedure was implemented
(see \cite{berges05} for details.) The counterterm
$\delta\lambda_2$ displayed in Fig.~\ref{Fig:cts} is decomposed into the sum of two terms $\delta\lambda_2^{\rm BPHZ}$ and $\Delta\lambda_2$. $\delta\lambda_2^{\rm BPHZ}$ is used to absorb the divergence of
the bubble integral present in $\Lambda_2^{(\rm v)}(0,p).$ The
remaining finite part of the bubble integral is denoted by 'F'. In
order to obtain the second line of the figure we localized all the
subdivergences. Then, we use the renormalization condition
(\ref{Eq:Gamma_cond}). To obtain the third line of the figure, a
cancellation between the third and fourth term of the second
line was exploited and the consequence of the renormalization
condition (\ref{Eq:Gamma_cond}) (which is $\lambda-\lambda_0=(\lambda_0
\lambda/2) \int_k \Gv(k),$ where $\lambda_0=\lambda+\delta\lambda_0$)
was rewritten in a diagrammatic form.  The remaining overall
divergences of the diagrams in the third line are removed by
$\Delta\lambda_2.$ Since the divergent integrals associated with these
diagrams are $T_d^{(0)}$ and $T_d^{(I)},$ keeping in mind that we also have
to add $\delta\lambda_2^{\rm BPHZ},$ from the condition of
vanishing of the square bracket in the last line of
Fig.~\ref{Fig:ren_with_cts} we obtain the expression (\ref{4-ct3}) of
$\delta\lambda_2$ determined previously.

At this point, we obtained the counterterms $\delta m_2^2, \delta \lambda_0$ and $\delta \lambda_2$. These are determined by
(\ref{4-ct1}), (\ref{4-ct2}) and (\ref{4-ct3}), respectively. Explicitly finite equations were also derived for $\Sigma^{(0)}$ and $\Sigma^{(r)}$ (and therefore for the complete self energy), these are (\ref{Eq:Sigma0-final}) and (\ref{Eq:sigma-r}).

\subsection{Renormalization of the equation of state}

We proceed with applying a similar procedure on the equation of state (\ref{Eq:EoS}). We use again the splitting of the
propagator introduced in the previously in (\ref{Eq:G_split})
and obtain
\bea
0&=&m^2+\delta m_0^2+\frac{\lambda+\delta\lambda_2}{2}\int_p \Gv(p)
-i\frac{\lambda^2}{6}\int_p\int_k \Gv(p+k) \Gv(p) \Gv(k)
\nonumber \\
&&+\frac{\lambda+\delta \lambda_4}{6}v^2
-i\frac{\lambda^2}{6}\int_p\int_k\Big(3\Gv(p)+\Gm(p)\Big) 
\Gm(p+k)\Gm(k)\nonumber\\
&&+\frac12\int_K\Lambda_2^{\vac}(0,k) \Gm(k).
\eea
Using (\ref{Eq:G-mat}) and (\ref{Eq:Sigma0}) in the second line, we obtain
\bea
0&=&m^2+\delta m_0^2+\frac{\lambda+\delta \lambda_2}{2}\int_p \Gv(p)
-i\frac{\lambda^2}{6}\int_p \int_k \Gv(p) \Gv(p+k)\Gv(k)
\nonumber\\
&&-\frac{i}{2}\Gamma_m^{\vac}\int_k\Lambda_2^{\vac}(0,k) \big(\Gv(k)\big)^2
+\frac{v^2}{6}\Gamma_4^{\vac}
+\frac12 \int_k \Gamma_2^{\vac}(0,k) \Gr(k)
\nonumber\\&&
-i\frac{\lambda^2}{6}\int_p \int_k
\Big(3\Gv(p)+\Gm(p)\Big)\Gm(p+k)\Gm(k),
\label{Eq:EoS-2}
\eea
where we used the expression of $\Gamma_2^{\vac}$ given in 
(\ref{Eq:Gamma-tilde-2}) and defined
\be
\Gamma_4^{\vac}=\lambda+ \delta \lambda_4-\frac32 i\int_k
\Lambda_2(0,k)\big(\Gv(k)\big)^2 \Gamma_2^{\vac}(0,k). 
\label{Eq:Gamma-4}
\ee
The integral in the expression above is divergent, and we impose the
renormalization condition
\bea
\Gamma_4^{\vac}=\lambda.
\eea
This leads to the equation
\bea
0=\delta\lambda_4+\frac{3}{2}\lambda(\lambda+\delta\lambda_2+\lambda^2
T_d^{(0)})\left(T_d^{(0)}+\lambda T_d^{(I)}\right)+\frac{3}{2}\lambda^3 
\left(T_d^{(I)}+\lambda T_d^{(I,2)}\right),
\label{4-ct4}
\eea
where
\bea
T_d^{(I,2)}=-i\int_p \big(G^{\vac}(p)\big)^2\big(I^{\vac}_F(p)\big)^2.
\label{Eq:TdI2}
\eea
Since the last but one term of (\ref{Eq:EoS-2}) is already finite in
view of (\ref{Eq:Gamma2_cond}) and (\ref{Eq:fin-Gamma2}), and the last
term is a finite integral, we are left with the divergences of the
first line and the first term of the second line. In order to complete
the renormalization we only have to find the quantity on which the
renormalization condition determining $\delta m_0^2$ can be imposed.
It is straightforward to check that 
\bea
\frac{\Delta^2 V_{\textnormal{2PI}}[v,G_v]}{\Delta v \Delta v}\bigg|_{v=0}&=&
m^2+\delta m_0^2+\frac{\lambda+\delta\lambda_2}{2}\int_p G_{v=0}(p)\nonumber\\
&&-i\frac{\lambda^2}{6}\int_p\int_k G_{v=0}(p+k) G_{v=0}(p) G_{v=0}(k),
\eea 
where $G_v$ on the left hand side is the $v$-dependent solution of the stationarity condition $\delta V_{\textnormal{2PI}}[v,G]/\delta G|_{G=G_v}=0$ and the subscript indicates
explicitly that it depends on the vacuum expectation value $v,$ 
so that the chain rule has to be applied when taking the derivatives
with respect to $v.$ With the splitting of the propagator $G_{v=0}$
into ``vacuum'', ``matter'', and ``regular'' parts, we can proceed
exactly as at the beginning of this subsection, but we have to
omit everywhere the terms proportional to $v^2$. Then, we have
\bea
\!\!\!\!\!\!\!\!\!\!\frac{\Delta^2 V_{\textnormal{2PI}}[v,G_v]}{\Delta v \Delta v}\bigg|_{v=0}
&\!\!\!\!\!\!\!\!=\!\!\!\!&m^2+\delta m_0^2+\frac{\lambda+\delta \lambda_2}{2}\int_p \Gv(p)
\nonumber\\
\!\!\!\!\!\!\!\!\!\!&&-i\frac{\lambda^2}{6}\int_p \int_k \Gv(p) \Gv(p+k)\Gv(k)
\nonumber\\
\!\!\!\!\!\!\!\!&&-\frac{i}{2}\Gamma_m^{\vac}\int_k\Lambda_2^{\vac}(0,k) \big(\Gv(k)\big)^2+\frac12 \int_k \Gamma_2^{\vac}(0,k) \Gr_{v=0}(k)
\nonumber\\
\!\!\!\!\!\!\!\!\!\!&&-i\frac{\lambda^2}{6}\int_p \int_k
\Big(3\Gv_{v=0}(p)+\Gm_{v=0}(p)\Big)\Gm_{v=0}(p+k)\Gm_{v=0}(k).
\label{Eq:magyarazat}
\eea 
The last term above is finite and since $\Gamma_2^{\vac}(0,k)$ was
already made finite by imposing the condition (\ref{Eq:Gamma2_cond}),
the last but one term above is also finite.  Therefore, in principle
we could choose our last renormalization condition as
$\frac{\Delta^2 V_{\textnormal{2PI}}[v,G_v]}{\Delta v \Delta v} \big|_\textrm{v=0}=m^2,$
but then $\delta m_0^2$ determined through this condition would differ
by a finite term from $\delta m_0^2$ given in \cite{patkos08}.
In order to obtain the desired expression for $\delta m_0^2$ we
choose the following condition
\be
\frac{\Delta^2 V_{\textnormal{2PI}}[v,G_v]}{\Delta v \Delta v} \bigg|_\textrm{v=0, overall}=m^2,\label{Eq:overall_cond2}
\ee 
where the concrete expression on the left hand side is given by the
right hand side of (\ref{Eq:magyarazat}) without the last two terms,
and the ``overall'' notation indicates that the divergence of
this term is the usual overall divergence, that is the divergence
which remains after all subdivergences are removed.
In other words, the quantity on the left hand
side of (\ref{Eq:overall_cond2}) is obtained formally by setting 
$v$ and $\Gr$ to zero after (\ref{Eq:G_split}), (\ref{Eq:G-mat}),
and (\ref{Eq:Sigma0_def}), (\ref{Eq:Sigma0}) 
are used in the explicit expression of 
$\frac{\Delta^2 V_{\textnormal{2PI}}[v,G_v]}{\Delta v \Delta v}.$
Then, it is easy to check that (\ref{Eq:overall_cond2}) gives
\bea
0&=&\delta m^2_0+\frac{1}{2}(\lambda+\delta\lambda_2)
\left[T_d^{(2)}+(m^2-M_0^2) T_d^{(0)}\right]\nonumber\\
&&+\frac{1}{2}\lambda^2(m^2-M_0^2)\left((T_d^{(0)})^2+T_d^{(I)}\right)+
\frac{1}{6}\lambda^2S_{0}(0),
\label{Eq:delta-m02}
\eea
where 
\bea
S_0(0)=-i\int_p\int_k G^{\vac}(p)G^{\vac}(k)G^{\vac}(p+k).
\label{Eq:S00}
\eea
The two renormalization conditions (\ref{Eq:Gamma-4}) and 
(\ref{Eq:delta-m02})
together with the expression of $\Gamma_2^{\vac}(0,k)$ given in
(\ref{Eq:fin-Gamma2}) lead to the explicitly finite version of the
equation of state 
\bea 0&=&m^2+\frac{\lambda}{6}v^2+\frac{\lambda}{2}
\int_P \Big(1+\lambda I_F^{\vac}(p)\Big)\Gr(p)
\nonumber\\ &&-i\frac{\lambda^2}{2}\int_p\int_k \Gv(p+k)\Gm(p)\Gm(k)\nonumber\\
&&-i\frac{\lambda^2}{6}\int_p \int_k \Gm(p+k)\Gm(p)\Gm(k).
\label{Eq:EoS-fin}
\eea

With this the renormalization is complete. We obtained equations for the remaining counterterms $\delta m_0^2$ and $\delta \lambda_4$, these are (\ref{Eq:delta-m02}) and (\ref{4-ct4}). The finite version (\ref{Eq:EoS-fin}) of the equation of state is also known.

The imposed renormalization conditions give exactly the same
expressions for these counterterms (see (\ref{4-ct1}), (\ref{4-ct2}), (\ref{4-ct3}), (\ref{4-ct4}) and (\ref{Eq:delta-m02})) introduced in the functional
(\ref{Eq:l2-2PI-Vct}) as those appearing in \cite{patkos08}. This
demonstrates the equivalence
between the two renormalization procedures. We note that even though
we formulated our set of renormalization conditions on the quantities
$\Gamma^{\vac}, \Gamma_2^{\vac}(0,p),$ $\Gamma_4^{\vac}$,
corresponding to the three four-point functions $\bar V, V$ and $\Gamma^{(4)}$ introduced in the last section of the previous chapter, and also
on the curvature $\frac{\Delta^2 V_{\textnormal{2PI}}[v,G_v]}{\Delta v\Delta v}$ and the self energy $\Sigma$, they do not look as natural as those imposed in
\cite{berges05,arrizabalaga06} at $T=T^*$ and in the symmetric
phase of the model ($v=0$). (This method was the one adopted to the renormalization of the complete 2PI-1/N approximation of the O(N) model in the previous chapter.) In our case $\Gv$ is not the full propagator at $v=0$ and in order to obtain given expressions of the
counterterms (which contains $\Gv$ in place of the full propagator), we
needed a specific way to formulate the renormalization conditions in
(\ref{Eq:overall_cond1}) and (\ref{Eq:overall_cond2}). We imposed these conditions in order to facilitate the comparison between the solution of the finite equations and those which use counterterms (see next section). On the other hand, as discussed in
Ref.~\cite{Reinosa:2011ut}, finding a set of natural renormalization
conditions, prescribing e.g. the mass defined from the
variational propagator and the curvature at the minimum of the
effective potential (as we know, these two quantities do not necessarily
coincide in a given truncation of the 2PI effective potential)
together with the value of the independent four-point functions is
problematic at $T=0,$ that is in the
broken symmetry phase, even at the Hartree approximation of the
2PI effective potential.

\section{Numerical algorithms and methods \label{sec:implement}}
Before discussing the algorithms and presenting the results we make
some general statements about the numerical method we use. 

Since we solve iteratively the set of coupled integral equations
consisting of the self-consistent propagator equation and the field
equation, we have to store and upgrade in each iteration step the
value of some functions on a grid and with the use of interpolation,
approximate the value of these functions at the points required by
the integration routine.  For this purpose we use an equidistant grid
for the modulus of the Euclidean four-momentum, the one-dimensional Akima spline
interpolation method and the numerical integration routines of the GNU
Scientific Library (GSL)~\cite{gsl}. When we solve the set of finite
equations, we store the ``regular'' and the ``matter'' part of the
propagator (see (\ref{Eq:G-mat})), while when solving the set of
equations with the counterterms it is the full self-energy which is
stored on the grid (see (\ref{Eq:sigma_decomp}) and
(\ref{Eq:sigmas})). In the first case the integrals are convergent in
the UV, but nevertheless, for the functions which are stored on the
grid we have to take into account the fact that there is a maximal
value of the modulus of the four-momentum. The convolution of these
functions should be calculated accordingly, using $\theta$-functions
to restrict the momenta of the propagators:
\bea
C_L[f,g](k_E)&\!\!\!=\!\!\!&\int \frac{d^4 p_E}{(2\pi)^4}f(|p_E|)g(|p_E+k_E|)
\theta(L-|k_E|) \theta(L-|k_E+p_E|)\nonumber\\
&\!\!\!=\!\!\!&\frac{1}{8\pi^3 k_E^2}
\int_0^L d p\,p\,f(p) \int_{|p-|k_E||}^{\textrm{min}(p+|k_E|,L)} dq\,q\,g(q)
\nonumber\\
&&\times\sqrt{4p^2 q^2-\big (p^2+q^2-k_E^2\big)^2},
\label{Eq:iden1}
\eea
where $\theta$ is the step function, for the moduli we introduced
$p=|p_E|,$ $q=|k_E+p_E|$. $L$ is the maximal value of the modulus
stored on the grid. The $q$-integral is a remnant of the angular
integration in a 4d spherical coordinate system and is obtained with
the change of variable $\cos\theta_1=(q^2-k^2-p^2)/(2 k p).$

For functions of the momentum with known analytical expressions (e.g. the vacuum propagator), there is no need to store them on the grid, and in consequence a simplification is encountered in the convolution involving such functions. In this case there is no need for the second $\theta$ function, the entire angular
integration can be done analytically, and with the change of variable
$t=\tan(\theta_1/2)$ we have
\bea
&&\int_0^\pi d\theta_1 \frac{\sin^2\theta_1}
{p_E^2+k_E^2 + 2 |p_E|\,|k_E|\cos\theta_1+M_0^2}=\nonumber\\
&&=\frac{\pi}{4 k_E^2 p_E^2} \left(
k_E^2+p_E^2+M_0^2-\sqrt{\big(k_E^2+p_E^2+M_0^2\big)^2-4 k_E^2 p_E^2}
\right).
\label{Eq:iden2}
\eea

In contradistinction to the previous type of functions, when counterterms are used, all momentum-dependent functions are cut at the maximal value of the modulus of the four-momenta, that is at the physical cutoff $\Lambda,$ irrespective of the fact that they are stored or not stored on the grid (in the latter case their expression is known analytically). We mention here that cutting the sum of momenta in the case of the bubble integral and as a consequence in all the double integrals encountered seems to be the cleanest way to
proceed. In fact, in the case of solving the set of finite
equations we could afford to simplify the calculation, by cutting only
the loop-momenta in the integrals involving the propagator
$\Gv(P)\equiv G_0(P)$ which is not stored on the grid, because the
integrals are all convergent in the UV, and in consequence the
corrections are suppressed. We will see indeed that the solution of
the finite equations nicely agrees with that obtained by solving the
equations with counterterms. Moreover, cutting only the loop-momenta,
when solving the set of equation containing the counterterms, produces
some small differences compared to the case when the sum of momenta is
also cut, at least up to the largest cutoff we investigated, and in
consequence is not a viable method for obtaining accurate numerical
results.

\subsection{Algorithm for solving the finite propagator equation and the equation of state}
\label{ss:Euclidean-algorithm}}

We will solve in Euclidean space the explicitly finite equation of state (\ref{Eq:EoS-fin}) and the propagator equation pieces (\ref{Eq:Sigma0-final}) and (\ref{Eq:sigma-r}) obtained before.
A Wick rotation is performed in every integral appearing in
our equations: with the continuation 
$k_0\to i k^4$ for example the tadpole integral reads 
$\int_K G(k_0,\k)\to i\int_{k_E} G(i k^4,\k)=\int_{k_E} \Delta(k_E)$
where the Euclidean four-momenta is $k_E=(\k,k^4)$, such that 
$k_E^2=\k^2 + k_4^2$ and the Euclidean propagator is defined as
$\Delta(k_E)=1/(k_E^2+\Sigma(k_E)).$

Using the expansion (\ref{Eq:prop}) of the propagator around the
vacuum propagator $\Dv(k_E)=1/(k_E^2+M_0^2)$ and the 
definition (\ref{Eq:G-mat}), the matter and regular parts of the 
Euclidean propagator reads
\bea
\Dm(k_E)&=&\frac{1}{k_E^2+M_0^2+\SigmaO(k_E)
+\Sigmar(k_E)}-\frac{1}{k_E^2+M_0^2},
\label{Eq:eucl-prop-m}
\\
\Dr(k_E)&=&\Dm(k_E)+\frac{\SigmaO(k_E)}{(k_E^2+M_0^2)^2}.
\label{Eq:eucl-prop-r}
\eea
From (\ref{Eq:Sigma0-final}) and (\ref{Eq:sigma-r}), the two pieces of the self-energy which appear above are given by 
\bea
\SigmaO(k_E)&=&m^2-M_0^2+\lambda\Big(1+\lambda I_F^{\vac}(k_E)\Big)\frac{v^2}{2}
+\frac{\lambda}{2}\int_{k_E}\Dr(k_E), 
\label{Eq:sigma0-eucl}
\\
\Sigmar(k_E)&=&
\lambda^2 v^2 I^\textrm{(mv)}(k_E)+\frac{\lambda^2v^2}{2} I^\textrm{(mm)}(k_E),
\label{Eq:sigmar-eucl}
\eea
where the ``matter-vacuum'' and ``matter-matter'' bubble integrals are
defined as
\begin{subequations}
\bea
I^\textrm{(mv)}(k_E)&=&-\int_{p_E}\Dm(p_E)\Dv(p_E+k_E),\\
I^\textrm{(mm)}(k_E)&=&-\int_{p_E}\Dm(p_E)\Dm(p_E+k_E).
\eea
\label{Eq:Imv_Imm}
\end{subequations}
The finite part of the vacuum bubble integral, $I_F^{\vac}$, appearing
in (\ref{Eq:sigma0-eucl}), is defined in (\ref{4-I-fin}).
After analytical continuation to Euclidean space, using (\ref{Eq:mid1}) and (\ref{Eq:Lfunc}) it has the following
expression (recall that $G_0\equiv \Gv$):
\be
I_F^{\vac}(k_E)=\frac{1}{16\pi^2}\Bigg[\sqrt{1+\frac{4M_0^2}{k_E^2}}
\ln\frac{\sqrt{1+\frac{4M_0^2}{k_E^2}}+1}
{\sqrt{1+\frac{4M_0^2}{k_E^2}}-1}-2\Bigg].
\label{Eq:IF_eucl}
\ee
Note that a self-consistent equation for the regular part $\Dr$ can be
obtained in principle if we substitute the expression of $\Dm$ from
(\ref{Eq:eucl-prop-m}) into (\ref{Eq:eucl-prop-r}) because $\SigmaO$
contains the integral of $\Dr$ and $\Dm$ appearing in the integrals of
$\Sigmar$ can be expressed in terms of $\Dr$ and $\SigmaO.$
Nevertheless, we store on the grid two quantities, $\Dm$ and $\Dr,$
which have to be solved simultaneously with the equation of state
(\ref{Eq:EoS-fin}), which in Euclidean space reads
\be
0=m^2+\frac{\lambda}{6}v^2+\frac{\lambda}{2}\int_{k_E} 
\Big(1+\lambda I_F^{\vac}(k_E)\Big)\Dr(k_E)+
\frac{\lambda^2}{6}\big(3 S^\textrm{(mmv)} + S^\textrm{(mmm)}\big),\\
\label{Eq:EoS-fin-eucl}
\ee
where the corresponding pieces of the setting-sun integral are defined
as
\begin{subequations}
\bea
S^\textrm{(mmv)}&=&-\int_{p_E}\int_{k_E}\Dm(p_E)
\Dm(k_E) \Dv(p_E+k_E), \\
S^\textrm{(mmm)}&=&-\int_{p_E}\int_{k_E} 
\Dm(p_E)\Dm(k_E) \Dm(p_E+k_E).
\eea
\end{subequations}
Using (\ref{Eq:iden2}) in the first term of (\ref{Eq:sigmar-eucl}) 
and for the setting-sun contribution to the equation of state 
(\ref{Eq:EoS-fin-eucl}) which contains one vacuum propagator we obtain
\bea
\label{Eq:I_mv}
I^\textrm{(mv)}(k_E)&=&
-\frac{1}{16\pi^2 k_E^2}\int_0^L d p\,p \,\Dm(p) \nonumber\\
&&\times\left(k_E^2+p^2+M_0^2- \sqrt{\big(k_E^2+p^2+M_0^2\big)^2-4k_E^2p^2}
\right), \\
S^\textrm{(mmv)}&=&
-\frac{1}{128\pi^4}\int_0^L d k\,k\,\Dm(k)
\int_0^L d p\,p\, \Dm(p) \nonumber\\
&&\times\left(k^2+p^2+M_0^2-\sqrt{\big(k^2+p^2+M_0^2\big)^2-4 k^2 p^2}\right). 
\eea
For the second term of (\ref{Eq:sigmar-eucl}) and the setting-sun 
contribution to (\ref{Eq:EoS-fin-eucl}) which contains only matter 
propagators we use (\ref{Eq:iden1}) and the results of the angular
integrations are
\bea
I^\textrm{(mm)}(k_E)&=&
-\frac{1}{8\pi^3 k_E^2}\int_0^L d p\,p\,\Dm(p)
\int_{|p-|k_E||}^{\textrm{min}(p+|k_E|,L)} 
d q\,q\,\Dm(q) \nonumber\\
&&\times\sqrt{4 p^2 q^2-\big(p^2+q^2-k_E^2\big)^2}, \\
S^\textrm{(mmm)}&=&-\frac{1}{64\pi^5}\int_0^L d k\,k\,\Dm(k)\int_0^L 
d p\,p\,\,\Dm(p)\nonumber\\
&&\times\int_{|p-k|}^{\textrm{min}(p+k,L)} d q\,q 
\sqrt{4 p^2q^2-(p^2+q^2-k^2)^2}
\Dm(q).\ \ \ 
\eea

Our iterative algorithm for solving the coupled set of equations for
$\Dr,$ $\Dm,$ and $v$ goes as follows. We start at {\it zeroth}
order by neglecting all the integrals in the self-energy and the field
equation, including the finite part of the bubble $I_F^\vac.$ By doing
so, from (\ref{Eq:EoS-fin-eucl}), (\ref{Eq:eucl-prop-m}), and
(\ref{Eq:eucl-prop-r}) we obtain:
\bea
&&v_0=\sqrt{-\frac{6m^2}{\lambda}},\ \ \quad 
\Dm_0(k_E)=-\frac{\left(m^2+\frac{\lambda}{2}v^2-M_0^2\right)}{(k_E^2+M_0^2)(k_E^2+m^2+\frac{\lambda}{2}v^2)},\nonumber\\
&&\Dr_0(k_E)=\frac{\left(m^2+\frac{\lambda}{2}v^2-M_0^2\right)^2}{(k_E^2+M_0^2)^2(k_E^2+m^2+\frac{\lambda}{2}v^2)}.
\eea
Then, starting from the {\it first} order, the iteration is done in two
steps.  At a generic order $n\ge 1$ we first upgrade
$\Dm_n$ and $\Dr_n$ on the grid using (\ref{Eq:eucl-prop-m}) and
(\ref{Eq:eucl-prop-r}), respectively, where the self-energy, which
includes now $I_F^\vac,$ is calculated with the lower order
quantities: $v_{n-1},$ $\Dm_{n-1},$ $\Dr_{n-1}.$ As a second step of
the {\it nth} iteration, we upgrade the vacuum expectation value by
calculating $v_{n}$ from (\ref{Eq:EoS-fin-eucl}) with the integrals
done with the already upgraded $\Dm_n$ and $\Dr_n.$ This procedure is
repeated until the iteration converges. \newline

\subsection{Algorithm for solving the propagator equation and equation of state using counterterms}

If we decide to solve the propagator equation and the equation of state using counterterms, whose role in this equations is to cancel the divergent part of the integrals, then we need a method to determine them. Such a method was developed in \cite{patkos08}, and we review it here because it will be slightly modified for numerical reasons.  In terms of the remaining finite part of the tadpole $T_{\rm F}[G]$ and bubble integral $I_{\rm F}(P,G)$ the propagator equation (\ref{Eq:full_prop}) reads
\be
i G^{-1}(p)=p^2-M^2-\frac{1}{2}\lambda^2 v^2 I_{\rm F}(p,G), \qquad
M^2=m^2+\frac{\lambda}{2}v^2+\frac{\lambda}{2}T_{\rm F}[G].
\ee
Using the expansion of $G$ around $G_0$
\be
G(p)=G_0(p)-i G_0^2(p)\left(
M^2-M_0^2+\frac{1}{2}\lambda^2 v^2 I_{\rm F}(p,G)\right)+\dots \,,
\ee
the tadpole, bubble, and setting-sun integrals can be explicitly
decomposed into divergent and finite parts (see Ref.~\cite{patkos08} 
for details):
\begin{subequations}
\bea
I(p,G)&=&T_d^{(0)}+I_{\rm F}(p,G),
\label{Eq:I_decomp}
\\
T[G]&=&T_d^{(2)}+(M^2-M_0^2) T_d^{(0)}+\frac{1}{2}\lambda^2 v^2 T_d^{(I)}
+T_{\rm F}[G],
\label{Eq:T_decomp}
\\
S(0,G)&=&S_0(0)+3 T_d^{(0)} T_{\rm F}[G]+3 (M^2 - M_0^2)\left(
\big(T_d^{(0)}\big)^2+T_d^{(I)}\right)
\nonumber\\
&&+\frac{3}{2}\lambda^2 v^2 \left(T_d^{(I)} T_d^{(0)}+T_d^{(I,2)}\right)
+S_{\rm F}(0,G),
\label{Eq:SS_decomp}
\eea
\label{Eq:TBSS_decomp}
\end{subequations}
\vspace{-2mm}   

\noindent
where the divergent integrals
$T_d^{(2)},T_d^{(0)},T_d^{(I)},T_d^{(I,2)}$ and $S_0(0)$ were already defined in (\ref{Eq:Td2}), (\ref{Eq:Td0}), (\ref{Eq:TdI}), (\ref{Eq:TdI2}) and (\ref{Eq:S00}), respectively.  Plugging the decomposed
expressions of the integrals (\ref{Eq:TBSS_decomp}) into the
propagator equation and the equation of state, after subtracting the corresponding explicitly finite equations from them, written in terms of the renormalized parameters $m^2,\lambda$ and the finite part of the
integrals, we obtain relations between the counterterms and the
divergent integrals. Requiring that the divergent coefficient of the $v^2,$ $v^0,$ and $T_F[G]$ vanish independently, we obtain the counterterms determined in \cite{patkos08}, which were also obtained from appropriate renormalization conditions previously (see (\ref{4-ct1}), (\ref{4-ct2}), (\ref{4-ct3}), (\ref{4-ct4}) and (\ref{Eq:delta-m02})). Actually, these requirements lead to six equations, instead of five. The extra condition arises when the outlined procedure is applied to the coefficient of $v^2$ in the equation of state. We obtain (\ref{Eq:delta-lambda-2b}), but as stated below this equation, its solution for $\delta\lambda_2$ coincides with the one obtained from (\ref{4-ct3}), as it should.

In Euclidean space these counterterms are functions of the 4d
rotational invariant cutoff $\Lambda$ appearing through the integrals
can be calculated explicitly. They read as follows:
\begin{subequations}
\bea
T_d^{(2)}&\!\!\!=\!\!\!&\frac{\Lambda^2}{16\pi^2}\left[1-
\frac{M_0^2}{\Lambda^2}\log\left(1+\frac{\Lambda^2}{M_0^2}\right)\right],
\\
T_d^{(0)}&\!\!\!=\!\!\!&\frac{1}{16\pi^2}\left[
\frac{\Lambda^2}{\Lambda^2+M_0^2}-\log\left(1+\frac{\Lambda^2}{M_0^2}\right)
\right],\\
T_d^{(I)}&\!\!\!=\!\!\!&\frac{1}{8\pi^2}\int_0^\Lambda d k\frac{k^3}{(k^2+M^2_0)^2}
C_\Lambda[\Dv,\Dv](k)-\big(T_d^{(0)}\big)^2,\\
T_d^{(I,2)}&\!\!\!=\!\!\!&-\frac{1}{8\pi^2} \int_0^\Lambda d k\frac{k^3}{(k^2+M^2_0)^2}
\Big(C_\Lambda[\Dv,\Dv](k)\Big)^2\nonumber\\
&&- 2 T_d^{(I)} T_d^{(0)} 
- \big(T_d^{(0)}\big)^3,\\
S_0(0)&\!\!\!=\!\!\!&-\frac{1}{8\pi^2}\int_0^\Lambda d k\frac{k^3}{(k^2+M_0^2)} C_\Lambda[\Dv,\Dv](k),
\eea
\label{Eq:div_ints_euclid}
\end{subequations}
\vspace{-2mm}   

\noindent
where for $C_\Lambda[\Dv,\Dv](k)$ we have to use (\ref{Eq:iden1})
with $\Dv(k)=1/(k^2+M_0^2)$ and with $L$ replaced by the physical
cutoff $\Lambda.$ 

The counterterms $\delta m_0^2, \delta m_2^2, \delta \lambda_0, \delta \lambda_2, \delta \lambda_4$ can be
used to solve iteratively the set of Euclidean equations
\begin{subequations}
\bea
\label{eq:Sigma}
\Sigma(p)&\!\!\!=\!\!\!&m^2+\delta m_2^2+\frac12(\lambda+\delta \lambda_2)v^2
+\frac12(\lambda+\delta \lambda_0)T(\Sigma)+
\frac12 \lambda^2 v^2 I(p,\Sigma), \\
\label{eq:EoS}
0&\!\!\!=\!\!\!&m^2+\delta m_0^2+\frac16(\lambda+\delta \lambda_4)v^2+
\frac12(\lambda+\delta \lambda_2)T(\Sigma)+\frac{\lambda^2}{6}S(0,\Sigma),
\eea
\end{subequations}
in which they have to cancel the divergences of tadpole, bubble
and setting-sun integrals, which in Euclidean space read as
\bea
&&T[\Sigma]=\frac{1}{8\pi^2}\int_0^\Lambda d k\, k^3 \Delta(k),\quad
I(p,\Sigma)=-C_\Lambda[\Delta,\Delta](p),\nonumber\\
&&S(0,\Sigma)=-\frac{1}{8\pi^2}\int_0^\Lambda d k\,k^3
\Delta(k) C_\Lambda[\Delta,\Delta](k).
\eea

However, since the counterterms are designed to cancel divergences of integrals produced by the full (i.e. iteratively converged) propagator, they cancel the complete cutoff dependence of $T[\Sigma],$ $I(p,\Sigma)$ and $S(0,\Sigma)$ only when the solution of this set of equations has converged. This is not a problem in itself, since, as it will be demonstrated, the iterative procedure converges, but the convergence of the solution of the set of equations is rather slow. It is possible to improve the iterative procedure if, instead of using the obtained counterterms, we rederive them at each order of the iteration using the procedure outlined above.  These counterterms, which come from the requirement to have finite equations at each order of the iteration, will evolve during the process of iteration toward the value of the counterterms already obtained. Since, as we will see, the intermediate counterterms are now environment dependent, that is they depend on $v$ and $T_F[\Sigma]$, this convergence can only occur, if $v$ and $\Sigma$ also converge to their appropriate values, which has to be checked numerically. The final counterterms then become environment independent and coincide with the ones determined by (\ref{4-ct1}), (\ref{4-ct2}), (\ref{4-ct3}), (\ref{4-ct4}) and (\ref{Eq:delta-m02}).

In what follows we present the improved iterative procedure which uses
evolving counterterms. At each order the iteration starts by the
upgrade of the self-energy, followed by the upgrade of the vacuum 
expectation value determined from the equation of state.

At {\it zeroth} order of the iteration the field and the self-energy
are equal to their tree-level expressions:
\bea
v_0^2=-\frac{6m^2}{\lambda},\qquad  
\Sigma_0(p)=m^2+\frac{\lambda}{2}v_0^2.
\eea
At this order, since there are no quantum fluctuations at all, there
are no divergences to cancel and in consequence all the counterterms 
are zero, that is
$\delta m_{0,0}^2=\delta m_{2,0}^2=\delta\lambda_{0,0}=\delta\lambda_{2,0}=
\delta\lambda_{4,0}=0,$
where the counterterms also carry the index of the iteration number
because all of them will change during the iteration process.

The general formulas for the {\it nth} order counterterms to be given
below encompass all orders with the following prescription: at
$n=0$ we start from the tree-level values, and formally
$T(\Sigma_n)=I(\Sigma_n)=0$ for $n\in\{-1,-2\}$ and $v_{-1}=v_0.$ Suppose that
the {\it (n-1)th} order of the iteration is done, i.e. $v_{n-1}^2$ and
$\Sigma_{n-1}(p)$ together with the corresponding counterterms are
already known, then at {\it nth} order we start by upgrading the
self-energy using (c.f. (\ref{eq:Sigma})):
\bea
\label{Eq:Sigma-itern}
\Sigma_n(p)&=&m^2+\delta m_{2,n}^2+\frac{\lambda}{2}v_{n-1}^2\nonumber\\
&&+\frac{\delta \lambda_{2,n}}{2}v_{n-2}^2+\frac{\lambda+\delta \lambda_{0,n}}{2}T(\Sigma_{n-1})+\frac{\lambda^2}{2}v_{n-1}^2 I(p,\Sigma_{n-1}).
\eea
The finite part of the tadpole and bubble are defined through
\bea
T(\Sigma_{n-1})&=&T_d^{(2)}+\Big(m^2+\frac{\lambda}{2}v_{n-2}^2+\frac{\lambda}{2}T_{F}(\Sigma_{n-2})-M_0^2\Big)T_d^{(0)}\nonumber\\
&&+\frac{\lambda^2 v_{n-2}^2}{2}T_d^{(I)}+T_F(\Sigma_{n-1}), 
\label{eq:tad_n-1}
\\
I(p,\Sigma_{n-1})&=&T_d^{(0)}+I_F(p,\Sigma_{n-1}).
\eea
The renormalization procedure requires the independent vanishing of the 
overall divergence, field-dependent subdivergence and $T_F$-dependent 
subdivergence:
\begin{subequations}
\label{Eq:cts-nb}
\bea
0&=&\delta m_{2,n}^2+\frac{\lambda+\delta\lambda_{0,n}}{2}\Big(T_d^{(2)}+(m^2-M_0^2)T_d^{(0)}\Big),\\
0&=&\delta \lambda_{2,n}
+\frac{\lambda}{2}T_d^{(0)}(\lambda+\delta \lambda_{0,n})
+\frac{\lambda^2}{2} T_d^{(I)} (\lambda+\delta \lambda_{0,n}) (1-\delta_{n1})\nonumber\\
&&+\lambda^2 T_d^{(0)}\frac{v_{n-1}^2}{v_{n-2}^2},\\
0&=&(\lambda+\delta \lambda_{0,n})\frac{\lambda}{2}T_d^{(0)}T_{F}(\Sigma_{n-2})
+\delta \lambda_{0,n}T_F(\Sigma_{n-1}).
\eea
\end{subequations}

The counterterms $\delta m_{2,n}^2$, $\delta \lambda_{2,n}$ and
$\delta \lambda_{0,n}$ are obtained from (\ref{Eq:cts-nb}), so that
$\Sigma_n(p)$ is finite and can be calculated. Note that for $n=1$ the
second term in the equation for $\delta\lambda_{2,n}$ vanishes as
it should because in this case there is no divergence due to the
bubble, that is there is no $T_d^{(I)}$ in $T(\Sigma_0)$, whose
expression is
\be
T(\Sigma_0)=T_d^{(2)}+\Big(m^2+\frac{\lambda}{2}v_0^2-M_0^2\Big)T_d^{(0)}+T_F(\Sigma_0).
\ee
Note also that for $n=1$ we have $\delta\lambda_{0,1}=0$, since 
$T_F(\Sigma_{-1})=0$ by convention.

We continue with the determination of $v_n^2$ from the equation of state, which at {\it nth} order of the iteration reads
\bea
\label{Eq:EoS-iter-n}
0&\!\!\!=\!\!\!&m^2+\delta m_{0,n}^2+\frac{\lambda}{6}v_n^2+\frac{\delta \lambda_{4,n}}{6}v_{n-1}^2+\frac{\lambda+\delta \tilde\lambda_{2,n}}{6}T(\Sigma_{n})+\frac{\lambda^2}{6}S(0,\Sigma_{n}),
\eea
where $T(\Sigma_n)$ is given by (\ref{eq:tad_n-1}) with $n-1\to n$ and
the setting-sun is
\bea
\!\!\!S(0,\Sigma_n)&\!\!\!=\!\!\!&S_0(0)+3 T_d^{(0)} T_{\rm F}[\Sigma_n]+3 \Big(m^2 
+\frac{\lambda}{2}v_{n-1}^2+\frac{\lambda}{2}T_{F}(\Sigma_{n-1})-M_0^2\Big)\nonumber\\
&\!\!\!\times\!\!\!&\left(\big(T_d^{(0)}\big)^2+T_d^{(I)}\right)+\frac{3}{2}\lambda^2 v^2_{n-1} \left(T_d^{(I)} T_d^{(0)}+T_d^{(I,2)}\right)
+S_{\rm F}(0,\Sigma_n).
\eea
Applying the same requirements as for the self-energy, we obtain:
\begin{subequations}
\label{Eq:ct-iter-n}
\bea
0&\!\!\!=\!\!\!&\delta m_{0,n}^2+\frac{\lambda+\delta \tilde\lambda_{2,n}}{2}\Big(T_d^{(2)}+(m^2-M_0^2)T_d^{(0)}\Big)\nonumber\\
&&+\frac{\lambda^2}{2}(m^2-M_0^2)\Big((T_d^{(0)})^2+T_d^{(I)}\Big)+\frac{\lambda^2}{6}S_{0}(0),\quad \\
0&\!\!\!=\!\!\!&\delta \lambda_{4,n}+\frac{3\lambda}{2}(\lambda+\delta \tilde\lambda_{2,n}+\lambda^2 T_d^{(0)})\Big(T_d^{(0)}+\lambda T_d^{(I)}\Big)\nonumber\\
&&+\frac{3\lambda^3}{2}\Big(T_d^{(I)}+\lambda T_d^{(I,2)}\Big), \\
0&\!\!\!=\!\!\!&\delta \tilde\lambda_{2,n}+\lambda^2 T_d^{(0)}+\frac{T_{F}(\Sigma_{n-1})}{T_F(\Sigma_{n})}\nonumber\\
&&\times\Big[\frac{\lambda}{2}(\lambda+\delta \tilde\lambda_{2,n})T_d^{(0)}+\frac{\lambda^3}{2}\Big((T_d^{(0)})^2+T_d^{(I)}\Big)\Big].
\eea
\end{subequations}
Note that a new counterterm $\delta \tilde{\lambda}_2$ has been
introduced instead of $\delta \lambda_2$, since in the $n$th iteration
$\delta \lambda_{2,n}$ does not eliminate properly the appropriate
subdivergences of the equation of state.  Extracting $\delta m_{0,n}^2$,
$\delta \lambda_{4,n}$ and $\delta\tilde \lambda_{2,n}$ from equations
(\ref{Eq:ct-iter-n}) we can obtain $v_n^2$ from
(\ref{Eq:EoS-iter-n}). The counterterm $\delta\tilde \lambda_{2,n}$
will be equal to $\delta\lambda_{2,n}$ determined from the propagator
equation only when the iteration converged, so that
$T_F(\Sigma_{n-1})/T_F(\Sigma_{n})\to 1.$ In this limit and when
$v_n/v_{n-1}\to 1$ the counterterms will converge to the values
obtained from the renormalization conditions. This happens only
asymptotically and in practice the iteration stops when the relative
change in the values of the field and the self-energy are smaller than
some given value dictated by our requirement of accuracy. The same
iterative procedure is used when the counterterms are not evolved
during the iteration, but used as determined from the renormalization conditions: we first upgrade the self-energy and
then the equation of state and check for their convergence.

\subsection{Numerical results}

In what follows we present the results on the numerical solution of
the propagator equation and the equation of state in Euclidean space.
As it was stressed, we wanted to build up a demanding numerical
framework, in which very highly accurate solutions can be obtained.
This concerns in particular the equations containing the counterterms
explicitly, since in these the increasing cutoff requires very large
cancellations, which is hard to be handled numerically. Previous attempts
on obtaining the cutoff independence of renormalized quantities for
large cut offs (e.g. \cite{arrizabalaga06}) could be only moderately
achieved, therefore such a numerical framework in question also
carries theoretical importance as well.

\begin{figure}[!t]
\begin{center}
\raisebox{0.35cm}{
\includegraphics[bb=316 65 440 291,scale=0.6]{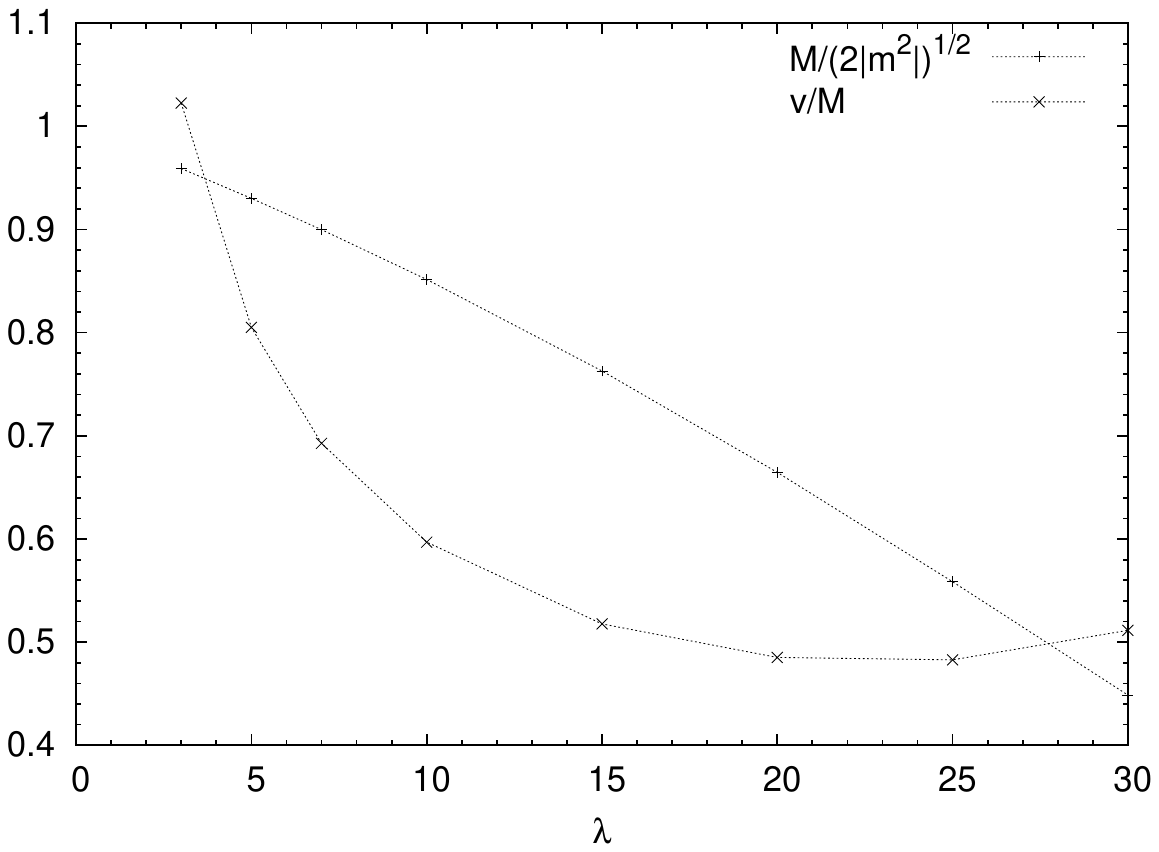}}
\includegraphics[bb=100 65 185 291,scale=0.642]{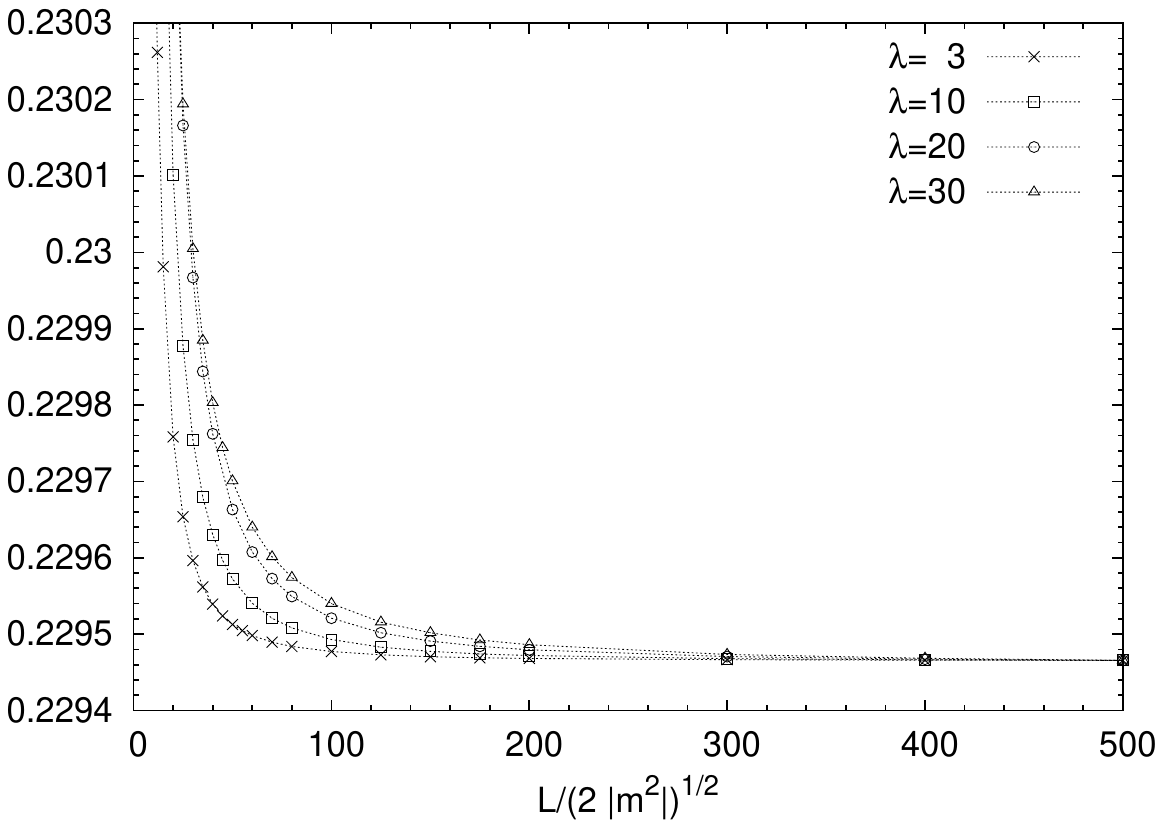}
\caption{The solution of the explicitly finite equations.  Left panel: $v/M$ and $M=\Sigma^{\frac{1}{2}}(p=0)$ as
  function of the coupling obtained for $L/\sqrt{2|m^2|}=500$.  Right
  panel: The converged value of $v$ as function of the maximal value
  of the modulus of the momentum stored on the grid for different
  values of the coupling constant $\lambda$.  An appropriate constant
  was subtracted from the values of $v$ obtained for $\lambda=3,$
  $10,$ and $20$ such that the resulting values coincide at
  $L/\sqrt{2|m^2|}=500$ with the value of $v$ obtained for
  $\lambda=30.$
\label{fig:L_dep}}
\end{center}
\end{figure}

When solving the explicitly finite set of equations the maximal value
of the modulus of the four-momentum stored on the grid is
$L/\sqrt{2|m^2|}=500,$ while in the case of solving equations with
counterterms the highest value of the modulus stored, that is the 4d
physical cutoff, is $\Lambda/\sqrt{2|m^2|}=200.$ In the numerics, we choose $|2m^2|$ as the unit mass, while the value of the renormalization scale is $M_0/\sqrt{2|m^2|}=2.1$. Actually, to have an idea on the physical scale, we can require to have $v/\Sigma^\frac{1}{2}(p=0)\simeq 0.66,$ which is the value of the ratio $f_\pi/m_\pi$ of the pion decay constant to the pion mass for $f_\pi=93$~MeV and $m_\pi=140$~MeV, and $\Sigma^\frac{1}{2}(p=0)=m_\pi.$ Then, from the left panel of Fig.~\ref{fig:L_dep} we see that the first requirement can be meet by choosing $\lambda\simeq 8,$ for which value the second one selects $\sqrt{2|m^2|}\simeq 158$~MeV. In units for which $m^2=-0.5$, the step size on the grid in general was $0.01.$ We have checked that by halving the step size the change in the results is comparable with or smaller than the precision required by our convergence criterion. The iteration was stopped when the change in the vacuum expectation value was smaller than $10^{-6}.$ The precision of the numerical integration routines used was higher than that, since we required relative error bounds between $10^{-6}$ and $10^{-8}$.

First, let us discuss the solution of the explicitly finite equations.
The field expectation value and the self energy at zero momentum can
be seen in Fig.~\ref{fig:L_dep}. In the right panel we see the converged
solution of the equation of state obtained for various values of the
coupling as function of $L,$ the maximal value of the modulus stored
on the grid.  As explained in the figure caption, $v$ was shifted
appropriately in order to make all four curves meet at
$L/\sqrt{2|m^2|}=500.$ In all cases a plateau can be observed for
increasing values of $L.$ However, with increasing value of the
coupling $\lambda$ the plateau starts at a higher $L.$ With the same
convergence criterion the number of iterations increases with
increasing $\lambda$ and for $\lambda=30$ we need twice as many
iterations until convergence occurs compared to the $\lambda=3$ case.
Interestingly, at a given $\lambda,$ the number of iterations does not
depend on $L.$

\begin{figure}[!t]
\begin{center}
\includegraphics[bb=160 55 315 301,scale=0.95]{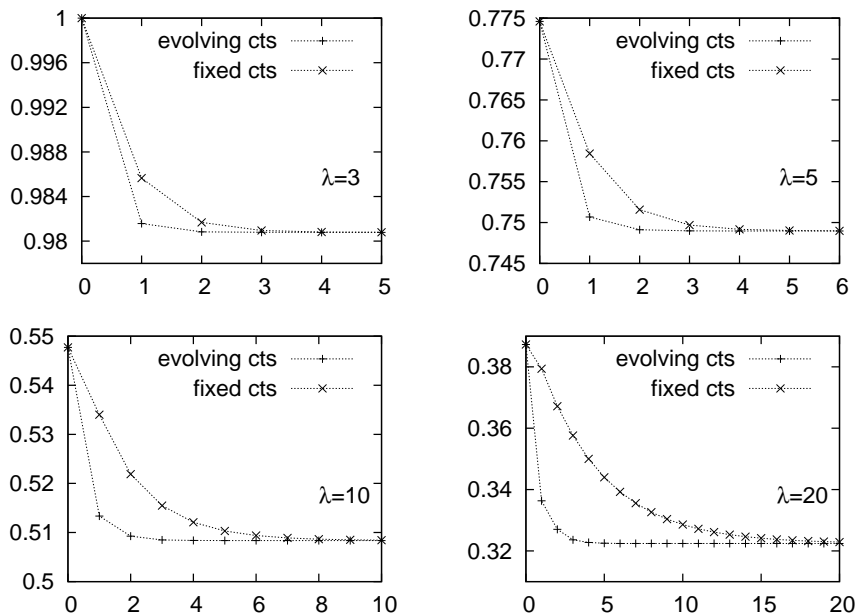}
\caption{A comparison between the convergence of the iterative
  algorithms using fixed and evolving counterterms for a fixed value of
  the physical cutoff: $\Lambda/\sqrt{2|m^2|}=200.$ The value
  of $v/\sqrt{2|m^2|}$ is shown as a function of the number of iterations for different values of the coupling.
\label{fig:comp1}}
\end{center}
\end{figure}

Next, let us focus on the numerical solution of the equations containing
explicitly the counterterms. We show, that we can get the very same results for the field expectation value and the self energy, as which obtained from the finite equations. We also quantify the extent of the
achievable improvement given by the use of the iteratively evolved
coupling and mass counterterms, as compared to the solution of the
equations which use the counterterms originally derived in \cite{patkos08}.
In order to see this, we compare in Fig.~\ref{fig:comp1} the number of
iterations needed for the convergence of $v$ at four different coupling
constants.  We use a reasonably large cutoff
$\Lambda/\sqrt{2|m^2|}=200$, around which the convergent results are
expected to show cutoff independence within some desired accuracy.
We observe the advantage of evolving the counterterms even for
lower values of the renormalized coupling. For larger values
($\lambda\gtrsim 10$) the improvement is significant, as we can see
from the large iteration number needed by the algorithm using fixed
counterterms.  Therefore, it is expected in general that solving
self-consistent equations iteratively in a fully non-perturbative
regime with fixed counterterms is not efficient from a numerical point
of view. For large couplings, $\lambda\gtrsim 10,$ the number of
iterations needed in case of using fixed counterterms are at least a
factor of two larger than the number of iterations needed for the
explicitly finite equations to converge.

\begin{figure}[!t]
\begin{center}
\includegraphics[bb=336 55 440 301,scale=0.58]{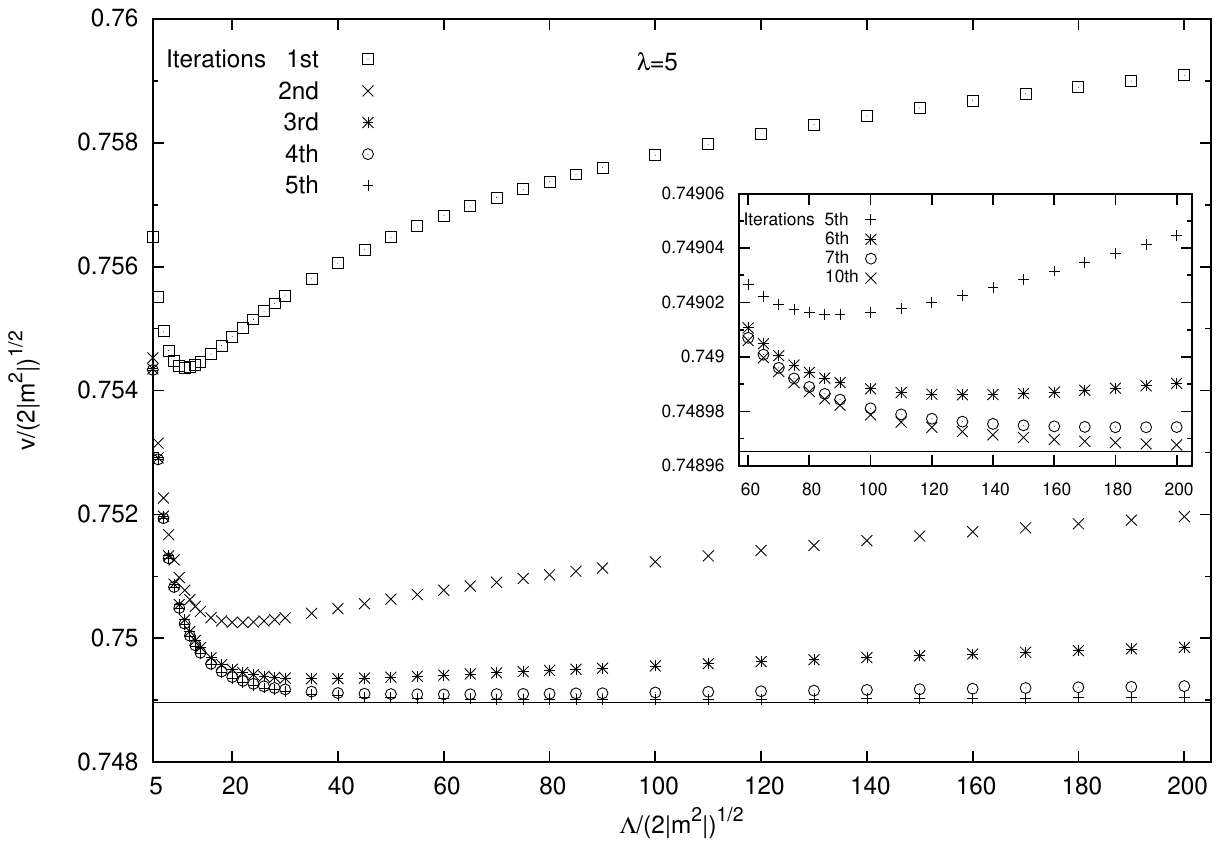}
\includegraphics[bb=80 55 125 301,scale=0.58]{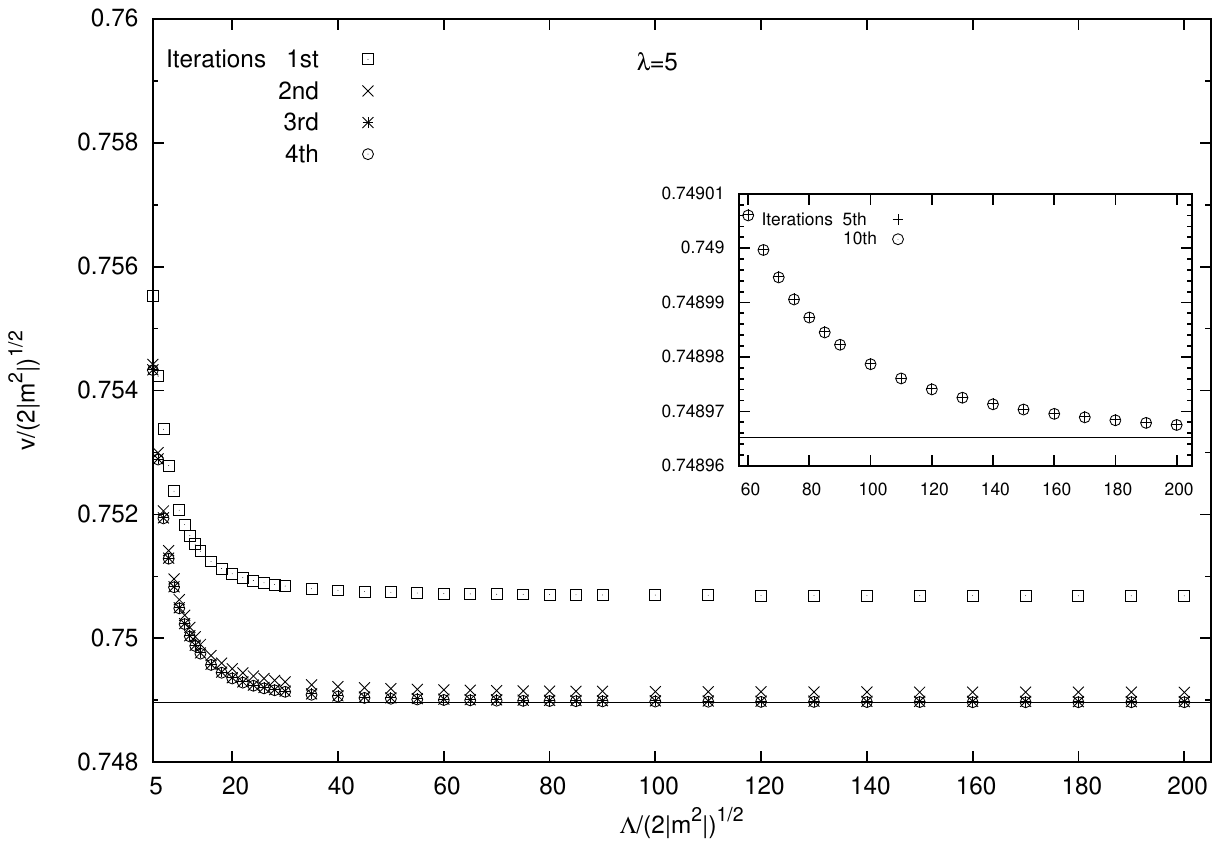}
\caption{
A comparison between the convergence rate of the iterative algorithms
using fixed (left panel) and evolving counterterms (right panel)
as reflected in the change of the vacuum expectation value at
different values of the physical cutoff $\Lambda$. The horizontal line
represents the solution of the explicitly finite system of equations
obtained with $L/\sqrt{2|m^2|}=500,$ where $L$ is the maximal value of
the modulus of the momentum stored on the grid. The insets show the
convergence at large iteration numbers.
\label{fig:comp2}}
\end{center}
\end{figure}

\begin{figure}[!t]
\begin{center}
\includegraphics[bb=145 55 315 301,scale=0.85]{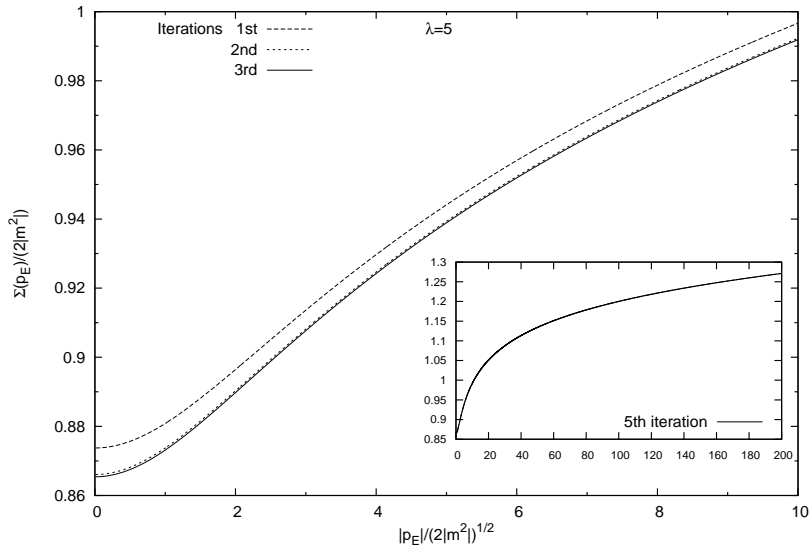}
\caption{
The convergence of the full self-energy obtained for $\lambda=5$ with
evolving counterterms for $\Lambda/\sqrt{2|m^2|}=200.$ The inset
shows the behaviour of the self-energy at large momenta.
\label{fig:SE}}
\end{center}
\end{figure}

In Fig.~\ref{fig:comp2} we compare for $\lambda=5$ the ways how a
cutoff independent solution is reached using fixed and evolving
counterterms.  The left panel shows that with increasing values of
$\Lambda$, a plateau in the $v(\Lambda)$ functions is approached
rather slowly, when the fixed counterterms of \cite{patkos08} are used,
despite the fact that the coupling is not large. Looking at the plot
with values of $\Lambda$ spanning a large range, after
several iterations a plateau is finally reached. However, the
structure in $v(\Lambda),$ better seen in the inset which uses a
smaller $\Lambda$-range, shows that the convergence is not uniform:
more iterations are needed for larger cutoffs to get close to the true
solution, that is the solution of the explicitly finite equations
(horizontal line).  This non-uniformity of the convergence in
$\Lambda$ is more pronounced for larger values of the coupling
constant. Contrary to this, the right panel of Fig.~\ref{fig:comp2}
shows a uniform convergence in $\Lambda$ when the iterative algorithm
uses evolving counterterms.  Since by the very construction of the
evolving counterterms a plateau is observed already for moderate
values of the cutoff at every order of the iteration, we see again
that with this improved approach the final, convergent result is
obtained in a more efficient way.

The convergence of the self-energy can be seen in Fig.~\ref{fig:SE}.
The function converges smoothly, i.e. for all momentum values the
self-energy reaches its converged value almost simultaneously. Note
however that, the convergence for $p_E\simeq 0$ is a slightly slower as
compared to larger values of the momenta.

\section{Concluding remarks}

In this chapter we presented an approach which can be used to obtain a
very accurate
numerical solution of the self-consistent propagator equation coupled
to the equation of state, both derived from the two-loop level
approximation of the 2PI effective action in the $\phi^4$ model. We
showed that the renormalization method developed in \cite{patkos08},
which removes the divergences using the minimal subtraction scheme
introduced in Chapter 3 is equivalent
to imposing nontrivial renormalization conditions on the self-energy,
the curvature of the effective potential and on kernels of
Bethe-Salpeter equations related to these quantities. The use of
renormalization conditions also allowed us to construct explicitly
finite propagator and field equations.

We compared the convergence of the iterative method applied to solve,
on the one hand, the explicitly finite equations and, on the other
hand, the equations which contains the explicitly calculated
counterterms of \cite{patkos08}. The very same results were obtained
numerically, which proves the correctness of our analytic study on
renormalization. It turned out that, especially for larger coupling
constants ($\lambda \gtrsim 10$), the convergence rate of the
algorithm which uses the counterterms is quite poor. This should not
come as a surprise given that the counterterms were determined based
on the asymptotic behavior of the propagator. Within an iterative
procedure the asymptotics develops progressively and is reached only
when the solution converged.  Only then the divergences present in the
equations match the counterterms constructed to cancel them, at
intermediate steps of the iteration process the use of fixed
counterterms results in an oversubtraction.  In order to cure this, we
managed to develop a different algorithm which rederives the
counterterms at every step of the iterative procedure from the
condition that the counterterms cancel the divergent part of the
integrals at every order of the iteration. This method, which uses
evolving counterterms turned out to be very effective in obtaining the
converged solution within a few iterative steps also for larger
couplings.

The presented approach and some of the methods applied in our present
work may be directly extended, first to a zero temperature numerical
solution of the model in Minkowski space, then to a finite
temperature study. It is expected that the
solution which can be obtained with the method described here could
serve as a good benchmark for other, even more effective methods used
in finite temperature studies (see e.g.
\cite{Berges:2004hn,arrizabalaga06}), and will represent a good basis
for obtaining highly accurate and/or numerically controlled solutions
of the 2PI approximations in more complicated theories as well.

\chapter{An approximate solution of the U(N)$\times$U(N) model in large-N}

Universal features of finite temperature and finite density variation of the QCD ground state realizing the $U(3)\times U(3)$ approximate chiral symmetry were investigated with the help of 
the corresponding meson model \cite{pisarski84}, introduced already in Chapter 2, for arbitrary flavor numbers. There were numerous attempts to treat quantitatively the finite temperature restoration of chiral symmetry of the strongly interacting matter in the framework of this model. A common goal of these investigations is to find out the nature of the phase transition, when the baryonic density is varied. Another central issue is the description of the quark mass dependence of the finite temperature symmetry restoration \cite{lenaghan00,roder03,herpay05}. For its investigation one couples constituent quarks carrying baryon number to the meson model \cite{kovacs08,schaefer09,bilic99,schaefer10}.

Optimistically one could say, that the location
of the characteristic points of the QCD phase diagram (i.e. the $T-\mu$ plane) determined with different variants of the model and in different approximations 
do agree with each other and with the results of the lattice field theoretical simulations within a factor of 2. In particular, improved agreement with lattice determinations of the QCD phase diagram were reported, when the Polyakov loop degree of freedom is coupled to the quark-meson model \cite{kahana08,mao10,gupta10}. In realistic approaches the effective models are all strongly coupled, therefore one usually experiences large variations in their predictions, when the simplest mean-field treatments are improved by taking into account quantum fluctuations of the mesons and the constituent quarks \cite{herbst10,marko10}. 

As discussed in Chapter 2, more recent investigations for the three-flavor meson model in the large-$N$ approximation are missing from the literature. Although the $N=\infty$ solution of the $O(N)$ model is the textbook example of the application of the $1/N$ expansion \cite{zinn-justin02}, the leading-order solution of the large-$N$ approximation of the $U(N)\times U(N)$ model is still unknown. There were various attempts to handle the model in the large-$N$ limit, but none of these went beyond the related $O(2N^2)$-symmetric nonlinearities (i.e. corresponding to the $\sim(s^a)^2+(\pi^a)^2$ part of the Lagrangian). A first step in this direction was made with the matrix generalization of the Hubbard-Stratonovich transformation involving the complete quartic potential of the $U(N)\times U(N)$ symmetric model \cite{frei90}. Based on this reformulation of the model, a saddle-point solution was constructed using an auxiliary matrix field \cite{meyer-ortmanns96}. It was {\it conjectured} that this solution describes the leading order behavior of the ground state for large flavor number. The approximate solution was applied directly to the thermodynamics of the three-flavor effective meson model (see also \cite{bilic99}). In the actual solution of the matrix saddle point equation the authors restricted their Ansatz to a condensate proportional to the unit matrix, which reduced once again the quartic potential to its $O(2N^2)$ invariant part. As a result, the part of the vacuum condensate pointing in the algebra space along the direction of the longest diagonal generator remained proportional to the corresponding component of the explicit symmetry breaking external source. To the author's actual knowledge, no published further step was attempted towards the exact large-$N$ analysis of the $U(N)\times U(N)$ symmetric matrix model.

The numerical details of the pattern of chiral symmetry breaking in theories of three-flavored strong interactions are essentially dictated by the observed mesonic spectra \cite{haymaker74,lenaghan00,black02,tornqvist99}. Potentials of the effective models are constructed with the requirement that in the ground state a condensate proportional  
to the $3\times 3$ unit matrix should arise spontaneously, and split the degeneracy of the parity partner states. This fundamental structure is slightly modified by an external source also proportional to the unit matrix, which generates mass for the pseudo-Goldstone pseudoscalar mesons. Guided by these aspects, in the following section first we will use the symmetry breaking pattern $U(N)\times U(N) \rightarrow U(N)$, corresponding to a condensate of the dynamical field which is proportional to the unit matrix. However, a possible parametrization of the linear sigma model has to reflect also some realistic mass information concerning the pseudoscalar sector. This leads us in the second half of this chapter to a more general condensate when $U(N)\times U(N) \rightarrow U(N-1)$ breaking can be realized. This is induced by the presence of the part of the external field proportional to the longest diagonal generator and results the kaon-pion mass splitting via the formation of a supplementary condensate along the same direction of the Lie algebra. 
It is a rather nontrivial question if this component of the condensate
arises fully from the external field, or can evolve spontaneously.

The goal of this chapter is to describe an approximate leading order ($N=\infty$) solution of the linear sigma model with $U(N)\times U(N)$ symmetry, and to obtain the ground state at zero temperature and density. Although the solution takes into account only two-loop contributions of the 2PI effective action, it definitely goes beyond the restricted $O(2N^2)$ symmetric solution. For the restriction of the contributing diagrams it exploits in addition to the large-$N$ expansion an assumption in the mass spectra of the model, concerning the relation of the scalar and pseudoscalar sector. The region in the coupling space, where this assumption is valid can be estimated by determining the spectra from the approximate solution self-consistently. In the first part of the chapter we also present the construction of the renormalized version of this solution in some details, and provide an illustrative investigation of its range of validity.

The idea to impose an extra assumption on the mass spectra stems from the way one deals with the 4 scalars and 4 pseudoscalars defining the $U(2)\times U(2)$ symmetric meson model. As we have seen in Chapter 2, there one simply omits half of the fields, i.e. the 3 components of the scalar-isovector triplet and the pseudoscalar-isoscalar singlet with reference to their higher mass. We shall assume an analogous feature to occur in the spectra of the leading order solution of the $U(N)\times U(N)$ meson model at large-$N$.
Then, introducing a more general symmetry breaking, there will be
splittings in the pseudoscalar masses as well. Based on the ``heaviness'' assumption concerning the scalar mesons, in a first approximation to the $N=\infty$ solution we retain only the quantum fluctuations of the light pion fields.  We do not attempt the anomalous realization of the $U_A(1)$ symmetry, therefore in the
first limiting case all pseudoscalars will have the same mass. 

After finding the propagators of the scalar fields, we can calculate corrections to the pion propagators and condensates arising from the heavy scalar fluctuations. It is possible then to upgrade also the scalar propagators and in principle, we can iterate this procedure until the solution of the large-$N$ $2$-loop 2PI approximation is reached. It will be demonstrated, that the pion fields still obey Goldstone's theorem after the scalar corrections are included. We shall explore the divergence structure of the equation of state, the pion self-energy and the saddle-point equations (SPEs), and determine the counterterm pieces in the effective action necessary for their renormalization.    

For the second part of the chapter we address the question of the determining the ground state of the system. We will describe a general symmetry breaking, then turn to the case corresponding to the $U(N)\times U(N) \longrightarrow U(N-1)$ pattern. Then we construct the $1$-loop effective potential and show that in a large part of the parameter space there are two local minima of the effective potential, in which non-trivial $n$-scalings are found corresponding to the field expectation values. This will result that instead of the usual power series in $n^2$, we will also found some parts of the effective potential which are of odd powers of $n$.

\section{Renormalized solution under the assumption of heavy scalars}

Let us recall the $U(n)\times U(n)$ symmetric Lagrangian from (\ref{2-L_ch2}) and extend it with the explicit symmetry breaking term in the zeroth scalar direction. Taking into account the $n$ scalings of the couplings, we have
\bea
\label{5-Lagr}
{\cal L}&=&\frac12\Big(\partial_{\mu}s^a\partial^{\mu}s^a+\partial_{\mu}\pi^a\partial^{\mu}\pi^a-m^2(s^as^a+\pi^a\pi^a)\Big)\nonumber\\
&-&\frac{g_1}{4n^2}\Big(s^as^a+\pi^a\pi^a\Big)^2-\frac{g_2}{2n}U^aU^a+h_0s^0.
\eea
Before proceeding, we introduce two auxiliary (composite) fields representing the combinations of $U^a=\frac12 d_{abc} (s^bs^c+\pi^b\pi^c)-f_{abc}s^b\pi^c$ and $s^as^a+\pi^a\pi^a$. This can be achieved by adding the following constraints to the Lagrangian:        
\be        
\Delta        
{\cal L}=-\frac{1}{2}\left(X-i\sqrt{\frac{g_1}{2n^2}}(s^as^a+\pi^a\pi^a)        
\right)^2-\frac{1}{2}\left(Y^a-i\sqrt{\frac{g_2}{n}}U^a\right)^2,        
\ee        
and treat $X$ and $Y^a$ as variables.
In the sum ${\cal L}\equiv L+\Delta L$, the $U(n)\times U(n)\rightarrow U(n)$ symmetry breaking pattern corresponds to the shifts    
\be        
s^a\rightarrow s^a+\sqrt{2n^2}v\delta^{a0},\qquad U^a\rightarrow        
U^a+2\sqrt{n}vs^a+n\sqrt{2n}\delta^{a0}v^2.        
\ee        
After the introduction of the auxiliary field variables and        
the shifts, the full Lagrangian has the following form:        
\bea        
L&=&        
\frac{1}{2}\Big[(\partial_\mu s^a)^2+(\partial_\mu\pi^a)^2-m^2\Big(2n^2v^2
+  2\sqrt{2n^2}vs^0+(s^a)^2+(\pi^a)^2\Big)\Big]+\nonumber\\        
&-&\frac{1}{2}X^2-\frac{1}{2}(Y^a)^2+i\sqrt{\frac{g_1}{2n^2}}X\Big(2n^2v^2+        
2\sqrt{2n^2}vs^0+(s^a)^2+(\pi^a)^2\Big)\nonumber\\        
&+&i\sqrt{\frac{g_2}{n}}Y^a(U^a+2\sqrt{n}vs^a+\sqrt{2n^3}\delta^{a0}v^2)+\sqrt{2n^2}h (s^0+\sqrt{2n^2}v),        
\label{Eq:shifted-aux-L}        
\eea          
where $\sqrt{2n^2}h=h_0$. Next, we shortly describe the assumed structure of the solution and introduce some notations.    
The classical constraint equations show that the background $v$        
induces nonzero $X,Y^0$ values and introduces mixing of the 
pair $s^0,X$ and also of $s^a,Y^a$ for every value of the index $a$. 
We construct correspondingly a quantum solution, where the saddle-point        
values of $X, Y^0$ are nonzero, $Y^a=0, a\neq 0$, and only        
the following 2-point functions do not vanish:        
\bea        
&\displaystyle        
[G_{s^0X},\quad G_{s^0Y^0},\quad G_{XY^0},\quad G_{s^0s^0},\quad        
G_{Y^0Y^0},\quad G_{XX}],\nonumber\\        
&        
\displaystyle        
[G_{s^us^v, u=v\neq 0},\quad G_{s^uY^v, u=v\neq 0},        
\quad G_{Y^uY^v, u=v\neq 0}], \qquad G_{\pi^u\pi^v, u=v}.        
\eea      
For arbitrary fields $A(x)$ and $B(y)$, the 2-point function $G_{AB}$ refers to $\bra{0}T(A(x)B(y))\ket{0}$. As in the previous chapters, we represent these propagators in Fourier space.  
The sets in square brackets form mixing sets of fields: there is a 3-dimensional mixing sector, and        
there are $n^2-1$ identical copies of 2-dimensional mixing two-point functions. The $\pi$ sector is diagonal.          
The equations below show degeneracy of the        
$2$-point functions for $u=v\neq 0$, therefore it is convenient to introduce the following short-hand notations:        
\bea        
&        
\displaystyle        
G_{s^0s^0}\equiv G_{s^0}, \quad G_{XX}\equiv G_{X}, \quad G_{Y^0Y^0}\equiv G_{Y^0}, \quad G_{\pi^0\pi^0}\equiv G_{\pi^0},        
\nonumber\\        
&        
\displaystyle        
\quad G_{\pi^u\pi^u}\equiv G_{\pi},\quad G_{s^us^u}\equiv G_s,        
\quad G_{s^uY^u}\equiv G_{sY},\quad G_{Y^uY^u}\equiv G_Y,\qquad        
u\neq 0.        
\eea          
\subsection{Approximate 2PI effective potential}

\begin{figure}[!t]
\centerline{ 
\includegraphics[bb=369 612 326 699]{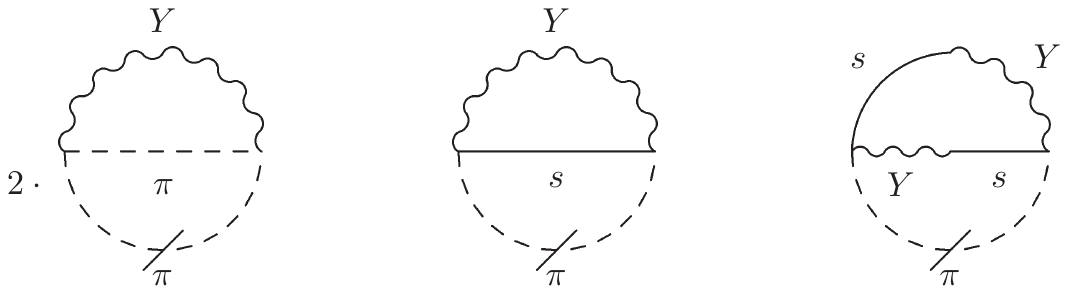}}
\caption{Vacuum diagrams contributing to the leading order equation of $G_\pi$ (cf. the ``setting sun'' integrals of (\ref{Eq:eff-pot-rescaled}) containing $G_\pi$ and proportional to $n^2$). One pion line is cut corresponding to the variation of the 2PI effective potential with respect to the pion propagator. The factor of 2 in front of the first diagram indicate the 2 ways of cutting the $\pi$ lines.}
\label{fig5_1}
\end{figure}

In this subsection a 2PI effective potential is given, from which equations of the $1$- and $2$-point
functions can be obtained by functional differentiation. The functional depends on the background $v$, the        
auxiliary (composite) fields ($X,Y^0$) and the 2-point     
functions. We use the same approximation as in Chapter 3: only the setting-sun type of diagrams are kept.
Since we have more fields and vertices, there are a lot more of these diagrams. Since we perform a {\it leading order}
large-$n$ analysis, only those parts of the expressions are needed, which contribute to the leading order of
the equations. For example on Fig. {\ref{fig5_1}} those diagrams are indicated which are needed for the pion propagator equation.

We introduce the following notations:
\bea
{\cal G}_{(s,Y)}=
\begin{pmatrix}
G_s & G_{sY} \\
G_{sY} & G_Y \\
\end{pmatrix},\quad {\cal G}_{(X,Y^0,s^0)}=
\begin{pmatrix}
G_X & G_{XY^0} & G_{s^0X} \\
G_{XY^0} & G_{Y^0} & G_{s^0Y^0} \\
G_{s^0X} & G_{s^0Y^0} & G_{s^0} \\
\end{pmatrix}.
\eea
The tree level versions of these quantities are:
\bea    
\displaystyle        
i{\cal D}_{(s,Y)}^{-1}=        
\begin{pmatrix}        
~iD_0^{-1}~ &        
~2i\sqrt{g_2}v~\\        
~2i\sqrt{g_2}v~        
 & ~-1~        
\end{pmatrix}, \quad       
\displaystyle   
i{\cal D}_{(X,Y^0,s^0)}^{-1}=  
\begin{pmatrix}        
~-1~ & 0 & ~2i\sqrt{g_1}v~ \\        
0 & -1 & ~2i\sqrt{g_2}v~ \\        
 ~2i\sqrt{g_1}v~ & ~2i\sqrt{g_2}v~ & iD_0^{-1} \\       
\end{pmatrix},      
\nonumber       
\eea     
\vspace{-0.7cm}
\bea
\label{Eq:treelevel} 
\eea
where we used the notation 
\bea
\label{5-M2def}
iD_0^{-1}(p)=p^2-m^2+\frac{i}{n}(\sqrt{2g_1}X+\sqrt{2g_2}Y^0)\equiv p^2-M^2,
\eea  
which turns out to be the tree-level pion propagator:
\bea
iD^{-1}_{\pi}=iD^{-1}_{\pi^0}=iD^{-1}_0.
\eea

The approximate 2PI effective potential without counterterms reads as:
\bea        
V_{\textnormal{2PI}}&\!\!\!=\!\!\!&n^2(m^2v^2-2hv)+\frac12\Big(X^2+(Y^a)^2\Big)-        
in(\sqrt{2g_1}X+\sqrt{2g_2}Y^0)v^2\nonumber\\        
&\!\!\!-\!\!\!&i\frac{n^2}{2}\int_k\Big(\ln G_{\pi}^{-1}(k)+D_\pi^{-1}(k)G_{\pi}(k)\Big)-i\frac{1}{2}\int_k\Big(\ln G_{\pi_0}^{-1}(k)+D_{\pi^0}^{-1}(k)G_{\pi^0}(k)\Big)\nonumber\\
&\!\!\!-\!\!\!&i\frac{n^2}{2}\int_k\textrm{Tr}[\ln{\cal G}_{(s,Y)}^{-1}(k)+    
{\cal D}_{(s,Y)}^{-1}(k){\cal G}_{(s,Y)}(k)]             
\nonumber\\
&\!\!\!-\!\!\!&i\frac{1}{2}\int_k\textrm{Tr}[\ln{\cal G}_{(X,Y^0,s^0)}^{-1}(k)+        
{\cal D}_{(X,Y^0,s^0)}^{-1}(k){\cal G}_{(X,Y^0,s^0)}(k)]             
\nonumber\\
&\!\!\!+\!\!\!&i\frac{n^2}{4}g_2\int_k\int_p\Big[\Big(G_{\pi}(k)G_{\pi}(k+p)+        
G_s(k)G_s(k+p)\Big)G_{Y}(p)\nonumber\\
&\!\!\!+\!\!\!&i\frac{n^2}{2}g_2 \int_k\int_p G_{sY}(k)G_{sY}(k+p)G_s(p)\nonumber\\        
&\!\!\!+\!\!\!&i\frac{n^2}{2}g_2 \int_k\int_p\Big(G_{Y}(k)G_s(k+p)-G_{sY}(k)      
G_{sY}(k+p)\Big)G_{\pi}(p)\nonumber\\
&\!\!\!+\!\!\!&i\frac{1}{2}g_1\int_k\int_p\Big(G_s(k)G_s(k+p)+G_{\pi}(k)G_{\pi}(k+p)\Big)G_{X}(p)
\nonumber\\
&\!\!\!+\!\!\!&i\sqrt{g_1g_2}\int_k\int_p\Big(G_s(k)G_s(k+p)+G_{\pi}(k)G_{\pi}(k+p)\Big)G_{XY^0}(p)
\nonumber\\
&\!\!\!+\!\!\!&i\frac{1}{2}g_2\int_k\int_p\Big(G_s(k)G_s(k+p)+G_{\pi}(k)G_{\pi}(k+p)\Big)G_{Y^0}(p)
\nonumber\\
&\!\!\!+\!\!\!&2i\sqrt{g_1g_2}\int_k\int_p G_{sY}(k)G_{s}(k+p)G_{s^0X}(p)+2ig_2\int_k\int_p G_{sY}(k)G_{s}(k+p)G_{s^0Y^0}(p)
\nonumber\\
&\!\!\!+\!\!\!&i\frac{1}{2}g_2\int_k\int_p\Big(G_s(k)G_Y(k+p)+G_{sY}(k)G_{sY}(k+p)\Big)G_{s^0}(p)
\nonumber\\
&\!\!\!+\!\!\!&ig_2\int_k\int_p G_{Y}(k)G_{\pi}(k+p)G_{\pi^0}(p).      
\label{Eq:eff-pot-rescaled}        
\eea     
        
Note, that since the $\pi$ and $s-Y$ sectors are of multiplicity ($n^2-1$), some
${\cal{O}}(1)$ terms in (\ref{Eq:eff-pot-rescaled}) are not relevant for their leading order equations. Also, there are more of ${\cal{O}}(1)$ terms of the complete effective potential, which contribute to the $\pi$ and $s-Y$ sector only at NLO, these are omitted from (\ref{Eq:eff-pot-rescaled}). However, for the remaining fields it is necessary to take into account the proper ${\cal{O}}(1)$ terms (otherwise we would have trivial equations for the corresponding propagators), which are therefore included in (\ref{Eq:eff-pot-rescaled}). This means that
(\ref{Eq:eff-pot-rescaled}) is {\it{not}} a large-$n$ expanded 2PI
effective potential truncated at 2-loop level. It is a potential from which leading order equations can be
derived in each sector of the theory. Note, that the structure appearing in the seventh line corresponds to
a setting-sun diagram with antisymmetrized vertex functions (i.e. $\sim f_{abc}$).    

We stressed in Chapter 3 (during the renormalization of the $O(N)$ model), that the setting-sun approximation of the (unrenormalized)
2PI effective potential corresponds to a specific truncation of the tower of the Dyson-Schwinger equations. Here an
analogous statement can be made. If we close the tower at the level of the $3$-point functions and substitute
their classical value to the equations of the $1$- and $2$-point functions, we obtain exactly the same equations
which can be derived from (\ref{Eq:eff-pot-rescaled}). These nonzero classical $3$-point couplings are:
\bea        
\Gamma_{X\pi^a\pi^b}(x,y,z)&\!\!\!=\!\!\!&\Gamma_{Xs^as^b}(x,y,z)=i\frac{\sqrt{2g_1}}{n}\delta_{ab}        
\delta(x-y)\delta(x-z),\nonumber\\        
\Gamma_{Y^as^bs^c}(x,y,z)&\!\!\!=\!\!\!&\Gamma_{Y^a\pi^b\pi^c}(x,y,z)=i\sqrt{\frac{g_2}{n}}        
d_{abc}\delta(x-y)\delta(x-z),\nonumber\\      
\Gamma_{Y^as^b\pi^c}(x,y,z)&\!\!\!=\!\!\!&-i\sqrt{\frac{g_2}{n}}f_{abc}\delta(x-y)\delta(x-z).     
\label{Eq:bare-vertex}  
\eea  
       
The derivatives of the unrenormalized effective potential with respect to the $1$-point functions are:
\bea       
\frac{\delta V_{\textnormal 2PI}}{\delta X}&\!\!\!=\!\!\!&X-in\sqrt{2g_1}v^2+i\sqrt{\frac{g_1}{2}}n\left(\int_k G_s (k)+\int_k G_{\pi}(k)\right),        
\nonumber\\        
\frac{\delta V_{\textnormal 2PI}}{\delta Y^0}&\!\!\!=\!\!\!&Y^0-in\sqrt{2g_2}v^2+i\sqrt{\frac{g_2}{2}}n\left(\int_k G_s (k)+\int_k G_{\pi}(k)\right),     
\nonumber\\
\frac{\delta V_{\textnormal 2PI}}{\delta v}&\!\!\!=\!\!\!&2n^2(M^2v-h)-2i\sqrt{g_2}n^2\int_k G_{sY}(k).     
\label{Eq:1pointeq}         
\eea     
When we look for the physical values of the $1$-point functions (mean fields), counterterms must be included and the derivatives must be set to zero.
In this case, comparing the solutions of the equations for the two auxiliary fields, we find the relation:        
\be        
\sqrt{g_1}X=\sqrt{g_2}Y^0,      
\label{Eq:saddle-relation}
\ee       
which should also be valid after renormalization.

Now we turn to the propagator equations. Zeros of the derivatives of the effective potential with respect to the $2$-point functions give the appropriate equations. In the pseudoscalar sector the following equations are found at leading order (the abbreviated notation $\int G_{[.]}G_{[.]}\equiv \int_k G_{[.]}(k) G_{[.]}(p+k)$ is used):  
\bea        
iG_{\pi^0}^{-1}&=&iD_0^{-1}-2ig_2\int G_Y G_{\pi},\nonumber\\        
iG_{\pi}^{-1}&=&iD_0^{-1}-ig_2\int G_{Y}(G_s+G_{\pi})+ig_2\int G_{sY}G_{sY}.  
\label{Eq:pionprop}  
\eea     
We note the potential violation of Goldstone's theorem when comparing       
$G_{\pi}(k=0)$ or $G_{\pi^0}(k=0)$ with the equation of state. Nevertheless,     
our forthcoming approximate solution obeys this theorem due to a rather    
nontrivial relation between the relevant tadpole and bubble contributions.      
The coupling between $s^a-Y^a$ can be seen explicitly in the structure of the corresponding sector, where (if $a\neq 0$)        
we find $n^2-1$ identical mixing $2\times 2$ equations:  
\bea        
iG_{s}^{-1}&\!\!\!=\!\!\!&iD_0^{-1}-ig_2\int G_{sY}G_{sY}-ig_2\int G_{Y}(G_{s}+G_{\pi}), \nonumber\\
iG_{Y}^{-1}&\!\!\!=\!\!\!&-1-ig_2\int G_s G_{\pi}-i\frac{g_2}{2}\int(G_sG_s+G_{\pi}G_{\pi}), \nonumber\\        
iG_{sY}^{-1}&\!\!\!=\!\!\!&2i\sqrt{g_2}v-ig_2\int G_{sY}(G_s-G_{\pi}).  
\label{Eq:sYprop}  
\eea         
In the $3\times 3$ mixing sector of $(X,Y^0,s^0)$, the following 6        
equations are obtained:        
\bea        
\!\!\!\!\!\!\!\!\!\!iG_{X}^{-1}&\!\!\!=\!\!\!&-1-ig_1\int(G_sG_s+G_{\pi}G_{\pi}), \quad iG_{XY^0}^{-1}=-i\sqrt{g_1g_2}\int(G_sG_s+G_{\pi}G_{\pi}), \nonumber\\         
\!\!\!\!\!\!\!\!\!\!iG_{s^0X}^{-1}&\!\!\!=\!\!\!&2i\sqrt{g_1}v-2i\sqrt{g_1g_2}\int G_sG_{sY}, \quad iG_{Y^0}^{-1}=-1-ig_2 \int(G_sG_s+G_{\pi}G_{\pi}), \nonumber\\      
\!\!\!\!\!\!\!\!\!\!iG^{-1}_{s^0Y^0}&\!\!\!=\!\!\!&2i\sqrt{g_2}v-2ig_2\int G_s G_{sY}, \!\!\!\!\quad iG^{-1}_{s^0}=iD_0^{-1}-2ig_2\int(G_{s}G_{Y}+G_{sY}G_{sY}).  
\label{Eq:s0XY0prop}  
\eea          

\subsection{Counterterm functional}

In view of the equations for the $1$-point functions and the propagators, we collect all the counterterms, which should ensure the finiteness of these equations. As before, we introduce the most general counterterm functional, which is allowed by 2PI formalism. 

In this case it means that we can introduce counterterms to all independent pieces appearing in the mean-field part of the effective potential and independently of them also to the terms showing up in the inverse tree-level propagators. Note, that we might have to introduce counterterms also to pieces which are chosen to have an apparently fixed numerical (zero or unity) renormalized coefficient. In this sense the most general form which we will need and is allowed by the structure of $V_{\textnormal 2PI}$ is the following:     
\bea    
V_{\textnormal ct}&\!\!\!=\!\!\!&n^2\delta m^2_0v^2+\frac{1}{2}\Big(\delta_x g_1X^2+\delta_y g_2(Y^0)^2\Big)\nonumber\\    
&&+\delta_{xy}XY_0-i\sqrt{2n^2}(\delta_{y0}\sqrt{g_2}Y^0+\delta_{x0}\sqrt{g_1}X)\nonumber\\    
&&-i\int_k \left[\frac{n^2}{2}\delta_{YY}G_{Y}(k)+\delta_{XX}G_{X}(k)+\delta_{Y^0Y^0}G_{Y^0}(k)+2\delta_{XY^0}G_{XY^0}(k)\right]  
\nonumber\\  
&&-\frac{n^2}{2}\int_k\left[(\delta Z_\pi k^2-\delta
m^2_\pi+i\frac{\sqrt{2g_1}}{n}\delta_{x\pi}X+i\frac{\sqrt{2g_2}}{n}\delta_{y\pi}Y^0)G_\pi(k)\right]\nonumber\\
&&-\frac{n^2}{2}\int_k\left[(\delta Z_s k^2-\delta
m^2_s+i\frac{\sqrt{2g_1}}{n}\delta_{xs}X+i\frac{\sqrt{2g_2}}{n}\delta_{ys}Y^0)G_s(k)\right].  
\label{Eq:counterfunct}    
\eea    
Note, that the self-energy corrections to $G_{sY}^{-1}$ turn out to be finite, therefore no counterterm is needed there. In the second line, countercouplings proportional to $XY^0, Y^0, X$ are also needed despite the fact that the corresponding renormalized values are chosen to be zero. Similarly, countercouplings in the third line belong to the contribution of the (pure) auxiliary propagators occurring in terms of the type ${\textrm {Tr}} {\cal D}^{-1}{\cal G}$, in which all their renormalized coefficients are fixed to unity except the coefficient of $G_{XY^0}$, which is zero. In the fourth line, countercouplings are introduced into the same type of terms, they appear in $D^{-1}_{s}$ and $D^{-1}_\pi$. As we know from Chapter 2 and 3, in the exact solution of the theory only unique quadratic and quartic counterterms should occur, but in any finite order of a 2PI approximation we have the freedom to choose the countercouplings in the counterterm functional independently \cite{berges05}. For the determination of (\ref{Eq:counterfunct}) we should analyze the divergence structure of the integrals appearing in the propagator equations.  

From now on, we make an assumption on the mass hierarchy of the model. We assume that the scalar sector is considerably heavier than      
the pionic, which simplifies the solution of the        
limiting form of the coupled $1$- and $2$-point equations valid at $n\rightarrow \infty$. In addition, we also assume that the scalar masses are      
considerably larger than the amplitude of the symmetry breaking vacuum      
condensate.  Since each component of the propagators in the mixed sectors will have a common denominator displaying the corresponding heavy mass, the only consequent way to neglect the heavy sector in a first approximation is to erase the bubble contributions containing at least one component of ${\cal G}_{(s,Y)}$ or ${\cal G}_{(X,Y^0,s^0)}$. Then only bubble diagrams exclusively built with $G_{\pi}$ are included. This means that in (\ref{Eq:pionprop}) all bubbles are suppressed, while in (\ref{Eq:sYprop}) and (\ref{Eq:s0XY0prop}) only pure pion bubbles remain.
In this case all $n^2$ pion propagators are equal to their tree-level value, and therefore have the same mass $M^2$ (check (\ref{Eq:pionprop})).

The equations of the $1$-point functions are not that simple, they also need renormalization. In the forthcoming procedure we use the scheme introduced in Chapter 3, i.e. we expand the propagators around an auxiliary perturbative propagator $G_0$ with mass $M_0$ and identify and subtract the divergences in terms of the integrals of $G_0$. The explicit form of the saddle-point equations can be written with the help of splitting the pion tadpole into finite $(T_\pi^F)$ and divergent parts as (see (\ref{Eq:tadpole}))     
\be     
\int_k G_{\pi}(k)\approx \int_k D_0(k) = T_\pi^F+T_d^{(2)}+(M^2-M_0^2)T_d^{(0)}.     
\ee  
With the use of the counterterms, it becomes the following:   
\begin{subequations} 
\label{Eq:xeq_ren}  
\bea        
\!\!\!\!\!\!\!\frac{x}{g_1}=i\Big(2v^2+T_d^{(2)}+(M^2-M_0^2)T_d^{(0)}+T_\pi^F\Big)-   
\Big(\delta_xx+\delta_{xy}\frac{1}{\sqrt{g_1g_2}}y^0\Big)+2i\delta_{x0}, \\         
\!\!\!\!\!\!\!\frac{y^0}{g_2}=    
i\Big(2v^2+T_d^{(2)}+(M^2-M_0^2)T_d^{(0)}+T_\pi^F\Big)    
-\Big(\delta_yy^0    
+\delta_{xy}\frac{1}{\sqrt{g_1g_2}}x\Big)+2i\delta_{y0}.       
\eea       
\end{subequations} 
Here we introduced the notations
\bea
\label{5-x-y10-def}
x=\frac{\sqrt{2g_1}}{n}X, \qquad y^0=\frac{\sqrt{2g_2}}{n}Y^0,
\eea
and we note that the counterterms $\delta_{xv}, \delta_{x\pi}, \delta_{xs}, \delta_{yv}, \delta_{y\pi}$ and $\delta_{ys}$ can be chosen zero,
therefore the sake of transparency we did not write down them
explicitly in (\ref{Eq:xeq_ren}).
Taking into account the definition of $M^2$,        
the following counterterms renormalize both equations:        
\bea       
\hspace{-0.2cm}\delta_x=\delta_y=T_d^{(0)},\qquad      
\delta_{xy}=\sqrt{g_1g_2}T_d^{(0)}, \nonumber\\
\delta_{x0}=\delta_{y0}=-\frac12\left(T_d^{(2)}+(m^2-M_0^2)T_d^{(0)}\right).
\label{Eq:cterms-one-point}        
\eea       
Because of the omission of the last (``heavy'') term in the third equation of ({\ref{Eq:1pointeq}}), 
the equation of state does not require any extra counterterm.        
The finite equations for the 1-point functions read as follows ($\delta m_0^2$ does not receive any contribution at this level):        
\be        
\frac{x}{g_1}=i(2v^2+T_\pi^F),        
\qquad \frac{y^0}{g_2}=i(2v^2+T_\pi^F),\qquad M^2v=h.    
\label{Eq:1-point-eqs}  
\ee  
        
In the similar approximation, the $2\times 2$ propagator matrix of the $(s,Y)$ sector simplifies to        
\be        
\displaystyle        
i{\cal G}_{(s,Y)}^{-1}(k)=        
\begin{pmatrix}        
~iD_0^{-1}(k)~ &        
~2i\sqrt{g_2}v~\\       
~2i\sqrt{g_2}v~        
 & ~-1+i\delta_{YY}+\frac{g_2}{2}I_{\pi}(k)~        
\end{pmatrix},        
\label{Eq:sy-sector-noscalar}        
\ee        
where $I_\pi$ is the usual bubble integral: $I_{\pi}(k)=-i\int_pD_0(p)D_0(p+k)$ and we have also written down the contribution obtained from $V^{2PI}_{ct}$.    
By choosing    
\be    
\delta_{YY}=i\frac{g_2}{2}T_d^{(0)},
\ee    
we ensure the finiteness of the matrix elements.        
The squared scalar mass is determined by the zero of the determinant:        
\be        
\displaystyle        
M_{(s,Y)}^2=M^2+4g_2v^2\frac{1}{1-\frac{g_2}{2}I_{\pi}^F(k=M_{(s,Y)})},    
\label{Eq:sYgap}    
\ee      
where $I_\pi^F(k)=I_\pi(k)-T_d^{(0)}$, check (\ref{Eq:I_pi^F}).
The mass matrix of the $(X,Y^0,s^0)$ sector also becomes more transparent:        
\be        
i{\cal G}_{(X,Y^0,s^0)}^{-1}=        
\begin{pmatrix}        
~-1+2i\delta_{XX}+g_1I_{\pi}(k)~  & ~2i\delta_{XY^0}    
+\sqrt{g_1g_2}I_{\pi}(k)~ &        
~2i\sqrt{g_1}v~\\        
 ~2i\delta_{XY^0}+\sqrt{g_1g_2}I_{\pi}(k)~& ~-1+2i\delta_{Y^0Y^0}    
+g_2I_{\pi}(k)~ &        
~2i\sqrt{g_2}v~\\        
~2i\sqrt{g_1}v~& ~2i\sqrt{g_2}v~        
& ~iD_0^{-1}(k)~        
\end{pmatrix},        
\label{Eq:s0y0x-sector-noscalar}        
\ee        
which after introducing the obvious counterterms    
\be    
\delta_{XX}=\frac{i}{2}g_1T_d^{(0)},\qquad \delta_{XY^0}=\frac{i}{2}\sqrt{g_1g_2}T_d^{(0)},    
\qquad \delta_{Y^0Y^0}=\frac{i}{2}g_2T_d^{(0)}    
\ee    
leads to a determinant equation completely analogous to the        
previous one:        
\be        
\displaystyle        
M_{(X,Y^0,s^0)}^2=M^2+4(g_1+g_2)v^2\frac{1}{1-(g_1+g_2)I^F_{\pi}(k=M_{(X,Y^0,s^0)})}.  
\label{Eq:xY0s0gap}       
\ee        
We see, that the spectra resulting from the assumption we made for        
the mass hierarchy might be consistent with the outcome in the sense that      
the scalar sector (and the auxiliary fields hybridized with it) can be  
indeed heavier than the pionic one.       
      
In the next subsection we investigate if there is a region of the
parameter space, where the scalar masses become much heavier than
the pseudoscalars ensuring the self-consistency of our approximate
solution. 

\subsection{Validity of the heavy mass assumption}

First one has to note, that in the case when no explicit symmetry breaking term is added to the Lagrangian ($h=0$) the mass assumption is true, since our approximation preserves Goldstone's theorem and therefore makes the pions massless, which means that they
are "infinitely" lighter than the scalars.

The interesting case is when $h\neq 0$. Let us define $r_{s,Y}:=M_{(s,Y)}/M$ and $r_{X,Y^0,s^0}:=M_{(X,Y^0,s^0)}/M$. In order to obtain a proper region of the parameter space, we introduce a heaviness criterion: the scalars are heavy enough if relations $r_{s,Y}>r_0$, $r_{X,Y^0,s^0}>r_0$ hold simultaneously, where $r_0$ is a given
number. It is somewhat arbitrary what value to choose for $r_0$. In an exploratory study, we work with the convenient choice $r_0=2$, because the quantity $I_\pi^F(k,M)$ appearing in both gap equations ({\ref{Eq:sYgap}) and ({\ref{Eq:xY0s0gap}) develops an imaginary part just for $k^2>4M^2$ (two-pion threshold). Above the threshold the masses are defined as real parts of the complex solutions. Expressing $v$ from the equation of state in ({\ref{Eq:1-point-eqs}), the two relevant equations are the following:
\bea
\!\!\!\!\!\!\!\!\!M_{(s,Y)}^2&\!\!\!=\!\!\!&M^2+4g_2\frac{h^2}{M^4}\frac{1-\frac{g_2}{2}\Re I_{\pi}^F(M_{(s,Y)})}{[1-\frac{g_2}{2}\Re I_{\pi}^F(M_{(s,Y)})]^2+[\frac{g_2}{2}\Im I_{\pi}^F(M_{(s,Y)})]^2}, \\
\!\!\!\!\!\!\!\!\!M_{(X,Y^0,s^0)}^2&\!\!\!=\!\!\!&M^2+4g_{12}\frac{h^2}{M^4}\frac{1-g_{12}\Re I_{\pi}^F(M_{(X,Y^0,s^0)})}{[1-g_{12}\Re I_{\pi}^F(M_{(X,Y^0,s^0)}]^2+[g_{12}\Im I_{\pi}^F(M_{(X,Y^0,s^0)})]^2},
\label{Eq:scalar-gaps}
\eea  
where $g_{12}=g_1+g_2$ and $\Re$, $\Im$ correspond to the real and imaginary parts, respectively. The finite part of the bubble integral was given explicitly in (\ref{Eq:I_pi^F}) and (\ref{Eq:Lfunc}). With the use of ({\ref{Eq:1-point-eqs}) and ({\ref{Eq:tadpole}}), we get
\be
M^2=m^2+g_{12}\Big(\frac{2h^2}{M^4}+\frac{M^2}{16\pi^2}\log \frac{M^2}{M_0^2}-\frac{M^2-M_0^2}{16\pi^2}\Big),
\label{Eq:pion-gap}
\ee
where $M_0$ is the renormalization scale. It is convenient to express all masses in proportion to the absolute value of the renormalized mass $(\sqrt{|m^2|})$. For a fixed $M_0$, ({\ref{Eq:pion-gap}}) determines $M^2/|m^2|$ as a function of $h/|m|^3,g_1,g_2$. 
Plugging it into the scalar gap equations, they
can be solved for $M_{(s,Y)}/\sqrt{|m^2|}$ and $M_{(X,Y^0,s^0)}/\sqrt{|m^2|}$. Then, we can trace out the region, where the heaviness criterion is fulfilled. This region is the part of the positive $g_1-g_2$ octant (the stability region of the $n\rightarrow\infty$ theory) below the surface displayed in Fig. \ref{Fig5_2}. As expected, the projection of the allowed region onto the $g_1-g_2$ plane shrinks for increasing value of $h/|m|^3$. We have varied $M_0/\sqrt{|m^2|}$ in the interval $(1,10)$ and a mild displacement of the allowed region was observed. This change can be balanced by an appropriate renormalization group transformation of the quartic couplings, i.e. using $g_i=g_i(M_0)$.

\begin{figure}[!t]
\centerline{ 
\includegraphics[bb=324 92 450 479,scale=0.5]{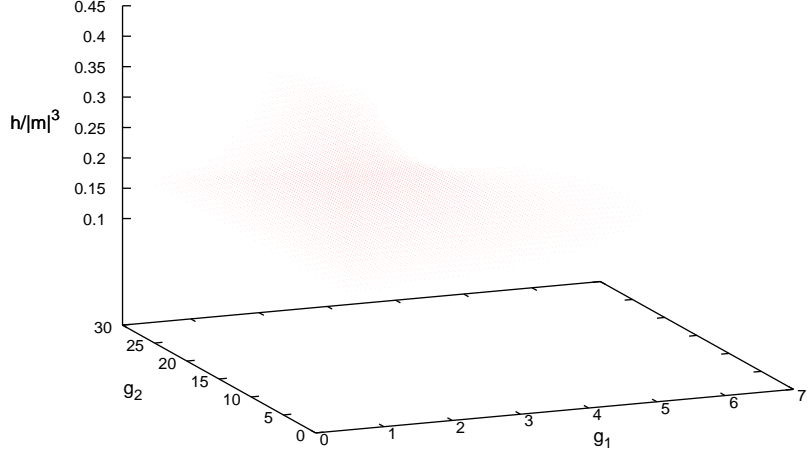}}
\caption{Region of the parameter space, where the mass assumption holds (using renormalization scale $M_0/\sqrt{|m^2|}=1.0$)}
\label{Fig5_2}
\end{figure} 

\subsection{Heavy scalar corrections} 
     
Using the propagators of the coupled $(s,Y)$-sector, we can     
start to systematically take into account the effect of the heavy      
degrees of freedom on the pion propagator, the EoS and the SPEs. 
In particular we now invoke the tadpole and bubble contributions to these equations evaluated with ({\ref{Eq:sy-sector-noscalar}}).   
For this we write explicitly the components of the heavy scalar      
propagator matrix:    
\bea           
G_{s}(k)&\!\!\!=\!\!\!&\frac{i}{d(k)}\left(1-\frac{g_2}{2}I_{\pi}^F(k)\right),\quad G_{Y}(k)=-\frac{i}{d(k)}(k^2-M^2),\nonumber\\     
G_{sY}(k)&\!\!\!=\!\!\!&-\frac{1}{d(k)}2\sqrt{g_2}v,\nonumber\\          
d(k)&\!\!\!=\!\!\!&\left(1-\frac{g_2}{2}I_{\pi}^F(k)\right)(k^2-M^2)-4g_2v^2.      
\eea                                         
With the help of these expressions, we readily write the scalar tadpole contributions to the SPEs.       
The correction of the EoS is of imminent interest, since it has      
an important role in the discussion of the validity of Goldstone's      
theorem. Using the expression of $G_{sY}$, we find the unrenormalized equation of state:      
\be\displaystyle      
M^2+2g_2\int_p\frac{i}{d(p)}=\frac{h}{v}.  
\label{Eq:unren-eos} 
\ee      
Now we proceed with the unrenormalized pion propagators:      
\bea      
\!\!\!\!\!\!\!\!iG_{\pi^0}^{-1}(k)&\!\!\!=\!\!\!&k^2-M^2-2g_2\int_p\frac{i}{d(p)}\frac{p^2-M^2}    
{(k-p)^2-M^2},      
\nonumber\\      
\!\!\!\!\!\!\!\!iG_{\pi}^{-1}(k)&\!\!\!=\!\!\!&k^2-M^2-g_2\int_p\frac{i}{d(p)}      
\frac{p^2-M^2}{(k-p)^2-M^2}\nonumber\\      
&&+4g_2^2v^2\int_p\frac{i}{d(p)d(k-p)}-ig_2\int_p      
\frac{(p^2-M^2)(1-g_2I_{\pi}^F(k-p)/2)}{d(p)d(k-p)}.      
\label{Eq:pion-props}
\eea      
When we set $k=0$, both propagator equations turn out to be the  
same. Comparing to the equation of state we  
find, that the approximation fulfills the Goldstone theorem
characterizing the $U(n)\times U(n)\rightarrow U(n)$ symmetry
breaking. It can be shown, that both equations receive the same counterterm contributions, therefore the theorem remains valid after renormalization.

Since the tadpoles of $G_s$ and $G_{sY}$ play a role in the EoS and
the SPEs, it is worthwhile to investigate their corrections, although
these contributions should be considered only as NNLO ``heavy'' corrections.
Substituting the leading order propagators into the second term on the
right hand side of the equation of $G_{sY}^{-1}$ (\ref{Eq:sYprop}), we find for the $\Sigma_{sY}$ self-energy:
\be
\Sigma_{sY}(p)\equiv ig_2\int_k G_{sY}(p-k)\Big(G_s(k)-G_{\pi}(k)\Big)=-8g_2^2\sqrt{g_2}v^3\int_k\frac{1}{d(p-k)d(k)(k^2-M^2)},
\ee 
which is actually finite.          
The heavy correction to $G_s^{-1}$ is very similar to that of $G^{-1}_{\pi}$ in (\ref{Eq:pion-props}):
\bea         
\!\!\!\!\!\!\!iG_{s}^{-1}(k)&\!\!\!=\!\!\!&k^2-M^2-g_2\int_p\frac{i}{d(p)}      
\frac{p^2-M^2}{(k-p)^2-M^2}\nonumber\\      
\!\!\!\!\!\!\!&&-4g_2^2v^2\int_p\frac{i}{d(p)d(k-p)}-ig_2\int_p      
\frac{(p^2-M^2)(1-g_2I_{\pi}^F(k-p)/2)}{d(p)d(k-p)}.      
\label{Eq:s-prop}
\eea 

The scalar corrections are taken into account also in the SPEs. They appear partly directly via the scalar tadpole, 
and also by the scalar correction of the pion tadpole. The structure of    
the two SPE's in ({\ref{Eq:1pointeq}}) is identical in view of
(\ref{Eq:saddle-relation}), therefore it is sufficient to investigate
the equation, which determines $x$. The unrenormalized form of the
saddle-point equation reads as follows:    
\begin{equation}    
0=\frac{x}{g_1}-2iv^2-i\int_k\big(G_s(k)+G_{\pi}(k)\big).  
\label{Eq:ren-spex}    
\end{equation}
The expression of the scalar tadpole is readily written down.           
The pion tadpole is expanded to linear order in the $\Sigma_{\pi}$ self-energy contribution:
\be
-i\int_kG_\pi(k)\approx -i\int_kD_0(k)+\int_k\frac{1}{(k^2-M^2)^2}\Sigma_{\pi},
\label{Eq:pion_prop-corr}
\ee
where
\be
\Sigma_\pi(k)=g_2\int_p\frac{i}{d(p)}\left[\frac{p^2-M^2}
{(k-p)^2-M^2}
-\frac{4g_2v^2}{d(k-p)}+\frac{(p^2-M^2)(1-g_2I^F_\pi(k-p)/2)}{d(k-p)}\right].
\ee

One can achieve the explicit (resummed) renormalization of the equations (\ref{Eq:unren-eos}), (\ref{Eq:pion-props}), (\ref{Eq:s-prop}) and (\ref{Eq:ren-spex}), which will not
be discussed here. In this procedure the remaining counterterms $\delta m_0^2, \delta_{xv}, \delta_{x\pi}, \delta_{xs}, \delta_{yv}, \delta_{y\pi}, \delta_{ys}$ and $\delta Z_{\pi}, \delta Z_s$ are determined. In a careful analysis, we find the following relations between them:
\bea
\delta m_0^2=\delta m_\pi^2=\delta m_s^2, \quad \delta Z_\pi=\delta Z_s, \nonumber\\
\delta_{xv}=\delta_{x\pi}=\delta_{xs}=\delta_{yv}=\delta_{y\pi}=\delta_{ys}.
\eea
The reader can find the corresponding calculations in details in one
of the papers (Phys. Rev. D{\bf 82}: 045011, 2010), whereupon this thesis is based on.

\section{Discussion of a more general breaking of the symmetry}

In this section we develop the previously introduced approximate large-$n$ solution with a more general condensate. In this way we can investigate the stability of the $U(n)\times U(n)\longrightarrow U(n)$ symmetry breaking pattern. Also we are motivated by possible phenomenological applications announced in the beginning of the chapter. 

It turns out that for a more general description of the symmetry breaking, it is convenient to introduce not two, but three auxiliary fields. In the previous section the auxiliary field $Y^a$ did not have a definite parity, however in the generalized analysis below it is worthwhile to work with quantities with a definite reflection property. Using the Jacobi identity, we easily recognize the decomposition
\bea
\label{5-Jacobi}
U^aU^a=U_1^aU_1^a+U_2^aU_2^a,
\eea
where $U_1^a=\frac12d_{abc}(s^bs^c+\pi^b\pi^c)$ and $U_2^a=f_{abc}s^b\pi^c$. Exploiting this in the Lagrangian (\ref{5-Lagr}), every piece of it can be expressed with fields with definite parity. Then, $U_1^a$ and $U_2^a$
are perfect combinations in addition to $s^as^a+\pi^a\pi^a$ to be replaced by auxiliary fields. It will turn out that one of them mixes with the scalar (as in the previous section), and the other with
the pseudoscalar elementary fields. The composite pseudoscalar (i.e. corresponding to $U_2^a$) is assumed to be similarly light as the elementary
$\pi^a$, therefore its dynamics will be included into our analysis, while the auxiliary fields corresponding to $U_1^a$ and $s^as^a+\pi^a\pi^a$ are classified to be ``heavy'', therefore their fluctuations will be omitted from the contributions to the equation of state and the saddle point
equations. In the present refined analysis, the mixing in the pseudoscalar sector results in a spectrum, where
also heavier excitations (``kaons'') might appear naturally. The self-consistency of this solution requires that
even the heavier pseudoscalars should be considerably lighter than the lightest scalar excitation.

\subsection{Introducing three auxiliary fields}

Using (\ref{5-Jacobi}), the Lagrangian (\ref{5-Lagr}) with a more general explicit symmetry breaking term reads as:
\bea
L&=&\frac{1}{2}[(\partial_\mu s^a)^2+(\partial_\mu\pi^a)^2-m^2((s^a)^2+(\pi^a)^2)]+\sqrt{2n^2}h_qs^q\nonumber\\        
&-&\frac{g_1}{4n^2}\Big((s^a)^2+(\pi^a)^2\Big)^2-\frac{g_2}{2n}[(U_1^a)^2+(U_2^a)^2],         
\eea
where ``q'' indices refer to the diagonal generators of $U(n)$. Then, three auxiliary composite fields are introduced by adding the following constraints to the Lagrangian:        
\bea        
\Delta L&\!\!\!=\!\!\!&-\frac{1}{2}\left(X-i\sqrt{\frac{g_1}{2n^2}}((s^a)^2+(\pi^a)^2)        
\right)^2\nonumber\\
&&-\frac{1}{2}\left(Y^a_1-i\sqrt{\frac{g_2}{n}}U^a_1\right)^2-\frac{1}{2}\left(Y^a_2-i\sqrt{\frac{g_2}{n}}U^a_2\right)^2.        
\eea
In the sum ${\cal L}\equiv L+\Delta L$ a general symmetry breaking pattern corresponding to diagonal directions
in the algebra space is realized through the shifts    
\bea
&        
s^a\rightarrow s^a+\sqrt{2n^2}v_q\delta^{qa},\qquad U_2^a\rightarrow U^a_2+n\sqrt{2}v_qf_{aqc}\pi^c,\nonumber\\
&
U^a_1\rightarrow        
U^a_1+n\sqrt{2}d_{abq}s^bv_q+n^2d_{apq}v_pv_q.        
\eea
The indices ``q'' correspond again to the diagonal generators. The index $q=0$ refers to the unit
matrix and $q=1,...,n-1$ refer to the elements of the Cartan subalgebra of $SU(n)$.    
  
After the introduction of the auxiliary field variables and        
the shifts, the full Lagrangian has the following form:        
\bea        
{\cal L}&\!\!\!=\!\!\!&        
\frac{1}{2}\Big[(\partial_\mu s^a)^2+(\partial_\mu\pi^a)^2-m^2\Big(2n^2v_q^2
+2\sqrt{2n^2}v_qs^q+(s^a)^2+(\pi^a)^2\Big)\Big]\nonumber\\
&&-\frac{1}{2}\left(X^2+(Y^a_1)^2+(Y_2^a)^2\right)\nonumber\\
&&+\sqrt{2n^2}h_q(s^q+\sqrt{2n^2}v_q)+i\sqrt{\frac{g_1}{2n^2}}X\Big(2n^2v_q^2        
+2\sqrt{2n^2}v_qs^q+(s^a)^2+(\pi^a)^2\Big)\nonumber\\        
&&+i\sqrt{\frac{g_2}{n}}[Y^a_1U^a_1+n\sqrt{2}v_qd_{abq}Y_1^as^b+Y_1^an^2d_{apq}v_qv_p]\nonumber\\
&&+i\sqrt{\frac{g_2}{n}}\left[Y_2^aU_2^a+n\sqrt{2}v_qf_{aqc}Y_2^a\pi^c\right].        
\label{Eq:shifted-aux-L}        
\eea              
The background $v_q$ introduces mixing of the 
pair $s^a,Y_1^a$ for every value of the index $a$. In addition
it induces nonzero values for $Y_1^q$ and also a $(2n+1)$-dimensional mixing sector is being formed for the propagators of these following degrees of freedom: $s^q,X,Y_1^q$.
On the other hand pions will mix (for some index values at least) with the composite field $Y_2^a$, which has negative parity. Since the ground state of strong
interactions respects parity, we expect to have a quantum solution of the model,
where $Y_2^a=0$. Note that if $Y_2^a$ had nonzero value, scalar and pseudoscalar fields would mix and the analysis of the mass spectrum would be a lot more complicated. Instead of going through the most general analysis, we assume from the start that $\pi^a$ and $s^a$ fields do not mix, and below we show that this is consistent
with $Y_2^a=0$.

We construct correspondingly a quantum solution in the same approximation as before (i.e. using a large-$n$
approximated 2PI effective action and neglecting heavy excitations), where the saddle point        
values of $X, Y_1^q$ are non-zero, $Y_2^a=0$, and the 2-point functions corresponding to the above
mentioned coupled (mixing) sectors are assumed not to vanish.
Our final goal is to construct the ground state of the system (which is determined by the equations of
the $1$-point functions), and since according to the heaviness assumption, in a first approximation
only the quantum fluctuations of the light ($\pi-Y_2$) fields enter to the equations of
state of the condensate and the saddle point equations, therefore in this section we 
fully avoid the analysis of the effect of the more complex symmetry breaking pattern on
the scalar sector. We note however, that due to the fact that now three auxiliary fields have been introduced,
the same approximation of the 2PI effective action introduced in the previous section leads to slightly
different equations in the scalar sector than the ones found in (\ref{Eq:sYprop}) and (\ref{Eq:s0XY0prop}).
We will not reanalyze these objects as it was done there, but simply expect that the heaviness
criterion remains valid in the essential part of the previously explored parameter space. For any specific phenomenological
applications it would be recommended to investigate the correctness of this expectation, however in this study
we limit ourselves to determine the features of the ground state of the system.

The $\pi$-sector breaks up into two separate groups of fields. The first group contains pions which mix
with each other. From now on, we say that an index is {\it diagonal}, if it corresponds to a diagonal generator. Our first observation is that only pions with diagonal indices can mix with each other. This can be seen using some useful identities of the Appendix, or can be understood as follows. 

Since a potential mixing could arise from the term $Y_1^a d_{abc}$ (i.e. $Y_1^aU_1^a$ in (\ref{Eq:shifted-aux-L})), assuming that $Y_1^a\neq 0$, if $a$ refers to a diagonal generator (this will be checked later), then the first index
of $d_{abc}$ must be also diagonal. If the other two indices ($b$ and $c$) are not diagonal, then they are necessary equal, since (check the Appendix) $d_{abc}\sim \Re \Tr (\lambda^a\lambda^b\lambda^c)$ (here $\lambda$'s are the generalized Gell-Mann matrices), therefore $\lambda^b \lambda^c$ must be a real and diagonal matrix.
From the generalization of the Pauli algebra (see the Appendix) this requirement is fulfilled only when $b=c$. This indeed shows that only diagonal pions can mix with each other. This mixing is determined by the specific pattern of the symmetry breaking and the corresponding structure constants can be found in the Appendix.

The second group of pions are those which can mix with $Y_2$. Here we find that only pions with non-diagonal indices
can enter the game and each of these can mix with exactly one $Y_2^a$. This means that these are disjoint
from the previous pions and form $2\times 2$ blocks. Again, using the identities for the structure constants given in the Appendix, one can immediately see this, however for the reader's convenience, we also give a short argument here. 

The explanation is very similar as the previous one, but now the mixing comes from $v_a f_{uav}$ (see the last term of (\ref{Eq:shifted-aux-L})), where $a\neq 0$ is a diagonal index. It is easy to see that independently of the value of $a$, we always get nonzero $f_{uav}$ for the same appropriately chosen pairs of $u-v$ indices. In other words, if we have chosen a ``$u$'' index, for a nonzero $f_{uav}$, we must point to the same ``$v$'' index independently of $a$. This is due to the fact (see the Appendix), that $f_{uav}\sim i\Tr \Big([\lambda^u,\lambda^v]\lambda^a\Big)$, which says that if $\lambda^a$ is diagonal, then the commutator $[\lambda^u,\lambda^v]$ must be also diagonal and purely imaginary (it is possible however that even in this case the matrix in the argument of the trace gives zero). This immediately fixes both $u$ and $v$, since they must correspond to the non-diagonal generators, which are the generalized Pauli matrices (i.e. generators with indices $(x,jk), (y,jk)$, for the notation see the Appendix).

We introduce notations for the propagators of fields described above. For the diagonal pions we have a huge $n\times n$ matrix, denoted by $\big({\cal G}^{\diag}_{\pi}\big)_{ab}\equiv G_{\pi^a \pi^b}$. The same quantity for the non-mixing $Y_2$'s are $\big({\cal G}^{\diag}_{Y_2}\big)_{ab}\equiv G_{Y_2^a Y_2^b}$. The inverses of these at tree level multiplied by $i$ are
\begin{subequations}
\bea
\label{5-pion_diag}
i\big({\cal D}_{\pi}^{\diag}\big)^{-1}_{ab}(p)&\!\!\!=\!\!\!&(p^2-m^2+i\sqrt{\frac{2g_1}{n^2}}X)\delta_{ab}+i\sqrt{\frac{g_2}{n}}d_{abc}Y_1^c\\
i\big({\cal D}_{Y_2}^{\diag}\big)^{-1}_{ab}(p)&\!\!\!=\!\!\!&-\delta_{ab}.
\eea
\end{subequations}
The $2\times 2$ blocks of non-diagonal pions and the corresponding $Y_2^a$ fields are denoted by:
\bea
{\cal G}_{\pi Y_2}^{(u,v)}=
\begin{pmatrix}
G_{\pi^u \pi^u} & G_{\pi^u Y_2^v} \\
G_{Y_2^v \pi^u} & G_{Y_2^v Y_2^v} \\
\end{pmatrix},
\eea
where $u=(x,jk)$ and $v=(y,jk)$ or vice versa. Since there are $n(n-1)/2$ possible choices for $u$ and $v$, taking also account the possible interchanging between them, we have a number of $n(n-1)$, in general different sectors of this type. The inverse matrix at tree level multiplied by $i$ is:
\bea
\label{5-tree-piY2}
i\big({\cal D}_{\pi Y_2}^{(u,v)}\big)^{-1}=
\begin{pmatrix}
p^2-m^2+i\sqrt{\frac{2g_1}{n^2}}X+i\sqrt{\frac{g_2}{n}}d_{auu}Y_1^a & i\sqrt{2g_2n}f_{vQu}v_Q \\
i\sqrt{2g_2n}f_{vQu}v_Q & -1 \\
\end{pmatrix}.
\eea
We see that sectors connected via the interchanging of the indices are almost the same, only the off-diagonal elements differ due to a minus sign (this comes from the total antisymmetry of the structure constants $f_{vQu}$ and the fact that $u$ and $v$ are not independent, therefore $d_{auu}=d_{avv}$).

We needed the expressions of the inverse tree level propagators, since they explicitly appear in the $1$-loop part of the 2PI effective potential. As it was already stressed, now we would like to focus only on the pseudoscalar sector, therefore pieces corresponding exclusively to the scalar sector are not present in the expression below. The approximate 2PI effective potential without counterterms looks as:
\bea
V_{\textnormal 2PI}&=&n^2m^2(v_a)^2+\frac12X^2+\frac12(Y_1^a)^2+\frac12(Y_2^a)-2n^2h_av_a\nonumber\\
&-&i\sqrt{\frac{2g_1}{n^2}}Xn^2(v^a)^2-i\sqrt{\frac{g_2}{n}}n^2Y_1^ad_{abc}v_bv_c\nonumber\\
&-&\frac{i}{2}\sum_{(u,v)}\int_k \Tr \log \Big({\cal G}_{\pi Y_2}^{(u,v)}\Big)^{-1}-\frac{i}{2}\int_k
\Tr \log \Big( {\cal G}_{\pi}^{\diag} \Big)^{-1}\nonumber\\
&-&\frac{i}{2}\int_k\Tr \log \Big( {\cal G}_{Y_2}^{\diag} \Big)^{-1}-\frac{i}{2}\sum_{(u,v)}\int_k \Tr \big({\cal D}_{\pi Y_2}^{(u,v)}\big)^{-1}{\cal G}_{\pi Y_2}^{(u,v)}\nonumber\\
&-&\frac{i}{2}\int_k \Tr \big({\cal D}_{\pi}^{\diag}\big)^{-1}{\cal G}^{\diag}_{\pi}-
\frac{i}{2}\int_k \Tr \big({\cal D}_{Y_2}^{\diag}\big)^{-1}{\cal G}^{\diag}_{Y_2} \nonumber\\
&+& i\frac{g_2}{n}f_{ai\alpha}f_{bj\beta}\int_p\int_k G_{Y_2^aY_2^b}(p)G_{s_i s_j}(k)G_{\pi_{\alpha}\pi_{\beta}}(p+k)\nonumber\\
&+& i\sqrt{\frac{2g_1g_2}{n^3}}f_{ai\alpha}\delta_{\beta\gamma}\int_p\int_kG_{\pi_\alpha\pi_\gamma}(p)G_{s_iX}(k)G_{Y_2^a\pi_\beta}(p+k)\nonumber\\
&+& i\frac{g_2}{n}f_{ai\alpha}d_{b\beta\gamma}\int_p\int_k G_{\pi_\alpha\pi_\gamma}(p)G_{s_iY_1^b}(k)G_{Y_2^a\pi^\beta}(p+k)\nonumber\\
&+& i\sqrt{\frac{g_1g_2}{2n^3}}d_{a\alpha\beta}\delta_{\gamma\delta}\int_p\int_k G_{\pi_\gamma\pi_\alpha}(p)G_{\pi_\delta\pi_\beta}(k)G_{XY_1^a}(p+k)\nonumber\\
&+& i\frac{g_1}{2n^2}\delta_{\alpha\beta}\delta_{\gamma\delta}\int_p\int_k G_{XX}(p)G_{\pi_\alpha\pi_\delta}(k)G_{\pi_\gamma\pi_\beta}(p+k)\nonumber\\
&+& i\frac{g_2}{4n}d_{a\alpha\beta}d_{b\gamma\delta}\int_p\int_k G_{Y_1^aY_1^b}(p)G_{\pi_\alpha\pi_\gamma}(k)G_{\pi_\beta\pi_\gamma}(p+k)\nonumber\\
&+& i\frac{g_2}{2n}f_{ai\alpha}f_{bj\beta}\int_p\int_k G_{s_is_j}(p)G_{Y_2^a\pi_\beta}(k)G_{Y_2^a\pi_\alpha}(p+k).
\label{5-2PI_3aux}
\eea

\subsection{Equations in the general framework}

From the effective potential (\ref{5-2PI_3aux}) we can derive equations for the $1$-point functions and for the propagators of the
$\pi-Y_2$ sector. The last seven lines correspond to the setting sun diagrams which give contributions to this
sector. In these expressions we did not simplify the index notations as we did in (\ref{Eq:eff-pot-rescaled}),
since we will not make use of these expressions in the actual equations. This is due to our heavy scalar
assumption: checking the setting sun diagrams in (\ref{5-2PI_3aux}), we see that they contain at least one
scalar (heavy) propagator, therefore in a first approximation the quantum corrections of the pseudoscalar 
propagators can be neglected. It means that the 2-loop truncated effective potential together with the heavy
scalar assumption leads to tree level propagators in the pseudoscalar sector:
\bea
{\cal G}^{\diag}_{\pi} \approx {\cal D}_{\pi}^{\diag}, \qquad {\cal G}^{\diag}_{Y_2} \approx {\cal D}_{Y_2}^{\diag}, \qquad 
{\cal G}_{\pi Y_2}^{(u,v)} \approx {\cal D}_{\pi Y_2}^{(u,v)}.
\label{5-Gapprox}
\eea

Now let us work out the equations for the $1$-point functions. The saddle point equations of the auxiliary variables are:
\begin{subequations}
\bea
0&\!\!\!=\!\!\!&\frac{\delta V_{\textnormal 2PI}}{\delta X}=X-i\sqrt{2n^2g_1}(v^a)^2-i\sqrt{\frac{g_1}{2n^2}}\int_p G_{\pi^a \pi^a},\\
\label{5-y1a}
0&\!\!\!=\!\!\!&\frac{\delta V_{\textnormal 2PI}}{\delta Y_1^a}=Y_1^a-in\sqrt{g_2n}d_{abc}v_bv_c-\frac{i}{2}\sqrt{\frac{g_2}{n}}d_{abc}\int_p G_{\pi^b\pi^c},\\
0&\!\!\!=\!\!\!&\frac{\delta V_{\textnormal 2PI}}{\delta Y_2^a}=Y_2^a.
\eea
\end{subequations}
Note that for $Y_2$ we obtained a trivial relation, since only scalar-pseudoscalar mixing propagator
$G_{s^a \pi^b}$ could have given contribution, which was previously excluded from the solution on the
basis of the parity conservation of strong interactions.  This shows that the $Y_2^a=0$ assumption is consistent.
Analyzing (\ref{5-y1a}) we obtain that for non-diagonal ``$a$'' indices, the right hand side is identically equal
to $Y_1^a$. This means that $Y_1$ inherits the structure of $v$, i.e. only diagonal indexed field expectation
values are non-zero. For the equations of state we obtain
\bea
0=\frac{\delta V_{\textnormal 2PI}}{\delta v^a}&\!\!\!=\!\!\!&2n^2(m^2v_a-h_a)-2i\sqrt{\frac{2g_1}{n^2}}Xn^2v^a\nonumber\\
&&-2in\sqrt{ng_2}d_{abc}v_bY_1^c\nonumber-i\sqrt{2g_2n}f_{vau}\int_p G_{Y_2^v\pi^u}(p).
\eea
It is convenient to treat the zeroth index independently. For the saddle point equations we get
\begin{subequations}
\label{5-eq-X-Y}
\bea
X&\!\!\!=\!\!\!&i\sqrt{2n^2g_1}(v_0^2+v_Q^2)+i\sqrt{\frac{g_1}{2n^2}} \int_p \big(G_{\pi_0\pi_0}(p)+G_{\pi_Q\pi_Q}(p)\big),\\
Y_1^0&\!\!\!=\!\!\!&i\sqrt{2n^2g_2}(v_0^2+v_Q^2)+i\sqrt{\frac{g_2}{2n^2}} \int_p \big(G_{\pi_0\pi_0}(p)+G_{\pi_Q\pi_Q}(p)\big),\\
Y_1^Q&\!\!\!=\!\!\!&i\sqrt{\frac{g_2}{4n}}\Big(4n\sqrt{2n}v_0v_Q+2n^2d_{QBC}v_Bv_C\nonumber\\
&&+d_{Quu}\int_p G_{\pi_u\pi_u}(p)+2\sqrt{\frac{2}{n}}\int_p G_{\pi_0\pi_Q}+d_{QBC}\int_p G_{\pi_B\pi_C}(p)\Big).
\eea
\end{subequations}
Here capital letters refer to nonzero diagonal indices, while ``$u$'' corresponds to non-diagonal ones. With the use of the previously introduced
definition of $M^2$ (check (\ref{5-M2def})), the equations of state simplify to
\begin{subequations}
\label{5-eqos}
\bea
0&\!\!\!=\!\!\!&\sqrt{2}n(h_0-M^2v_0)+2i\sqrt{g_2}Y_1^Qv_Q,\\
0&\!\!\!=\!\!\!&\sqrt{2}n(h_Q-M^2v_Q)+2i\sqrt{g_2}Y_1^Qv_0+i\sqrt{2g_2n}d_{QBC}Y_1^Bv_C \nonumber\\
&&+i\sqrt{\frac{g_2}{n}}f_{vQu}\int_p G_{Y_2^v\pi^u}(p).
\eea
\end{subequations}

In order to solve these equations and to determine the ground state of the model, we need to obtain the propagators $G_{Y_2^v\pi^u}$.
We saw that our approximation led to tree-level propagators in the pseudoscalar sector, therefore we only need to invert
the $2\times 2$ matrices given in (\ref{5-tree-piY2}). We obtain
\bea
\label{5-mixpion}
{\cal G}_{\pi Y_2}^{(u,v)}(p)&\!\!\!=\!\!\!&\frac{1}{p^2-M_{(u,v)}^2}
\begin{pmatrix}
i & \sqrt{2g_2n}f_{ubv}v_b \\
\sqrt{2g_2n}f_{ubv}v_b & (p^2-m^2+2i\sqrt{\frac{g_1}{2n^2}}X)/i+\sqrt{g_2}{n}d_{auu}Y_1^a
\end{pmatrix},\nonumber\\
\eea
where
\bea
M_{(u,v)}^2=m^2-2i\sqrt{\frac{g_1}{2n^2}}X-i\sqrt{\frac{g_2}{n}}d_{auu}Y_1^a-2g_2n(f_{vbu}v_b)^2,
\eea
which is symmetric under the change $u\leftrightarrow v$ (recall that $d_{auu}=d_{avv}$).

Before we choose a specific symmetry breaking pattern, we have to analyze the scaling behavior of the quantities, since we would like to work in the large-$n$ approximation. Let us therefore introduce the following rescaled variables:
\bea
\label{5-rescaled-var}
\chi_Q=\sqrt{n}v_Q,\quad y_1^Q=Y_1^Q/\sqrt{n},\quad x=X/n,\quad y_1^0=Y_1^0/n.
\eea
Note that the definitions of $x$ and $y_1^0$ somewhat differ from the ones introduced in the previous section, check (\ref{5-x-y10-def}). Using these notations, we approximate our equations at LO of the large-$n$ limit. For the equations of state
we have
\begin{subequations}
\label{5-eqos2}
\bea
0&=&h_0-M^2v_0+i\sqrt{2g_2}y_1^Q\chi_Q/n,\\
0&=&h_Q-M^2\chi_Q+i\sqrt{2g_2}y_1^Qv_0+i\sqrt{g_2}d_{QBC}y_1^B\chi_C\nonumber\\
&&-g_2\sum_{(u,v)}\frac{f_{vQu}f_{vAu}}{n}\chi_A T(M^2_{(u,v)}).
\eea
\end{subequations}
Note that we have also rescaled the diagonal indexed external fields: $h_Q\rightarrow h_Q/\sqrt{n}$. With the
new variables $M^2$ can be written as $M^2=m^2-i(\sqrt{2g_1}x+\sqrt{2g_2}y_1^0)$ and $T(\mu^2)=\int_p i/(p^2-\mu^2)$
is the usual tadpole integral.

The saddle point equations read as:
\begin{subequations}
\label{5-eq-X-Y2}
\bea
y_1^0&\!\!\!=\!\!\!&i\sqrt{2g_2}\Big(v_0^2+\chi_A\chi_A/n+\frac{1}{2n^2}\sum_{(u,v)}T(M_{(u,v)}^2)\Big),\\
y_1^Q&\!\!\!=\!\!\!&i\sqrt{g_2}\Big(2\sqrt{2}v_0\chi_Q+d_{QBC}\chi_B\chi_C\nonumber\\
&&+\frac{1}{2n}d_{QBC}\int_p G_{\pi^B\pi^C}(p)+\frac{1}{2n}\sum_{(u,v)}d_{Quu}T(M_{(u,v)}^2)\Big),
\eea
\end{subequations}
with $\sqrt{g_2}x=\sqrt{g_1}y_1^0$.

From now on, we choose a symmetry breaking pattern, where $v_Q=v_8\delta_{Q8}$. The index ``8'' refers to the longest diagonal generator(!) of the $U(n)$ group, reminding to the $n=3$ case. We stress that this is just a notation, our analysis is a large-$n$ study and has nothing to do with the $U(3)$ case.

\section{Ground state of the system with a breaking in the 0-8 sector}

First we discuss the mixings when the only spontaneously broken direction beyond the zeroth one is the ``8'', corresponding to
the longest diagonal generator. Recalling (\ref{5-pion_diag}), we
assume that $Y_1$ inherits further the structure of $v$ (it will be checked later) and therefore $Y_1^a=Y_1^0\delta_{a0}+Y_1^8\delta_{a8}$. Then we obtain
\bea
i\big({\cal D}_\pi\big)^{-1}_{ab}(p)=(p^2-M^2)\delta_{ab}+i\sqrt{g_2} d_{8ab}y_1^8.
\eea
Using the Appendix, when $A,B\neq 0,8$ are diagonal indices, then
\bea
\label{5-diag_d}
d_{8AB}=\sqrt{\frac{2}{n(n-1)}}\delta_{AB}.
\eea
Also,
\bea
d_{800}=0, \qquad d_{880}=\sqrt{\frac{2}{n}}, \qquad d_{888}=-(n-2)\sqrt{\frac{2}{n(n-1)}}.
\eea
This shows that the pion mixing is reduced to the $0-8$ sector.

Now we turn to the non-diagonal pions, which can mix with $Y_2$. Recalling (\ref{5-tree-piY2}), we need $f_{u8v}\neq 0$ (since $f_{u0v}=0$). Using the Appendix, the only possible choice is to have $u=(x,jn)$ with $v=(y,jn)$ or vice versa.

We have $(n-1)$ choices for the value of $j$, and since the interchange of $u\leftrightarrow v$ doubles the number of nonzero structure constants, therefore we have $2(n-1)$ pions which mix in the large-$n$ limit with the corresponding $Y_2$ fields. These pions will be called the {\it kaons}.
Using the Appendix, independently of the $u,v$ values in question (i.e. $(x,jn), (y,jn)$), we obtain 
\bea
f_{u8v}=-\frac{1}{\sqrt{2}}\sqrt{\frac{n}{n-1}}. 
\eea
The $d_{8uv}$ values for the non-diagonal indices are as follows. For indices which do not count in the mixing with $Y_2$ (i.e. $u,v=(x,jk),(y,jk)$ where $k\neq n$):
\bea
\label{5-nonmix-d}
d_{8uv}=\sqrt{\frac{2}{n(n-1)}}\delta_{uv},
\eea
and for the ones which mix (i.e. $u,v=(x,jn),(y,jn)$):
\bea
d_{8uv}=-\frac{(n-2)}{\sqrt{2n(n-1)}}\delta_{uv}.
\eea
This shows that apart from the kaons, the ``88'' and the ``00'' pions, every other pions are identical. With the help of (\ref{5-diag_d}) and (\ref{5-nonmix-d}), we find ($a,b\neq(x,jn),(y,jn),8,0$)
\bea
i\big({\cal G}_\pi\big)_{ab}^{-1}(p)=(p^2-M^2+i\frac{\sqrt{2g_2}}{n}y_1^8)\delta_{ab}\equiv (p^2-M^2-\frac{1}{n}\Delta M^2_\pi)\delta_{ab},
\label{5-pionmassNLO}
\eea
where we have written down the NLO correction to the mass, which is needed for the NLO construction of the effective potential (see later). Recalling (\ref{5-mixpion}), for the kaon-$Y_2$ sector we obtain:
\bea
{\cal G}_{K Y_2}(p)&\!\!\!=\!\!\!&\frac{1}{p^2-M_K^2}
\begin{pmatrix}
i & \pm\sqrt{g_2}\chi_8 \\
\pm\sqrt{g_2}\chi_8 & (p^2-M^2)/i-\sqrt{\frac{g_2}{2}}y_1^8
\end{pmatrix},\nonumber\\
\eea
where
\bea
M_K^2=M^2+i\sqrt{\frac{g_2}{2}}y_1^8+g_2\chi_8^2.
\eea
Note that we have changed $\pi$ to $K$ in the notation of the propagator matrix, and erased the indices of it, these should be thought as appropriate $(u,v)$ pairs regarding to the previous analysis. The $\pm$ sign in the off-diagonal elements signals the doubling which corresponds to interchanging of $u\leftrightarrow v$ indices. This however has no effect in the mass $M_K$, as it was already stated in the general analysis.

Now let us see how the equations of state and the saddle point equations simplify. Using the rescaled variables introduced in (\ref{5-rescaled-var}), sticking to the large-$n$ limit, from (\ref{5-eqos2}) and (\ref{5-eq-X-Y2}) we have
\bea
\label{5-v0LO}
0&=&h_0-M^2v_0,\\
\label{5-chi8LO}
0&=&h_8-M^2\chi_8+i\sqrt{2g_2}y_1^8(v_0-\chi_8)-g_2\chi_8T(M_K^2),\\
\label{5-xLO}
x&=&i\sqrt{2g_1}\big(v_0^2+\frac12T(M^2)\big),\\
\label{5-y10LO}
y_1^0&=&i\sqrt{2g_2}\big(v_0^2+\frac12T(M^2)\big),\\
\label{5-y1Q}
y_1^Q&=&i\sqrt{g_2}(2\sqrt{2}v_0\chi_Q+d_{Q88}\chi_8^2)\nonumber\\
&&+\frac{i\sqrt{g_2}}{2n}\Big(\sum_{(u,v)}d_{Quu}T(M_{(u,v)}^2)+\sum_{A\neq8} d_{QAA} T(M^2)\Big).
\eea
Again, capital letters ($Q,A\neq 0$) correspond to diagonal indices.
First we should prove that $y_1^Q$ inherits the structure of $v$, i.e. only $Q=8$ gives nontrivial equation. The first two terms of
(\ref{5-y1Q}) are obviously zero for $Q\neq 8$, therefore we have
to analyze the last two. In the third term, almost every pion has $M^2$ mass, only the kaons are exceptions. In the last term, if we include $A=8$ into the sum, the error we commit counts only at NLO. In this case we have
\bea
\sum_{(u,v)}d_{Quu}T(M_{(u,v)}^2)&\!\!\!+\!\!\!&\sum_{A} d_{QAA} T(M^2)\nonumber\\
&\!\!\!=\!\!\!&\sum_{u\,=\kaon}d_{Quu}\big(T(M_K^2)-T(M^2)\big)+\sum_a d_{Qaa} T(M^2),
\eea
where in the second sum of the right hand side the summation goes over every possible index. Since $\sum_a d_{Qaa}=0$ for every $Q$, this last term cancels and we must analyze the sum $\sum_{u\,=\kaon} d_{Quu}$. Let us assume that
$Q\neq 8$. Using the Appendix, we can easily obtain that
\bea
\sum_{u\,=\kaon} d_{Quu}=0.
\eea
This proves that $y_1^Q$ is nontrivial only for $Q=8$. In the latter case the sum is
\bea
\sum_{u\,=\kaon} d_{8uu}=-(n-1)(n-2)\sqrt{\frac{2}{n(n-1)}}\longrightarrow -\sqrt{2}n,
\eea
therefore the equation of $y_1^8$ is
\bea
\label{5-y18LO}
y_1^8=i\sqrt{2g_2}\big(2v_0\chi_8-\chi_8^2-\frac12 T(M_K^2)+\frac12 T(M^2)\big).
\eea
We can solve the equations (\ref{5-chi8LO}), (\ref{5-y18LO}) for the "8" components $\chi_8,y_1^8$ with $v_0, y_1^0, x$ taken from their respective equations (\ref{5-v0LO}), (\ref{5-xLO}), (\ref{5-y10LO})
for a set of $h_0, g_1, g_2$. 
This latter part of the calculation coincides with the solution presented in the first part of the chapter where only $v_0$-condensate was considered.

Equations (\ref{5-chi8LO}), (\ref{5-xLO}), (\ref{5-y10LO}) and (\ref{5-y18LO}) are divergent, therefore need renormalization. We would like to have the very same equations, but with the tadpole integrals substituted with their finite part in the renormalization scheme we used before. In this one $T_F(\mu^2)=T(\mu^2)-T_d^{(2)}-(\mu^2-M_0^2)T_d^{(0)}$, check (\ref{Eq:Tds}) and (\ref{Eq:tadpole}). We can apply the method described in details in Chapter 3: identify first the divergent pieces in the equations, then integrate back to obtain the appropriate counterterm functional of the 2PI effective potential. Without going into the details, we just state that every divergent equation can be made finite in the previously mentioned scheme with the following counterterm functional:
\bea
V_{ct}&\!\!\!=\!\!\!&i\frac{n^2}{2}\int_p\ln (-p^2+M_0^2)\nonumber\\
&\!\!\!-\!\!\!&\frac{n^2}{2}\left[(M^2-m^2)(T_d^{(2)}-M_0^2T_d^{(0)})+\frac{1}{2}(M^4-m^4)T_d^{(0)}\right]\nonumber\\
&\!\!\!-\!\!\!&n\left[\left(\frac{1}{2}\Delta
M_\pi^2+M_K^2-M^2\right)(T_d^{(2)}+(M^2-M_0^2)T_d^{(0)})\right]\nonumber\\
&\!\!\!+\!\!\!&\frac{n}{2}(M^2_K-M^2)^2T_d^{(0)},
\label{5-Eq:ct}
\eea
Although (\ref{5-Eq:ct}}) is written in a rather compact form, it is a function of $v_0,\chi_8,y_1^0,y_1^8,x$ and is completely allowed by the formal requirements of renormalizability \cite{berges05}.
Note also the important simplification $\Delta M_\pi^2/2+M_K^2-M^2=g_2\chi_8^2$. The first term of (\ref{5-Eq:ct}) is field independent, therefore gives no contribution to the equations for the 1-point functions. It serves only for eliminating the overall divergence of the effective potential generated by the zero point fluctuations of the pions.

Plugging (\ref{5-Gapprox}) to (\ref{5-2PI_3aux}), we obtain the 1PI effective potential. The obtained functional depends also on the auxiliary variables and as it was already stressed, we include only the quantum fluctuations of the light particles (i.e. pions and kaons). Since every setting sun in (\ref{5-2PI_3aux}) contains at least one heavy scalar line, we neglect these diagrams. The corresponding formal expression is the following with ${\cal O}(n)$ accuracy:
\bea
V&=&V_{cl}+V_{quant},\nonumber\\
V_{cl}&=&n^2\left[M^2v_0^2+\frac{1}{2}(x^2+(y_1^0)^2)-2h_0v_0\right]\nonumber\\
&&+n\left[M^2\chi_8^2+\frac{1}{2}(y_1^8)^2-2 h_8\chi_8-i\sqrt{2g_2}y_1^8\chi_8(2v_0-\chi_8)\right],
\nonumber\\V_{quant}&=&-\frac{i}{2}\left[(n^2-2n)\int_p \ln(-p^2+M_\pi^2)+2n\int_p\ln(-p^2+M_K^2)\right],
\label{Eq:eff-pot}
\eea
where $V_{quant}$ is the standard expression of the one-loop part of the effective potential. It is important to note that for the accurate evaluation of the quantum contribution of the pions, we have to make use of the ${\cal O}(n)$ accurate value of the pion mass, that is $M_\pi^2=M^2+\frac{1}{n}\Delta M_\pi^2$, see (\ref{5-pionmassNLO}). The divergences of $V_{quant}$ are eliminated by $V_{ct}$.

The renormalized effective potential ($V_R\equiv V+V_{ct}$) in the renormalization scheme defined through $V_{ct}$ specified in (\ref{5-Eq:ct}) above looks like
\bea
V_{R}&=&n^2\left[M^2v_0^2+\frac{1}{2}(x^2+(y_1^0)^2)-2h_0v_0\right]\nonumber\\
&+&n\left[M^2\chi_8^2+\frac{1}{2}(y_1^8)^2-2 h_8\chi_8-i\sqrt{2g_2}y_1^8\chi_8(2v_0-\chi_8)\right]\nonumber\\
&+&\frac{n^2}{32\pi^2}\left[\frac{3}{4}(M_0^4-M^4)+M_0^2(M^2-M_0^2)+\frac{1}{2}M^4\ln\frac{M^2}{M_0^2}\right]\nonumber\\
&+&\frac{n}{16\pi^2}\Big[\frac{1}{2}(\Delta M_\pi^2-M^2)M^2\ln\frac{M^2}{M_0^2}
+\frac{1}{2}M_K^4\ln\frac{M_K^2}{M_0^2}+g_2\chi_8^2M_0^2\nonumber\\
&-&\frac{1}{2}\Delta M_\pi^2M^2+\frac{3}{4}(M^4-M_K^4)\Big].
\label{Eq:eff-pot}
\eea
The fact that the $"8"$ condensate contributes only to ${\cal O}(n)$ 
means that the $v_8$ contribution will be subleading relative to that of $v_0$. Still it belongs to the leading large $n$ piece, since the NLO contributions which were omitted from the very start would contribute ${\cal O}(n^0)$. 

\subsection{Numerical results}

We have explored the $g_1, g_2$ plane and in a large region we found three solutions for $\chi_8, y_1^8$ in a fixed background of $v_0,x,y$ values. It turned out that one of them is the trivial solution (when $h_8=0$), and the other two represent a local minimum and maximum of the effective potential. In order to select the true ground state of the system between the trivial and non-trivial local minima, we must look for the energetically more favorable one. It is clear that we must investigate only the $\chi_8, y_1^8$ dependent parts of the effective potential. We can immediately see that only the ${\cal{O}}(n)$ part of $V_R$ is what we have to study. This part coincides with the set of the ($\chi_8, y_1^8$) dependent terms. It is important to stress that due to this fortunate effect, the selection of the ground state does not depend on the explicit value of $n$, since it appears as an overall multiplicative factor. In the forthcoming figures we refer to $V^{(n)}$ as the ${\cal{O}}(n)$ part of the full effective potential divided by this overall $n$ factor.

In the first part of this chapter we mapped out the region of the three-dimensional parameter space $(g_1,g_2,h_0)$, where the scalar masses turned out to be at least twice as large as the mass of the pseudo-Goldstone fields (pions). In this part of the investigations, relative to the case of the pure $v_0$ background, we have chosen rather low values of $h_0$ (all dimensional quantities are measured in proportion to the appropriate powers of the absolute value of the renormalized mass). The results below are obtained in those regions of the parameter space, where we found the heavy scalar assumption valid (i.e. where the scalar masses are at least twice as large as the pion mass).

Let us start the discussion with the mass splitting in the $\pi$-sector, in particular the kaon/pion mass ratio. 
The most detailed study was achieved for $h_8=0$, addressing the question of the $v_8$ (i.e. $\chi_8$)-dependence of the effective potential after eliminating the $y_1^8$ dependence of $V_R$ with the help of (\ref{5-y18LO}). We scanned through the allowed region at different values of $h_0$. As announced, in addition to the trivial solution $v_8=y_1^8=0$, also a non-trivial (meta)stable solution (local minimum of the effective potential) of (\ref{5-chi8LO}) and (\ref{5-y18LO}) with positive squared masses of all $\pi$-fields was found in a large
part of the region allowed by the heavy scalar mass assumption. In Fig. \ref{Fig5_3} the ratio $M_K^2/M_\pi^2$ is displayed over the $(g_1,g_2)$-plane for 2 different
values of $h_0$. We see, that in a quite large region of the allowed parameter space the mass ratio varies mildly, but the kaons are getting
heavier when $h_0$ is lowered. Also it can be observed, that for small $g_1$ values increasing $g_2$ induces very large kaon mass values, while for larger $g_1$ the same does not lead to significant changes of the mass ratio. 

\begin{figure}
\includegraphics[bb=85 90 490 380,scale=0.63]{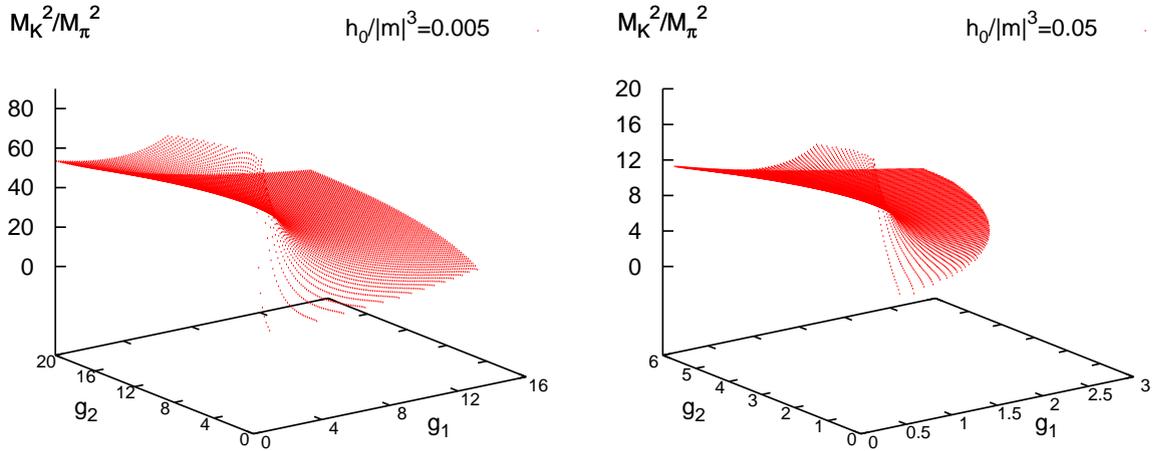}
\caption{Kaon/pion squared mass ratio as a function of the couplings ($g_1,g_2$) for two different $h_0$ values. Kaons become lighter as $h_0$ increases, at the same time the region where non-trivial solutions are found narrows.}
\label{Fig5_3}
\end{figure}

\begin{figure}
\includegraphics[bb=-100 50 50 580,scale=0.4]{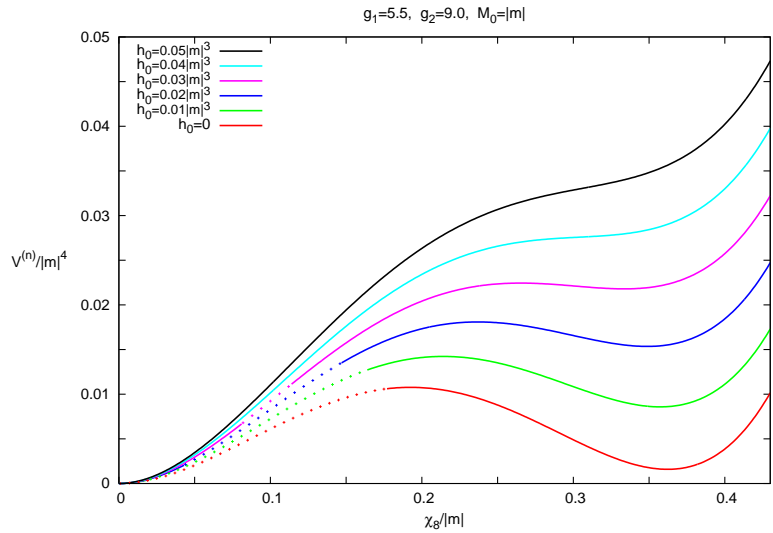}
\caption{The ${\cal{O}}(n)$ part of the quantum effective potential divided by $n$ as a function of the $\chi_8$ condensate. The real part of the potential is plotted with dotted lines for $\chi_8$ values where some negative mass squares lead to complex values of the potential. The curves from top to bottom follow the order of the $h_0$ values indicated in the upper-left-hand corner.}
\label{Fig5_4}
\end{figure}

The ${\cal O}(n)$ part of the full renormalized effective potential $V_R$ could be plotted in an extended interval around the non-trivial minimum, where all mass-squares are positive. 
An example is shown in Fig. \ref{Fig5_4}, where we see that the energy density of the non-trivial solution grows monotonically with $h_0$ increasing. (As already mentioned, the ${\cal{O}}(n)$ part of the full potential divided by $n$ is denoted as $V^{(n)}$, and the same holds for the classical part as $V_{cl}^{(n)}$.) There is a critical
value $h_0^c(g_1,g_2)$, where the non-trivial local minimum disappears. For example we can verify on Fig. \ref{Fig5_3} that for the $g_1,g_2$ values of Fig. \ref{Fig5_4}, there does exist a non-trivial solution of the equations for $h_0=0.005$ but not for $h_0=0.05$. This circumstance explains why we used much smaller values of $h_0$ compared to the first part of this chapter.

The dotted pieces of the curves represent the real part of the effective potential in regions where it becomes complex due to the fact that one of the mass-squares becomes negative. This phenomenon is a well-known feature of the loop expansion \cite{rivers,fujimoto83,raifear86}. Although the effective potential should be a real quantity by construction, in regions where the classical potential is chosen to be non-convex, the loop expansion breaks down and produces an imaginary part. Mathematically, a non-convex classical potential can be understood as an analytic continuation of a convex classical potential through the mass term:
$m \rightarrow im$. The loop-expansion creates a divergent asymptotic series, but there is no such theorem which would state that the analytically continued series corresponds term-by-term to the series of the analytically continued effective potential. In other words, the analytic continuation and the expansion in $\hbar$ are not interchangeable operations. From a mathematical point of view, this is the origin of the appearance of the imaginary part. Nevertheless, the naive calculation leading to complex effective potentials has a physical meaning. First, it turns out that if we want to understand the effective potential at a $\phi_c$ point as the expectation value of the minimum energy density of the class of quantum states in which the field expectation value is $\phi_c$ (which property can be derived from the usual Legendre-transform definition of the effective potential), then we have to take the {\it real part} of the loop-expansion and apply Maxwell's construction. Furthermore, Weinberg and Wu showed \cite{weinberg87} that even the naively calculated real part in itself contains physics: it is the minimum energy density of a class of quantum states with a $\phi_c$ field expectation value together with the more restrictive constraint that their wave functionals are concentrated on the uniform configuration $\phi_c$. The state chosen this way may differ from the state which {\it minimizes} the energy density of states with a $\phi_c$ field expectation value. In this case the state in question is unstable and the {\it imaginary part} of the potential can be interpreted as half the decay rate per unit volume. 

We have also plotted the curve
of the classical part of the ${\cal{O}}(n)$ piece of the potential in a typical point of the coupling space for two values of $h_0$, which demonstrates that with the present renormalization scale $M_0$ the quantum fluctuations only moderately modify the value of the potential, see Fig. \ref{Fig5_5}.
\begin{figure}
\includegraphics[bb=15 50 90 420,scale=0.5]{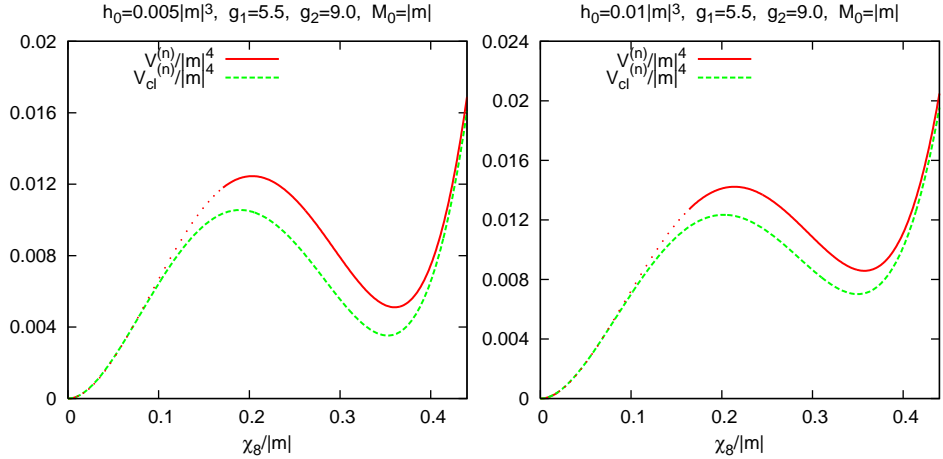}
\caption{The effect of the quantum fluctuations is only moderate, as indicated by comparing the $V_{cl}^{(n)}$  (dashed lines) with $V^{(n)}$. The dotted part displays the real part of $V^{(n)}$, where it becomes complex.}
\label{Fig5_5}
\end{figure}
In the whole allowed region of the coupling space we found for the potential in the non-trivial minimum positive
values, therefore we conclude that with $h_8=0$, there is no $h_0,v_0, M^2$ combination which would induce spontaneously nonzero $v_8$ condensate into the true ground state of the system.
The effect of the existence of such a metastable state might still show up in out-of equilibrium situations.

In a number of energetically "promising"
coupling points we have studied the effect of the application of a non-zero $h_8$ on the effective potential.
In these cases also the trivial solution shifts in proportion of the external field, and the effective potential can be displayed in a finite, but small region
around the shifted trivial minimum too (i.e. no negative mass squares appear in its neighborhood). If $h_0$ is small, the value of the effective potential in the non-trivial minimum varies faster with $h_8$ than near the shifted trivial minimum, therefore we find a critical external field $h_8^c(g_1,g_2)$, which drives the minimum characterized by a large $v_8$ value to be the true ground
state of the system. In Fig. \ref{Fig5_6} we show the typical variation of the potential as a function of $h_8$ for the value of $(g_1,g_2,h_0)$ chosen as in the left panel of Fig. \ref{Fig5_5}.
\begin{figure}
\includegraphics[bb=-73 50 90 580,scale=0.42]{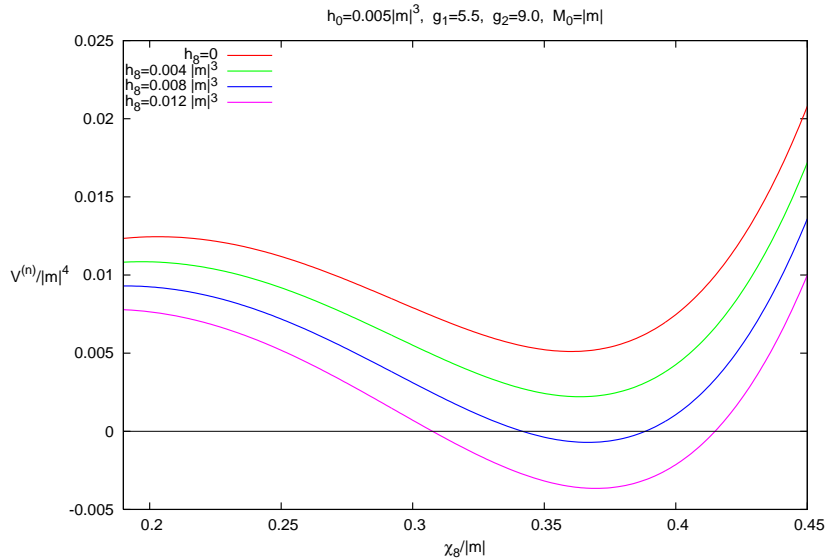}
\caption{Smooth variation of $V^{(n)}$ with increasing $h_8$ displaying the symmetry breaking pattern $U_V(n)\rightarrow U_V(n-1)$. The curves from top to bottom follow the order of the applied $h_8$ values indicated in the upper-left-hand corner.}
\label{Fig5_6}
\end{figure}
\begin{figure}
\includegraphics[bb=43 90 290 580,scale=0.6]{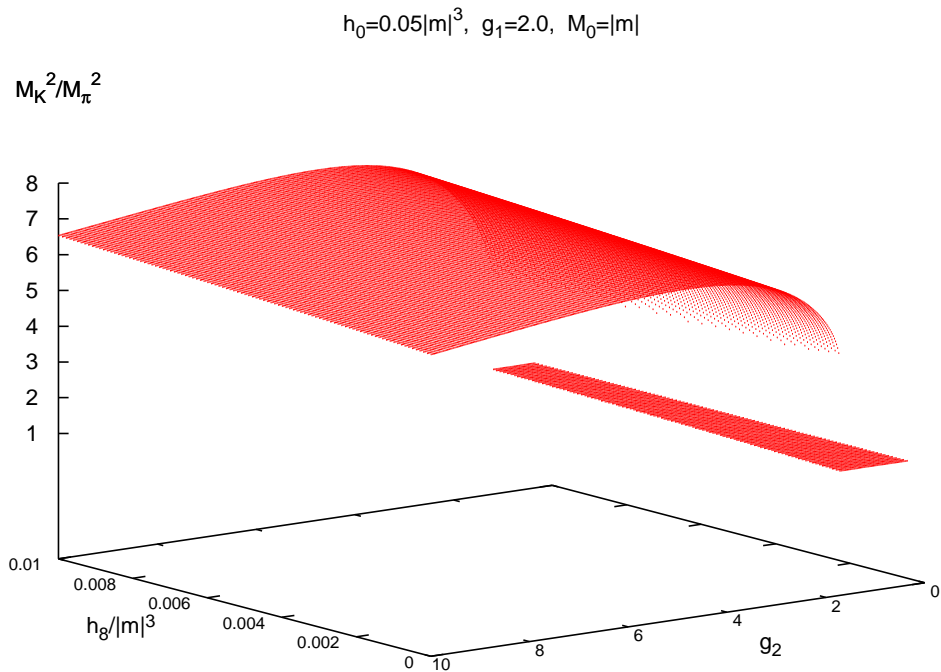}
\caption{The variation of $M_K^2/M_\pi^2$ with $h_0$ and $g_2$. The relative variation of the mass ratio is very small with respect to $h_8$. Note that the
abrupt jump downwards is occurring when the non-trivial minimum is missing.}
\label{Fig5_7}
\end{figure}
In Fig. \ref{Fig5_7} we display the variation of the ratio $M_K^2/M_\pi^2$ as a function of $h_8$ and $g_2$ for given $g_1$ and $h_0$ values as taken in the non-trivial minimum. We see that apart from a jump to unity happening at lower $g_2$ values, the ratio is not varying too much. This ratio equals unity in the region where the non-trivial minimum is missing. Choosing higher values for $h_0$ involves smaller kaon masses (as it can also be seen in Fig. \ref{Fig5_3}), which points to the direction in the parameter
space with a chance of phenomenological applications. We also note, that choosing other $g_1$ values, the shape of the figure changes only mildly. The figure also shows that the mass ratio is increasing linearly with $h_8$ at fixed $g_2$, however this change is almost negligible compared to its full magnitude, i.e. the relative change is very small. This means that the kaon mass is induced dominantly by the action of the $v_0$ condensate, and not by the explicit symmetry breaking proportional to $h_8$. This circumstance is rather different from what comes in the $n=3$ flavor models from various perturbative discussions.

\section{Concluding remarks}

In this chapter, a leading order approximate large-$n$ solution of the $U(n)\times U(n)$ model was presented in 
auxiliary field formalism. In the first part of the chapter a certain symmetry breaking pattern was
assumed, which reduces the $U(n)\times U(n)$ symmetry of the Lagrangian to $U(n)$. During the
construction of the solution, a light pseudoscalar/heavy scalar hierarchy of the spectra was
assumed. The renormalization of the saddle point equations, the equation of state and the propagator
equations was performed
with explicit counterterm construction, which was followed by an investigation of the
consistency of the additional heavy scalar assumption. It turned out that in a large region of the
parameter space the existence of {\it heavy} scalars occurs. It was also shown, that the proposed solution explicitly fulfills Goldstone's theorem.

In the second part of the chapter we examined the possibility of a more general symmetry breaking,
in particular when the remaining $U(n)$ symmetry is reduced further to $U(n-1)$. In this case new solutions were found.
After renormalization, it was shown that in the direction which corresponds to ``8'' in the
$SU(3)$ classification, in addition to the trivial minimum, a metastable local minimum of the
renormalized effective potential appears, even without introducing explicit symmetry breaking in
the corresponding direction. This means that a condensate can arise spontaneously in the direction
in question, which breaks the remaining $U(n)$ symmetry to $U(n-1)$. It turned out that
the application of a moderate external field transforms the nontrivial minimum into the true ground state
(i.e. the global minimum of the effective potential) of the system. 

The different ground states are
characterized by different pseudoscalar excitation spectra, also explicitly constructed. The mass
ratio of the heavier modes (``kaons'') and the remaining pseudo-Goldstone modes was found
sensitive mainly only to the external field in the zeroth direction (which is proportional to the unit matrix),
i.e. to the symmetry breaking pattern of the starting solution. If it turns out that phenomenology can be built
on the nontrivial vacuum condensate, this observation would imply also rather different finite temperature
behavior compared to the perturbative descriptions. 

Although the transition of the condensate in the zeroth direction in itself would be much similar to that
of the $O(2n^2)$ symmetric model, the insensitivity of the location of the new nontrivial minimum with respect
to the external field in the ``8'' direction (see Fig. \ref{Fig5_7}) hints a first-order transition. We
observed that it is not the external field but rather the condensate in the zeroth direction which determines
the strength of the symmetry breaking to $U(n-1)$. The tightly connected variation of the two
condensates probably leads then to a phase transition with both fields going through a discontinuous change
of similar amplitude. This expectation complies well with renormalization group considerations which classify
the restoration of the broken symmetry in this model quite differently than in models displaying symmetries
of the orthogonal group \cite{paterson81,pisarski84}. An obvious next step is the study of the finite
temperature symmetry restoration from this nontrivial ground state, but this is not included in this thesis.

Further consolidation of the discovered structure should require the investigation of the fluctuation effect of
the heavy scalar fields. On a general basis, one expects the modification of the different coefficients in
different EoS, SPEs, and self-energies by terms of ${\cal O}(M_\pi^2/M_s^2)$ (where $M_s^2$ is one of the
scalar masses), which would just slightly deform the coupling region where the nontrivial vacuum structure
is self-consistently present.

A final goal could be to make use of this solution at $n=3$, when also the contributions from the effective
term reflecting the axial anomaly (the 't Hooft determinant) would be included perturbatively using the
large-$n$ propagators of the different fields. The comparison with the treatments of the $n=3$ three flavored
meson model in different perturbative calculations \cite{kovacs08,schaefer09,schaefer10,dumm05,dumm06} and
functional approaches, which go beyond perturbation theory \cite{braun04,jaeckel04}, partly based on the
exact renormalization group techniques \cite{berges02,gies06,schaefer08}, could give some insight into the
importance of strong coupling effects. 

\clearemptydoublepage

\chapter*{Conclusions}
\addcontentsline{toc}{chapter}{Conclusions}

\lhead[\let\uppercase\relax\bf{\small{\thepage}}]{\let\uppercase\relax\bf{\footnotesize{Conclusions}}}
\rhead[\let\uppercase\relax\bf{\footnotesize{Conclusions}}]{\let\uppercase\relax\bf{\small{\thepage}}}

In this thesis we applied 2PI formalism to various scalar quantum field theories, at zero temperature. The main scope of the work was to investigate the renormalizability of the partially resummed perturbative series induced by different 2PI approximations. Beyond the theoretical realization, the study of the numerical features of 2PI renormalization was also a major motivation of this work. In a specific theory ($U(N)\times U(N)$ model) we also managed to present new ground state solutions of the 2PI equations, which could lead to phenomenological applications.

First we dealt with the $O(N)$ symmetric model in auxiliary field formulation. We constructed the
2PI effective potential at next-to-leading order of the $1/N$ expansion, and showed its renormalizability with
explicit counterterm construction. We found that the renormalization can be done without any restriction on the background fields, stated earlier in \cite{andersen04,andersen08}. We also eliminated the auxiliary field to get the next-to-leading order
renormalized 2PI effective potential in terms of the original variables of the model. During this step we could see in
a transparent way how the partial series of counterterm diagrams
are resummed in the original formulation of the model. The renormalization of the 2PI effective potential formulated with an auxiliary field led to renormalized functionals after the elimination of any subset of the variables. 

We also discussed algorithms of the numerical implementation of 2PI renormalization. We employed the one
component $\phi^4$ theory at $2$-loop level of the 2PI effective potential. Our main goal was to implement
very accurate numerical realizations of the renormalization procedure, and to quantify the rate of convergence
of the numerical solutions. In order to develop different algorithms, we used different theoretical approaches
to obtain renormalized equations. We pointed out that counterterms of
\cite{patkos08} are equivalent with appropriate renormalization conditions imposed on the $2$- and $4$-point
functions. We developed accurate numerics regarding both variants of the renormalization programme and found the very same
result numerically, which confirmed our theoretical considerations. The method using the exact counterterms
turned out to be inconvenient at large coupling constants: solving the equations required large computational resources. We managed to overcome this with a new algorithm which did not work
with fixed counterterms, but iteratively developed them together with the equations. This led to a significant
improvement in the convergence rate, therefore the method is expected to be a valuable tool for solving bare (i.e. counterterm included) equations of strongly coupled theories.

In the last chapter, we worked on the solution of the $U(N)\times U(N)$ meson model in the large-$N$ limit. We built up the
2PI effective potential at $2$-loop level using auxiliary fields. We presented a solution of the
coupled set of equations of the $1$- and $2$-point functions in the broken phase of a one component
condensate realizing the $U(N)\times U(N) \longrightarrow U(N)$ symmetry breaking. Our solution involved an additional
assumption on the scalar/pseudoscalar masses: the scalars are considered to be much heavier than the
pseudoscalars. Due to this, we neglected terms (diagrams) of the equations, which contained at least one
scalar propagator. This approximate solution was the first large-$N$ study of the model, which
went beyond the Ans{\"a}tze related to the $O(2N^2)$ symmetric piece of the Lagrangian of the model. The renormalizability of the solution was
presented with explicit counterterm construction and we showed that it fulfills Goldstone's theorem. We also solved the scalar propagator equations numerically. With this we demonstrated that in a large part of the coupling space, the heavy scalar assumption is valid, confirming the self-consistency of our approximate solution.

We also discussed the stability of the symmetry breaking with respect to a more general condensate. After deriving equations in the same approximation for a general breaking of the symmetry, the study of the renormalized effective potential showed that the
field equations have further solutions in certain part of the parameter space. One of them
turned out to be a new local minimum of the effective potential, which is a metastable vacuum even without
the existence of an explicit symmetry breaking term. However, moderate application of a conjugate source, the metastable vacuum transforms into the global minimum, i.e. the true ground state of the
system. This nontrivial structure of the effective potential is a signal of a more completed treatment of the
model in the sense that, our solution dealt with the full $U(N)\times U(N)$ structure, not just the
partial $O(2N^2)$ nonlinearities, which had been a typical approximation before. Our analysis hinted
a first order finite temperature transition in the cases where the nontrivial minimum is the true ground state. This is
expected on general grounds of the renormalization group analysis in the large-$N$ limit \cite{pisarski84}, however until now no
publications are known which could constructively reproduce this conjecture.

\renewcommand{\theequation}{A\arabic{equation}}
\chapter*{Appendix}
\addcontentsline{toc}{chapter}{Appendix}

\setcounter{equation}{0}

\lhead[\let\uppercase\relax\bf{\small{\thepage}}]{\let\uppercase\relax\bf{\footnotesize{Appendix}}}
\rhead[\let\uppercase\relax\bf{\footnotesize{Appendix}}]{\let\uppercase\relax\bf{\small{\thepage}}}
\section*{Structure of the U(N) algebra}

The $U(n)$ algebra is $n^2$ dimensional, the generators are denoted by $T^a$. They are traceless and normalized in a way that $\Tr (T^aT^b)=\delta_{ab}/2$. The generalized Gell-Mann matrices $\lambda^a$ are defined as
\bea
T^a=\frac{\lambda^a}{2},
\eea
therefore $\Tr(\lambda^a\lambda^b)=2\delta_{ab}$. The generators can be classified into two sets: there are diagonal, and non-diagonal ones.
There are $n$ diagonal Gell-Mann matrices:
\bea
\lambda^{0}&\!\!\!=\!\!\!&\sqrt{\frac{2}{n}}
\left( \begin{array}{cccc}
1 & & & \\
& 1 & & \\
& & ... & \\
& & & 1 \\
\end{array} \right), 
\qquad \lambda^{1}=
\left( \begin{array}{cccc}
1 & & & \\
& -1 & & \\
& & 0 & \\
& & & ... \\
\end{array} \right), \qquad... 
\nonumber\\ &&\lambda^{n-1}=\sqrt{\frac{2}{n(n-1)}}
\left( \begin{array}{cccc}
1 & & & \\
& 1 & & \\
& & ... & \\
& & & -(n-1) \\
\end{array} \right).
\eea
We note that in the Chapter 5, we use the ``8'' index to $\lambda^{n-1}$. The number of the non-diagonal ones is $n(n-1)$ and they read as
\bea
\label{app-nondiag}
\big(\lambda^{(x,jk)}\big)_{ab}=\delta_{ak}\delta_{bj}+\delta_{aj}\delta_{bk}, \qquad \big(\lambda^{(y,jk)}\big)_{ab}=i\delta_{ak}\delta_{bj}-i\delta_{aj}\delta_{bk},
\eea
where we introduced the compact index notations $(x,jk)$ and $(y,jk)$ ($j<k$). This arises from the fact that the matrices defined in (\ref{app-nondiag}) are related to the $\sigma_x$ and $\sigma_y$ Pauli matrices: they are mainly the same but the elements of them appear in the $j-k$ row-column.

The structure constants of the algebra are defined through the relation
\bea
\lambda^a\lambda^b=(d_{abc}+if_{abc})\lambda^c,
\eea
where $d_{abc}$ is totally symmetric and $f_{abc}$ is totally antisymmetric in their indices. From this we have
\begin{subequations}
\label{app-struct}
\bea
[\lambda^a,\lambda^b]&\!\!\!=\!\!\!&2if_{abc}\lambda^c \qquad \!\!\Longrightarrow \qquad \Tr\big[[\lambda^a,\lambda^b]\lambda^c\big]=4if_{abc}, \\
\{\lambda^a,\lambda^b\}&\!\!\!=\!\!\!&2d_{abc}\lambda^c \qquad \Longrightarrow \qquad \Tr \big[\{\lambda^a,\lambda^b\}\lambda^c\big]=4d_{abc}.
\eea
\end{subequations}
Here $[.,.]$ and $\{.,.\}$ refer to commutation and anticommutation, respectively. Alternatively, we can write
\begin{subequations}
\label{app-struct2}
\bea
f_{abc}=\frac12\Im \Tr(\lambda^a\lambda^b\lambda^c), \\
d_{abc}=\frac12\Re \Tr(\lambda^a\lambda^b\lambda^c).
\eea
\end{subequations}
We can immediately see that
\bea
d_{0ab}=\sqrt{\frac{2}{n}}\delta_{ab}, \qquad f_{0ab}=0,
\eea
and in can easily be proved that
\bea
\sum_{i} d_{aii}=\sqrt{2n^3}\delta_{a0}.
\eea
Now we list some useful identities which can be derived from the properties (\ref{app-struct}) or (\ref{app-struct2}). We assume that $Q\neq 0$ corresponds to a diagonal generator. Then, for the antisymmetric structure constants the following can be derived:
\bea
f_{(x,jk),Q,(y,jk)}=
\begin{cases}
\sqrt{\frac{Q}{2(Q+1)}}, \!\! \qquad \qquad \qquad j=Q+1<k, \\
-\frac{1}{\sqrt{2Q(Q+1)}}, \!\!\qquad \qquad \quad j<Q+1<k, \\
-\sqrt{\frac{Q+1}{2Q}}, \qquad \qquad \qquad j<Q+1=k, \\
0, \qquad \qquad \qquad \qquad \quad j<k<Q+1.
\end{cases}
\eea
A special case when $Q=n-1$, that is ``8'', then only the third and fourth lines make sense. The nonzero values are:
\bea
f_{(x,jn),8,(y,jn)}\equiv f_{(x,jn),n-1,(y,jn)}=-\sqrt{\frac{n}{2(n-1)}}.
\eea
We also note that when $j\neq j'$ and $k\neq k'$,
\begin{subequations}
\bea
f_{(x,jk),Q,(x,j',k')}=f_{(y,jk),Q,(y,j',k')}=0, \\
f_{(x,jk),Q,(y,j',k')}=f_{(y,jk),Q,(x,j',k')}=0.
\eea
\end{subequations}
For indices corresponding fully to diagonal generators, the symmetric tensor looks like ($Q,P,T\neq 0$):
\bea
d_{QPT}=\begin{cases}
0 \qquad \qquad \qquad \qquad \qquad Q<P<T, \\ 
0 \qquad \qquad \qquad \qquad \qquad Q<P=T, \\
\sqrt{\frac{2}{T(T+1)}} \qquad \qquad \qquad \!\quad Q=P<T, \\
-(Q-1)\sqrt{\frac{2}{Q(Q+1)}} \qquad \quad \!\!Q=P=T.
\end{cases}
\eea
Finally, we have ($Q\neq 0$ diagonal)
\bea
d_{Q,(x,jk),(x,jk)}=d_{Q,(y,jk),(y,jk)}=\begin{cases}
0, \qquad \qquad \qquad \qquad \quad \!Q+1<j<k, \\
-\sqrt{\frac{Q}{2(Q+1)}}, \qquad \qquad \quad Q+1=j<k, \\
\frac{1}{\sqrt{2Q(Q+1)}}, \qquad \qquad \qquad \!\!j<Q+1<k, \\
-\frac{Q-1}{\sqrt{2Q(Q+1)}}, \qquad \qquad \quad \!j<Q+1=k, \\
\sqrt{\frac{2}{Q(Q+1)}}, \qquad \qquad \qquad \!j<k<Q+1,
\end{cases}
\eea
and
\bea
d_{Q,(x,jk),(x,j',k')}=d_{Q,(y,jk),(y,j',k')}=d_{Q,(y,jk),(x,j',k')}=0.
\eea
\clearemptydoublepage

\renewcommand{\headrulewidth}{0pt}
\lhead[\let\uppercase\relax]{\let\uppercase\relax}
\rhead[\let\uppercase\relax]{\let\uppercase\relax}
\cfoot{}

\pagestyle{plain}

\end{document}